\documentclass[article]{emulateapj}
\usepackage{epsfig,lscape}

 \newcommand{\hi}{\mbox{\rm \ion{H}{1}}}
\newcommand{\hii}{\mbox{\rm \ion{H}{2}}} \newcommand{\htwo}{\mbox{\rm H$_2$}}
 
\newcommand{\kmpers}{\mbox{km~s$^{-1}$}}

\newcommand{\xcounits}{\mbox{cm$^{-2}$ (K km s$^{-1}$)$^{-1}$}}

\newcommand{\xco}{\ensuremath{X_{\rm CO}}}

\newcommand{\ML}{\ensuremath{\Upsilon_{\star}^K}}
\newcommand{\mlunit}{\ensuremath{M_\odot/L_{\odot,K}}}

\shorttitle{The Star Formation Efficiency in Nearby Galaxies}
\shortauthors{Leroy et al.}  \slugcomment{Accepted for publication in
  \emph{The Astronomical Journal}}

\begin{document}
\title{The Star Formation Efficiency in Nearby Galaxies: Measuring Where Gas
  Forms Stars Effectively}

\author{Adam K. Leroy\altaffilmark{1}, Fabian Walter\altaffilmark{1}, Elias
  Brinks\altaffilmark{2}, Frank Bigiel\altaffilmark{1}, W.J.G. de
  Blok\altaffilmark{3,4}, Barry Madore\altaffilmark{5}, M. D.
  Thornley\altaffilmark{6}}

\altaffiltext{1}{Max-Planck-Institut f{\"u}r Astronomie, K{\"o}nigstuhl 17,
  D-69117, Heidelberg, Germany}
\altaffiltext{2}{Centre for Astrophysics Research, University of
  Hertfordshire, Hatfield AL10 9AB, U.K.}
\altaffiltext{3}{Research School of Astronomy \& Astrophysics, Mount Stromlo
  Observatory, Cotter Road, Weston ACT 2611, Australia}
\altaffiltext{4}{Department of Astronomy, University of Cape Town, Private Bag
  X3, Rondebosch 7701, South Africa}
\altaffiltext{5}{Observatories of the Carnegie Institution of Washington,
  Pasadena, CA 91101, USA}
\altaffiltext{6}{Department of Physics and Astronomy, Bucknell University,
  Lewisburg, PA, 17837, USA}

\begin{abstract}
  We measure the star formation efficiency (SFE), the star formation rate per
  unit gas, in 23 nearby galaxies and compare it to expectations from proposed
  star formation laws and thresholds. We use \hi\ maps from THINGS and derive
  \htwo\ maps from CO measured by HERACLES and BIMA SONG. We estimate the star
  formation rate by combining GALEX FUV maps and SINGS 24$\mu$m maps, infer
  stellar surface density profiles from SINGS $3.6\mu$m data, and use
  kinematics from THINGS.  We measure the SFE as a function of: the free--fall
  and orbital timescales; midplane gas pressure; stability of the gas disk to
  collapse (including the effects of stars); the ability of perturbations to
  grow despite shear; and the ability of a cold phase to form. In spirals, the
  SFE of H$_2$ alone is nearly constant at $5.25 \pm 2.5 \times
  10^{-10}$~yr$^{-1}$ (equivalent to an \htwo\ depletion time of $1.9 \times
  10^9$~yr) as a function of all of these variables at our 800~pc resolution.
  Where the ISM is mostly \hi , on the other hand, the SFE decreases with
  increasing radius in both spiral and dwarf galaxies, a decline reasonably
  described by an exponential with scale length $0.2$--$0.25~r_{25}$. We
  interpret this decline as a strong dependence of GMC formation on
  environment. The ratio of molecular to atomic gas appears to be a smooth
  function of radius, stellar surface density, and pressure spanning from the
  \htwo --dominated to \hi --dominated ISM. The radial decline in SFE is too
  steep to be reproduced only by increases in the free--fall time or orbital
  time. Thresholds for large--scale instability suggest that our disks are
  stable or marginally stable and do not show a clear link to the declining
  SFE. We suggest that ISM physics below the scales that we observe --- phase
  balance in the \hi , \htwo\ formation and destruction, and stellar feedback
  --- governs the formation of GMCs from \hi .
\end{abstract}


\section{Introduction}
\label{INTRO}

In nearby galaxies, the star formation rate (SFR) is observed to
correlate spatially with the distribution of neutral gas, at least to
first order. This is observed using a variety of SFR and gas tracers,
but the quantitative relationship between the two remains poorly
understood. Although it is common to relate SFR to gas surface density
via a power law, the relationship is often more complex. The same
surface density of gas can correspond to dramatically different SFRs
depending on whether it is found in a spiral or irregular galaxy or in
the inner or outer part of a galactic disk. Such variations have
spurred suggestions that the local potential well, pressure, coriolis
forces, chemical enrichment, or shear may regulate the formation of
stars from the neutral interstellar medium (ISM).

In this paper, we compare a suite of proposed star formation laws and
thresholds to observations. In this way, we seek to improve observational
constraints on theories of galactic-scale star formation. Such theories are
relevant to galaxy evolution at all redshifts, but must be tested mainly in
nearby galaxies, where observations have the spatial resolution and
sensitivity to map star formation to local conditions. An equally important
goal is to calibrate and test empirical star formation recipes. In lieu of a
strict theory of star formation, such recipes remain indispensable input for
galaxy modeling, particularly because star formation takes place mostly below
the resolution of cosmological simulations. This requires the implementation
of ``subgrid'' models that map local conditions to the SFR
\citep[e.g.,][]{SPRINGEL03}.

Our analysis is based on the highest quality data available for a significant
sample of nearby galaxies: \hi\ maps from The \hi\ Nearby Galaxy Survey
\citep[THINGS,][]{WALTER08}, far ultraviolet (FUV) maps from the GALEX Nearby
Galaxies Survey \citep[][]{GILDEPAZ07}, infrared (IR) data from the {\em
  Spitzer} Infrared Nearby Galaxies Survey \citep[SINGS,][]{KENNICUTT03}, CO
$1\rightarrow0$ maps from the BIMA Survey of Nearby Galaxies \citep[BIMA
SONG,][]{HELFER03} and CO $2\rightarrow1$ maps from the HERA CO-Line
Extragalactic Survey \citep[HERACLES][]{LEROY08}.  This combination yields
sensitive, spatially resolved measurements of kinematics, gas surface density,
stellar surface density, and SFR surface density across the entire optical
disks of 23 spiral and irregular galaxies.

The topic of star formation in galaxies is closely linked to that of giant
molecular cloud (GMC) formation. In the Milky Way, most star formation takes
place in GMCs, which are predominantly molecular, gravitationally bound clouds
with typical masses $\sim 10^5$ -- $10^6$~M$_{\odot}$ \citep[][]{BLITZ93}.
Similar clouds dominate the molecular ISM in Local Group galaxies
\citep[e.g.,][]{FUKUI99,ENGARGIOLA03}. If the same is true in other galaxies,
then a close association between GMCs and star formation would be expected to
be a general feature of our data. \citet{BIGIEL08} study the relationship
between atomic hydrogen (\hi), molecular gas (\htwo), and the SFR in the same
data used here. Working at a resolution of 750~pc, they do not resolve
individual GMCs, but do find that a single power law with index $n = 1.0 \pm
0.2$ relates H$_2$ and SFR surface density over the optical disks of spirals.
This suggests that as in the Milky Way, a key prerequisite to forming stars is
the formation of GMCs (or at least H$_2$).

\citet{BIGIEL08} find no similar trend relating \hi\ and SFR. Instead the
ratios of \htwo -to-\hi\ and SFR-to-\hi\ vary strongly within and among
galaxies. GMC formation, therefore, appears to be a function of local
conditions.  Here we investigate this dependence. We focus on where the ISM
can form gravitationally bound, predominantly molecular structures, i.e., the
``star formation threshold,'' and investigate how the molecular fraction of
the ISM varies with local conditions. In equilibrium, the fraction of the ISM
in GMCs may be set by the timescale over which these structures form.
Therefore we also consider suggested timescales for the formation of GMCs and
compare them to observations.

Maps with good spatial coverage and sensitivity are critical to distinguish
between the various proposed thresholds and timescales. Perhaps {\em the} key
observation to test theories of galactic--scale star formation is that the
star formation per unit gas mass decreases in the outer disks of spiral and
irregular galaxies \citep[e.g.,][]{KENNICUTT89,MARTIN01,THORNLEY06}. The
details of this decrease vary with the specifics of the observations. For
example, \citet{MARTIN01} observed a sharp drop in the distribution of \hii\
regions, while UV maps suggest a steady decline \citep{BOISSIER07}, but it is
without dispute that the SFR per unit gas mass does indeed decline \citep[see
also][]{WONG02}. Maps with good spatial extent contain both regions where GMC
formation proceeds efficiently and regions where it is suppressed.  Including
both \hi -rich dwarf galaxies and \htwo -dominated spirals offers a similar
contrast.

In \S \ref{SECT_BACKGROUND}, we present a set of star formation laws and
thresholds that we will compare to observations. We phrase these in terms of
the star formation per unit neutral gas, which we call the ``star formation
efficiency.'' This quantity, the inverse of the gas depletion time, removes
the basic scaling between stars and gas and measures how effectively each
parcel of the ISM forms stars.

In \S \ref{SECT_DATA}, we briefly describe our sample, data, and methodology.
In order to focus the main part of the paper on analysis, we defer most
detailed discussion of data and methodology to the appendices.

In \S \ref{SECT_RESULTS} we look at how the star formation efficiency relates
to other basic quantities (\S \ref{SECT_SFEOBS}), proposed laws (\S
\ref{SECT_SFELAW}), and thresholds (\S \ref{SECT_SFETHRESH}) described in \S
\ref{SECT_BACKGROUND}. In \S \ref{SECT_DISCUSSION} we analyze and interpret
these results. In \S \ref{SECT_RECIPE}, we illustrate our conclusions by
comparing predictions for the star formation efficiency to observations. In \S
\ref{CONCLUSIONS}, we summarize our results.

Appendices \ref{GASAPP} -- \ref{SFRAPP} contain all the information required
to reproduce our calculations, including descriptions of the data and how we
convert from observables to physical quantities. We present our data as an
electronic table of radial profiles described in Appendix \ref{RPROFAPP} and
as maps and plotted profiles for each galaxy in Appendix \ref{ATLASAPP}.

\section{Background}
\label{SECT_BACKGROUND}

Following, e.g., \citet[][]{KENNICUTT89}, we break the topic of star formation
in galaxies into two parts. Where star formation is widespread, we refer to
the quantitative relationship between neutral gas and the SFR as the star
formation {\em law}. To predict the SFR over an entire galactic disk, it is
also necessary to know which gas is actively forming stars. This topic is
often phrased as the star formation {\em threshold}, but may be more generally
thought of as the problem of where a cold phase ($n \sim 4$--$80$~cm$^{-3}$,
$T \sim 50$--$200$~K) or gravitationally bound clouds can form; both are
thought to be prerequisites to star formation. We give a brief background on
both laws and thresholds, first noting that neither term is strictly accurate:
``laws'' here refer to observed (or predicted) correlations and the
``threshold'' is probably a smooth variation from non-star forming to actively
star forming gas.

We cast this discussion in terms of the {\em star formation efficiency} (SFE).
There are many definitions for the SFE, but throughout this paper we use the
term only to refer to the star formation rate surface density per unit neutral
gas surface density along a line of sight, i.e., SFE=$\Sigma_{\rm SFR} /
\Sigma_{\rm gas}$ with units of yr$^{-1}$. We will also discuss SFE~(\htwo )
which refers to the SFR per unit \htwo\ ($\Sigma_{\rm SFR}$/$\Sigma_{\rm
  H2}$), and SFE~(\hi) ($\Sigma_{\rm SFR}$/$\Sigma_{\rm HI}$). The SFE is the
inverse of the gas depletion time, the time required for present day star
formation to consume the gas reservoir. It represents a combination of the
real timescale for neutral gas to form stars and the fraction of gas that ends
up in stars, e.g., if 1\% of the gas is converted to stars every $10^7$~yr,
the SFE$=10^{-9}$~yr$^{-1}$.  Because it is normalized by $\Sigma_{\rm gas}$,
the SFE is more useful than $\Sigma_{\rm SFR}$ alone to identify where
conditions are conducive to star formation (i.e., where gas is ``good at
forming stars'').

As we describe proposed laws (\S \ref{SECT_LAWBACK}) and thresholds (\S
\ref{SECT_THRESHBACK}), we present quantitative forms for each that can be
compared to the observed SFE. Table \ref{IDEATAB} collects these expressions,
which we compare to observations in \S \ref{SECT_RESULTS}.

\begin{deluxetable*}{l l l}
  \tabletypesize{\small} \tablewidth{0pt} \tablecolumns{9}
  \tablecaption{\label{IDEATAB} Star Formation Laws and Thresholds}
  \tablehead{\colhead{Theory} & \colhead{Form} & \colhead{Observables}}
  \startdata
  \multicolumn{3}{c}{Star Formation Laws} \\
  \hline
  disk free--fall time & & \\
  $\ldots$ fixed scale height & ${\rm SFE} \propto \Sigma_{\rm gas}^{0.5}$ & $\Sigma_{\rm gas}$ \\
  $\ldots$ variable scale height & ${\rm SFE~or~}R_{\rm mol} \propto
  \frac{\Sigma_{\rm gas}}{\sigma_{\rm g}}~\left(1 +
    \frac{\Sigma_*}{\Sigma_{\rm gas}} \frac{\sigma_{\rm
        g}}{\sigma_{*,z}}\right)^{0.5}$ & $\Sigma_{\rm gas}$, $\Sigma_*$, $\sigma_g$, $\sigma_*$ \\
  orbital timescale & ${\rm SFE~or~}R_{\rm mol} \propto \tau_{\rm orb}^{-1} =
  \frac{v (r_{\rm gal})}{2 \pi r_{\rm gal}}$ & $v(r_{\rm gal})$ \\
  cloud-cloud collisions & ${\rm SFE} \propto \tau_{\rm orb}^{-1}~Q_{\rm
    gas}^{-1}~\left(1 - 0.7~\beta\right)$ & $v(r_{\rm gal})$ \\
  fixed GMC efficiency & ${\rm SFE} = {\rm SFE} \left({\rm H}_2\right)
  \frac{R_{\rm mol}}{R_{\rm mol} + 1}$ & $\Sigma_{\rm H2}$ \\
  pressure and ISM phase & $R_{\rm mol} \propto \left(\Sigma_{\rm gas}~\left(
      \Sigma_{\rm gas} + \frac{\sigma_{\rm g}}{\sigma_{*,z}} \Sigma_*
    \right)~P_0^{-1} \right)^{1.2}$ & $\Sigma_{\rm gas}$, $\Sigma_*$,
  $\sigma_{\rm g}$, $\sigma_*$ \\
  \hline
  \multicolumn{3}{c}{Star Formation Thresholds} \\
  \hline
  gravitational instability & & \\
  $\ldots$ in the gas disk & $Q_{\rm gas} =
  \left(\frac{\sigma_g~\kappa}{\pi~G~\Sigma_{\rm gas}}\right) < 1$ &
  $\Sigma_{\rm gas}$, $\sigma_{\rm g}$, $v(r_{\rm gal})$ \\
  $\ldots$ in a disk of gas and stars & $Q_{\rm stars+gas} =
  \left(\frac{2}{Q_{\rm stars}} \frac{q}{1+q^2} + \frac{2}{Q_{\rm
        gas}}~R~\frac{q}{1 + q^2 R^2}\right)^{-1} < 1$ & $\Sigma_{\rm gas}$,
  $\Sigma_*$, $\sigma_{\rm g}$, $\sigma_*$, $v(r_{\rm gal})$ \\
  competition with shear & $\Sigma_{\rm gas} > \frac{2.5~A~\sigma_g}{\pi~G}$ &
  $\Sigma_{\rm gas}$, $\sigma_{\rm g}$, $v(r_{\rm gal})$ \\
  cold gas phase & $\Sigma_{\rm gas} > 6.1~{\rm M}_{\odot}~{\rm
    pc}^{-2}~f_{\rm g}^{0.3}~Z^{-0.3} ~I^{0.23}$ & $\Sigma_{\rm gas}$,
  $\Sigma_*$, $Z$, $I$
  \enddata
\end{deluxetable*}

\subsection{Star Formation Laws}
\label{SECT_LAWBACK}

A star formation law should predict the SFE from local conditions. Here we
describe three proposals for the limiting timescale over which gas forms
stars: the free-fall timescale in the gas disk, the orbital timescale, and the
characteristic timescale for cloud-cloud collisions. We also describe
proposals that GMCs form stars with a fixed SFE and that the midplane gas
pressure regulates the fraction of the ISM in the molecular phase. We present
each proposal as a prediction for the SFE in terms of observables. These
appear together in the upper part of Table \ref{IDEATAB}. We expect a
successful star formation law to reproduce the observed SFE (in practice,
combined with an empirical calibration).

\subsubsection{Disk Free-Fall Time With Fixed Scale Height}

The most common formulation of the star formation law is a power law relating
gas and star formation (surface) densities following
\citet{SCHMIDT59,SCHMIDT63}. \citet{KENNICUTT89,KENNICUTT98A} calibrated this
law in its observable (surface density) form. Averaging over the star-forming
disks of spiral and starburst galaxies, he found

\begin{equation}
\label{KLAW}
\Sigma_{\rm{SFR}} \propto \Sigma_{\rm{gas}}^{1.4}~,
\end{equation}

\noindent often referred to as the ``Kennicutt--Schmidt law.''

The exponent in Equation \ref{KLAW}, $n \approx 1.5$, can be approximately
explained by arguing that stars form with a characteristic timescale equal to
the free--fall time in the gas disk, which in turn depends inversely on the
square root of the gas volume density, $\tau_{\rm ff} \propto \rho_{\rm
  gas}^{-0.5}$ \citep[e.g.,][]{MADORE77}. For a fixed scale height $\rho_{\rm
  gas} \propto \Sigma_{\rm gas}$ and $\Sigma_{\rm SFR} \propto \Sigma_{\rm
  gas}^{1.5}$. The first star formation law that we consider is thus

\begin{equation} 
{\rm SFE} \propto \Sigma_{\rm gas}^{0.5}~,
\end{equation}

\noindent which is approximately Equation \ref{KLAW}.

\subsubsection{Disk Free-Fall Time With Variable Scale Height}

If the scale height is not fixed, but instead set by hydrostatic equilibrium
in the disk, then

\begin{equation}
  \label{DISKFFEQ}
  \tau_{\rm ff} \propto \frac{1}{\sqrt{\rho_{\rm mp, gas}}} \propto
  \frac{\sigma_{\rm g}}{\Sigma_{\rm gas}~\sqrt{1 + \frac{\Sigma_*}{\Sigma_{\rm
          gas}} \frac{\sigma_{\rm g}}{\sigma_{*,z}}}}
\end{equation}

\noindent where $\sigma_{\rm g}$ and $\sigma_{*,z}$ are the (vertical)
velocity dispersions of gas and stars, $\Sigma_{\rm gas}$ and $\Sigma_{\rm *}$
are the surface densities of the same, and $\rho_{\rm mp, gas}$ is the
midplane gas density. Equation \ref{DISKFFEQ} combines the expression for
midplane density from \citet[][their Equation 34]{KRUMHOLZ05} and midplane gas
pressure from \citet[][his Equation 11, used to calculate
$\phi_P$]{ELMEGREEN89}. The second star formation law that we consider is

\begin{equation}
  \label{SFEFF}
  {\rm SFE} \propto \tau_{\rm ff}^{-1} \propto \frac{\Sigma_{\rm gas}}{\sigma_{\rm g}}~\left(1 + \frac{\Sigma_*}{\Sigma_{\rm gas}} \frac{\sigma_{\rm g}}{\sigma_{*,z}}\right)^{0.5}~,
\end{equation}

\noindent which incorporates variations in the scale height and thus gas
volume density with a changing potential well.

\subsubsection{Orbital Timescale}

It is also common to equate the timescale for star formation and the orbital
timescale \citep[e.g.,][]{SILK97,ELMEGREEN97}. \citet{KENNICUTT98A} and
\citet{WONG02} found that such a formulation performs as well as Equation
\ref{KLAW}. In this case

\begin{equation}
  \label{TORBEQ}
  {\rm SFE} \propto \tau_{\rm orb}^{-1} = \frac{\Omega}{2 \pi} =
  \frac{v(r_{\rm gal})}{2\pi r_{\rm gal}}~.
\end{equation}

\noindent where $v(r_{\rm gal})$ is the rotational velocity at a
galactocentric radius $r_{\rm gal}$ and $\Omega$ is the corresponding angular
velocity.

\subsubsection{Cloud--Cloud Collisions}

\citet{TAN00} suggested that the rate of collisions between gravitationally
bound clouds sets the timescale for star formation so that

\begin{equation} {\rm SFE} \propto \tau_{\rm orb}^{-1}~Q_{\rm
    gas}^{-1}~\left(1 - 0.7 \beta \right)~.
\end{equation}

\noindent where $Q_{\rm gas}$, defined below, measures gravitational
instability in the disk and $\beta = d \log~v(r_{\rm gal}) / d \log~r_{\rm
  gal}$ is the logarithmic derivative of the rotation curve. The dependence on
$\beta$ reflects the importance of galactic shear in setting the frequency of
cloud-cloud collisions. In the limit $\beta = 0$ (a flat rotation curve) this
prescription reduces to essentially Equation \ref{TORBEQ}; for $\beta = 1$
(solid body rotation) the SFE is depressed by the absence of shear.

\subsubsection{Fixed GMC Efficiency}
\label{GMCFFSECT}

If the SFE of an individual GMC depends on its intrinsic properties and if
these properties are not themselves strong functions of environment or cloud
formation, then we expect a fixed SFR per unit molecular gas, SFE~(\htwo ).
\citet{KRUMHOLZ05} posited such a case, arguing that the SFE of a GMC depends
on the free--fall time in the cloud, itself only a weak function of cloud mass
in the Milky Way \citep{SOLOMON87}. \citet{BIGIEL08} found support for this
idea. Studying the same data used here, they derived a linear relationship
between $\Sigma_{\rm H2}$ and $\Sigma_{\rm SFR}$ on scales of 750~pc.

SFE~(\htwo ) is likely to appear constant if: the scaling relations and mass
spectrum (i.e., the intrinsic properties) of GMCs are approximately universal,
the gas pressure is low enough that GMCs are largely decoupled from the rest
of the ISM, individual resolution elements contain at least a few GMCs, and
the properties of a cloud regulate its ability to form stars \citep[\S
\ref{FIXEDHTWO_SECT} and][]{BIGIEL08}. This is the fifth star formation law
that we consider, that star formation in spiral galaxies occurs mostly in GMCs
and that once such clouds are formed, they have approximately uniform
properties so that

\begin{equation} 
\label{SFECONST}
{\rm SFE}~\left( {\rm H}_2 \right) = {\rm constant}~,
\end{equation}

\noindent which we can convert to the SFE of the total gas given $R_{\rm mol}
= \Sigma_{\rm H2} / \Sigma_{\rm HI}$, the ratio of \htwo\ to \hi\ gas.  Then

\begin{equation}
\label{FIXEDSFEEQ}
{\rm SFE} = {\rm SFE}~\left( {\rm H}_2 \right) \frac{R_{\rm mol}}{R_{\rm mol}+1}
\end{equation}

\noindent or if we measure only $\Sigma_{\rm HI}$ (as is the case in dwarfs),
then ${\rm SFE}~\left( {\rm HI} \right) = {\rm SFE}~\left( {\rm H}_2
\right)~R_{\rm mol}$.

The balance between GMC/\htwo\ formation and destruction will set $R_{\rm mol}
= \Sigma_{\rm H2}/\Sigma_{\rm HI}$. If GMCs with fixed lifetime form over a
free fall time or orbital time then $R_{\rm mol} \propto \tau_{\rm ff}^{-1}$
or $R_{\rm mol} \propto \tau_{\rm orb}^{-1}$ (\S \ref{GMCFORM_SECT}), which we
have noted in Table \ref{IDEATAB}. Combined with Equation \ref{FIXEDSFEEQ}, an
expression for $R_{\rm mol}$ predicts the SFE.

\subsubsection{Pressure and Phase of the ISM}

\citet{WONG02}, \citet{BLITZ04}, and \citet[][]{BLITZ06} explicitly
consider $R_{\rm mol}$. Following \citet{ELMEGREEN89} and
\citet{ELMEGREEN94}, they identify pressure as the critical quantity
that sets the ability of the ISM to form \htwo . They show that the
midplane hydrostatic gas pressure, $P_{\rm h}$, correlates with this
ratio in the inner parts of spiral galaxies.

Pressure, which is directly proportional to the gas volume density, should
affect both the rate of \htwo\ formation/destruction and the likelihood of a
gravitationally unstable overdensity condensing out of a turbulent ISM
\citep{ELMEGREEN89,ELMEGREEN94}.  \citet{ELMEGREEN89} gives the following
expression for $P_{\rm h}$,

\begin{equation}
\label{PRESSEQ}
  P_{\rm h} \approx \frac{\pi}{2}~G~\Sigma_{\rm gas}~\left( \Sigma_{\rm gas} +
    \frac{\sigma_{\rm g}}{\sigma_{*,z}} \Sigma_* \right)~,
\end{equation}

\noindent and \citet{ELMEGREEN93} predicted that the fraction of gas in the
molecular phase depends on both $P_{\rm h}$ and the interstellar radiation
field, $j$, via $R_{\rm mol} \propto P^{2.2}~j^{-1}$. If $\Sigma_{\rm SFR}
\propto \Sigma_{\rm H2}$ and we make the simple assumption that $j \propto
\Sigma_{\rm SFR}$ then \citet{ELMEGREEN93} predicts

\begin{equation}
\label{PRESSLAW} 
R_{\rm mol} \propto ~P_{h}^{1.2}~{\rm~or~} \Sigma_{\rm H2} = \Sigma_{\rm HI}
P_{h}^{1.2}~,
\end{equation}

\noindent which combines with Equation \ref{FIXEDSFEEQ} to predict the SFE.

\citet{WONG02} and \citet{BLITZ06} found observational support for Equation
\ref{PRESSLAW}. Using a modified Equation \ref{PRESSEQ} appropriate where
$\Sigma_* \gtrsim \Sigma_{\rm gas}$, \citet{BLITZ06} fit a power law of the
form

\begin{equation}
\label{RMOLPRESS}
R_{\rm mol} = \frac{\Sigma_{\rm{H2}}}{\Sigma_{\rm{HI}}} = \left(
  \frac{P_{h}}{P_0} \right)^{\alpha}~,
\end{equation}

\noindent finding $P_0 = 4.3 \times 10^4$~cm$^{-3}$~K, the observed pressure
where the ISM is equal parts \hi\ and \htwo , and a best--fit exponent $\alpha
= 0.92$. \citet{WONG02} found $\alpha = 0.8$. \citet{ROBERTSON08} recently
found support from simulations for $\alpha \sim 0.9$.

\subsection{Star Formation Thresholds}
\label{SECT_THRESHBACK}

We have described suggestions for the efficiency with which gas form stars,
but not {\em whether} gas forms stars. A ``star formation threshold'' is often
invoked to accompany a star formation law. This is a criterion designed to
address the question ``which gas is actively forming stars?'' or ``where can
the ISM form gravitationally bound, molecular clouds?'' and proposed
thresholds have mostly focused on the existence of gravitational or thermal
instability in the gas disk.

A common way to treat the issue of thresholds is to formulate a critical gas
surface density, $\Sigma_{\rm crit}$, that is a function of local conditions
--- kinematics, stellar surface density, or metallicity. If $\Sigma_{\rm gas}$
is below $\Sigma_{\rm crit}$, star formation is expected to be suppressed; we
refer to such regions as ``subcritical.''  Where the gas surface density is
above the critical surface density, star formation is expected to be
widespread.  We refer to such regions as ``supercritical.''

In practice, we expect to observe a drop in the SFE associated with the
transition from super- to subcritical. We do not necessarily expect ${\rm SFE}
= 0$ in subcritical regions. Even with excellent resolution, a line of sight
through a galaxy probes a range of physical conditions.  At our working
resolution of 400 -- 800~pc, each resolution element encompasses a wide range
of local conditions. Within a subcritical resolution element, star formation
may still occur in isolated pockets that locally meet the threshold criterion.

Expressions for star formation thresholds are collected in the lower part of
Table \ref{IDEATAB}.

\subsubsection{Gravitational Instability}
\label{GRAV_SECT}

\citet{KENNICUTT89}, \citet{KENNICUTT98A}, and \citet{MARTIN01} argued that
star formation is only widespread where the gas disk is unstable against large
scale collapse.  Following \citet{TOOMRE64}, the condition for instability in
a thin gas disk is

\begin{equation}
\label{QGAS}
Q_{\rm gas} = \frac{\sigma_g~\kappa}{\pi~G~\Sigma_{\rm gas}}~< 1~.
\end{equation}

\noindent where $\sigma_g$ is the gas velocity dispersion, $G$ is the
gravitational constant, and $\kappa$ is the epicyclic frequency, calculated
via

\begin{equation}
  \kappa = 1.41~\frac{v(r_{\rm gal})}{r_{\rm gal}}~\sqrt{1+\beta}~,
\end{equation}

\noindent where $\beta = d \log~v(r_{\rm gal}) / d \log~r_{\rm gal}$.  

\citet{MARTIN01} found that \hii\ regions are common where $\Sigma_{\rm gas}$
exceeds a critical surface density derived following Equation \ref{QGAS},

\begin{equation}
\Sigma_{{\rm crit,}Q} = \alpha_Q~\frac{\sigma_g~\kappa}{\pi~G}~.
\end{equation}

\noindent In regions where $\Sigma_{\rm gas}$ is above this threshold, gas is
unstable against large scale collapse, which leads to star formation. Below
the threshold, Coriolis forces counteract the self-gravity of the gas and
suppress cloud/star formation. The factor $\alpha_Q$ is an empirical
calibration, the observed average value of $1/Q_{\rm gas}$ at the star
formation threshold. For an ideal thin gas disk, the condition for gas to be
unstable to collapse is $\alpha_Q > 1$. At the edge of star forming disks,
\citet{KENNICUTT89} found $\alpha_Q = 0.63$ and \citet{MARTIN01} found
$\alpha_Q = 0.69$ ($Q_{\rm gas} \sim 1.5$).

\citet{KENNICUTT89} and \citet{MARTIN01} mention the influence of stars as a
possible cause for $Q_{\rm gas} > 1$ at the star formation threshold.
\citet{HUNTER98} present an in-depth discussion of how several factors
influence $\alpha_Q$, e.g., stars and viscosity lower it, while the thickness
of the gas disk raises it.  \citet{KIM01,KIM07} argue based on simulations
that the observed threshold corresponds to the onset of nonlinear,
non-axisymmetric instabilities. \citet{SCHAYE04} and \citet{DEBLOK06} suggest
a different explanation, that $\alpha_Q \ne 1$ because $\sigma_g$ has been
systematically mishandled; they point out that $\sigma_g$ measured from 21-cm
emission will overestimate the true velocity dispersion of gas in a cold
phase.

The stellar potential well may substantially affect the stability of the gas
disk.  \citet{RAFIKOV01} extended work by \citet{JOG84} to provide a
straightforward way to calculate the instability of a gas disk in the presence
of a collisionless stellar disk. \citeauthor{RAFIKOV01} defined

\begin{equation}
\label{QSTAREQ}
Q_{\rm stars} = \frac{\sigma_{*,r}~\kappa}{\pi~G~\Sigma_*}
\end{equation}

\noindent where $\sigma_{*,r}$ is the (radial) velocity dispersion of stars
and $\Sigma_*$ is the stellar mass surface density. The condition for
instability in the gas disk is

\begin{equation}
  \frac{1}{Q_{\rm stars + gas}} = \frac{2}{Q_{\rm stars}} \frac{q}{1+q^2} + \frac{2}{Q_{\rm
      gas}}~R~\frac{q}{1 + q^2 R^2} > 1~,
\end{equation}

\noindent where $q = k \sigma_{*,r} / \kappa$, with $k$ the wavenumber of the
instability being considered, and $R = \sigma_{\rm g}/\sigma_{*,r}$. The
minimum value of $Q_{\rm stars + gas}$ indicates whether the gas disk is
unstable to large scale collapse. In our sample, typical values of $q$
correspond to wavelengths $\lambda = 2 \pi / k \approx 1$--$5$~kpc.

\citet[][]{HUNTER98} and \citet{BLITZ04} observed strong correlations between
star and GMC formation and the distribution of stars, consistent with stellar
gravity playing a key role in star formation.  \citet{YANG07} recently showed
that $Q_{\rm stars + gas}$ does an excellent job of predicting the location of
star formation in the Large Magellanic Cloud and \citet{BOISSIER03} showed
that including stars improves the correspondence between $Q$ and star
formation in disk galaxies.  \citet{LI05,LI06} found the same results from
numerical simulations of disk galaxies, i.e., that stability against large
scale collapse depends critically on the stellar potential well, with star
formation where $Q_{\rm stars+gas} \lesssim 1.6$.

\subsubsection{Galactic Shear}
\label{SECT_BACKSHEAR}

Motivated by the failure of the Toomre $Q_{\rm gas}$ threshold in dwarf
irregular galaxies, \citet{HUNTER98} suggested that collecting the material
for cloud formation may be easier than implied by $Q_{\rm gas}$, e.g., through
the aid of magnetic fields \citep[see also][]{KIM01}. They hypothesize that
the destructive influence of galactic shear may instead limit where GMCs can
form and describe a threshold that depends on the ability of clouds to form in
the time allowed by shear.

This threshold is based on the local shear rate, described by Oort's A
constant

\begin{equation}
\label{SHEARONE}
A = - 0.5~r_{\rm gal}~\frac{d\Omega}{dr_{\rm gal}}~.
\end{equation}

\noindent Substituting $\Omega = v(r_{\rm gal}) / r_{\rm gal}$,

\begin{equation}
  A = 0.5~\left( \frac{v(r_{\rm gal})}{r_{\rm gal}} - \frac{dv(r_{\rm
        gal})}{dr_{\rm gal}} \right) = 0.5~\frac{v(r_{\rm gal})}{r_{\rm gal}}~\left( 1 - \beta \right)
\end{equation}

\noindent Then the threshold has the form

\begin{equation}
\label{SHEAREQ}
\Sigma_{\rm crit,A} = \frac{\alpha_A~\sigma_g~A}{\pi~G}~.
\end{equation}

\noindent \citet{HUNTER98} suggest $\alpha_{\rm A} = 2.5$, but this
normalization for $\Sigma_{\rm crit,A}$ is relatively uncertain. The value
chosen by \citet{HUNTER98} corresponds to perturbations growing by a factor of
$\sim 100$ during the time allowed by shear, which roughly matches both the
surface density contrast between $\Sigma_{\rm HI}$ and a GMC and the condition
$Q_{\rm gas} \lesssim 1$ where $dv(r_{\rm gal})/dr_{\rm gal} = 0$.

The practical advantage of shear over $Q_{\rm gas}$ is that shear is low in
dwarf galaxies and the inner disks of spiral galaxies ($\beta = 1$ for solid
body rotation), both locales where widespread star formation is observed. In
the outer disks of spiral galaxies --- where star formation cutoffs are
observed --- rotation curves tend to be flat ($\beta = 0$) so that
$\Sigma_{\rm crit, A}$ and $\Sigma_{\rm crit, Q}$ reduce to the same form.

\subsubsection{Formation of a Cold Phase}
\label{THERM_SECT}

The very long time needed to assemble a massive GMC from coagulation of
smaller clouds suggests that most GMCs in galaxy disks form ``top down''
\citep[e.g.,][]{MCKEE07}. However this does not necessarily require that the
whole gas disk to be unstable.  Where cold \hi\ is abundant, the lower
velocity dispersion associated with this phase may render the ISM locally
unstable \citep{SCHAYE04}, leading to the formation of GMCs and stars.

Therefore, instead of large-scale gravitational instability or cloud
destruction by shear, the ability to form a cold neutral medium
\citep[][]{MCKEE77,WOLFIRE03} may regulate GMC formation.  \citet{SCHAYE04}
argues based on modeling that near the cutoffs observed by \citet{MARTIN01}
gas becomes mostly cold \hi\ and H$_2$, $\sigma_g$ drops accordingly, and $Q$
becomes $<1$ in the cold gas.  In a similar vein, \citet{ELMEGREEN94} suggest
that the star formation efficiency in the outer parts of galaxies drops
because the pressure becomes too low to allow a cold phase to form even given
perturbations, e.g., from supernova shocks.  \citet{BRAUN97} found support for
this idea using 21--cm observations; he associated networks of high surface
brightness filaments with cold \hi\ and showed that these filaments are
pervasive across the star forming disk, but become less common at large radii
\citep[though work on THINGS by][calls this result into question]{USERO08}.

\citet{SCHAYE04} modeled the ISM to estimate where the average temperature
drops to $\approx 500$~K, the molecular fraction reaches $\approx 10^{-3}$,
and $Q_{\rm gas} \approx 1$; good indicators that cold \hi\ is common and
\htwo\ formation is efficient. These all occur where $\Sigma_{\rm gas}$
exceeds

\begin{equation}
\label{SCHAYEEQ}
\Sigma_{\rm S04} \approx \frac{6.1}{{\rm M}_{\odot}~{\rm pc}^{-2}}~f_{\rm
  g}^{0.3}~\left( \frac{Z}{0.1 Z_{\odot}}
\right)^{-0.3}~\left(\frac{I}{10^6~\rm{cm}^{-2}~{\rm s}^{-1} }\right)^{0.23},
\end{equation}

\noindent where $f_{\rm g} \approx \Sigma_{\rm gas} / ( \Sigma_{\rm gas} +
\Sigma_* )$ is the fraction of mass in gas (we assume a two-component disk),
$Z$ is the metallicity of the ISM, and $I$ is the flux of ionizing photons.
$\Sigma_{\rm S04}$ also depends on the ratio of thermal to turbulent pressure
and higher order terms not shown here. \citet{SCHAYE04} selects fiducial
values to match those expected in outer galaxy disks, but concludes that the
influence of $Z$, $f_{\rm g}$, and the radiation field is relatively small.
Most reasonable values yield $\Sigma_{{\rm S04}} \approx 3 -
10$~M$_{\odot}$~pc$^{-2}$.

\citet{SCHAYE04} argues that a simple column density threshold may work as
well as dynamical thresholds. This agrees with the observation by, e.g.,
\citet{SKILLMAN87} and \citet{DEBLOK06} that a simple \hi\ column density
threshold does a good job of predicting the location of star formation in
dwarf irregulars. This threshold, $\Sigma_{\rm HI} \approx
10$~M$_{\odot}$~pc$^{-2}$, also corresponds to the surface density above which
\hi\ is observed to saturate \citep[][]{MARTIN01,WONG02,BIGIEL08}; that is,
gas in excess of this surface density in spiral galaxies is in the molecular
phase.

\section{Data}
\label{SECT_DATA}

The right hand column of Table \ref{IDEATAB} lists the observables required to
evaluate each law or threshold. We require estimates of: the surface density
of atomic gas ($\Sigma_{\rm HI}$), molecular gas ($\Sigma_{\rm H2}$), star
formation rate ($\Sigma_{\rm SFR}$), and stellar mass ($\Sigma_{\rm *}$), the
velocity dispersions of gas and stars ($\sigma_{\rm gas}$ and $\sigma_*$), and
the rotation curve ($v(r_{\rm gal})$). Estimates of the metallicity await
future work.

\subsection{The Sample}
\label{SAMPLE_SECT}

\begin{deluxetable}{l c c c l}
  \tabletypesize{\small} \tablewidth{0pt} \tablecolumns{5}
  \tablecaption{\label{SAMPLETAB} Sample Galaxies}
  \tablehead{\colhead{Galaxy\tablenotemark{a}} &
    \colhead{Res.\tablenotemark{b}} & \colhead{CO} & \colhead{Rotation} &
    \colhead{Also in} \\
    & ($\arcsec$) & & Curve\tablenotemark{c} & sample of\tablenotemark{d} }
  \startdata
  DDO 154 & 19 & \nodata & dB & \nodata \\
  Ho I & 21 & \nodata & T & \nodata \\
  Ho II & 24 & \nodata & T & \nodata \\
  IC 2574 & 21 & \nodata & dB & \nodata \\
  NGC 4214\tablenotemark{e} & 28 & \nodata & T & \nodata \\
  NGC 2976 & 23 & \nodata & dB & \nodata \\
  NGC 4449\tablenotemark{e} & 20 & \nodata & \nodata & \nodata \\
  NGC 3077\tablenotemark{e} & 22 & \nodata & \nodata & \nodata \\
  NGC 7793 & 21 & \nodata & dB & \nodata \\
  NGC  925 & 9 & \nodata & dB & 1, 2, 4 \\
  NGC 2403 & 26 & \nodata & dB & 1, 2, 4 \\
  \hline
  NGC  628 & 23 & HERACLES & T & 1, 2 \\
  NGC 3198 & 12 & HERACLES & dB & \nodata \\
  NGC 3184 & 15 & HERACLES & T & \nodata \\
  NGC 4736 & 35 & HERACLES & dB &  1, 2, 3, 5 \\
  NGC 3351 & 16 & HERACLES & T & \nodata \\
  NGC 6946 & 28 & HERACLES & dB & 2 \\
  NGC 3627 & 18 & BIMA SONG & dB & 5 \\
  NGC 5194 & 21 & BIMA SONG & T & 2, 4, 5 \\
  NGC 3521 & 15 & HERACLES & dB & 5 \\
  NGC 2841 & 12 & HERACLES & dB & 1, 2 \\
  NGC 5055 & 16 & HERACLES & dB & 2, 3, 5 \\
  NGC 7331 & 11 & HERACLES & dB & 2, 5
\enddata
\tablenotetext{a}{In order of increasing stellar mass.}
\tablenotetext{b}{Angular resolution to match working spatial resolution in
  the subsample, 400 ~pc for dwarf galaxies and and 800~pc for spirals.}
\tablenotetext{c}{Rotation curve data: dB = \citet{DEBLOK08}; T = only THINGS
  first moment \citep{WALTER08}}
\tablenotetext{d}{1: \citet{KENNICUTT89}; 2: \citet{MARTIN01}; 3:
  \citet{WONG02}; 4: \citet{BOISSIER03}; 5: \citet{BLITZ06}}
\tablenotetext{e}{IR data from {\em Spitzer} archive (not SINGS).}
\end{deluxetable}

We assemble maps and radial profiles of the necessary quantities in 23 nearby,
star--forming galaxies that we list in order of increasing stellar mass in
Table \ref{SAMPLETAB}. These are galaxies for which we could compile the
necessary data, which means the overlap of THINGS, SINGS, the GALEX NGS, and
(for spirals) either BIMA SONG or HERACLES.

We work with two subsamples: 11 \hi -dominated, low-mass galaxies and 12 large
spiral galaxies. In Table \ref{SAMPLETAB}, the galaxies that we classify
``dwarf galaxies'' lie above the horizontal dividing line.  These have
rotation velocities $v_{\rm rot} \lesssim 125$~km~s$^{-1}$, stellar masses
$M_* \lesssim 10^{10}$~M$_\odot$, and $M_B \gtrsim -20$~mag. The galaxies that
we label ``spirals'' lie below the dividing line and have $v_{\rm rot} \gtrsim
125$~km~s$^{-1}$, $M_* \gtrsim 10^{10}$~M$_\odot$, and $M_B \lesssim -20$~mag.

This division allows us to explore two distinct regimes in parallel. Compared
to their larger cousins, dwarf galaxies have low metallicities, intense
radiation fields, lower galactic shear, and weak or absent spiral structure.
Metallicity, in particular, should have a strong effect on the thermal balance
of the ISM. In lieu of direct measurements, separating the sample in this way
allows us to assess its impact.

We treat the two subsamples slightly differently in two ways. First, we place
data for spirals at a common spatial resolution of 800~pc and data for dwarf
galaxies at 400~pc. The spirals in our sample are farther away than the dwarf
galaxies with larger physical radii, and this approach ensures a good number
of resolution elements across each galaxy and a fairly uniform angular
resolution of $\sim 20\arcsec$ (see Table \ref{SAMPLETAB}).

Second, we use CO maps combined with a constant CO-to-H$_2$ conversion factor,
\xco , to derive $\Sigma_{\rm H2}$ in spirals, while we treat the molecular
gas content of dwarf galaxies as unknown (see Appendix \ref{XCOAPP}). CO
emission in very low mass galaxies is usually weak or not detected
\citep[e.g.,][and see Table \ref{STRUCTURETAB}]{TAYLOR98,LEROY05} and its
interpretation is confused by potential variations in \xco . Because dwarf
galaxies lack \htwo-filled \hi\ depressions like those observed in the centers
of spirals, we expect $\Sigma_{\rm HI}$ to at least capture the basic
morphology of the total gas. Although we do not measure $\Sigma_{\rm H2}$ in
dwarf galaxies, we consider our results in light of the possibility of an
unseen reservoir of molecular gas (\S \ref{SECT_MISSINGH2}).

\subsection{Data to Physical Quantities}

\begin{deluxetable*}{l l l l l}
  \tabletypesize{\footnotesize} \tablewidth{0pt} \tablecolumns{9}
  \tablecaption{\label{OBSERVABLETAB} Data to Physical Quantities}
  \tablehead{\colhead{Quantity} & \colhead{Observation} & \colhead{Survey} &
    \colhead{Reference} & \colhead{Key Assumptions} } \startdata
  $\Sigma_{\rm HI}$ & 21-cm line & THINGS & \citet{WALTER08} \\
  $\Sigma_{\rm H2}$ (spirals only) & CO $2\rightarrow1$ & HERACLES &
  \citet{LEROY08} & fixed line ratio, CO-to-H$_2$ conversion  \\
  & CO $1\rightarrow0$ & BIMA SONG & \citet{HELFER03} & fixed CO-to-H$_2$
  conversion \\
  Unobscured $\Sigma_{\rm SFR}$ & FUV & GALEX NGS & \citet{GILDEPAZ07} \\
  Embedded $\Sigma_{\rm SFR}$ & 24$\mu$m & SINGS & \citet{KENNICUTT03} \\
  $\Sigma_{\rm *}$ & 3.6$\mu$m & SINGS & \citet{KENNICUTT03} &
  $\ML = 0.5$~$M_\odot/L_{\odot,K}$ \\
  Kinematics & 21-cm line & THINGS & \citet{DEBLOK08} & simple functional fit;
  fixed $\sigma_{\rm gas}$
  \enddata
\end{deluxetable*}

Appendices \ref{GASAPP} -- \ref{SFRAPP} explain in detail how we translate
observables into physical quantities. Here and in Table \ref{OBSERVABLETAB} we
summarize this mapping.

{\em Atomic Hydrogen Surface Density (Appendix \ref{GASAPP}):} We derive
atomic gas mass surface density, $\Sigma_{\rm HI}$, from 21-cm line integrated
intensity maps obtained by \citet{WALTER08} as part of the THINGS survey using
the Very Large Array\footnote{The VLA is operated by the National Radio
  Astronomy Observatory, which is a facility of the National Science
  Foundation operated under cooperative agreement by Associated Universities,
  Inc.}.  $\Sigma_{\rm HI}$ is corrected for inclination and includes a factor
of 1.36 to account for helium.

{\em Molecular Hydrogen Surface Density (Appendix \ref{GASAPP}):} In spirals,
we estimate the molecular gas mass surface density, $\Sigma_{\rm H2}$, from CO
line emission. For 10 galaxies we use data from HERACLES, a large program at
the IRAM\footnote{IRAM is supported by CNRS/INSU (France), the MPG (Germany)
  and the IGN (Spain)}~30--m telescope \citep{LEROY08} that used the HERA
focal plane array \citep{SCHUSTER04} to map a subsample of THINGS in the CO
$J=2\rightarrow1$ line. For NGC~3627 and NGC~5194, we use $J=1\rightarrow0$
line maps from the BIMA SONG survey \citep{HELFER03}.

We convert from CO line intensity to $\Sigma_{\rm H2}$ assuming a constant
CO-to-H$_2$ conversion factor appropriate for the solar neighborhood, $ \xco =
2\times10^{20}$~\xcounits , and a fixed line ratio $I_{\rm CO} (2\rightarrow1)
= 0.8~I_{\rm CO} (1\rightarrow0)$, typical of the disks of spiral galaxies.
We correct for the effects of inclination and include a factor of 1.36 to
reflect the presence of helium .

{\em Galactic Rotation (Appendix \ref{KINAPP}):} We fit a simple functional
form to the high quality rotation curves derived from THINGS by
\citet{DEBLOK08} and the THINGS first moment maps \citep{WALTER08}. These fits
yield smooth, well--behaved (analytic) derivatives and match the observations
well. Two galaxies (NGC~3077 and NGC~4449) have complex velocity fields that
require substantial effort to interpret and we omit them from analyses
requiring kinematics.

{\em Gas Velocity Dispersion (Appendix \ref{KINAPP}):} We assume a fixed gas
velocity dispersion, $\sigma_{\rm gas} = 11$~km~s$^{-1}$, a value motivated by
the THINGS second moment maps.

{\em Stellar Velocity Dispersion (Appendix \ref{KINAPP}):} We estimate the
vertical stellar velocity dispersion, $\sigma_{*,z}$, from hydrostatic
equilibrium, the assumption of an isothermal disk, and an estimated (radially
invariant) stellar scale height. We derive this scale height for each galaxy
from our measured stellar scale length and an average flattening ratio for
disk galaxies. We take the vertical and radial velocity dispersions to be
related by $\sigma_{*,z} = 0.6~\sigma_{*,r}$.

{\em Stellar Surface Density (Appendix \ref{STARAPP}):} We estimate the
stellar surface density, $\Sigma_*$, from {\em Spitzer} 3.6$\mu$m maps, mostly
from SINGS \citep{KENNICUTT03}. To avoid contamination by hot dust and
foreground stars, we construct radial profiles only, using the median
3.6$\mu$m intensity in each tilted ring.  We convert from 3.6$\mu$m intensity
to $\Sigma_*$ via an empirical $K$-to-3.6$\mu$m calibration and adopt a fixed
$K$-band mass-to-light ratio, $\ML = 0.5~M_\odot/L_{\odot,K}$.

{\em Star Formation Rate Surface Density (Appendix \ref{SFRAPP}):} We combine
FUV and 24$\mu$m maps to derive maps of $\Sigma_{\rm SFR}$; giving us a tracer
sensitive to both exposed and dust-embedded star formation.  The FUV data come
from the GALEX Nearby Galaxies Survey \citep{GILDEPAZ07} and the 24$\mu$m maps
are part of SINGS. Because this precise combination of data is new, Appendix
\ref{SFRAPP} includes an extended motivation for how we convert intensity to
$\Sigma_{\rm SFR}$.

\subsection{Properties of the Sample}

\begin{deluxetable*}{l c c c c c c c c c c c c c c c}
  \tabletypesize{\footnotesize} \tablewidth{0pt} \tablecolumns{16}
  \tablecaption{\label{STRUCTURETAB} Properties of Sample Galaxies}
  \tablehead{\colhead{Galaxy} & \colhead{Dist.} & \colhead{$i$} & \colhead{PA}
    & \colhead{Morph.} & \colhead{$M_{B}$} & \colhead{$r_{25}$}
    & \colhead{$v_{\rm flat}$} & \colhead{$l_{\rm flat}$} & \colhead{$\log
      M_*$} &  \colhead{$\log M_{\rm HI}$} & \colhead{$\log M_{\rm H2}$} & \colhead{${\rm SFR}$} &
    \colhead{$l_*$} & \colhead{$l_{\rm SFR}$} & \colhead{$l_{\rm CO}$} \\ &
    (Mpc) & ($\arcdeg$) & ($\arcdeg$) & & (mag) & (kpc) & (km s$^{-1}$) &
    (kpc) & (M$_\odot$) & (M$_\odot$) & (M$_\odot$) & (M$_\odot$ yr$^{-1}$) &
    (kpc) & (kpc) & (kpc) }
\startdata
DDO 154 &  4.3 &  66 & 230 & Irr & -14.4 &  1.2 &  50 &  2.0 &  7.1 &  8.7 & $\leq  6.8$ & 0.005 &  0.8 &  1.0 &  \nodata \\
Ho I &  3.8 &  12 &  50 & Irr & -14.9 &  1.8 &  53 &  0.4 &  7.4 &  8.3 & $\leq  7.2$ & 0.009 &  0.8 &  1.2 &  \nodata \\
Ho II &  3.4 &  41 & 177 & Irr & -16.9 &  3.7 &  36 &  0.6 &  8.3 &  8.9 & $\leq  7.6$ & 0.048 &  1.2 &  1.3 &  \nodata \\
IC 2574 &  4.0 &  53 &  56 & Irr & -18.0 &  7.5 & 134 & 12.9 &  8.7 &  9.3 & $\leq  7.9$ & 0.070 &  2.1 &  4.8 &  \nodata \\
NGC 4214 &  2.9 &  44 &  65 & Irr & -17.4 &  2.9 &  57 &  0.9 &  8.8 &  8.7 & $ 7.0$ & 0.107 &  0.7 &  0.5 &  \nodata \\
NGC 2976 &  3.6 &  65 & 335 & Sc & -17.8 &  3.8 &  92 &  1.2 &  9.1 &  8.3 & $ 7.8$ & 0.087 &  0.9 &  0.8 &  1.2 \\
NGC 4449 &  4.2 &  60 & 230 & Irr & -19.1 &  2.8 &  \nodata &  \nodata &  9.3 &  9.2 &  6.9\tablenotemark{a} & 0.371 &  0.9 &  0.8 &  \nodata \\
NGC 3077 &  3.8 &  46 &  45 & Sd & -17.7 &  3.0 &  \nodata &  \nodata &  9.3 &  9.1 &  6.5\tablenotemark{a} & 0.086 &  0.7 &  0.3 &  \nodata \\
NGC 7793 &  3.9 &  50 & 290 & Scd & -18.7 &  6.0 & 115 &  1.5 &  9.5 &  9.1 &  \nodata & 0.235 &  1.3 &  1.3 &  \nodata \\
NGC 2403 &  3.2 &  63 & 124 & SBc & -19.4 &  7.3 & 134 &  1.7 &  9.7 &  9.5 &  7.3 & 0.382 &  1.6 &  2.0 &  1.9 \\
NGC 0925 &  9.2 &  66 & 287 & SBcd & -20.0 & 14.2 & 136 &  6.5 &  9.9 &  9.8 & $ 8.4$ & 0.561 &  4.1 &  4.1 &  \nodata \\
 \hline
NGC 0628 &  7.3 &   7 &  20 & Sc & -20.0 & 10.4 & 217 &  0.8 & 10.1 &  9.7 &  9.0 & 0.807 &  2.3 &  2.4 &  2.4 \\
NGC 3198 & 13.8 &  72 & 215 & SBc & -20.7 & 13.0 & 150 &  2.8 & 10.1 & 10.1 &  8.8 & 0.931 &  3.2 &  3.4 &  2.7 \\
NGC 3184 & 11.1 &  16 & 179 & SBc & -19.9 & 11.9 & 210 &  2.8 & 10.3 &  9.6 &  9.2 & 0.901 &  2.4 &  2.8 &  2.9 \\
NGC 4736 &  4.7 &  41 & 296 & Sab & -20.0 &  5.3 & 156 &  0.2 & 10.3 &  8.7 &  8.6 & 0.481 &  1.1 &  0.9 &  0.8 \\
NGC 3351 & 10.1 &  41 & 192 & SBb & -19.7 & 10.6 & 196 &  0.7 & 10.4 &  9.2 &  9.0 & 0.940 &  2.2 &  1.8 &  2.5 \\
NGC 6946 &  5.9 &  33 & 243 & SBc & -20.9 &  9.8 & 186 &  1.4 & 10.5 &  9.8 &  9.6 & 3.239 &  2.5 &  2.7 &  1.9 \\
NGC 3627 &  9.3 &  62 & 173 & SBb & -20.8 & 13.9 & 192 &  1.2 & 10.6 &  9.0 &  9.1 & 2.217 &  2.8 &  1.9 &  2.2 \\
NGC 5194 &  8.0 &  20 & 172 & SBc & -21.1 &  9.0 & 219 &  0.8 & 10.6 &  9.5 &  9.4 & 3.125 &  2.8 &  2.4 &  2.3 \\
NGC 3521 & 10.7 &  73 & 340 & SBbc & -20.9 & 12.9 & 227 &  1.4 & 10.7 & 10.0 &  9.6 & 2.104 &  2.9 &  3.1 &  2.2 \\
NGC 2841 & 14.1 &  74 & 153 & Sb & -21.2 & 14.2 & 302 &  0.6 & 10.8 & 10.1 &  8.5 & 0.741 &  4.0 &  5.3 &  \nodata \\
NGC 5055 & 10.1 &  59 & 102 & Sbc & -20.6 & 17.4 & 192 &  0.7 & 10.8 & 10.1 &  9.7 & 2.123 &  3.2 &  3.1 &  3.1 \\
NGC 7331 & 14.7 &  76 & 168 & SAb & -21.7 & 19.6 & 244 &  1.3 & 10.9 & 10.1 &  9.7 & 2.987 &  3.3 &  4.5 &  3.1 \\

\enddata
\tablenotetext{a}{Unless noted $\log M_{\rm H2}$ comes from HERACLES
  \citet{LEROY08} or BIMA SONG \citep{HELFER03}. NGC~3077 is from
  \citet[][]{WALTER01}, NGC~4449 is from \citet{BOLATTO08}. Upper
  limits are at $5\sigma$ significance.}
\end{deluxetable*}

Table \ref{STRUCTURETAB} compiles the integrated properties of each galaxy in
our sample. Columns (1) -- (7) give basic parameters adopted from other
sources: the name of the galaxy; the distance, inclination, and position angle
\citep[][except that we adopt $i=20\arcdeg$ in M~51]{WALTER08}; and the
morphology, $B$-band isophotal radius at 25 mag arcsec$^{-2}$ ($r_{25}$), and
$B$-band absolute magnitude from LEDA \citep[][]{PRUGNIEL98}. Columns (8) and
(9) give $v_{\rm flat}$ and $l_{\rm flat}$, the free parameters for our
rotation curve fit (Appendix \ref{KINAPP}); from these two parameters one can
calculate $v~(r_{\rm gal})$ and $\beta$. Columns (10) -- (13) give the total
stellar mass, \hi\ mass, H$_2$ mass and SFR from integrating our data within
$1.5~r_{25}$.

Columns (14) -- (17) give scale lengths derived from exponential fits
to the $\Sigma_*$, $\Sigma_{\rm SFR}$, and $\Sigma_{\rm H2}$ (CO)
radial profiles.  The stellar scale lengths match those found by
\citet{TAMBURRO08} with 15\% scatter; they are $\sim 10\%$ shorter
than those found by \citet{REGAN01}, with RMS scatter of 20\%.  Our CO
scale lengths are taken from \citet{LEROY08}; these are $\sim 30\%$
shorter than those of \citet{REGAN01} on average.

\subsection{Methodology}

We work with maps of $\Sigma_{\rm HI}$, $\Sigma_{\rm H2}$ and $\Sigma_{\rm
  SFR}$ on the THINGS astrometric grid. All data are placed at a common
spatial resolution, $400$~pc for dwarf galaxies and $800$~pc for spirals; when
necessary, we use a Gaussian kernel to degrade our data to this resolution.
The convolution occurs before any deprojection and may be thought of as
placing each subsample at a single distance. Radial profiles of these maps and
$\Sigma_*$ appear in Appendix \ref{RPROFAPP}.

Using these data, we compute each quantity in Table \ref{IDEATAB} for
each pixel inside $1.2~r_{25}$ and derive radial profiles over the
same range following the methodology in Appendix
\ref{RPROFAPP}. Because we measure $\Sigma_*$ and $v(r_{\rm gal})$
only in radial profile, these maps are often a hybrid between radial
profiles and pixel--by--pixel measurements.

In \S \ref{SECT_RESULTS} -- \ref{SECT_RECIPE}, we analyze the combined
data set for the two subsamples and avoid discussing results for
individual galaxies. We refer readers interested in individual
galaxies to the Appendices. Appendix \ref{RPROFAPP} gives our radial
profile data and the atlas in Appendix \ref{ATLASAPP} shows maps of
$\Sigma_{\rm HI}$, $\Sigma_{\rm H2}$, total gas, unobscured
$\Sigma_{\rm SFR}$, dust-embedded $\Sigma_{\rm SFR}$, and total
$\Sigma_{\rm SFR}$, as well as profiles of the quantities in Table
\ref{IDEATAB}.

In keeping with our emphasis on the combined dataset, we default to quoting
the mean and $1\sigma$ scatter when we give uncertainties in parameters
derived from the ensemble of galaxies (we usually estimate the scatter using
the median absolute deviation to reduce sensitivity to outliers). We prefer
this approach to giving the uncertainty in the mean because we are usually
interested in how well a given number describes our whole sample, not how
precisely we have measured the mean.

\section{Results}
\label{SECT_RESULTS}

Here we present our main observational results, how the star formation
efficiency varies as a function of other quantities. We begin in \S
\ref{SECT_SFEOBS} by showing the SFE as a function of three basic parameters:
galactocentric radius, stellar surface density, and gas surface density. Then
in \S \ref{SECT_SFELAW}, we look at SFE as a function of the laws described in
\S \ref{SECT_LAWBACK}. Finally, in \S \ref{SECT_SFETHRESH} we show the SFE as
a function of the thresholds described in \S \ref{SECT_THRESHBACK}.

\begin{figure*}
\begin{center}
\plottwo{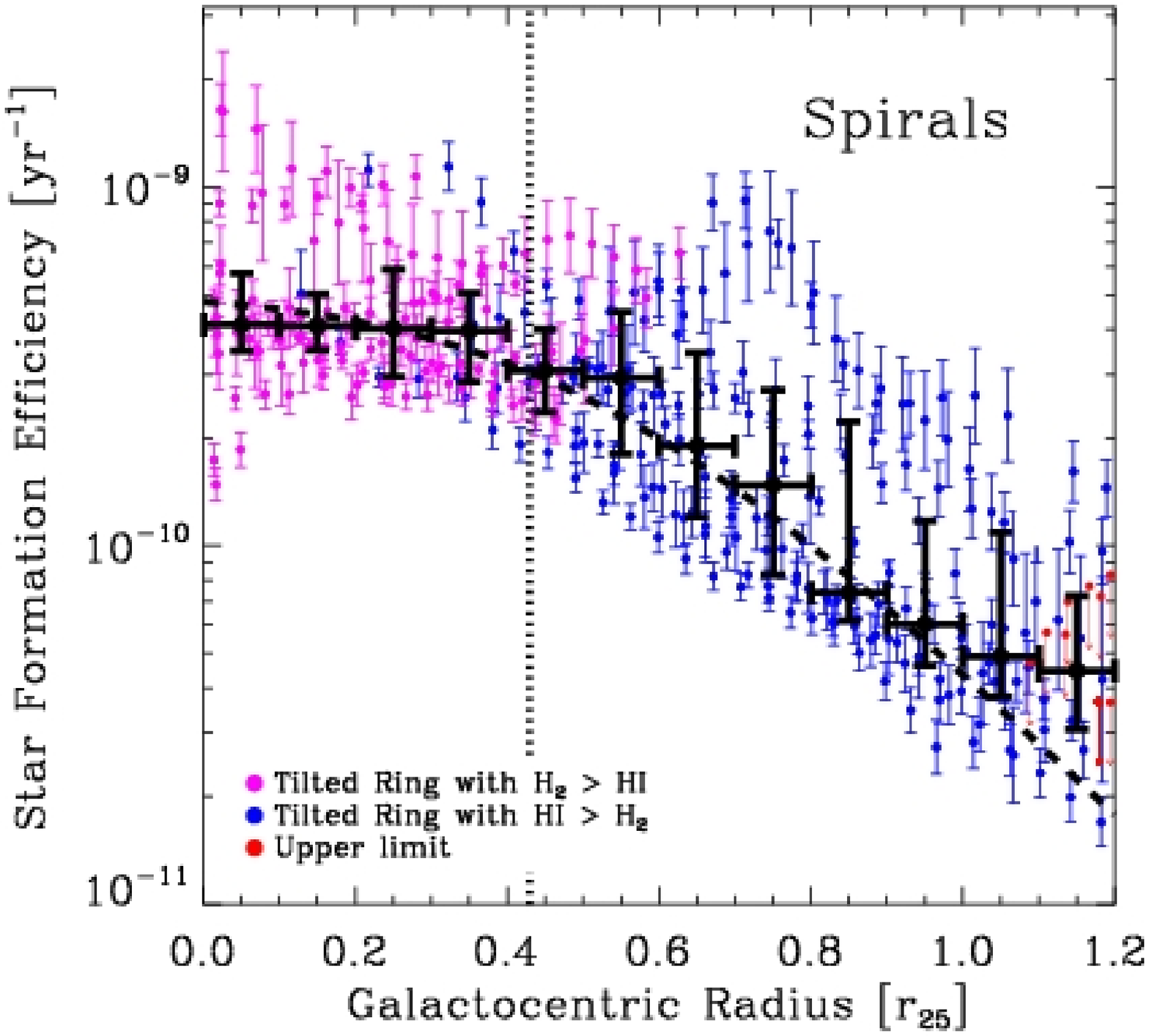}{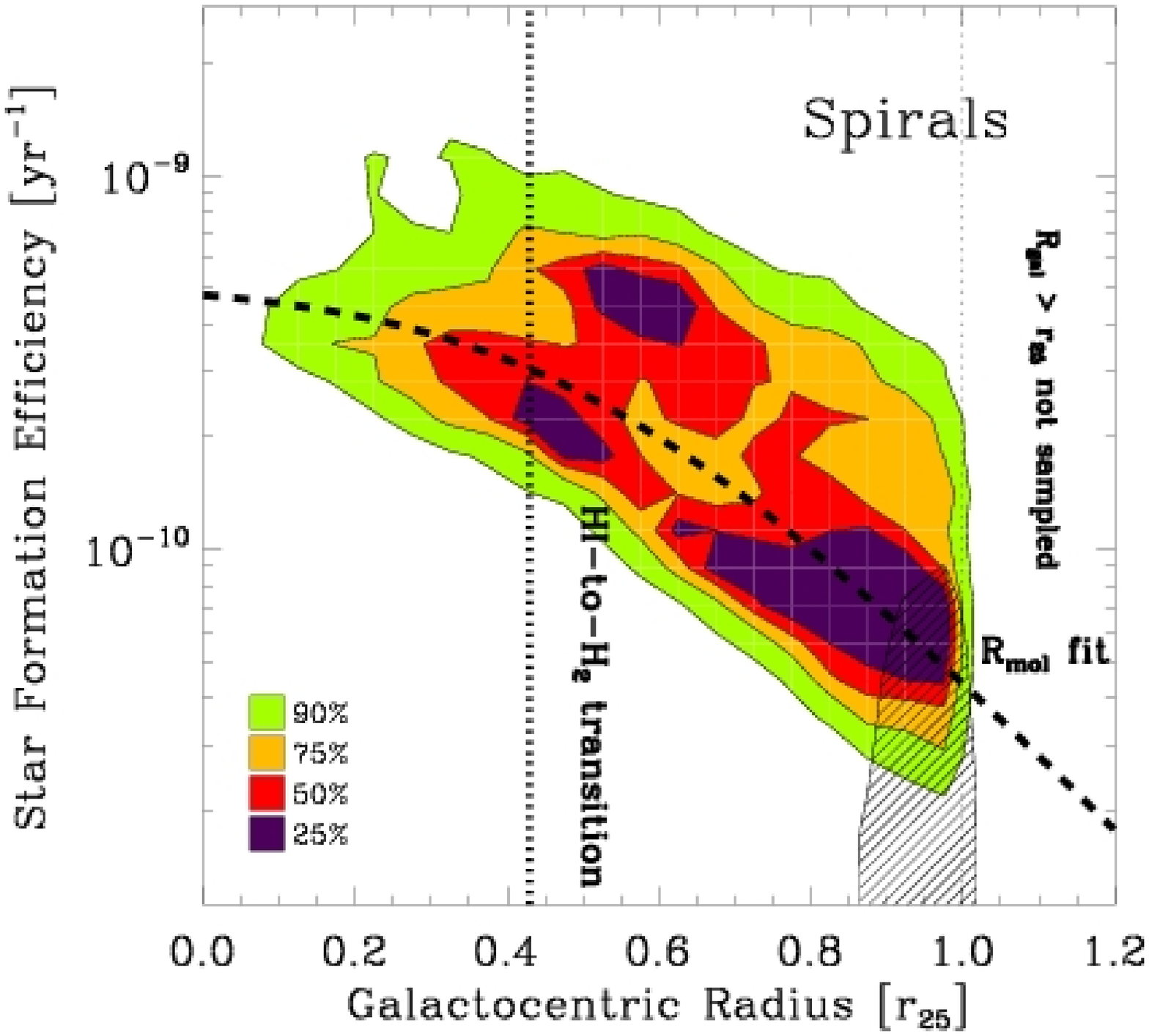}
\plottwo{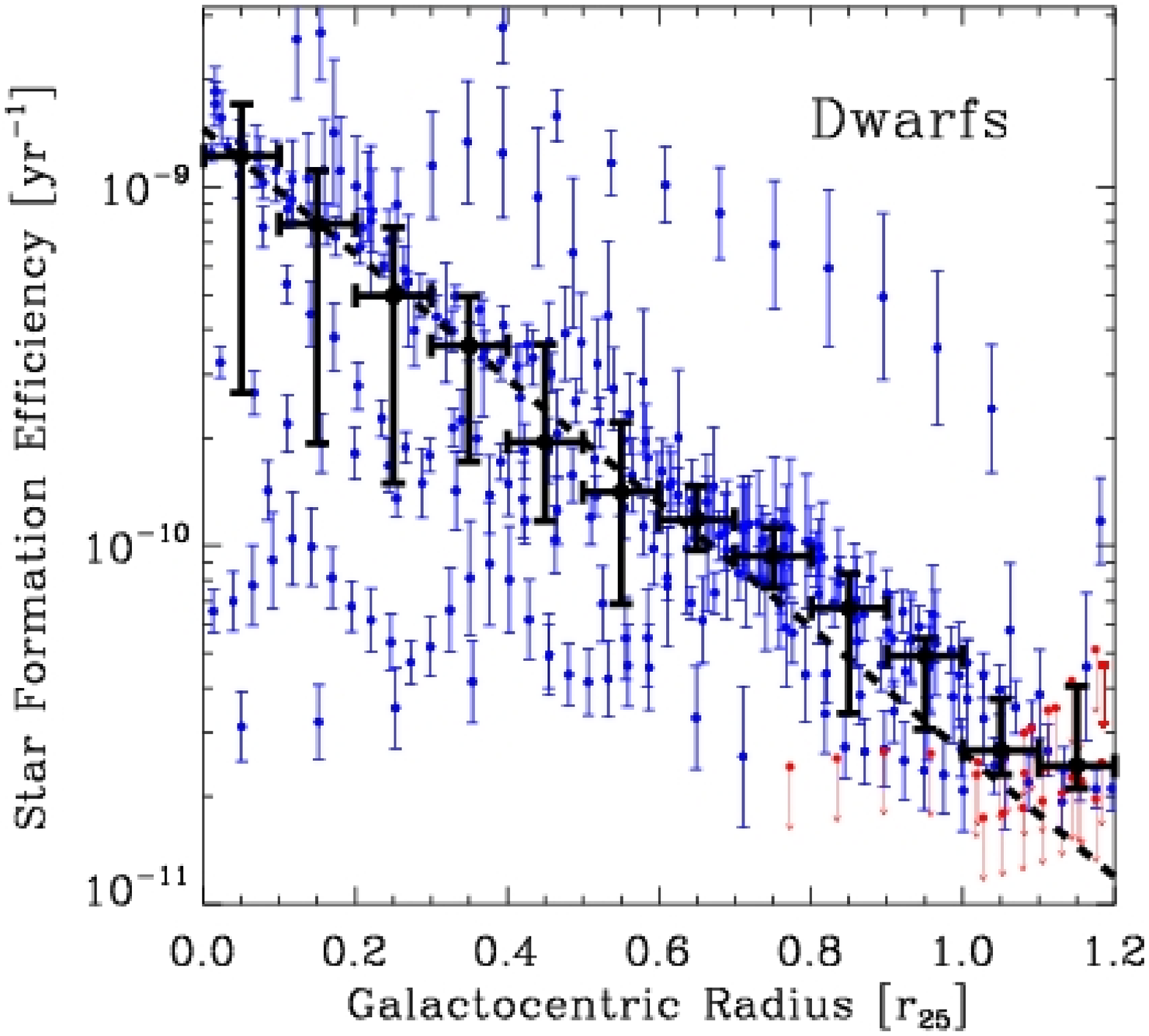}{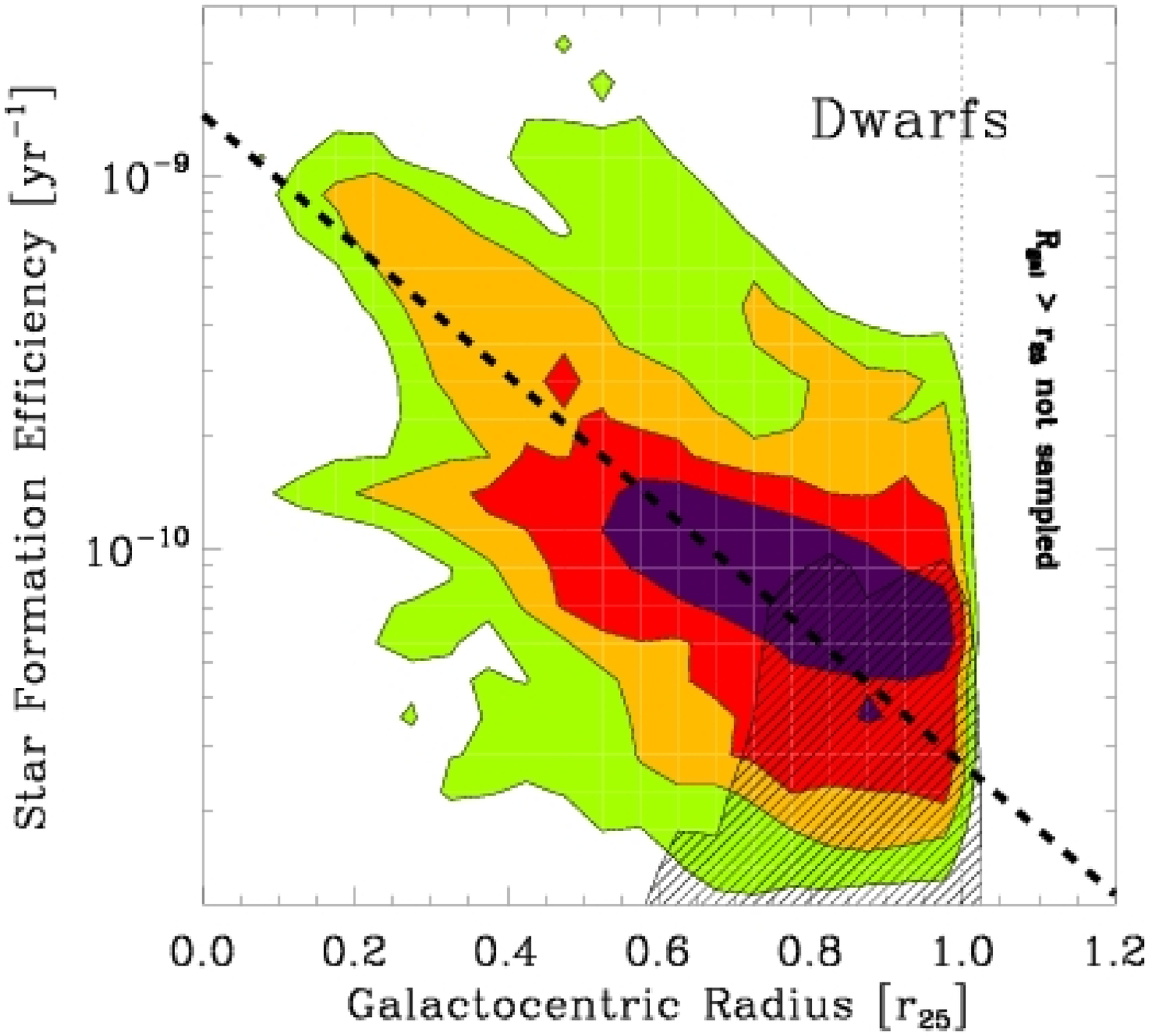}
\end{center}
\caption{\label{SFEVSRAD} Star formation efficiency as a function of
  galactocentric radius in spiral (top row) and dwarf (bottom row) galaxies.
  The left panels show results for radial profiles; each point shows the
  average SFE over a $10\arcsec$-wide tilted ring; magenta points are \htwo
  -dominated ($\Sigma_{\rm H2} > \Sigma_{\rm HI}$), blue points are \hi
  -dominated ($\Sigma_{\rm H2} < \Sigma_{\rm HI}$), and red arrows indicate
  upper limits.  The right panels show data for individual lines of sight. We
  give each galaxy equal weight and choose contours that include 90\%, 75\%,
  50\%, and 25\% of the data. The hatched regions indicate where we are
  incomplete.  The top panels show a nearly fixed SFE in \htwo -dominated
  galaxy centers (magenta). Where \hi\ dominates the ISM (blue), we observe
  the SFE to decline exponentially with radius; the thick dashed lines show
  fits of SFE to $r_{\rm gal}$ (Equations \ref{SFEFIT2} and
  \ref{DWARFSFEFIT1}). The vertical dotted line in the upper panels shows
  $r_{\rm gal}$ at the \hi -to-\htwo\ transition in spirals, $0.43 \pm
  0.18~r_{25}$ (\S \ref{TRANS_SECT}).}
\end{figure*}

We present these results as a series of plots that each show SFE as a function
of another quantity. These all follow the format seen in Figure
\ref{SFEVSRAD}, where we show SFE ($y$-axis) versus galactocentric radius
($x$-axis), normalized to the optical radius, $r_{25}$. We plot the subsamples
of spiral (top row) and dwarf galaxies (bottom row) separately.

On the left, we show results for radial profiles. Each point shows the average
SFE over one $10\arcsec$-wide tilted ring in one galaxy. The color indicates
whether the ISM averaged over the ring is mostly ($> 50\%$) \hi\ (blue) or
\htwo\ (magenta). Thick black crosses show all data binned into a single
trend. For each bin, we plot the median, $50\%$ range ($y$-error bar), and bin
width ($x$-error bar, here $0.1~r_{25}$).

On the right, we again show SFE as a function of radius, this time
calculated for each line of sight. We coadd all galaxies, giving equal
weight to each, and pick contours that contain 90\% (green), 75\%
(yellow), 50\% (red), and 25\% (purple) of the resulting data. Most
numerical results use the annuli, which are easier to work with; these
pixel-by-pixel plots verify that conclusions based on rings hold
pixel-by-pixel down to kiloparsec scales.

We do not analyze data with $\Sigma_{\rm gas} < 1$~M$_{\odot}$~pc$^{-2}$
because the SFE is not well-determined for low gas surface densities; that is,
we only address the question ``where there is gas, is it good at forming
stars?'' Data with $\Sigma_{\rm SFR} <
10^{-4}$~M$_{\odot}$~yr$^{-1}$~kpc$^{-2}$ are treated as upper limits. These
are red arrows in the radial profiles plots. In the pixel-by-pixel plots,
hatched regions show the area inhabited by 95\% of data with $\Sigma_{\rm SFR}
\leq 10^{-4}$~M$_{\odot}$~yr$^{-1}$~kpc$^{-2}$, i.e., the hatched regions
indicate the area where we are incomplete. In the pixel--by--pixel plots, we
include data out to $r_{25}$, while we plot radial profile data out to
$1.2~r_{25}$.

\subsection{SFE and Other Basic Quantities}
\label{SECT_SFEOBS}

\subsubsection{SFE and Radius}
\label{SECT_SFERAD}

We argued in \S \ref{INTRO} that a critical observation for theories of
galactic--scale star formation is that the SFE declines in the outer parts of
spiral galaxies. Figure \ref{SFEVSRAD} shows this via plots of SFE against
galactocentric radius (normalized to $r_{25}$) in our two subsamples.

In spiral galaxies (top row), the SFE is nearly constant where the ISM is
mostly \htwo\ (magenta), which agrees with our observation of a linear
relationship between $\Sigma_{\rm H2}$ and $\Sigma_{\rm SFR}$ in
\citet[][]{BIGIEL08}.  Typically, the ISM is equal parts \hi\ and \htwo\ at
$r_{\rm gal} = 0.43 \pm 0.18~r_{25}$ (\S \ref{TRANS_SECT}).  Outside this
transition, the SFE decreases steadily with increasing radius. This decline
continues to $r_{\rm gal} \gtrsim r_{25}$, the limit of our data.  This is
similar, though not identical, to the observation by \citet{KENNICUTT89} and
\citet{MARTIN01} that star formation is not widespread beyond a certain
radius.

The SFE in spirals can be reasonably described in two ways. First, a constant
SFE in the inner parts of galaxies followed by a break at $0.4~r_{25}$
(slightly inside the transition to a mostly-\hi\ ISM):

\begin{equation}
  \label{SFEFIT1}
  {\rm SFE} = 
  \left\{\begin{array}{cl} 4.3 \times 10^{-10} & r_{\rm gal} < 0.4 r_{25} \\
      2.2 \times 10^{-9}~\exp \left(\frac{-r_{\rm gal}}{0.25~r_{25}}\right) & r_{\rm gal} > 0.4 r_{25} \\ 
\end{array}~\right.
{\rm yr}^{-1}~.
\end{equation}

\noindent Alternatively, we can adopt Equation \ref{FIXEDSFEEQ}, appropriate
for a fixed SFE (\htwo), and derive the best-fit exponential relating $R_{\rm
  mol}$ to $r_{\rm gal}$,

\begin{eqnarray}
  \label{SFEFIT2}
  {\rm SFE} &=& 5.25 \times 10^{-10}~\frac{R_{\rm mol}}{R_{\rm mol} + 1}~{\rm yr}^{-1}\\
  \nonumber R_{\rm mol} &=& 10.6~\exp \left(-r_{\rm gal} / 0.21~r_{25}\right)~,
\end{eqnarray}

\noindent which appears as a thick dashed line in the upper panels of Figure
\ref{SFEVSRAD}. The two fits reproduce the observed SFE with similar accuracy;
the scatter about each is $\approx 0.26$~dex, slightly better than a factor of
2. 

In dwarf galaxies (lower panels), we observe a steady decline in the SFE with
increasing radius for all $r_{\rm gal}$, approximately described by

\begin{equation}
  \label{DWARFSFEFIT1}
  {\rm SFE} = 1.45 \times 10^{-9}~\exp \left( -r_{\rm gal} / 0.25 ~r_{25}
  \right)~{\rm yr}^{-1}~
\end{equation}

\noindent with $\sim 0.4$~dex scatter about the fit, i.e., a factor of 2--3.

In dwarfs, we take $\Sigma_{\rm gas} \approx \Sigma_{\rm HI}$, so that SFE$=
\Sigma_{\rm SFR} / \Sigma_{\rm HI}$. For comparison with Equation
\ref{SFEFIT2}, however, we rewrite Equation \ref{DWARFSFEFIT1} {\em assuming}
that SFE~(\htwo )$= 5.25 \times 10^{-10}$~yr$^{-1}$, the value measured in
spirals.  In terms of $R_{\rm mol}$, Equation \ref{DWARFSFEFIT1} becomes

\begin{eqnarray}
  \label{DWARFSFEFIT2}
  {\rm SFE} &=& \frac{\Sigma_{\rm SFR}}{\Sigma_{\rm HI}} = 5.25 \times 10^{-10}~R_{\rm mol}~{\rm yr}^{-1}\\
  \nonumber R_{\rm mol} &=& 2.76~\exp \left(-r_{\rm gal} / 0.25~r_{25}\right)~.
\end{eqnarray}

The outer parts of dwarfs, $r_{\rm gal} \gtrsim 0.4~r_{25}$, appear similar to
the outer disks of spiral galaxies in Figure \ref{SFEVSRAD}. Surprisingly,
however, we find the SFE to be {\em higher} in the central parts of dwarf
galaxies than in the molecular gas of spirals. A higher SFE in dwarf galaxies
is quite unexpected.  Their lower metallicities, more intense radiation
fields, and weaker potential wells should make gas {\em less} efficient at
forming stars. A simple explanation for the high observed SFE is the presence
of a significant amount of \htwo .  Figure \ref{SFEVSRAD} assumes that
$\Sigma_{\rm gas} \approx \Sigma_{\rm HI}$ in dwarfs. If we miss a significant
amount of \htwo\ along a line of sight, we will overestimate the SFE because
we underestimate $\Sigma_{\rm gas}$. We quantify the possibility of
substantial \htwo\ in dwarfs in \S \ref{SECT_MISSINGH2}, but the magnitude of
the effect can be read directly from Equation \ref{DWARFSFEFIT2}. At $r_{\rm
  gal} = 0$, if dwarf galaxies have the same SFE~(\htwo ) as spirals, $R_{\rm
  mol} \approx 2.76$, i.e., $\Sigma_{\rm H2} \approx 2.76~\Sigma_{\rm HI}$.

\subsubsection{SFE and Stellar Surface Density}
\label{SECT_SFESTARS}

\begin{figure}
 \plotone{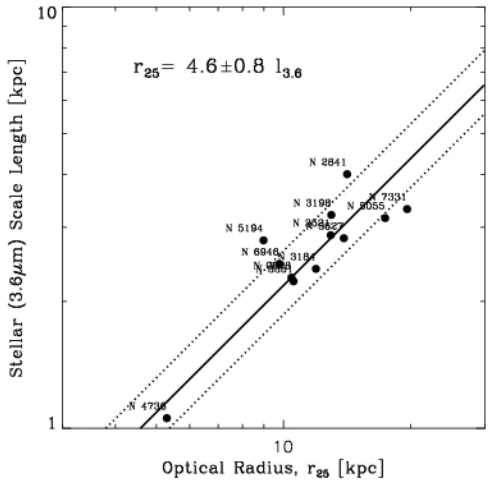}
 \caption{\label{SLVSR25} Stellar scale length, $l_*$, as a function of
   isophotal radius, $r_{25}$. Solid and dashed lines show $r_{25} = (4.6 \pm
   0.8)~l_*$.}
\end{figure}

Galactocentric radius is probably not intrinsically important to a local
process like star formation, but Figure \ref{SFEVSRAD} suggests that local
conditions covariant with radius have a large effect on the ability of gas to
form stars. The radius, $r_{25}$, that we use to normalize the $x$-axis is
defined by an optical isophote and thus measures stellar light. Therefore
$r_{25}$ is closely linked to the stellar distribution.

Figure \ref{SLVSR25} shows this link directly. We plot stellar scale length,
$l_*$, measured via an exponential fit to the $3.6\mu$m profile as a function
of $r_{25}$ for our spiral subsample. We see that $r_{25} = (4.6 \pm 0.8)~l_*$
and that we could have equivalently normalized the $x$-axis in Figure
\ref{SFEVSRAD} by $l_*$. We may suspect, then, that the stellar surface
density, $\Sigma_*$, underlies the well-defined relation between SFE and
$r_{\rm gal}$ observed in Figure \ref{SFEVSRAD}.

\begin{figure*}
  \begin{center}
    \plottwo{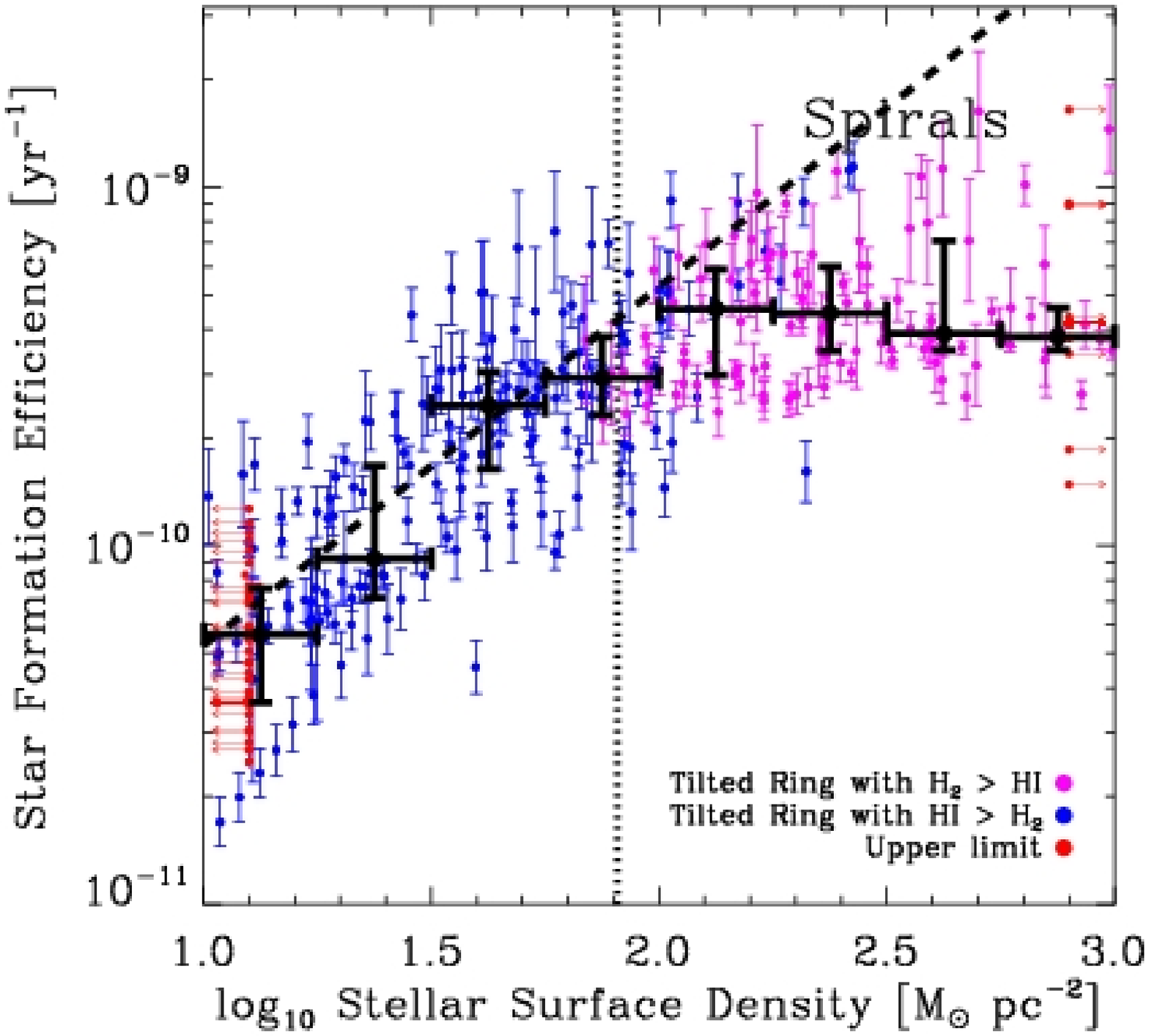}{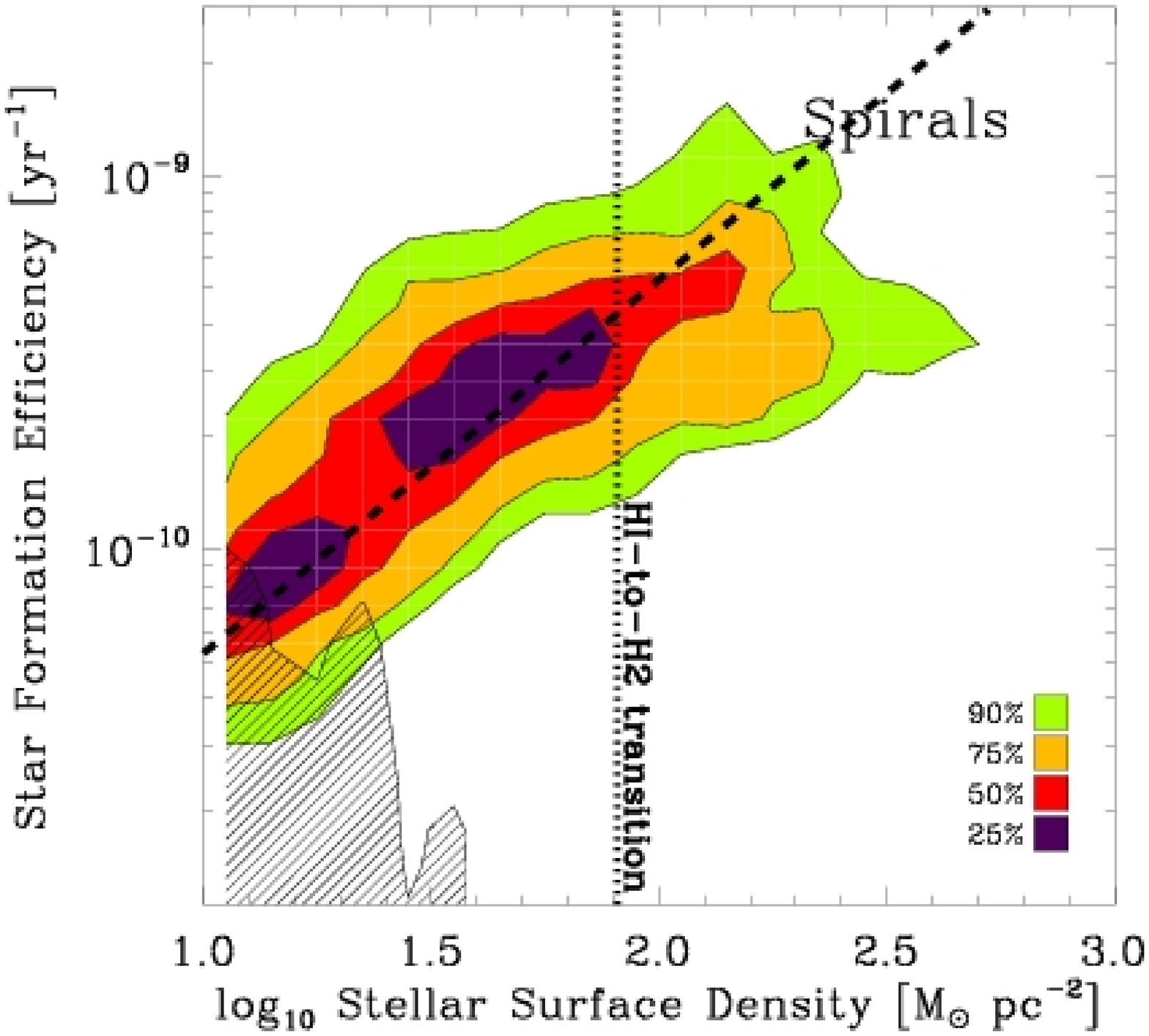}
    \plottwo{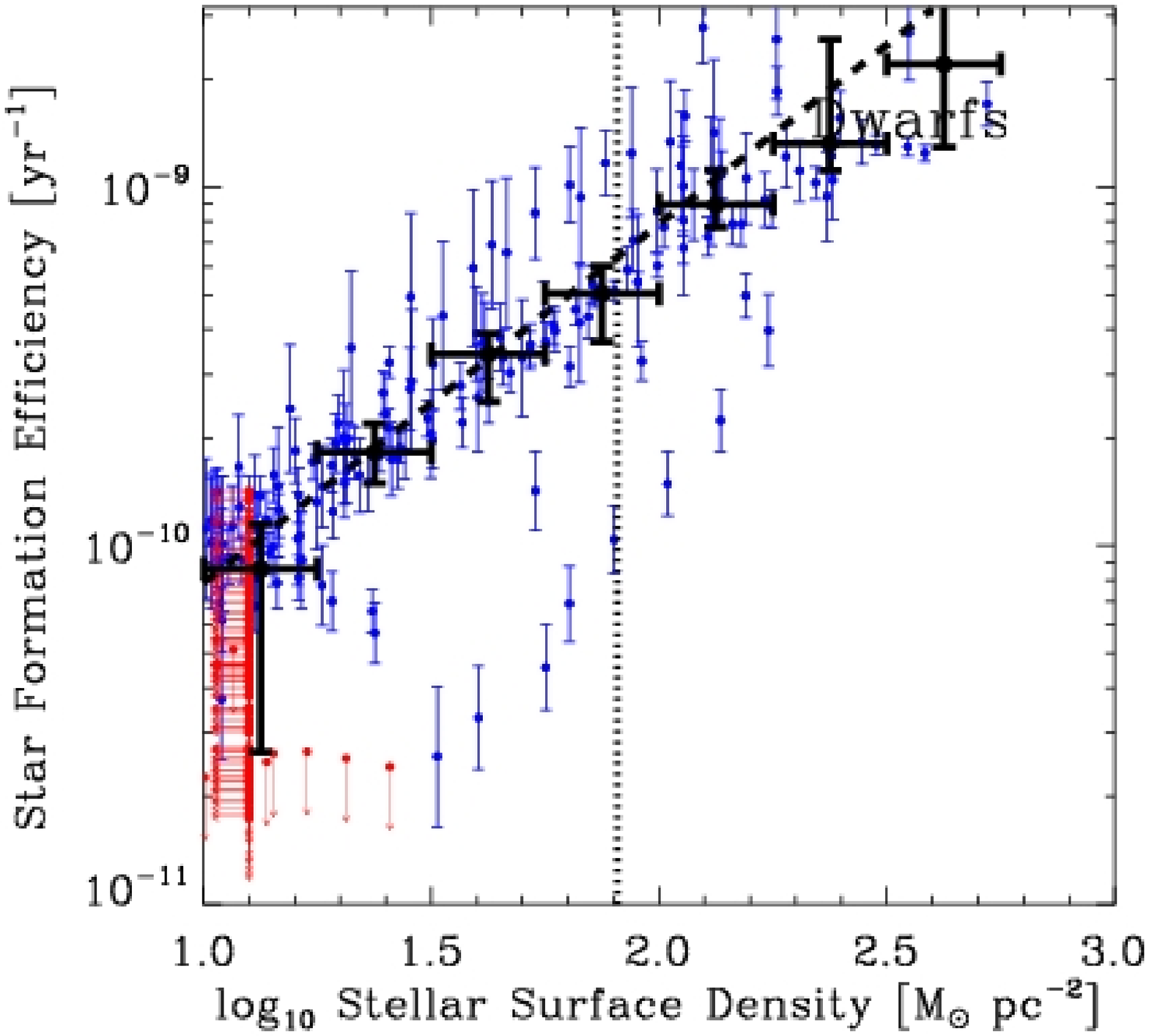}{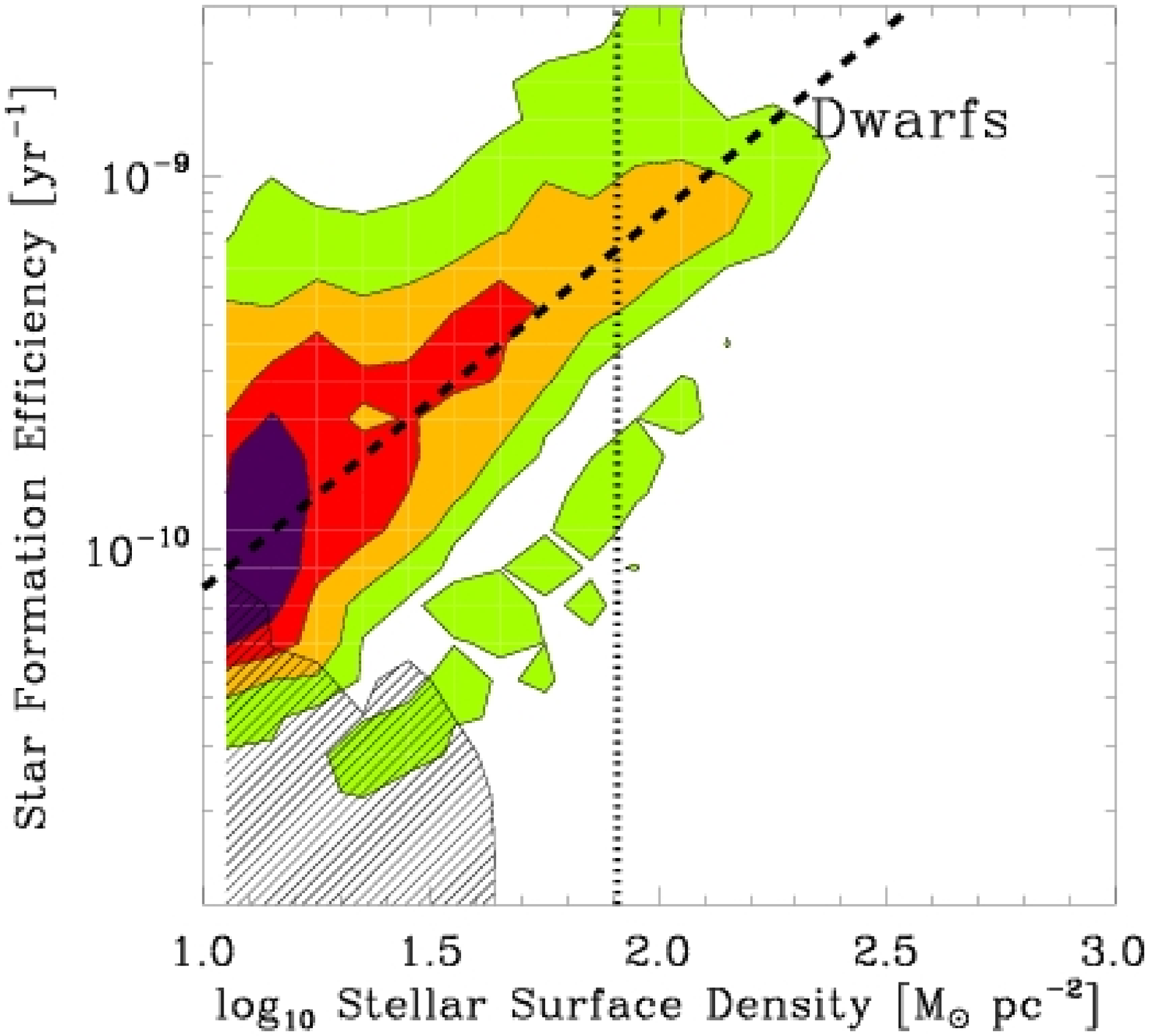}
  \end{center}
  \caption{\label{SFEVSSTARS} SFE as a function of stellar surface density,
    $\Sigma_*$, in spiral (top row) and dwarf (bottom row) galaxies.
    Conventions and symbols are as in Figure \ref{SFEVSRAD}. Dashed diagonal
    lines show the linear relationship between SFE and $\Sigma_*$ expected for
    the mean stellar assembly time and $\Sigma_{\rm gas}$ for each subsample.
    Vertical dotted lines show $\Sigma_*$ where the ISM is equal parts \hi\
    and \htwo\ in spirals (\S \ref{TRANS_SECT}), $\Sigma_* = 81 \pm
    25$~M$_\odot$~pc$^{-2}$.}
\end{figure*}

In Figure \ref{SFEVSSTARS}, we explore this connection by plotting SFE as a
function of $\Sigma_*$. In both spiral and dwarf galaxies, we see a nearly
linear relationship between SFE and $\Sigma_*$ where the ISM is \hi -dominated
(blue points).

A basic result of THINGS is that over the optical disk of most star forming
galaxies, the \hi\ surface density varies remarkably little \citep[Appendices
\ref{RPROFAPP} and \ref{ATLASAPP} and][]{WALTER08}. Inspecting our atlas, one
sees that $\Sigma_{\rm HI} \approx 6$~M$_{\odot}$~pc$^{-2}$ (within a factor
of 2) over a huge range of local conditions, including most of the optical
disk in most galaxies.  Because $\Sigma_{\rm gas}$ is nearly constant in the
\hi -dominated (blue) regime, ${\rm SFE}\propto\Sigma_*$ approximately defines
a line of fixed specific star formation rate (SSFR), i.e., star formation rate
per unit stellar mass.

The inverse of the SSFR is the stellar assembly time, $\tau_* = \Sigma_* /
\Sigma_{SFR}$. This is the time required for the present star formation rate
to build up the observed stellar disk.  In our spiral subsample, the mean
$\log_{10} \tau_* \approx 10.5 \pm 0.3$, i.e., $3.2 \times 10^{10}$ years or
slightly more than 2 Hubble times. Dwarf galaxies have shorter assembly times,
$\log_{10} \tau_{*}\approx 10.2 \pm 0.3$ years, about a Hubble time (dashed
lines in Figure \ref{SFEVSSTARS} show these values using average values of
$\Sigma_{\rm gas}$ for each subsample). Taking these numbers at face value,
dwarfs are forming stars at about their time-average rate, while spirals are
presently forming stars at just under half of their average rate.

We {\em only} observe SFE~$\propto\Sigma_*$ where the ISM is mostly \hi .
Where the ISM is mostly \htwo\ in spirals galaxies, we observe a constant SFE
at a range of $\Sigma_*$; similar to the constancy as a function of $r_{\rm
  gal}$ observed in the inner parts of spirals (Figure \ref{SFEVSRAD}). The
transition between these two regimes occurs at $\Sigma_* = 81 \pm
25$~M$_{\odot}$~pc$^{-2}$ (\S \ref{TRANS_SECT}) in spirals. In dwarfs, lines
of sight with $\Sigma_*$ above this transition value exhibit systematically
high SFE, lending further, albeit indirect, support to the idea that these
points correspond to unmeasured \htwo .

\begin{figure}
  \plotone{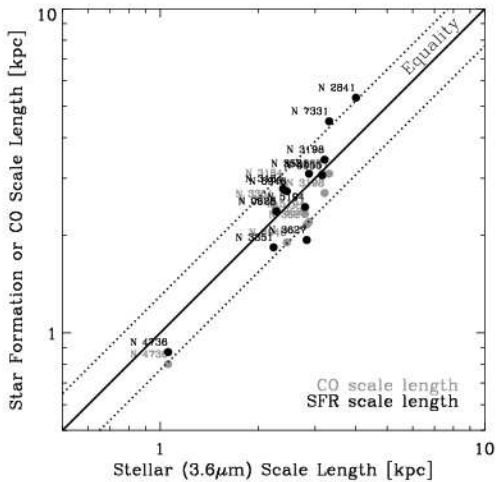}
  \caption{\label{SLVSSL} The scale lengths of star formation (black) and CO
    (gray) as a function of the stellar scale length ($x$-axis). All three
    scale lengths are similar, the dashed lines show slope unity and $\pm
    30\%$ (the approximate scatter in the data).}
\end{figure}

Figures \ref{SFEVSRAD} and \ref{SFEVSSTARS} show that where the ISM is mostly
\htwo , the star formation rate per unit {\em gas} (SFE) is nearly constant
and that where the ISM is mostly \hi , the star formation rate per unit {\em
  stellar mass} (SSFR) is nearly constant.  Together these observations
suggest that \htwo , stars, and star formation have similar structure with all
three embedded in a relatively flat distribution of \hi. Figure \ref{SLVSSL}
shows that the scale lengths of these three distributions are, in fact,
comparable.  The star formation rate (black) and CO (gray) scale lengths of
spiral galaxies are both roughly equal to the stellar scale length:

\begin{equation}
  l_{\rm CO} = (0.9 \pm 0.2)~l_* {\rm~and~} l_{\rm SFR} = (1 \pm 0.2)~l_*~.
\end{equation}

\noindent \citet{REGAN01} also found that $l_{\rm CO} \approx l_*$
comparing $K$-band maps to BIMA SONG and \citet{YOUNG95} found $l_{\rm
  CO} \approx 0.2~r_{25}$, which is almost identical to our $l_{\rm
  CO} \sim 0.9~l_*$ and $(4.6 \pm 0.8)~l_* = r_{25}$.

\subsubsection{SFE and Gas Surface Density}
\label{GAS_SECT}

\begin{figure*}
\begin{center}
\plottwo{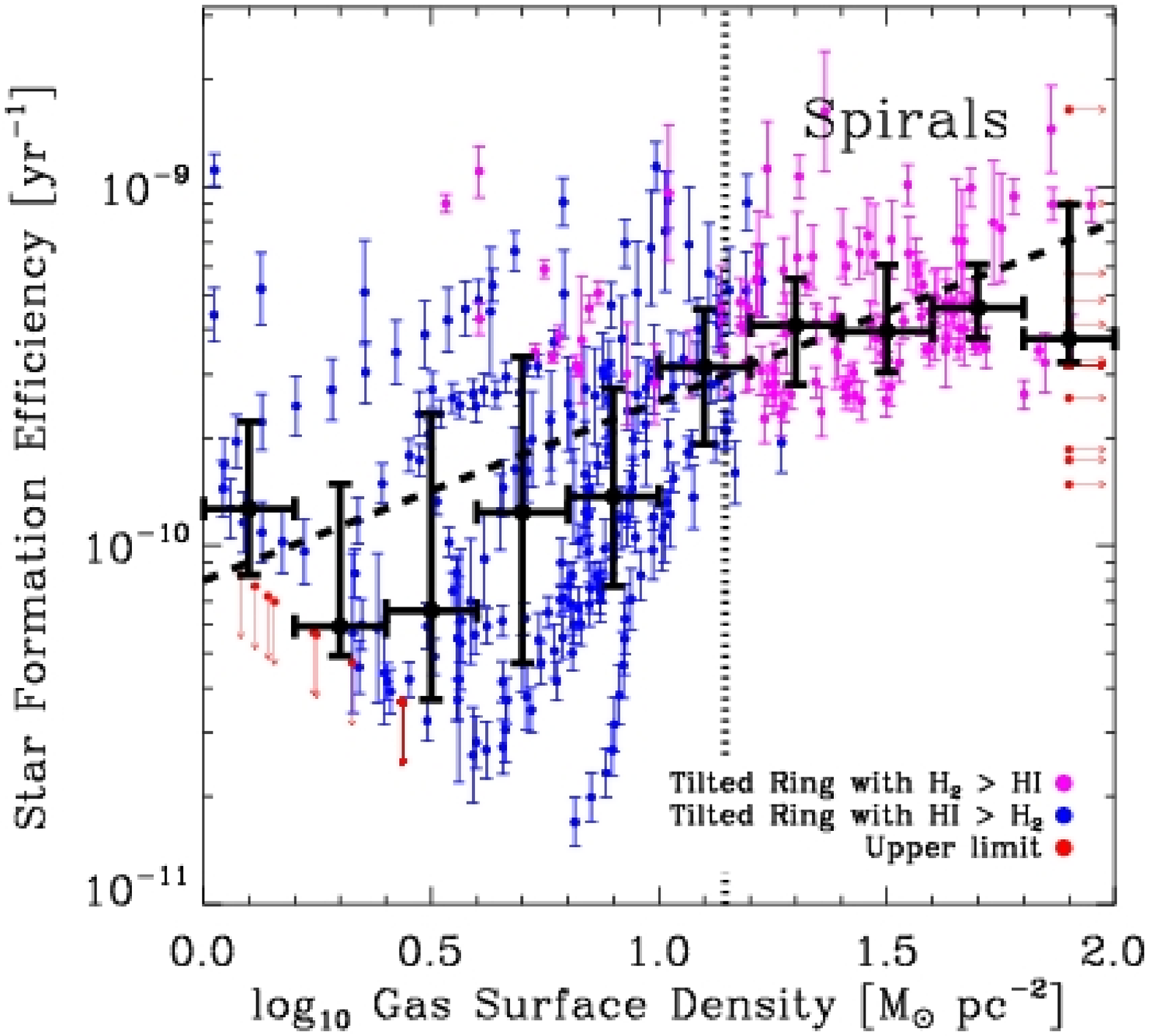}{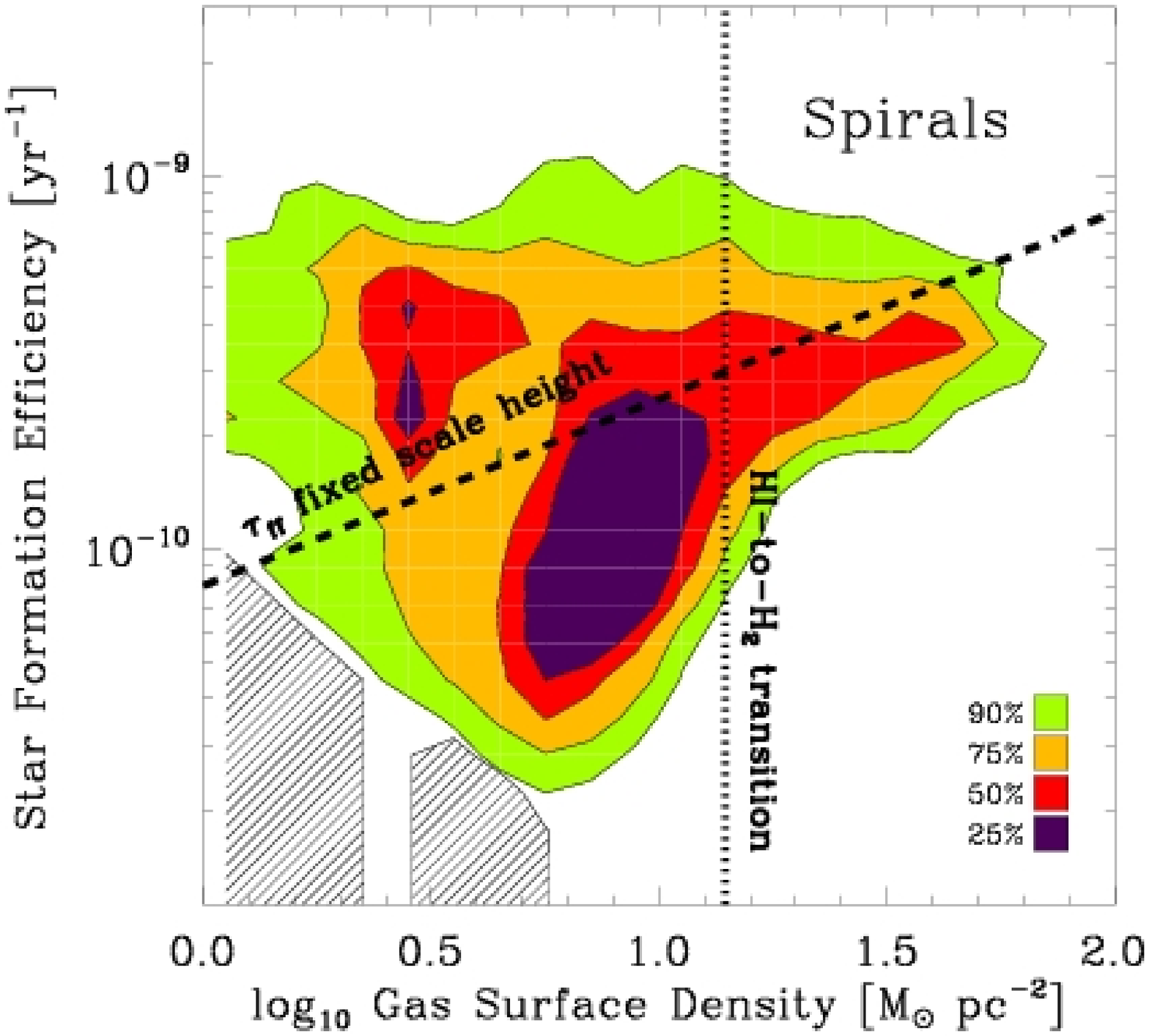}
\plottwo{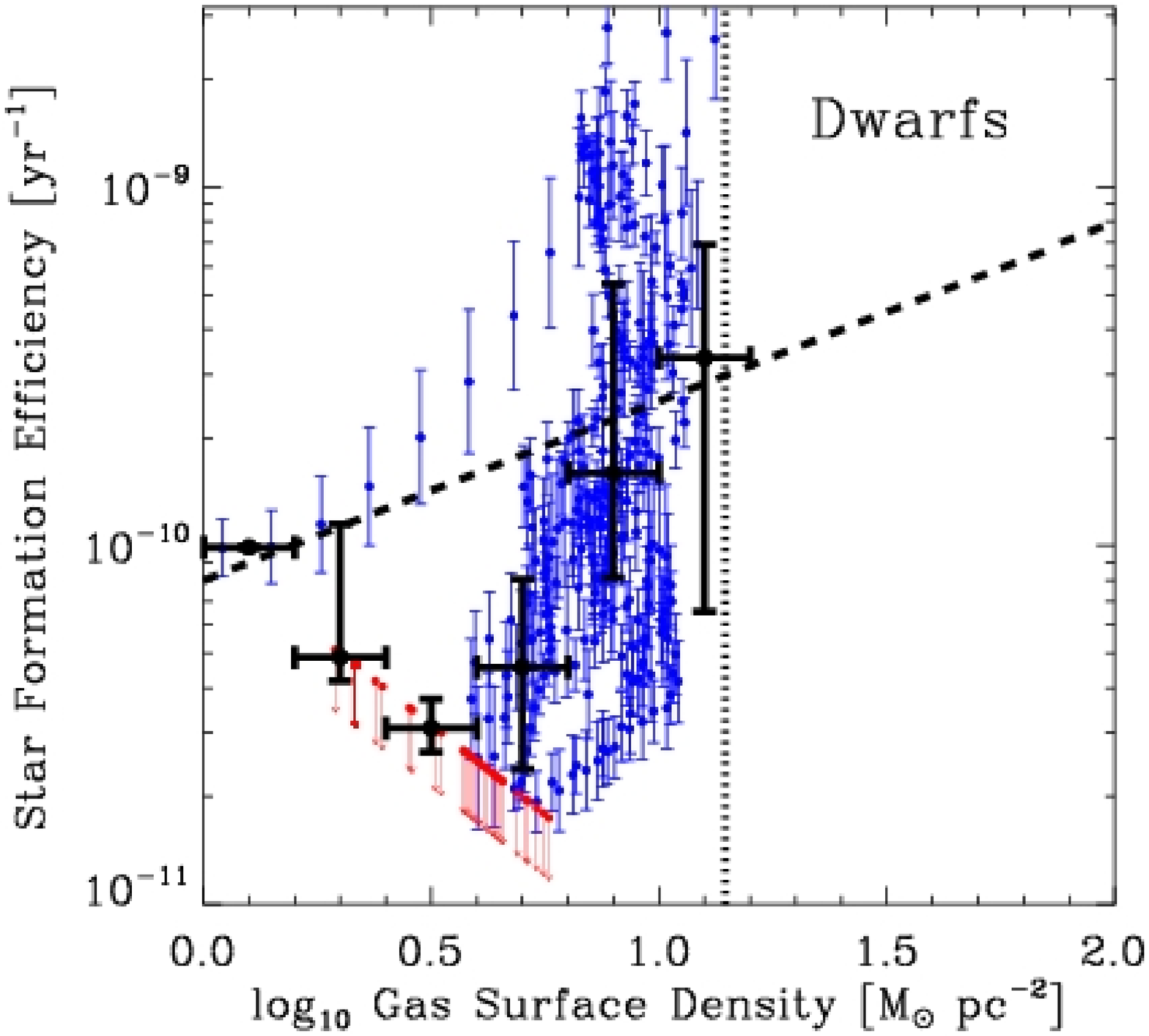}{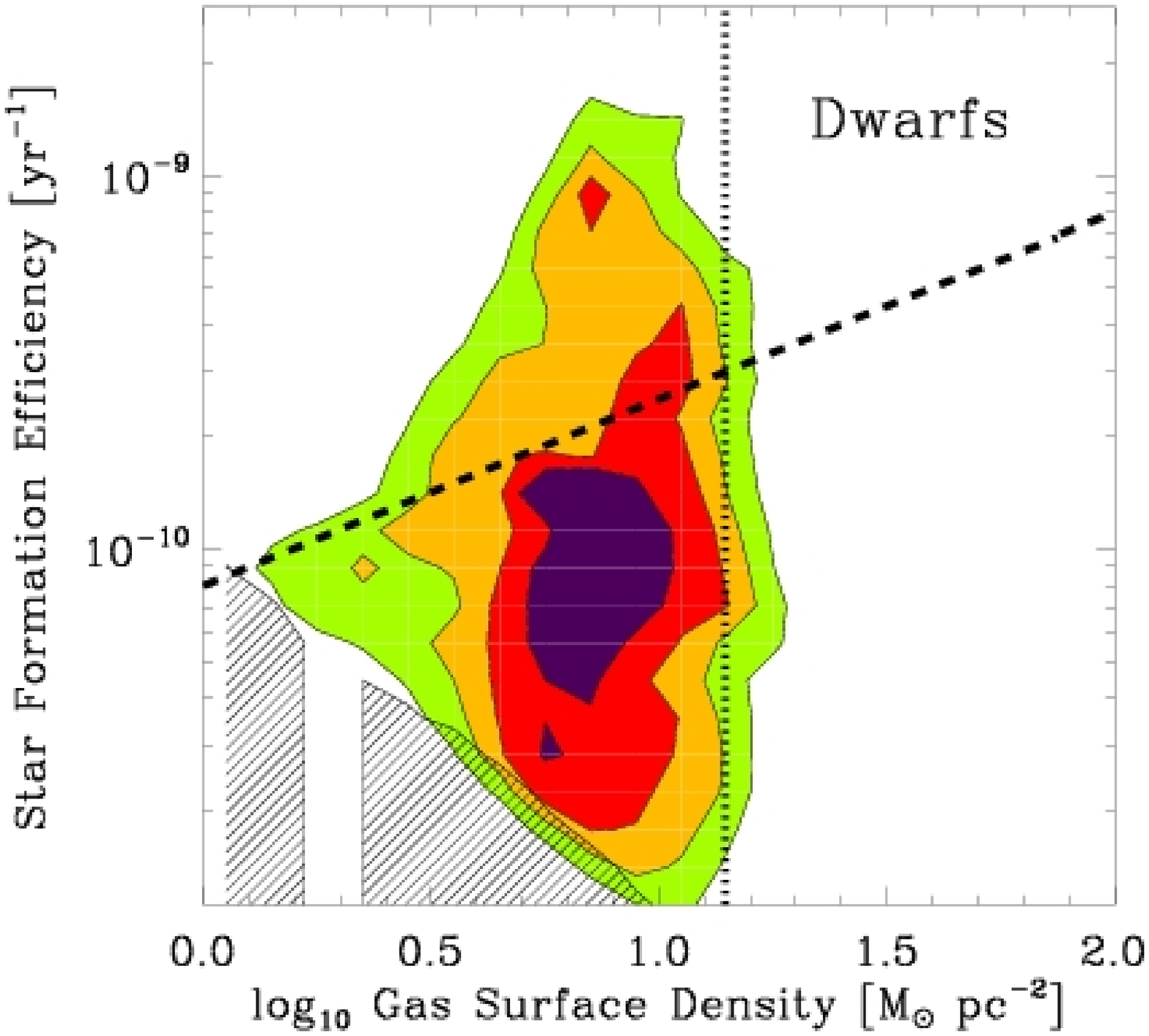}
\end{center}
\caption{\label{SFEVSGAS} SFE as a function of $\Sigma_{\rm gas}$ in
  spiral (top row) and dwarf (bottom row) galaxies. Conventions and
  symbols are the same as in Figure \ref{SFEVSRAD}. The vertical
  dotted line shows $\Sigma_{\rm gas}$ at the \hi -to-\htwo\
  transition in spirals (\S \ref{TRANS_SECT}), $\Sigma_{\rm gas} =
  14 \pm 6$~M$_{\odot}$~pc$^{-2}$.  The dashed line shows the SFE
  proportional to the free--fall time in a fixed scale height
  disk. Clearly the line cannot describe both high and low SFE data,
  even if the normalization is adjusted, and so changes in this
  timescale cannot drive the radial decline that we observe in the
  SFE.}
\end{figure*}

This link between $\Sigma_*$ and the SFE is somewhat surprising because it is
common to view $\Sigma_{\rm SFR}$, and thus the SFE, as set largely by
$\Sigma_{\rm gas}$ alone over much of the disk of a galaxy \citep[following,
e.g.,][]{KENNICUTT98A}. In Figure \ref{SFEVSGAS} we show this last slice
through SFR-stars-gas parameter space, plotting SFE as a function of
$\Sigma_{\rm gas}$.

As in Figures \ref{SFEVSRAD} and \ref{SFEVSSTARS}, we observe two distinct
regimes. In spirals, where $\Sigma_{\rm gas} > 14 \pm 6$~M$_\odot$~pc$^{-2}$
(\S \ref{TRANS_SECT}) the ISM is mostly \htwo\ and we observe a fixed SFE.
This $\Sigma_{\rm gas}$, shown by a vertical dotted line, corresponds
approximately to both $N(H) \sim 10^{21}$~cm$^{-2}$ star formation threshold
noted by \citet{SKILLMAN87} and the saturation value for \hi\ observed by,
e.g., \citet[][]{MARTIN01} and \citet{WONG02} \citep[and seen strikingly in
THINGS at $\Sigma_{\rm gas} = 12$~M$_{\odot}$~pc$^{-2}$ by][who quote
$\Sigma_{\rm gas} = 9$~M$_{\odot}$~pc$^{-2}$ but do not include
helium]{BIGIEL08}.

In contrast to $r_{\rm gal}$ and $\Sigma_*$, $\Sigma_{\rm gas}$ does not
exhibit a clear correlation with the SFE where the ISM is mostly \hi .
Instead, over the narrow range $\Sigma_{\rm gas} \approx
5$--$10$~M$_\odot$~pc$^{-2}$, the SFE varies from $\sim 3 \times 10^{-11}$ to
$10^{-9}$ yr$^{-1}$. We see little evidence that $\Sigma_{\rm HI}$ plays a
central role regulating the SFE in either spirals or dwarfs.  Rather, the most
striking observation in Figure \ref{SFEVSGAS} is that $\Sigma_{\rm HI}$
exhibits a narrow range of values over the optical disk and is therefore
itself likely subject to some kind of regulation.

The possibility of a missed reservoir of molecular gas in dwarfs is again
evident from the lower panels in Figure \ref{SFEVSGAS}. A subset of data has
SFE higher than that observed for \htwo\ in spirals and just to the left of
the \hi\ saturation value. If H$_2$ were added to these points, they would
move down (as the SFE decreases) and to the right (as $\Sigma_{\rm gas}$
increases), potentially yielding a data distribution similar to that we
observe in spirals.

\subsection{SFE and Star Formation Laws}
\label{SECT_SFELAW}

We now ask whether the star formation laws proposed in \S \ref{SECT_LAWBACK}
can explain the radial decline in SFE and whether SFE$~(\htwo )$, already
observed to be constant as a function of $r_{\rm gal}$, $\Sigma_*$, and
$\Sigma_{\rm gas}$ (but with some scatter), exhibits {\em any} kind of
systematic behavior. We compare the SFE to four quantities that drive the
predictions in Table~\ref{IDEATAB}: gas surface density (already seen in
Figure \ref{SFEVSGAS}), gas pressure (density), the orbital timescale, and the
derivative of the rotation curve, $\beta$.

\subsubsection{Free--Fall Time in a Fixed Scale Height Disk}
\label{FIXEDSH_SECT}

A dashed line in Figure \ref{SFEVSGAS} illustrates SFE~$\propto
\Sigma_{\rm gas}^{0.5}$, expected if the SFE is proportional to the
free-fall time in a fixed scale height gas disk \citep[similar to the
Kennicutt--Schmidt law,][]{KENNICUTT98A}. The normalization matches
the \htwo --dominated parts of spirals and roughly bisects the range
of SFE observed for dwarfs, but large areas of the disk have much
lower SFE than one would predict from this relation. Adjusting the
normalization can move the line up or down but cannot reproduce the
distribution of data observed in Figure \ref{SFEVSGAS}.

The culprit here is the small dynamic range in $\Sigma_{\rm
  HI}$. Because $\Sigma_{\rm HI}$ does not vary much across the disk,
while the SFE does, the free--fall time in a fixed scale height disk,
or any other weak dependence of SFE on $\Sigma_{\rm gas}$ alone,
cannot reproduce variations in the SFE where the ISM is mostly \hi .
A quantity other than $\Sigma_{\rm gas}$ must play an important role
at radii as low as $\sim 0.5~r_{25}$ \citep[a fact already recognized
by][among others]{KENNICUTT89}.

\subsubsection{Free--Fall Time in a Variable Scale Height Gas Disk; Pressure
  and ISM Phase}
\label{DENSITY_SECT}

\begin{figure*}
\begin{center}
\plottwo{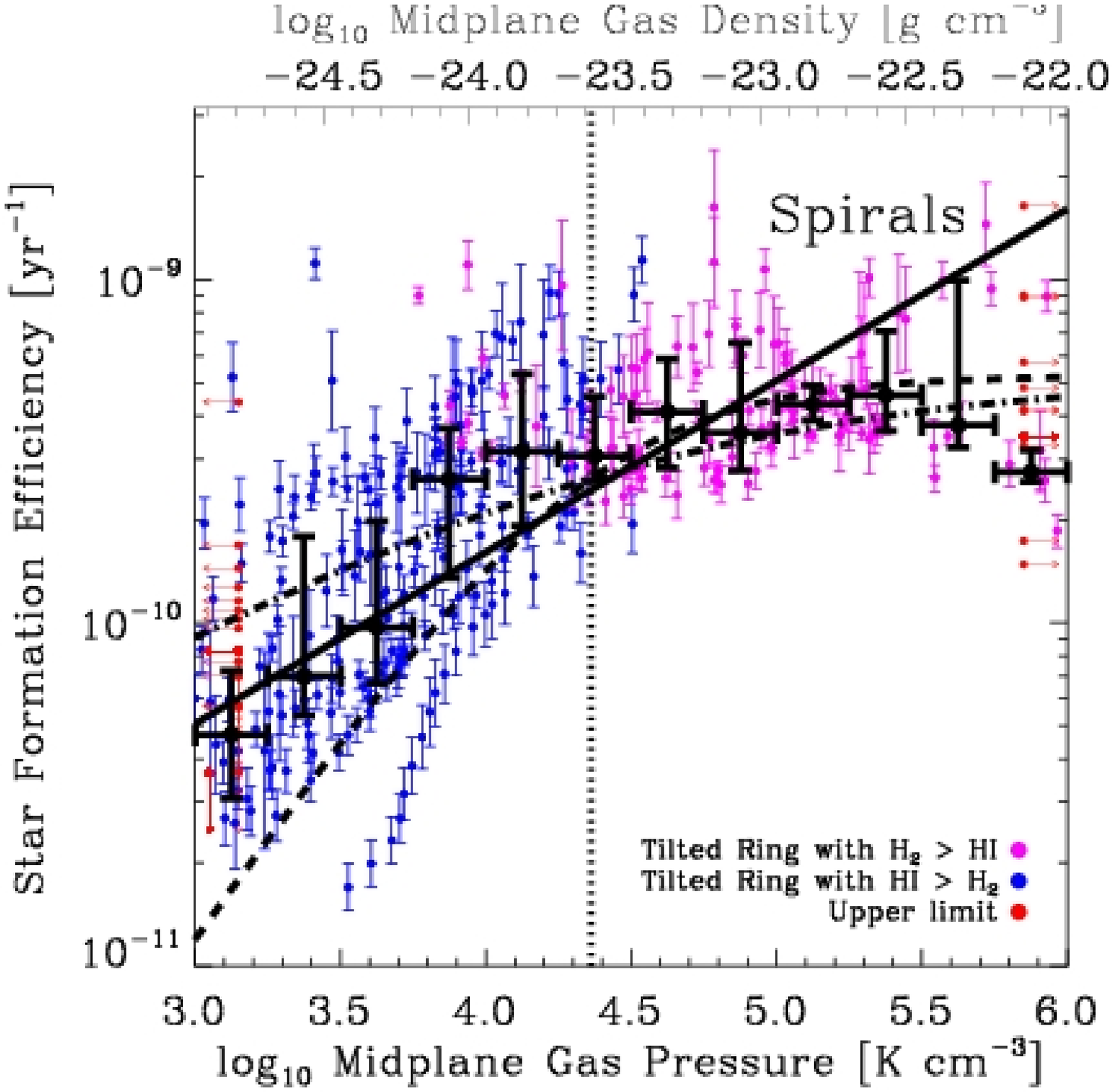}{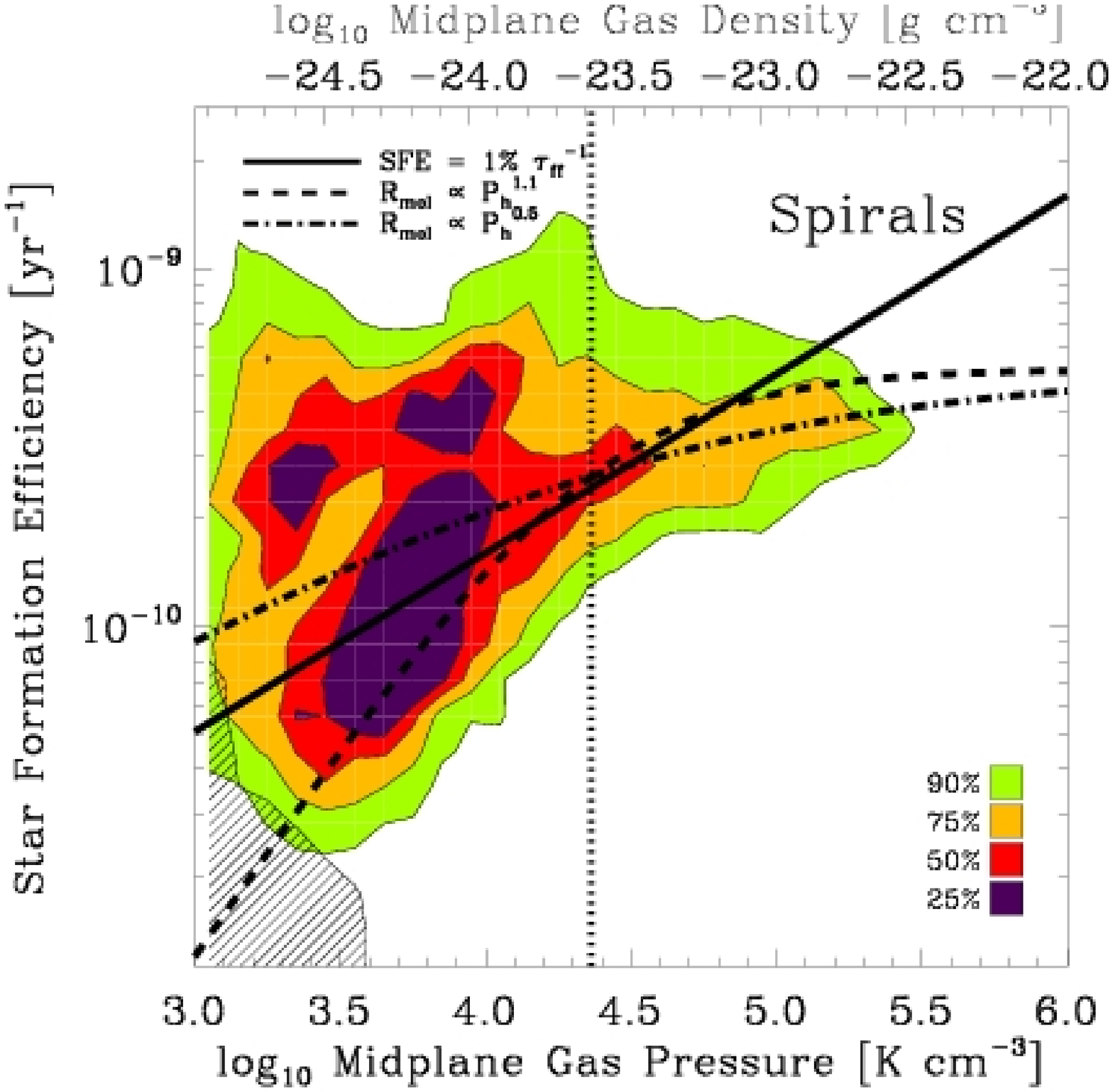}
\vspace{0.2in}
\plottwo{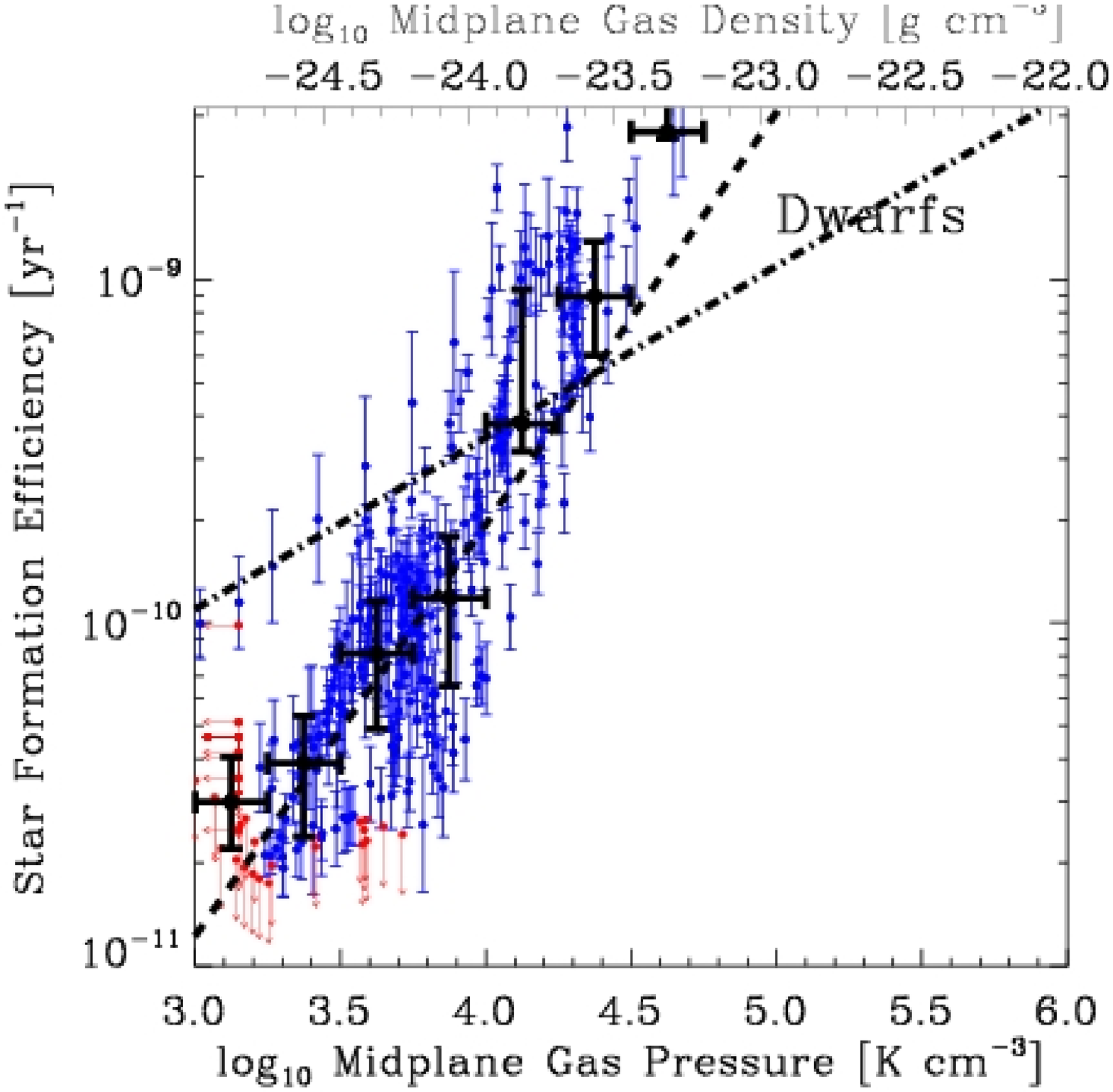}{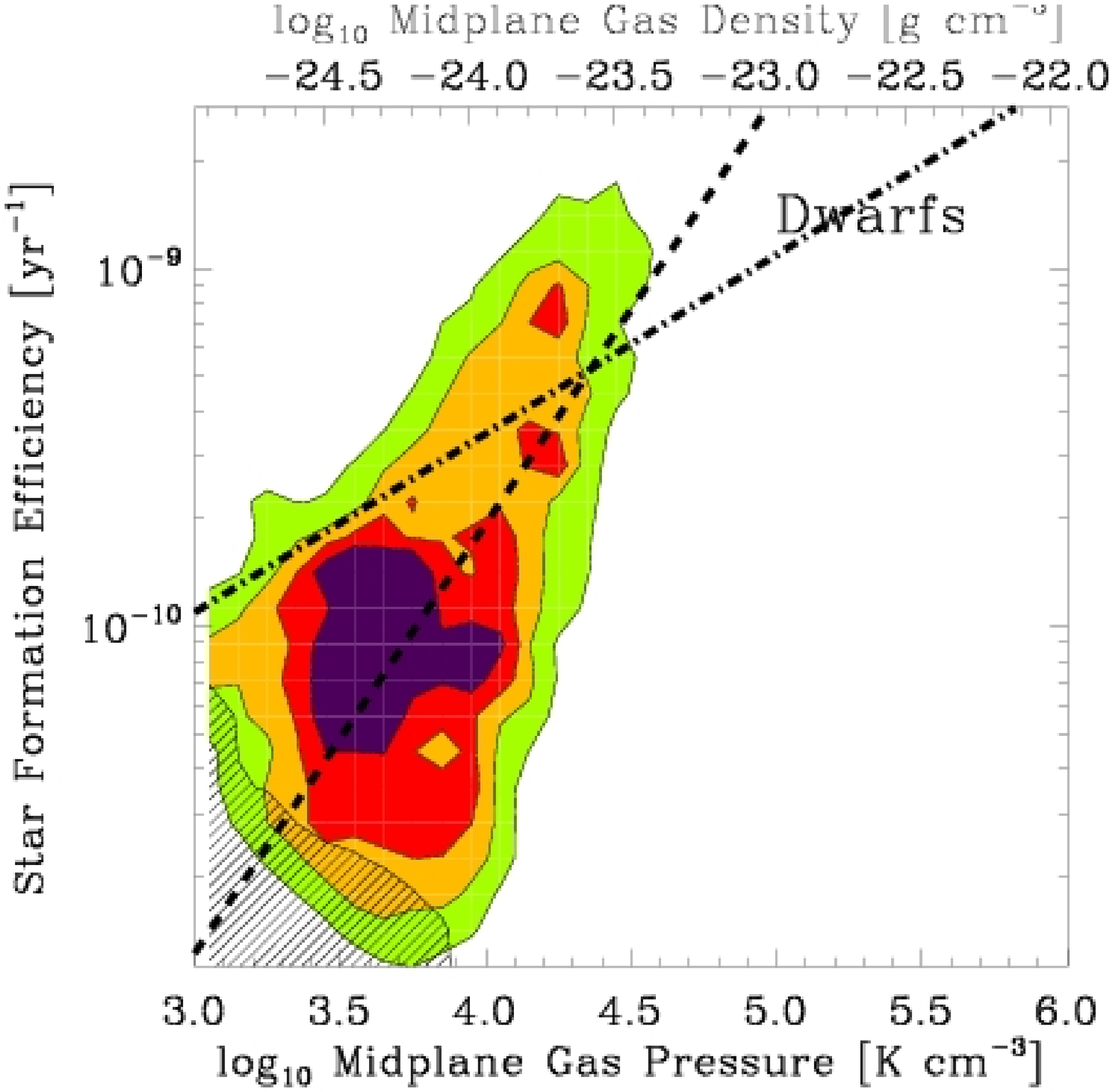}
\end{center}
\caption{\label{SFEVSPRESS} SFE as a function of midplane hydrostatic gas
  pressure, $P_{\rm h}$ (bottom $x$-axis) and equivalent volume density (top
  $x$-axis) in spiral (top row) and dwarf (bottom row) galaxies.  Conventions
  follow Figure \ref{SFEVSRAD}. The vertical line shows $P_{\rm h}$ at the \hi
  -to-\htwo\ transition in spirals, $\log_{10} P_{\rm h} / k_{\rm B}~[{\rm
    K~cm}^{-3}] \approx 4.36$. The solid line illustrates 1\% of gas converted
  to stars per disk free--fall time. Dash-dotted and dashed lines show $R_{\rm
    mol} \propto P_{\rm h}^{0.5}$ ($\tau_{\rm ff}^{-1}$) and $R_{\rm mol}
  \propto P_{\rm h}^{1.2}$ \citep{ELMEGREEN93}.  For our adopted $\sigma_{\rm
    gas} = 11$~km~s$^{-1}$ and including helium: $\rho \left[{\rm
      g~cm^{-3}}\right] = 1.14 \times 10^{-28}~(P_{\rm h} / k_{\rm B})
  \left[{\rm K~cm}^{-3}\right]$ and $n\left[{\rm cm}^{-3}\right] = 4.4 \times
  10^{23}~\rho \left[ {\rm g ~cm}^{-3} \right]$.}
\end{figure*}

We saw in \S \ref{SECT_SFEOBS} that where the ISM is mostly \hi , the SFE
correlates better with $\Sigma_*$ than with $\Sigma_{\rm gas}$. This might be
expected if the stellar potential well plays a central role in setting the
{\em volume} density of the gas, $\rho_{\rm gas}$, because $\Sigma_*$ varies
much more strongly with radius than $\Sigma_{\rm HI}$. In \S
\ref{SECT_BACKGROUND} we present two predictions relating SFE to $\rho_{\rm
  gas}$: that the timescale over which GMCs form depends on the $\tau_{\rm
  ff}$, the free--fall time in a gas disk with a scale height set by
hydrostatic equilibrium\footnote{Hereafter $\tau_{\rm ff}$ refers only to the
  free fall time in a gas disk with a scale height set by hydrostatic
  equilibrium.}, and that the ratio $R_{\rm mol} = \Sigma_{\rm H2}/\Sigma_{\rm
  HI}$ depends primarily on midplane gas pressure, $P_{\rm h}$.

Under our assumption of a fixed $\sigma_{\rm gas}$, $P_{\rm h} \propto
\rho_{\rm gas}$ and both predictions can be written as a power law
relating SFE or $R_{\rm mol}$ to $P_{\rm h}$. In Figure
\ref{SFEVSPRESS} we plot SFE as a function of $\rho_{\rm gas}$ and
$P_{\rm h}$ (top and bottom $x$-axis), estimated from hydrostatic
equilibrium (Equation \ref{PRESSEQ}).

Where the ISM is mostly \htwo\ (magenta points) in spirals (top row), we
observe no clear relationship between $P_{\rm h}$ and SFE, further evidence
that SFE~(\htwo ) is largely decoupled from global conditions of the ISM in
our data.

Where the ISM is mostly \hi\ (blue points) in dwarf galaxies and the outer
parts of spirals, the SFE correlates with $P_{\rm h}$. $P_{\rm h}$ predicts
the SFE notably better than $\Sigma_{\rm gas}$ in this regime, supporting the
idea that the volume density of gas (at least \hi ) is more relevant to star
formation than surface density. \citet{WONG02} and \citet{BLITZ06} observed a
continuous relationship between $R_{\rm mol}$ and $P_{\rm h}$, mostly where
$\Sigma_{\rm H2} \gtrsim \Sigma_{\rm HI}$. Figure \ref{SFEVSPRESS} suggests
that such a relationship extends well into the regime where \hi\ dominates the
ISM.

The solid line in Figure \ref{SFEVSPRESS} illustrates the case of 1\% of the
gas formed into stars per $\tau_{\rm ff}$ (SFE~$\propto \rho_{\rm
  gas}^{0.5}$), a typical value at the \hi --to--\htwo\ transition in spirals
(\S \ref{TRANS_SECT}). Adjusting the normalization slightly, such a line can
intersect both the high and low end of the observed SFE in spirals, but
predicts variations in SFE~(\htwo ) that we do not observe and is too shallow
to describe dwarf galaxies.

The dash--dotted line shows $R_{\rm mol} \propto \tau_{\rm ff}^{-1} \propto
P_{\rm h}^{0.5}$, expected for GMC formation over a free fall time. In dwarf
galaxies, where we take $\Sigma_{\rm gas} = \Sigma_{\rm HI}$, this is
equivalent too SFE$\propto \tau_{\rm ff}^{-1}$. This description can describe
spirals at high and intermediate $P_{\rm h}$, but is too shallow to capture
the drop in SFE at large radii in spirals and across dwarf galaxies. If
$\tau_{\rm ff}$ is the characteristic timescale for GMC formation, effects
other than just an increasing timescale must suppress cloud formation in these
regimes.

A dashed line shows the steeper dependence, $R_{\rm mol} \propto P_{\rm
  h}^{1.2}$, expected for low $R_{\rm mol}$ based on modeling by
\citet{ELMEGREEN93}. This may be a reasonable description of both spiral and
dwarf galaxies (note that at high SFE, $P_{\rm h}$ may be underestimated in
dwarf galaxies because we fail to account for \htwo ). We explore how $P_{\rm
  h}$ relates to $R_{\rm mol}$ more in \S \ref{SECT_DISCUSSION}.

\subsubsection{Orbital Timescale}
\label{KIN_SECT}

 \begin{figure*}
  \begin{center}
    \plottwo{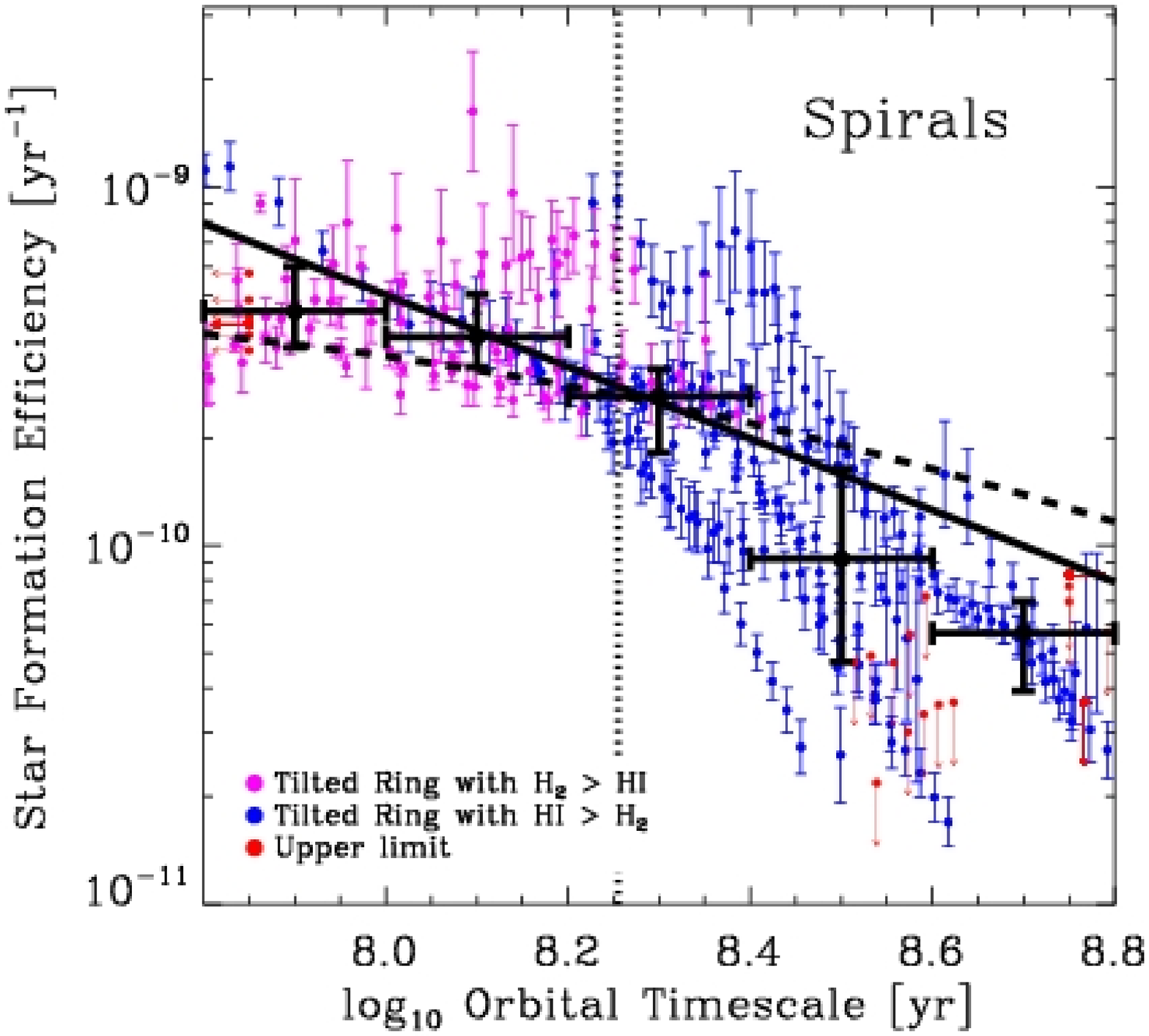}{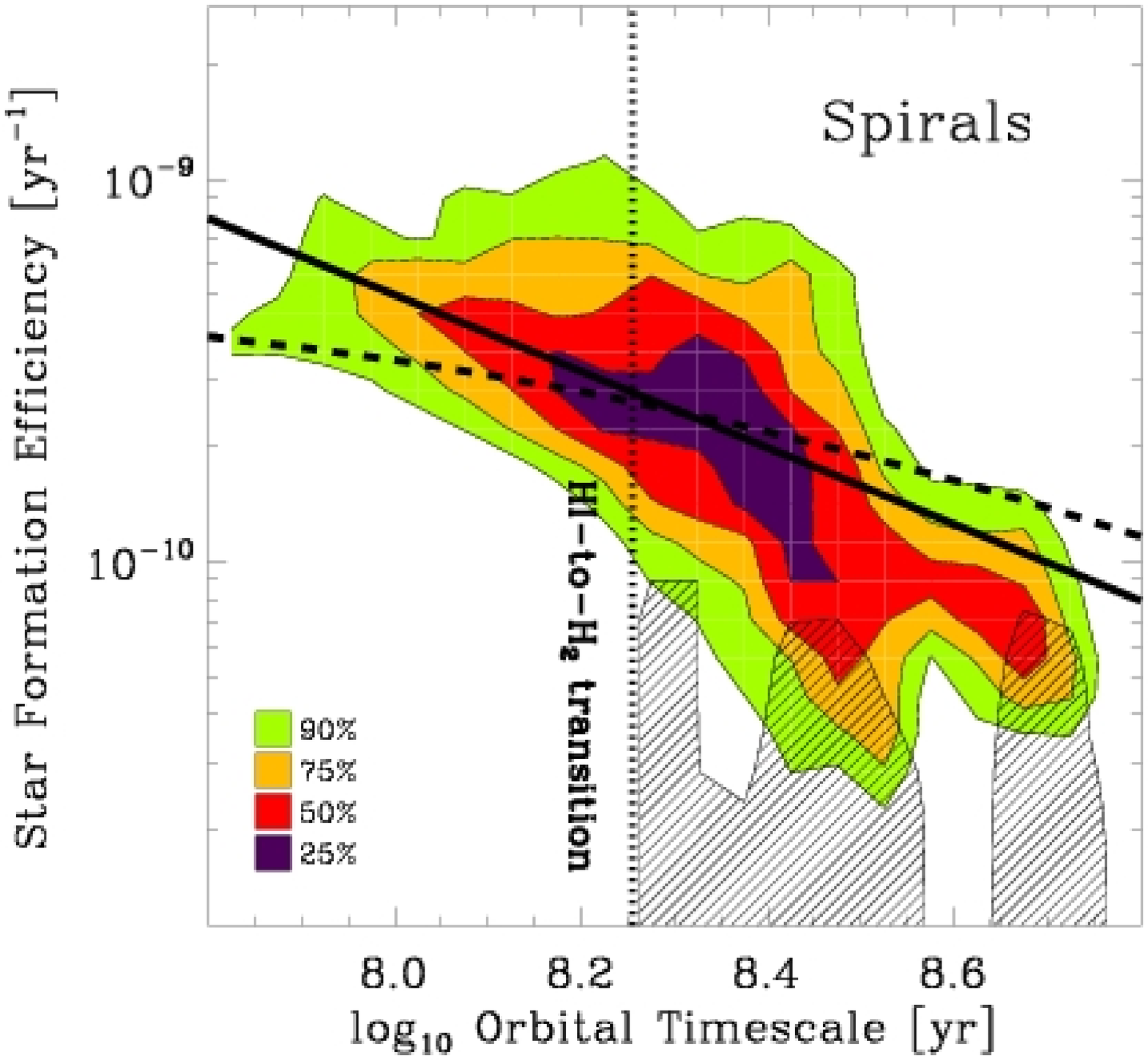}
    \plottwo{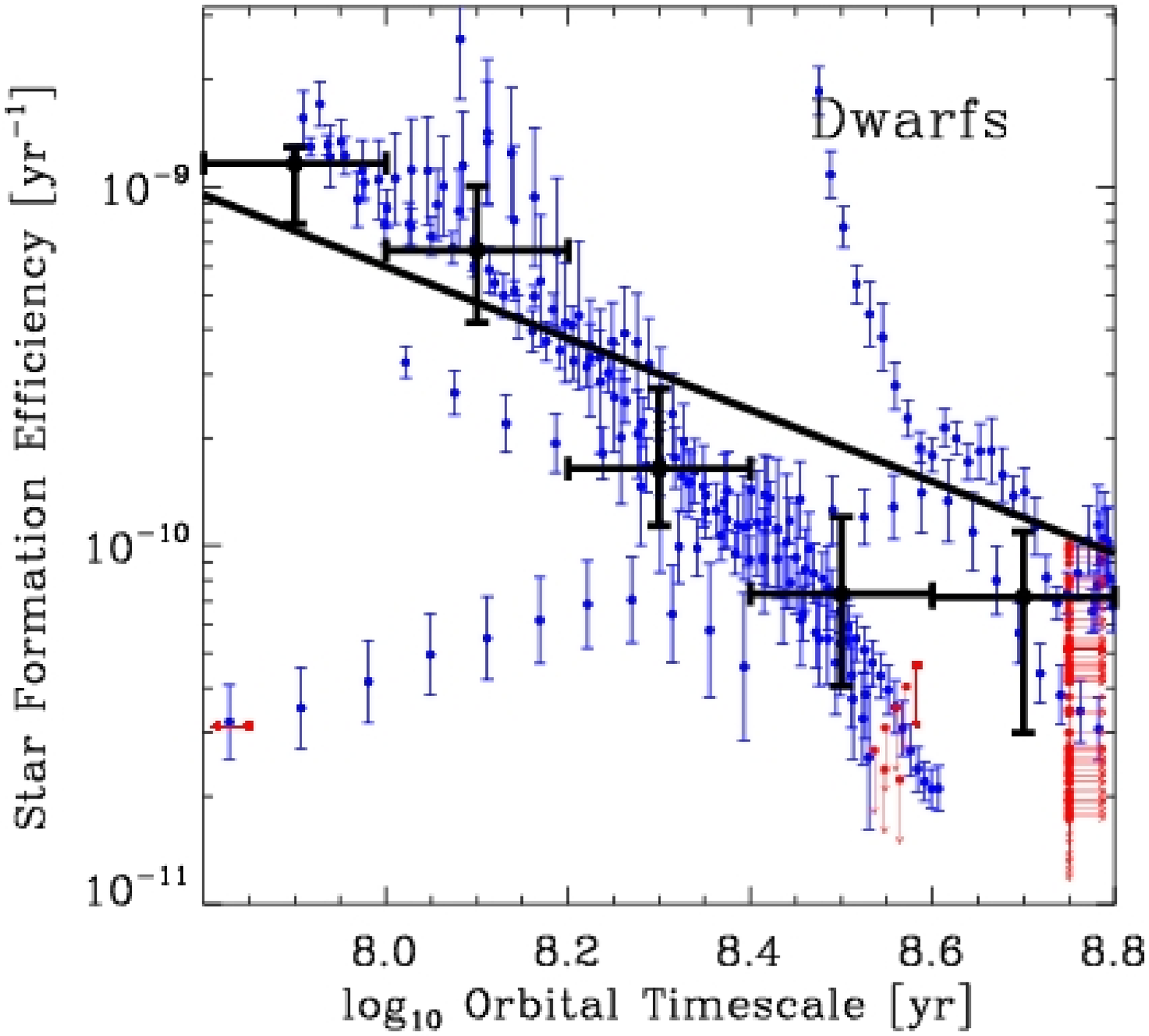}{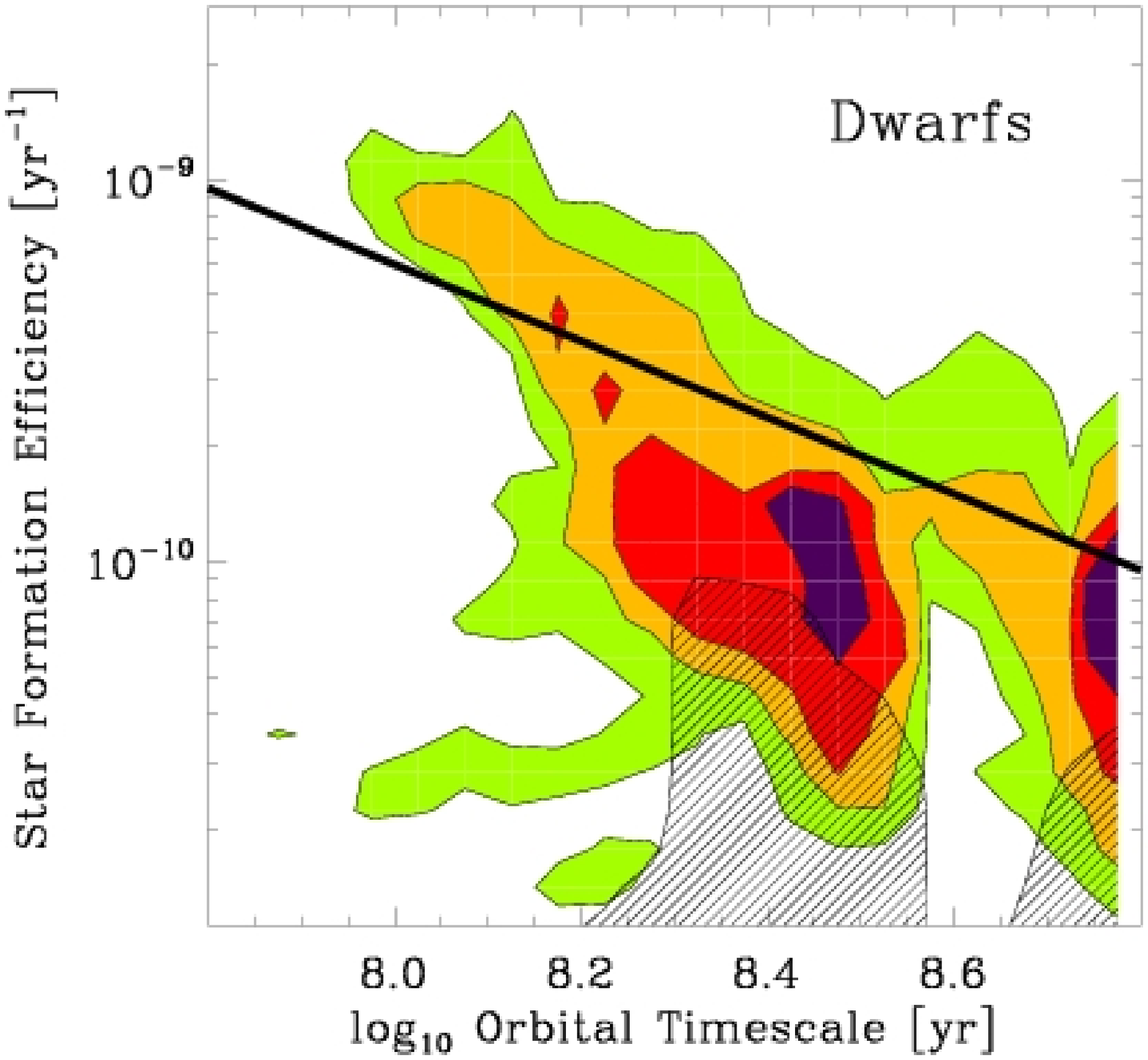}
  \end{center}
   \caption{\label{SFEVSTORB} SFE as a function of the orbital timescale,
     $\tau_{\rm orb}$, in spiral (top row) and dwarf (bottom row) galaxies,
     following the conventions from Figure \ref{SFEVSRAD}. The solid line
     shows $6\%$ of gas converted into stars per $\tau_{\rm orb}$. The dashed
     line shows the expected SFE if $R_{\rm mol} = \Sigma_{\rm H2} /
     \Sigma_{\rm HI} \propto \tau_{\rm orb}^{-1}$. The SFE is a well-defined
     function of $\tau_{\rm orb}$, but the decline in $\tau_{\rm orb}$ alone
     cannot reproduce the radial decline in SFE or $R_{\rm mol}$.}
\end{figure*}

The orbital timescale, $\tau_{\rm orb}$, varies strongly with radius and
\citet{KENNICUTT98A} found $\tau_{\rm orb}$ to be a good predictor of
disk--averaged SFE. In Figure \ref{SFEVSTORB}, we plot SFE as a function of
$\tau_{\rm orb}$ in our sample.

The solid line shows 6\% of the gas converted to stars per $\tau_{\rm orb}$
and is a reasonable match to spirals near the \hi --to--\htwo\ transition
(vertical dotted line). This value agrees with the range of efficiencies found
by \citet{WONG02} and with \citet{KENNICUTT98A}, who found $\approx 7\%$ of
gas converted to stars per $\tau_{\rm orb}$ averaged over galaxy disks
(converted to our adopted IMF). Like \citet{WONG02}, we do not observe a clear
correlation between SFE and $\tau_{\rm orb}$ where the ISM is mostly \htwo .

Where the ISM is mostly \hi\ (blue points), the SFE clearly anti-correlates
with $\tau_{\rm orb}$ in both spiral and dwarf galaxies. However, we do not
observe a constant efficiency per $\tau_{\rm orb}$. In both subsamples, SFE
drops faster than $\tau_{\rm orb}$ increases, so that data at large radii
(longer $\tau_{\rm orb}$, lower SFE) show lower efficiency per $\tau_{\rm
  orb}$ than those from inner galaxies. Although $\tau_{\rm orb}$ correlates
with the SFE, the drop in $\tau_{\rm orb}$ is not enough on its own to explain
the drop in SFE.

We reach the same conclusion if we posit that $\tau_{\rm orb}$ is the relevant
timescale for GMC formation, so that $R_{\rm mol} \propto \tau_{\rm
  orb}^{-1}$. The dashed lines in Figure \ref{SFEVSTORB} show this relation
combined with a fixed SFE~(\htwo ) and normalized to $R_{\rm mol} = 1$ at
$\tau_{\rm orb} = (1.8 \pm 0.4) \times 10^8$~years, which we observe at the
\hi --to--\htwo\ transition in spirals (\S \ref{TRANS_SECT}). This dependence
is even shallower than SFE $\propto \tau_{\rm orb}^{-1}$ and cannot reproduce
the SFE in both inner and outer disks by itself. If $\tau_{\rm orb}$ is the
relevant timescale for cloud formation, then the fraction of gas that is
actively forming stars must vary substantially between the middle and the edge
of the optical disk.

\subsubsection{Derivative of the Rotation Curve, $\beta$}

\begin{figure*}
  \begin{center}
  \plottwo{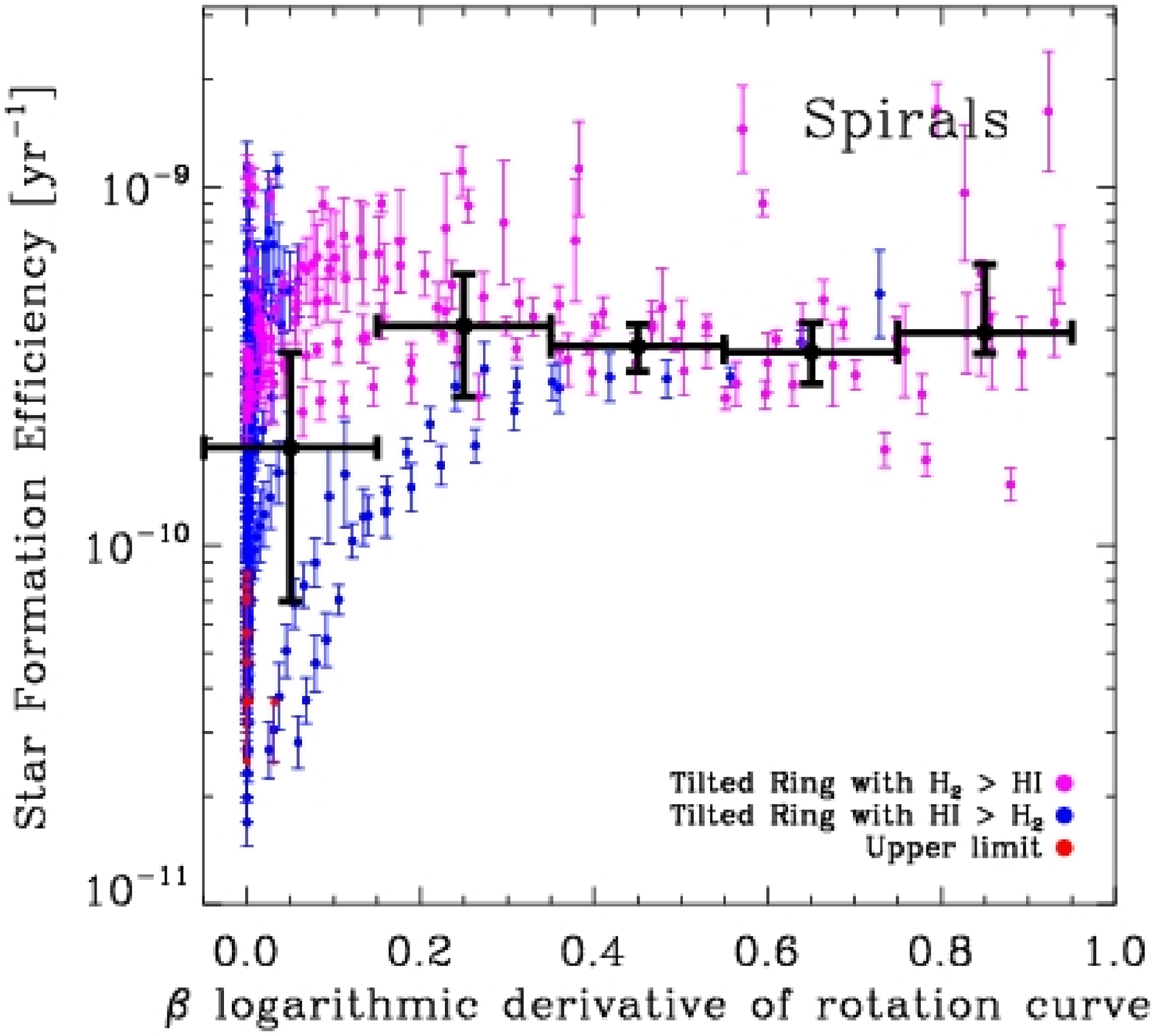}{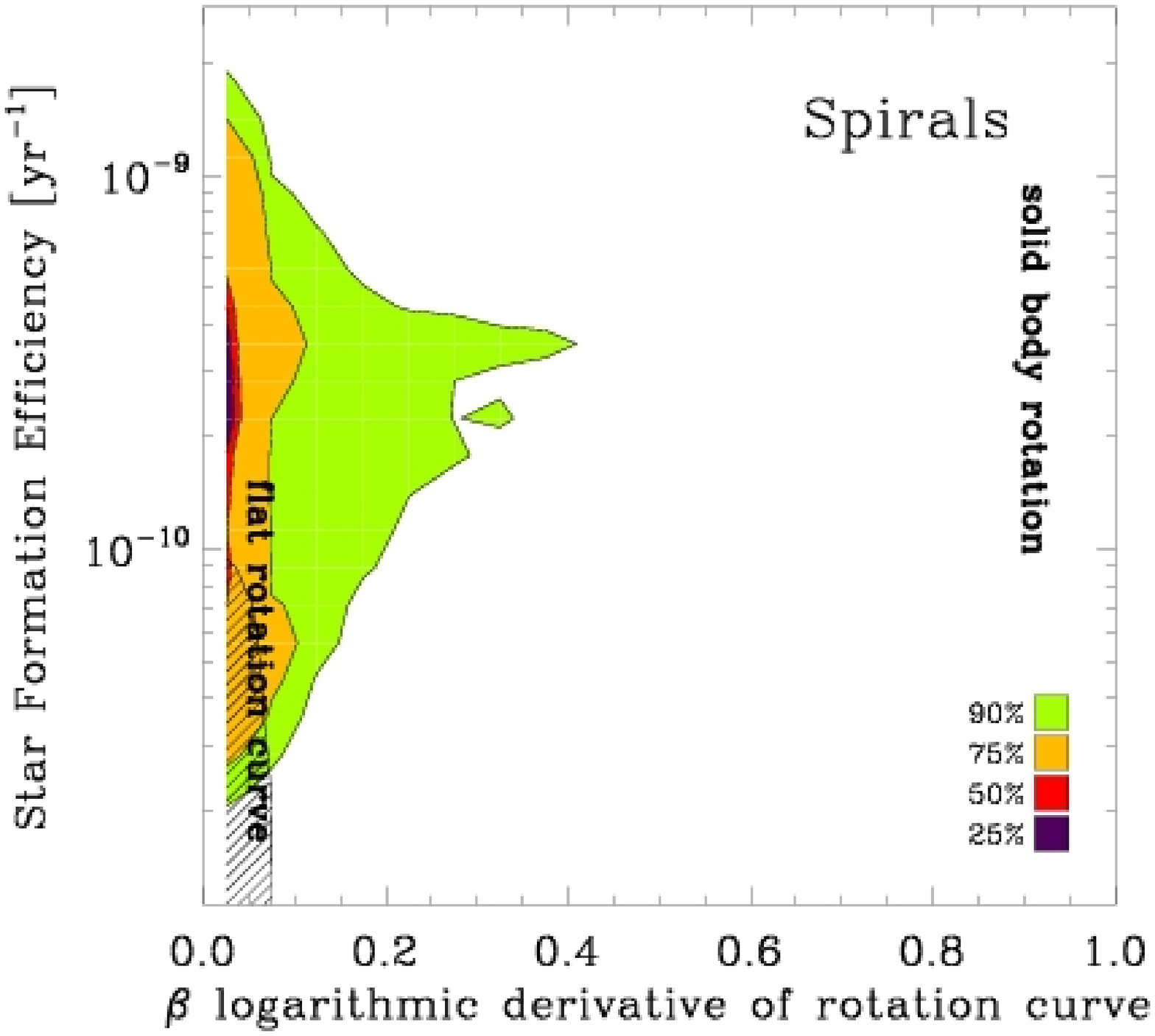}
  \plottwo{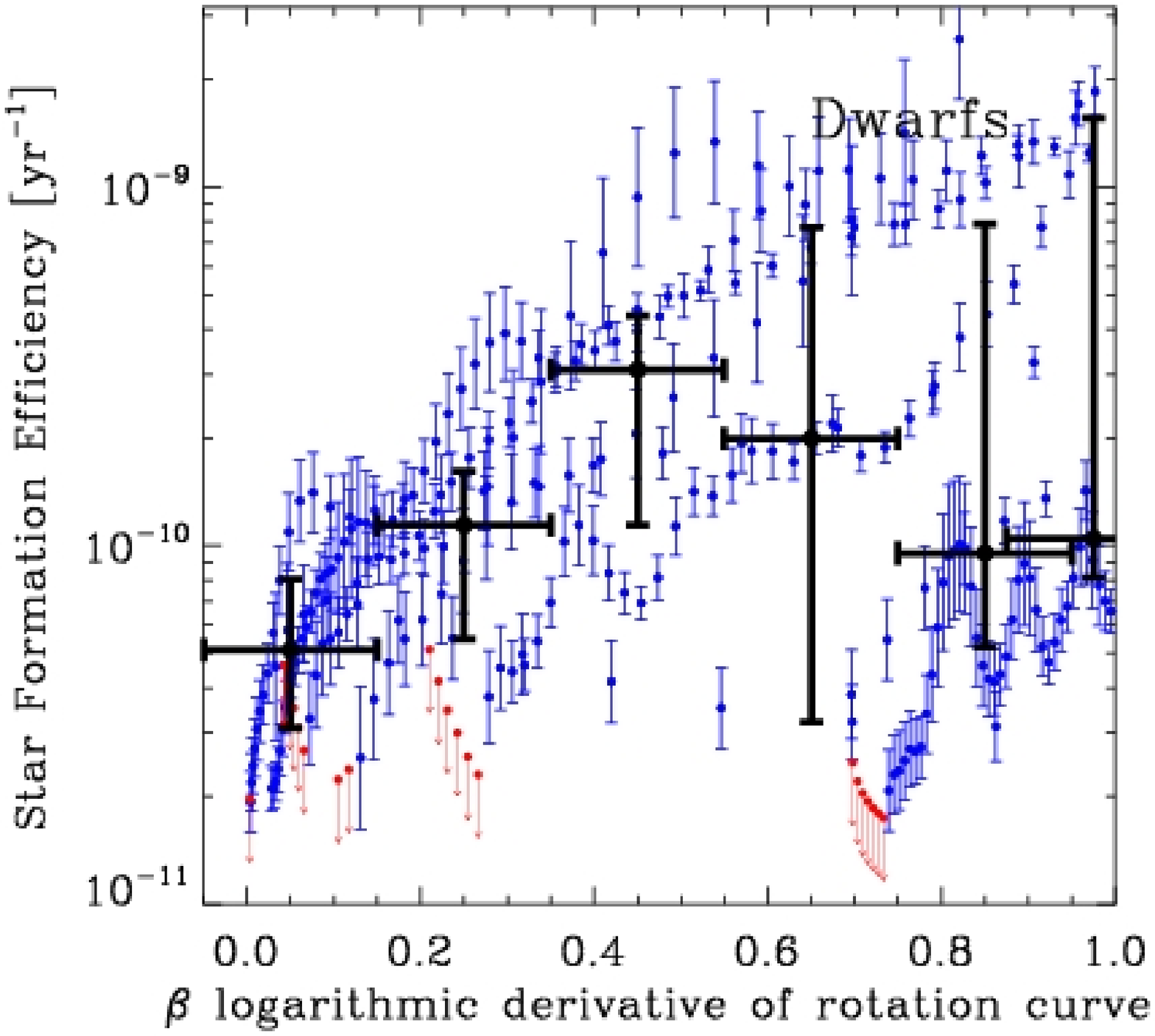}{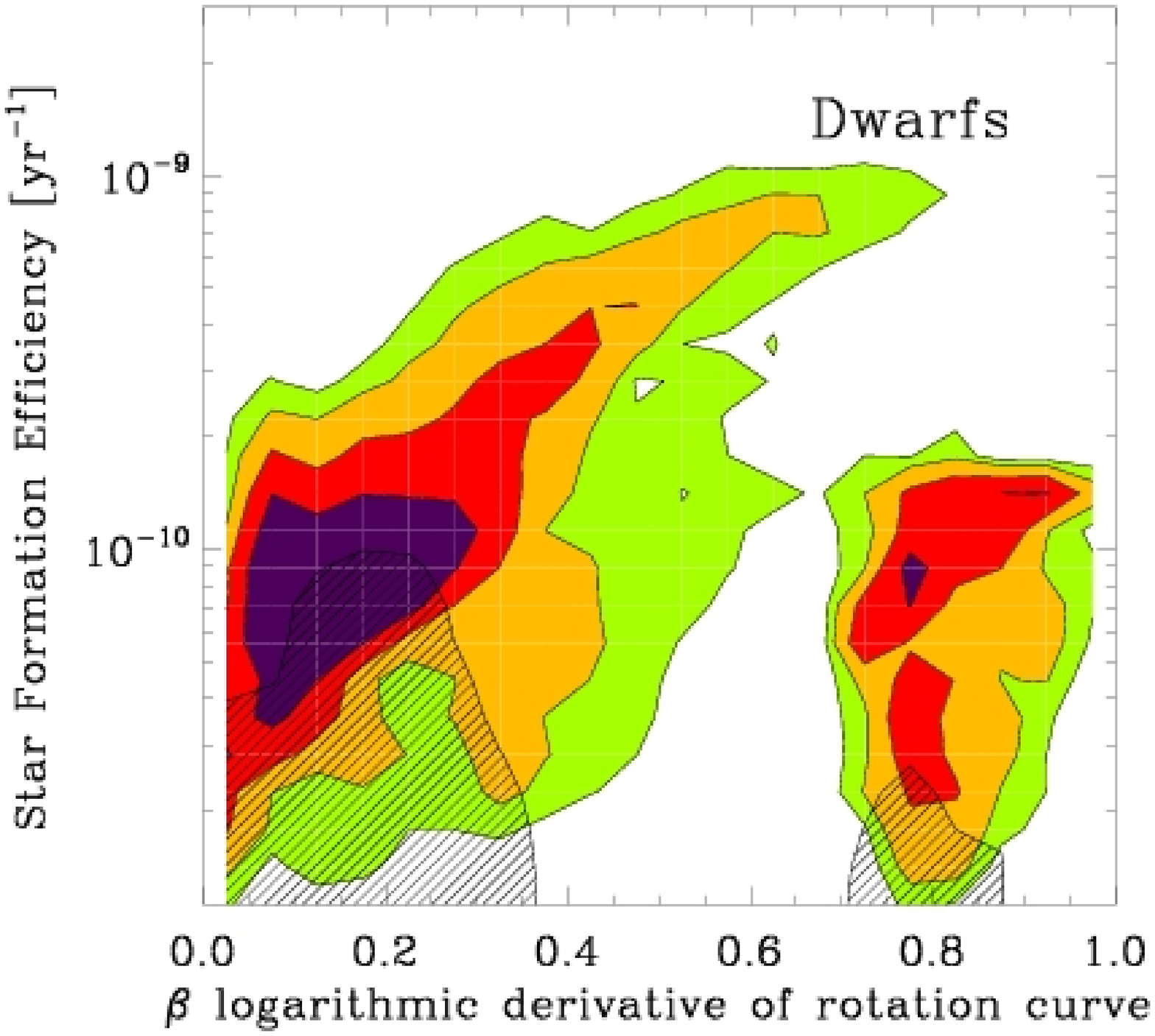}
  \end{center}
  \caption{\label{SFEVSBETA} SFE as a function of $\beta$, the logarithmic
    derivative of the rotation curve in spiral (top row) and dwarf (bottom
    row) galaxies. $\beta = 1$ for solid body rotation and $\beta = 0$ for a
    flat rotation curve. If collisions between GMCs were important to
    triggering star formation, we would expect the SFE in the \htwo\ dominated
    (magenta) parts of spirals to be higher for low $\beta$ (high shear),
    which is not apparent from the data.}
\end{figure*}

\citet{TAN00} suggests that cloud-cloud collisions regulate the
SFE. The characteristic timescale for such collisions is $\tau_{\rm
  orb}$ modified by the effects of galactic shear. We saw in Figure
\ref{SFEVSTORB} that the SFE of molecular gas is not a strong function
of $\tau_{\rm orb}$.  Therefore, in Figure \ref{SFEVSBETA}, we plot
the SFE as a function of $\beta$, the logarithmic derivative of the
rotation curve (we plot SFE against $Q_{\rm gas}$, the other component
of this timescale in \S \ref{QSECT}).  This isolates the effect of
differential rotation; $\beta = 0$ for a flat rotation curve and
$\beta = 1$ for solid body rotation (no shear).

Figure \ref{SFEVSBETA} shows a simple relationship between $\beta $ and SFE in
spirals: $\beta > 0$ is associated with high SFE. High $\beta$ occurs almost
exclusively at low radius (where the rotation curve rises steeply) and in
these regions the ISM is mostly \htwo\ with accordingly high SFE. On the other
hand, the outer disks of spirals have $\beta \sim 0$ and a wide range of SFE.
Beyond basic relationship, it is unclear that $\beta $ has utility predicting
the SFE. In particular, we see no clear relationship between SFE and $\beta $
where the ISM is mostly \htwo\ (magenta points). If collisions between bound
clouds regulate the SFE, we would expect an anti-correlation between $\beta $
and SFE because cloud collisions are more frequent in the presence of greater
shear.

In dwarf galaxies increasing $\beta$ corresponds mostly to increasing SFE.
This relationship has the sense of the shear threshold proposed by
\citet{HUNTER98}, that where rotation curves are nearly solid body low shear
allows clouds to form via instabilities aided by magnetic fields \citep[see
also][]{KIM01}. The rotation curves in dwarf galaxies rise more slowly than
those in spirals, leading to $\beta > 0$ over a larger range of radii in dwarf
galaxies and limiting $\beta = 0$ to the relative outskirts of the galaxy. A
positive correlation between $\beta$ and SFE is opposite the sense expected if
cloud collisions are important: at high $\beta$ collisions should be less
frequent.

\subsection{SFE and Thresholds}
\label{SECT_SFETHRESH}

\begin{figure*}
  \begin{center}
    \plottwo{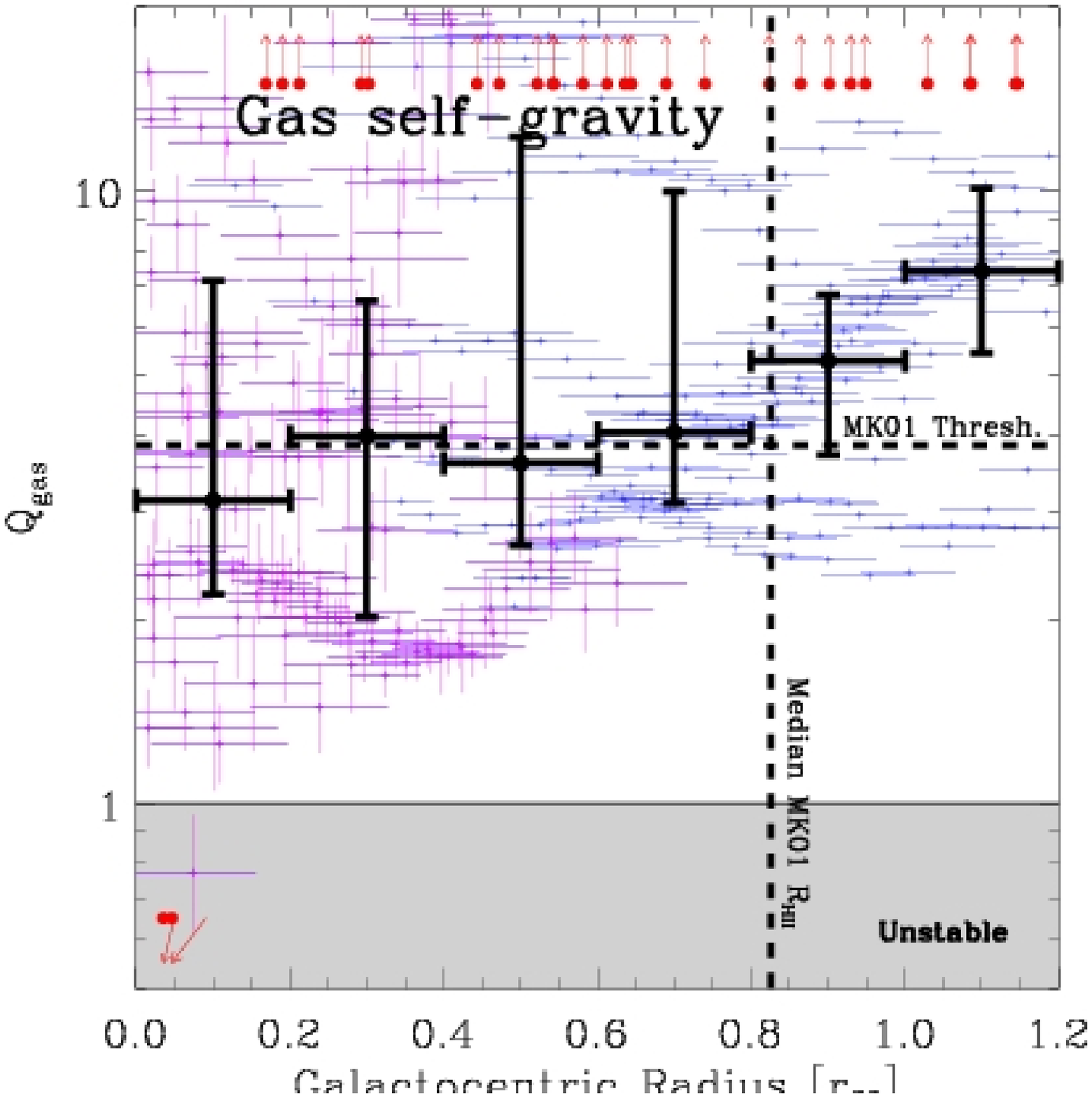}{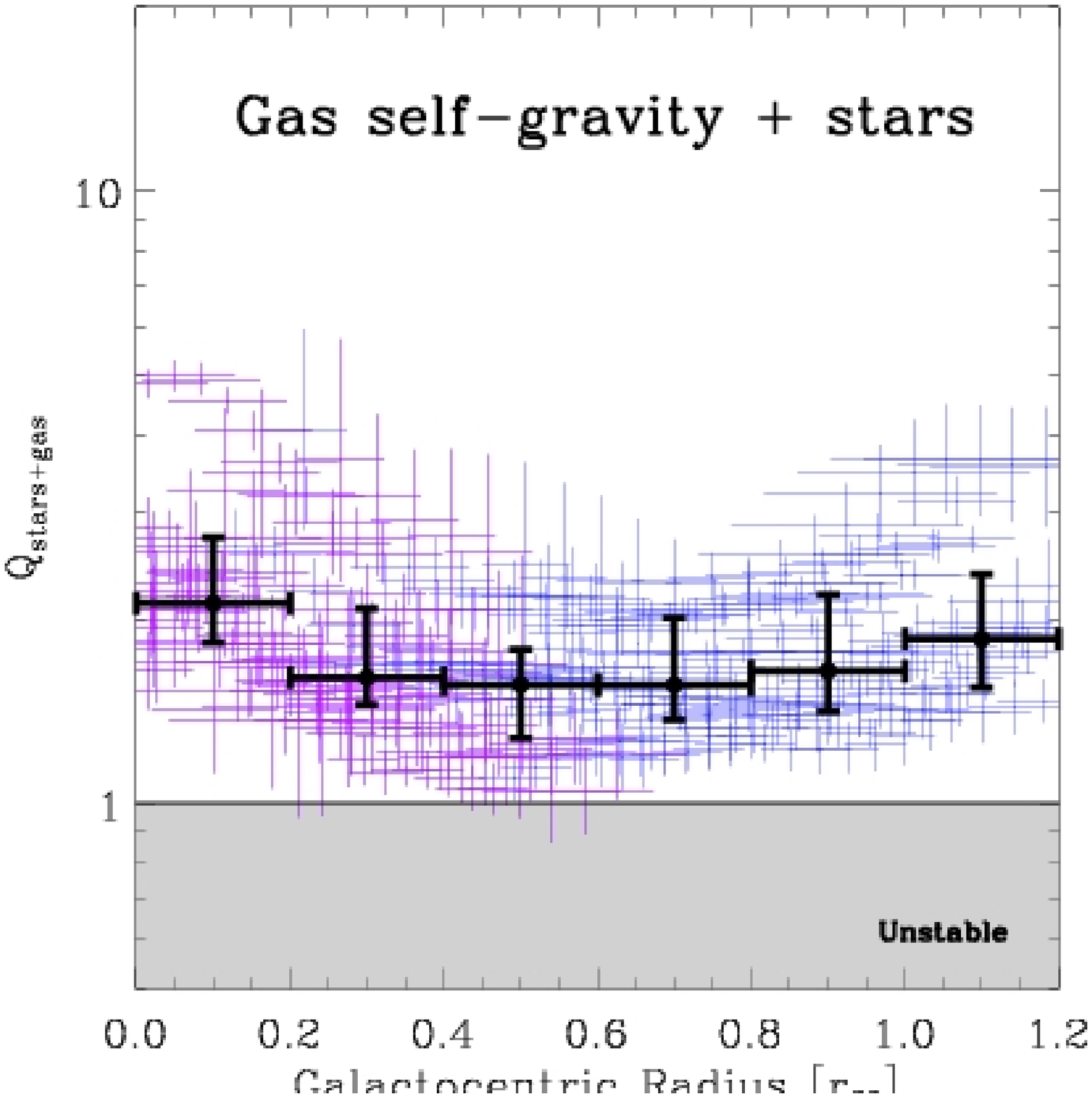}
    \plottwo{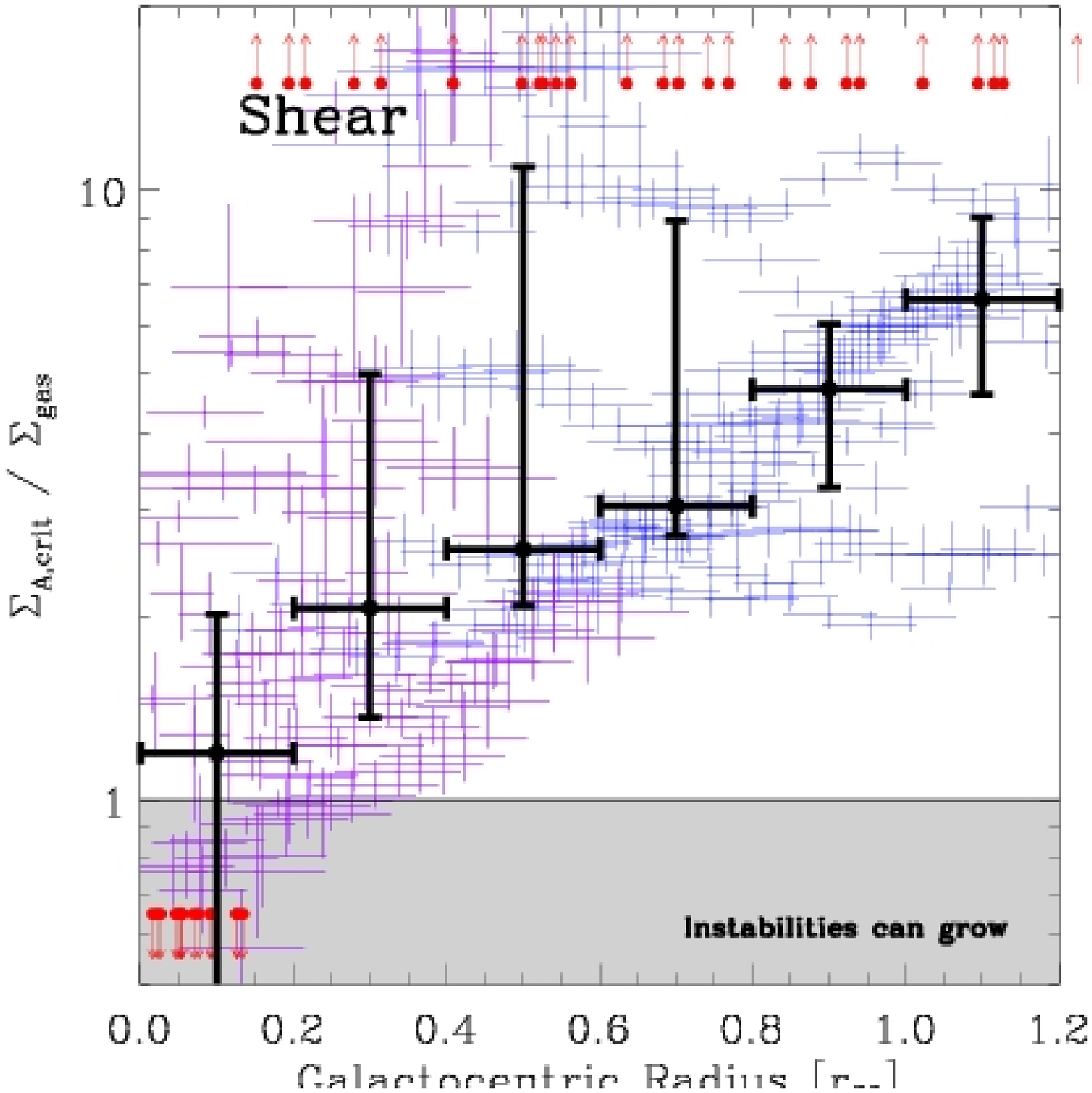}{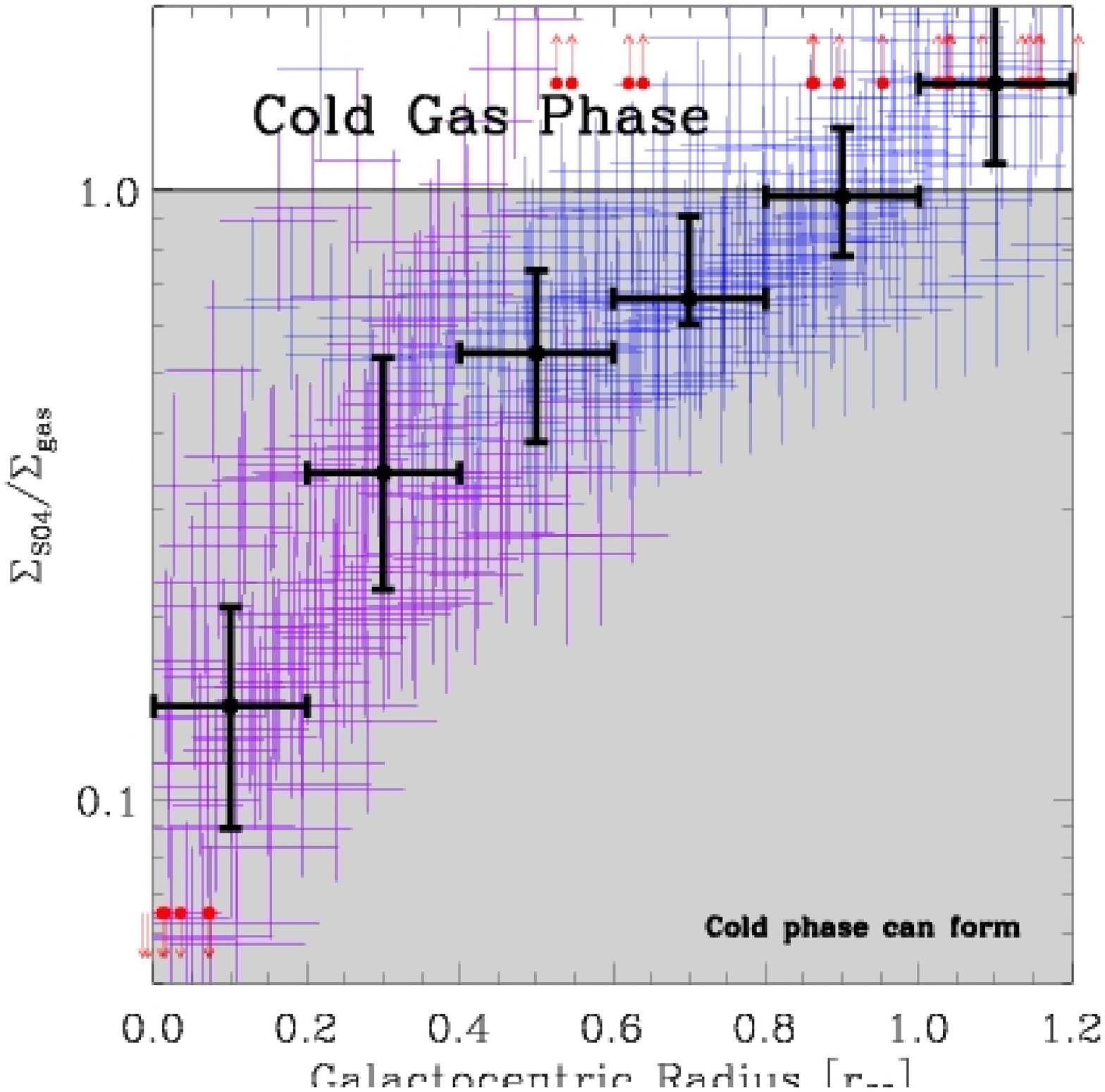}
  \end{center}
  \caption{\label{THRESHVSRADSPIRALS} Radial behavior of thresholds in spiral
    galaxies: ({\em top left}) gravitational instability due to gas self
    gravity; ({\em top right}) gravitational instability due to the
    combination of self--gravity and stellar gravity; ({\em bottom left})
    competition between cloud formation and destruction by shear; ({\em bottom
      right}) formation of a cold phase.  Each point shows average
    $\Sigma_{\rm crit}/\Sigma_{\rm gas}$ over one $10\arcsec$ tilted ring in
    one galaxy. In magenta rings, the ISM is mostly \htwo , in blue rings the
    ISM is mostly \hi .  Gray regions show the condition required for star
    formation.}
\end{figure*}

\begin{figure*}
  \begin{center}
    \plottwo{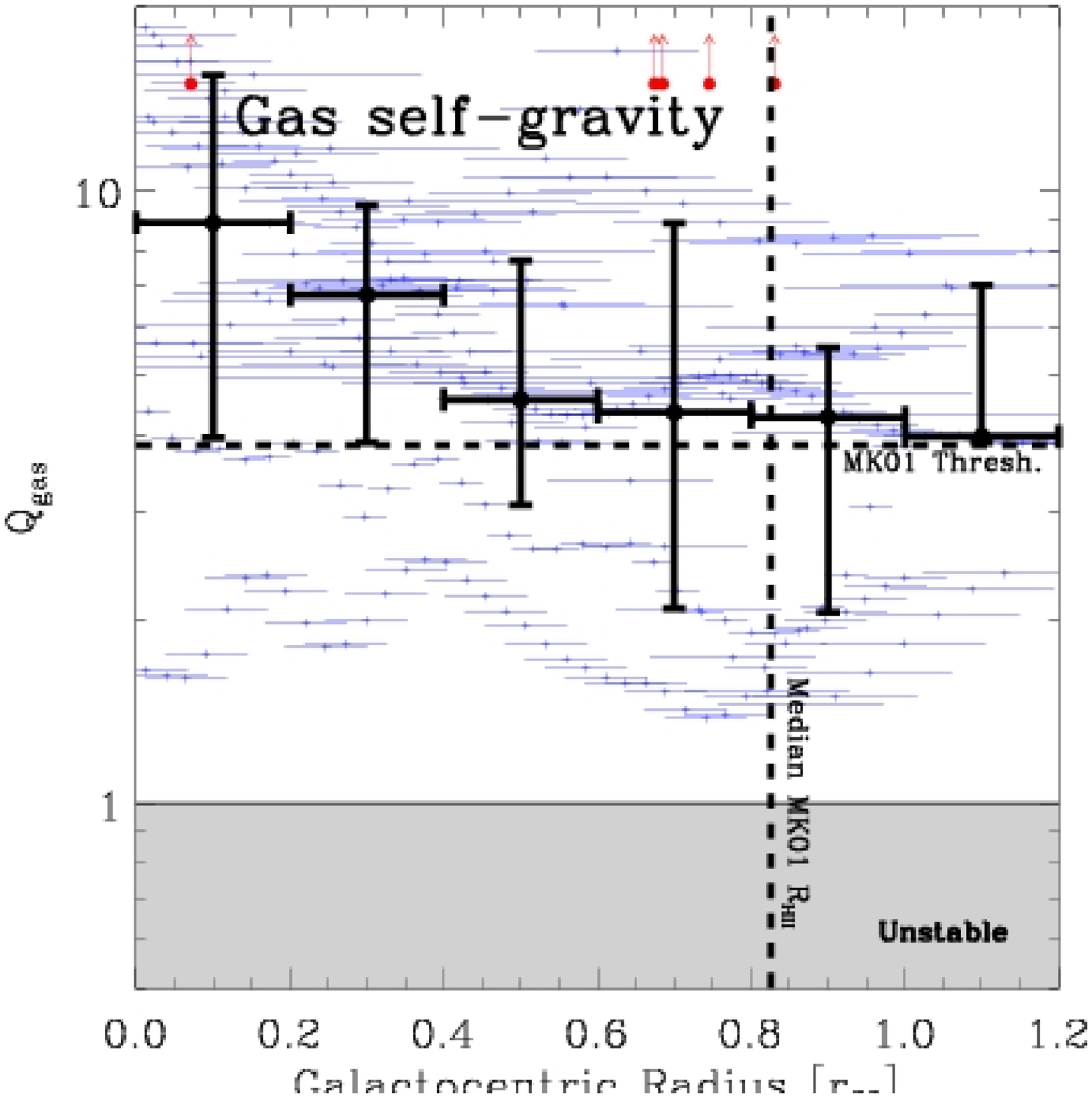}{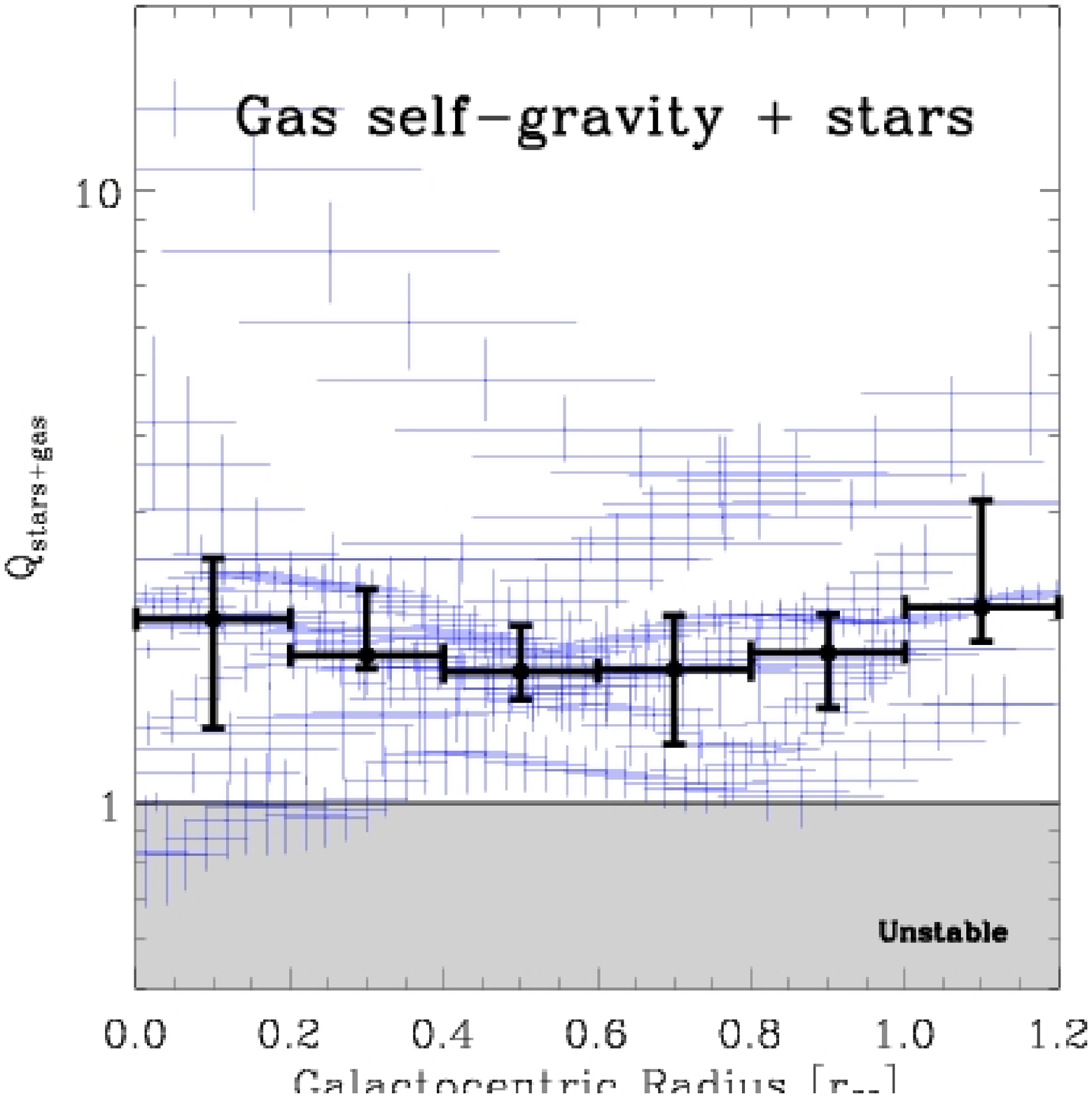}
    \plottwo{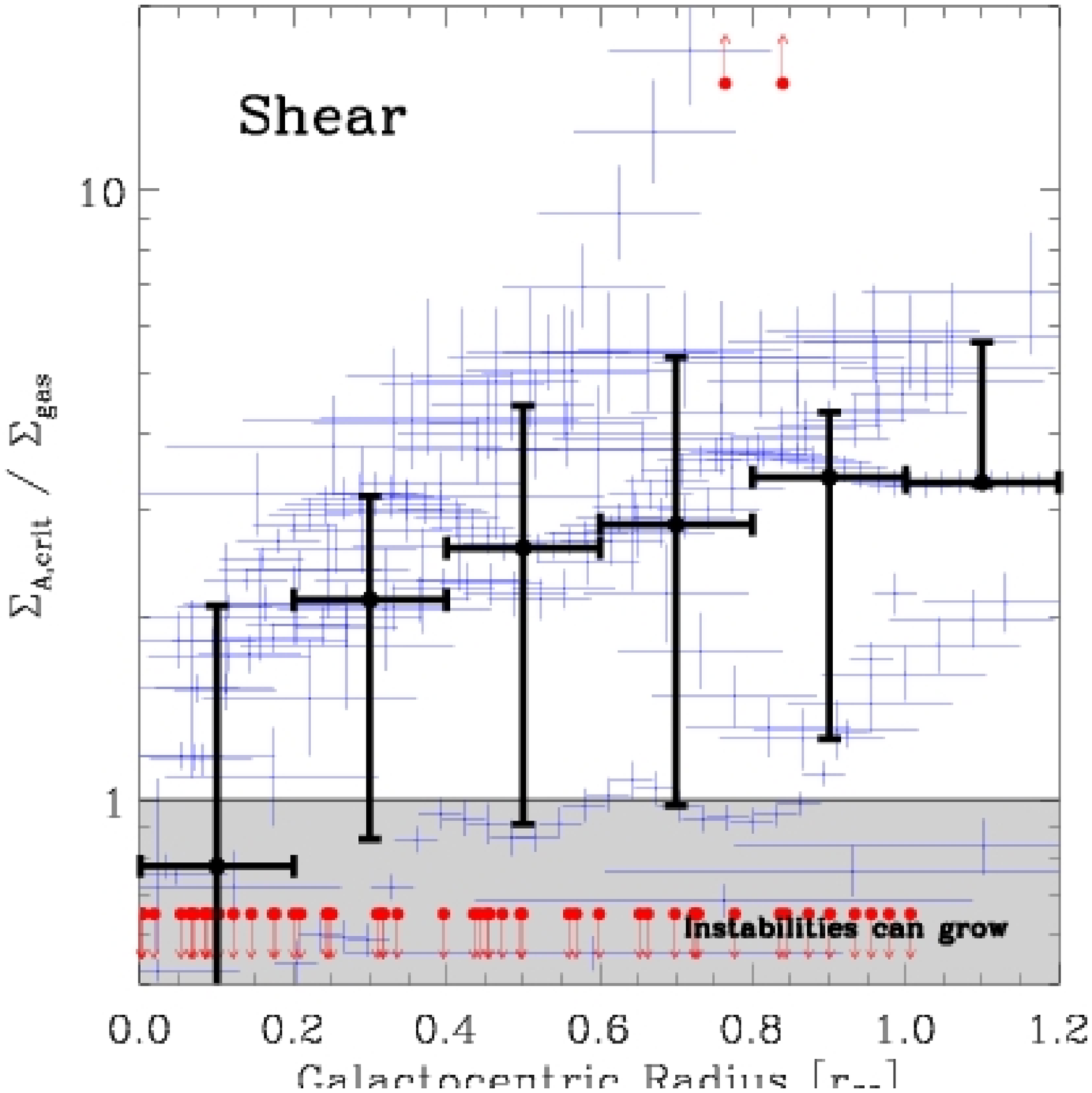}{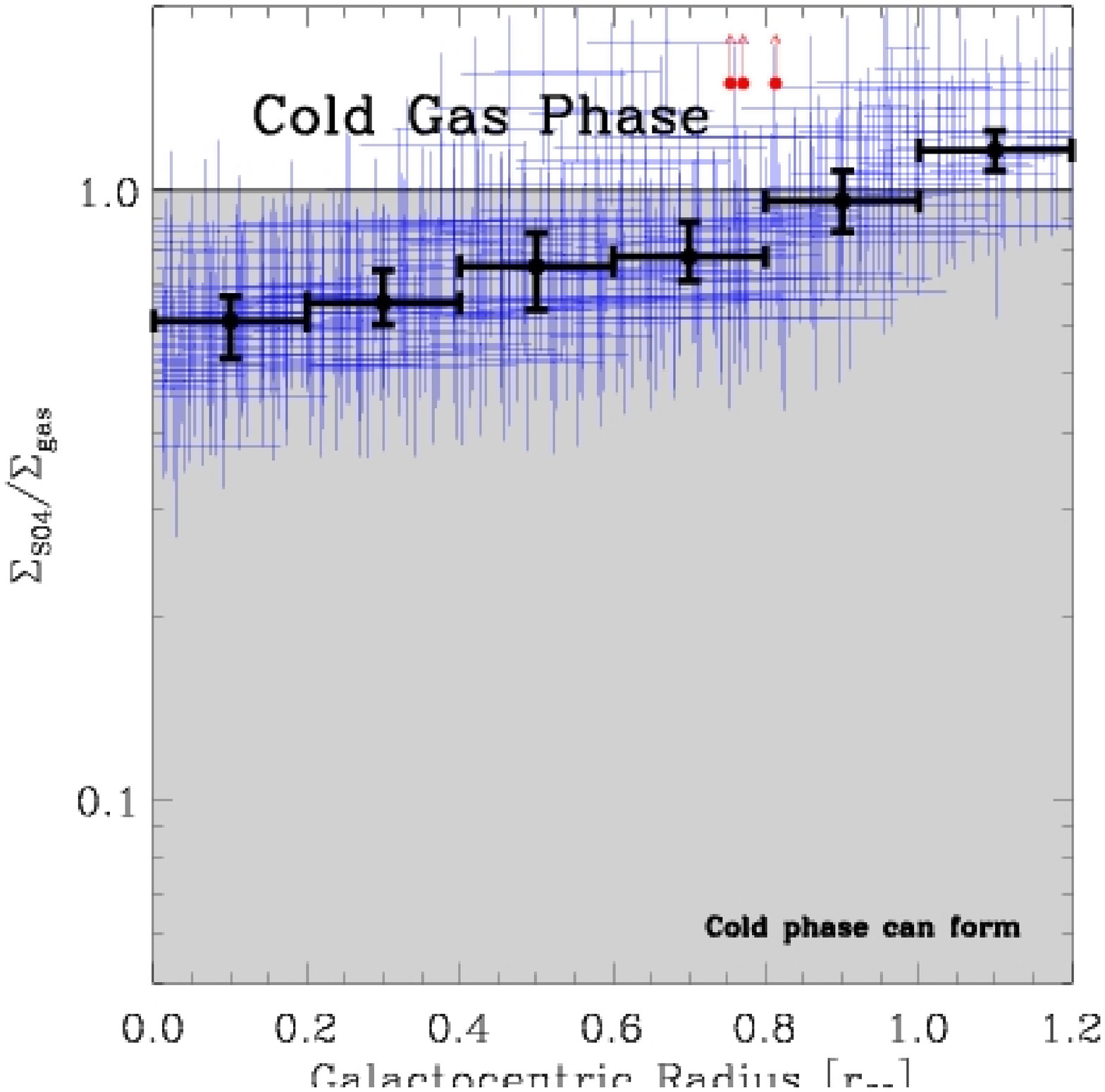}
  \end{center}
  \caption{\label{THRESHVSRADDWARFS} Radial behavior of thresholds in dwarf
    galaxies: ({\em top left}) gravitational instability due to gas self
    gravity; ({\em top right}) gravitational instability due to the
    combination of self--gravity and stellar gravity; ({\em bottom left})
    competition between cloud formation and destruction by shear; ({\em bottom
      right}) formation of a cold phase. Each point shows average $\Sigma_{\rm
      crit}/\Sigma_{\rm gas}$ over one $10\arcsec$ tilted ring in one galaxy.
    Gray regions show the condition required for star formation.}
\end{figure*}

The decline in the SFE where the ISM is mostly \hi\ is too dramatic to be
reproduced across our whole sample by changes in $\tau_{\rm orb}$ or
$\tau_{\rm ff}$ alone. This may be because at large radii a significant amount
of gas is simply unrelated to star formation. If the fraction of gas that is
unable to form GMCs increases with radius, the SFE will decline independent of
any change in GMC formation time. Here we consider the SFE as a function of
proposed star formation thresholds: gravitational instability in the gas alone
($Q_{\rm gas}$), in a disk of gas and stars ($Q_{\rm stars+gas}$), the ability
of instabilities to develop before shear destroys them, and the ability of a
cold gas phase to form.

First we plot each threshold as a function of galactocentric radius in spiral
(Figure \ref{THRESHVSRADSPIRALS}) and dwarf galaxies (Figure
\ref{THRESHVSRADDWARFS}).  Individual points correspond to averages over
$10\arcsec$--wide tilted rings. For magenta points $\Sigma_{\rm H2} >
\Sigma_{\rm HI}$ and for blue points $\Sigma_{\rm H2} < \Sigma_{\rm HI}$. The
gray region in each plot shows the nominal condition for instability, i.e.,
where we expect star formation to occur. Red arrows indicate data outside the
range of the plot.

We proceed creating plots like Figure \ref{SFEVSRAD} for each threshold and
comparing them to Figure \ref{THRESHVSRADSPIRALS}. We expect supercritical gas
to exhibit a (dramatically) higher SFE than subcritical gas, where star
formation proceeds only in isolated pockets or not at all.

\subsubsection{Gravitational Instability in the Gas Disk}
\label{QSECT}

\begin{figure*}
\begin{center}
\plottwo{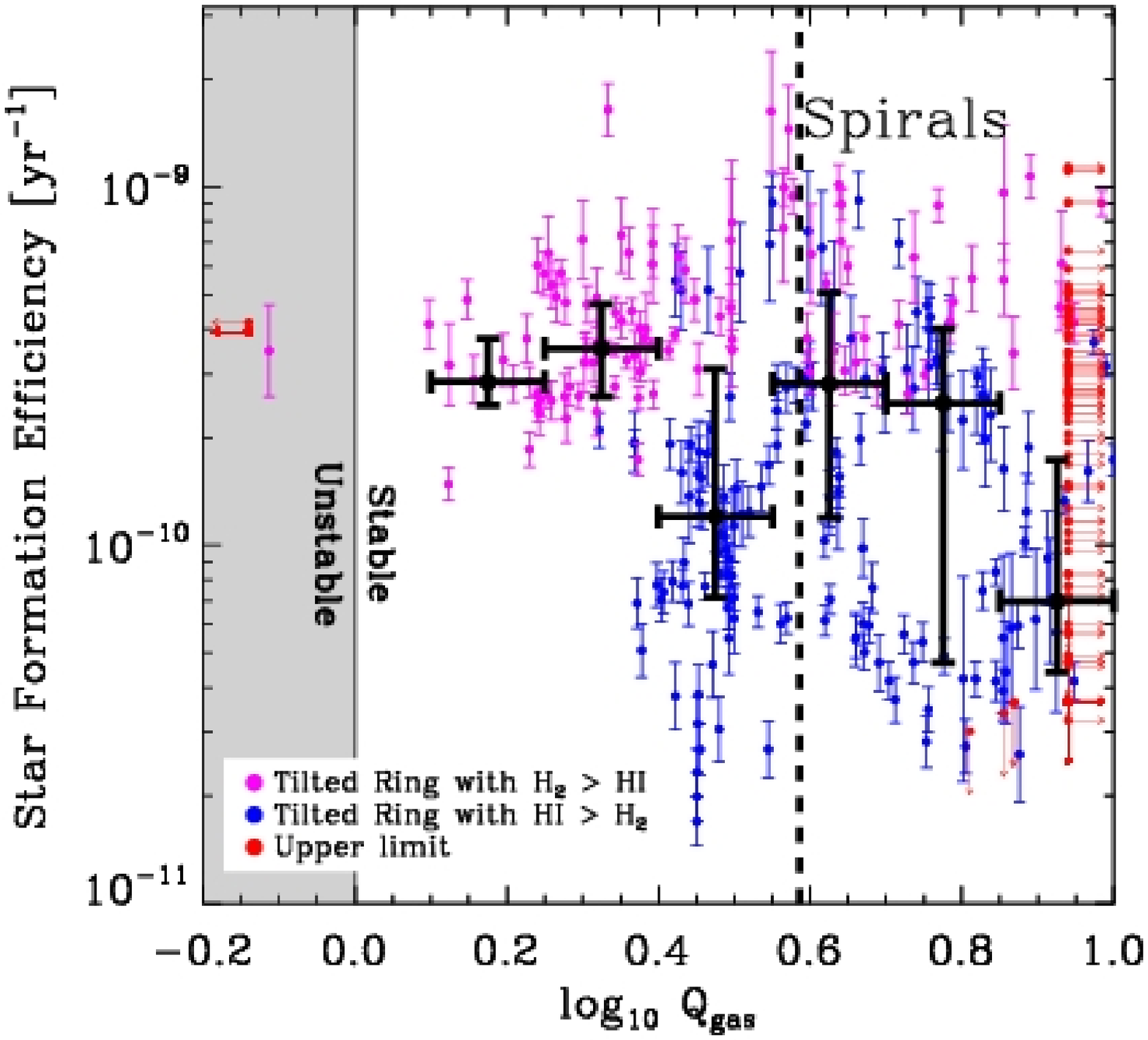}{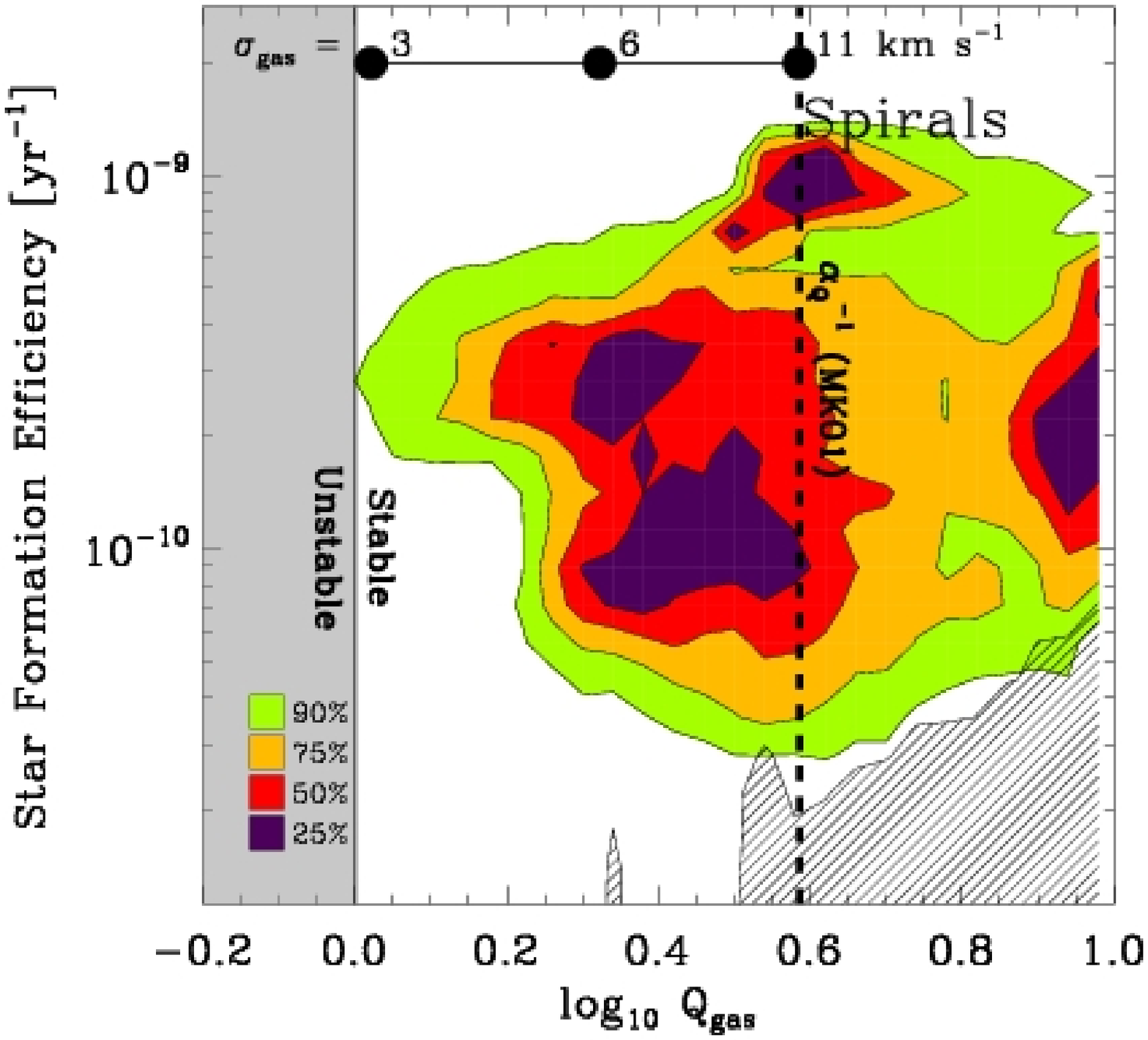}
\plottwo{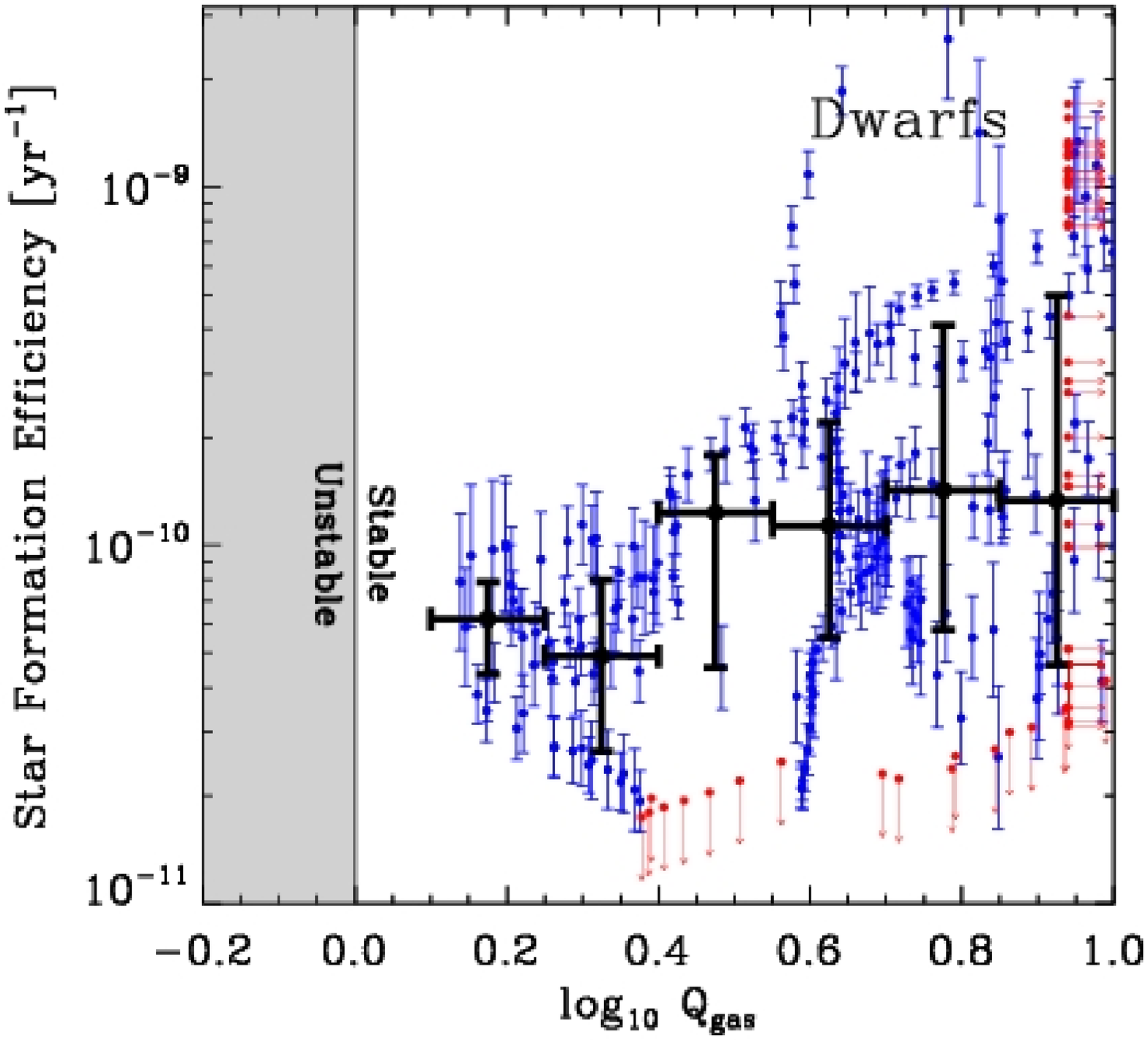}{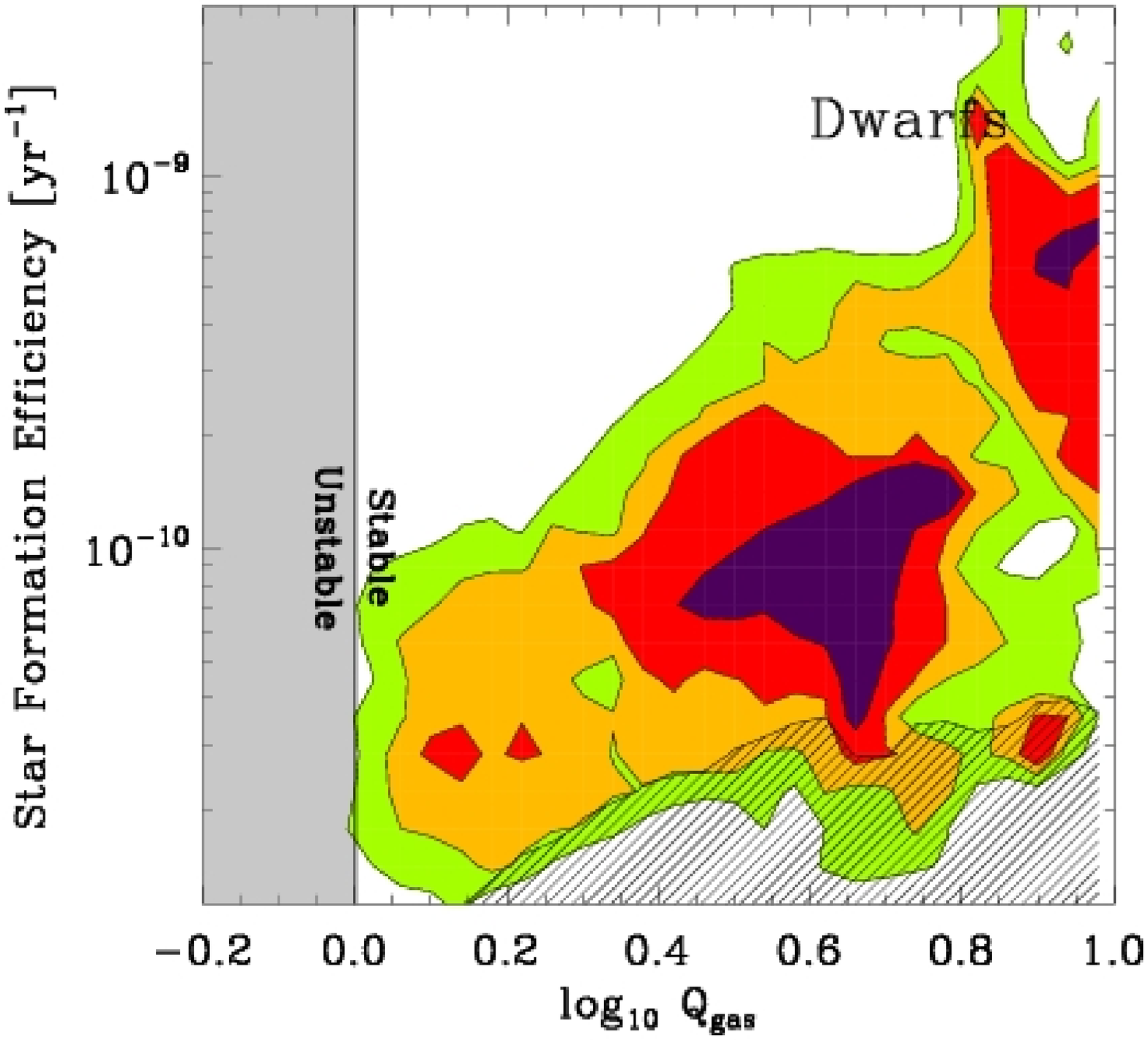}
\end{center}
\caption{\label{SFEVSQGAS} SFE as a function of $Q_{\rm gas}$, the Toomre $Q$
  parameter, which measures instability to axisymmetric collapse in a gas
  disk. Symbols and conventions follow Figure \ref{SFEVSRAD}. The gray region
  shows where instability is expected. A dashed line in the top right panel
  shows the $Q_{\rm gas}$ threshold derived from H$\alpha$ emission by
  \citet{MARTIN01} converted to our assumptions. In the same panel, we show
  the effect on $Q_{\rm gas}$ of changing $\sigma_g$ from our adopted
  11~km~s$^{-1}$ to 6~km~s$^{-1}$ and then 3~km~s$^{-1}$, expected for a cold
  phase.}
\end{figure*}

\begin{figure*}
  \begin{center}
    \plottwo{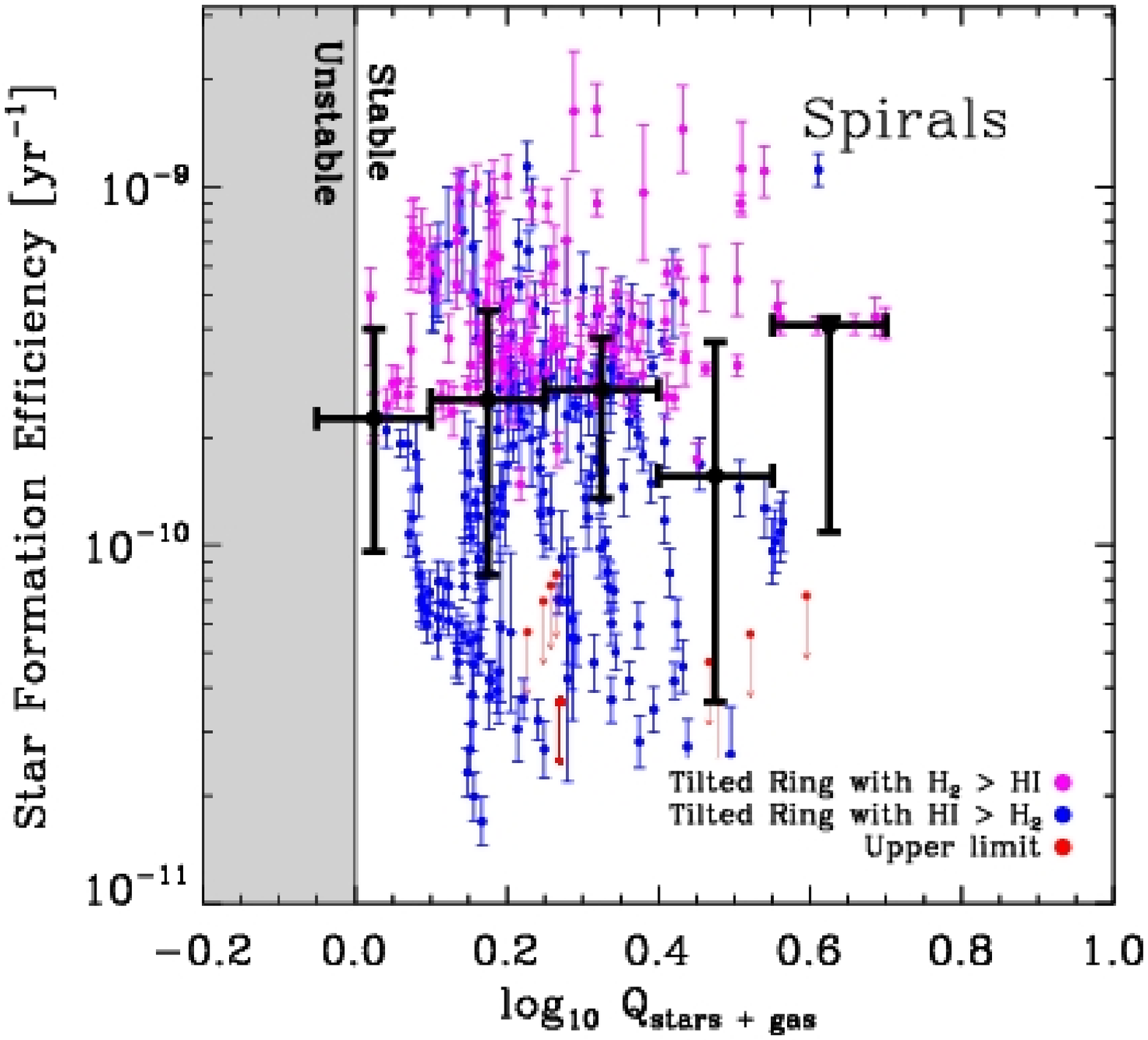}{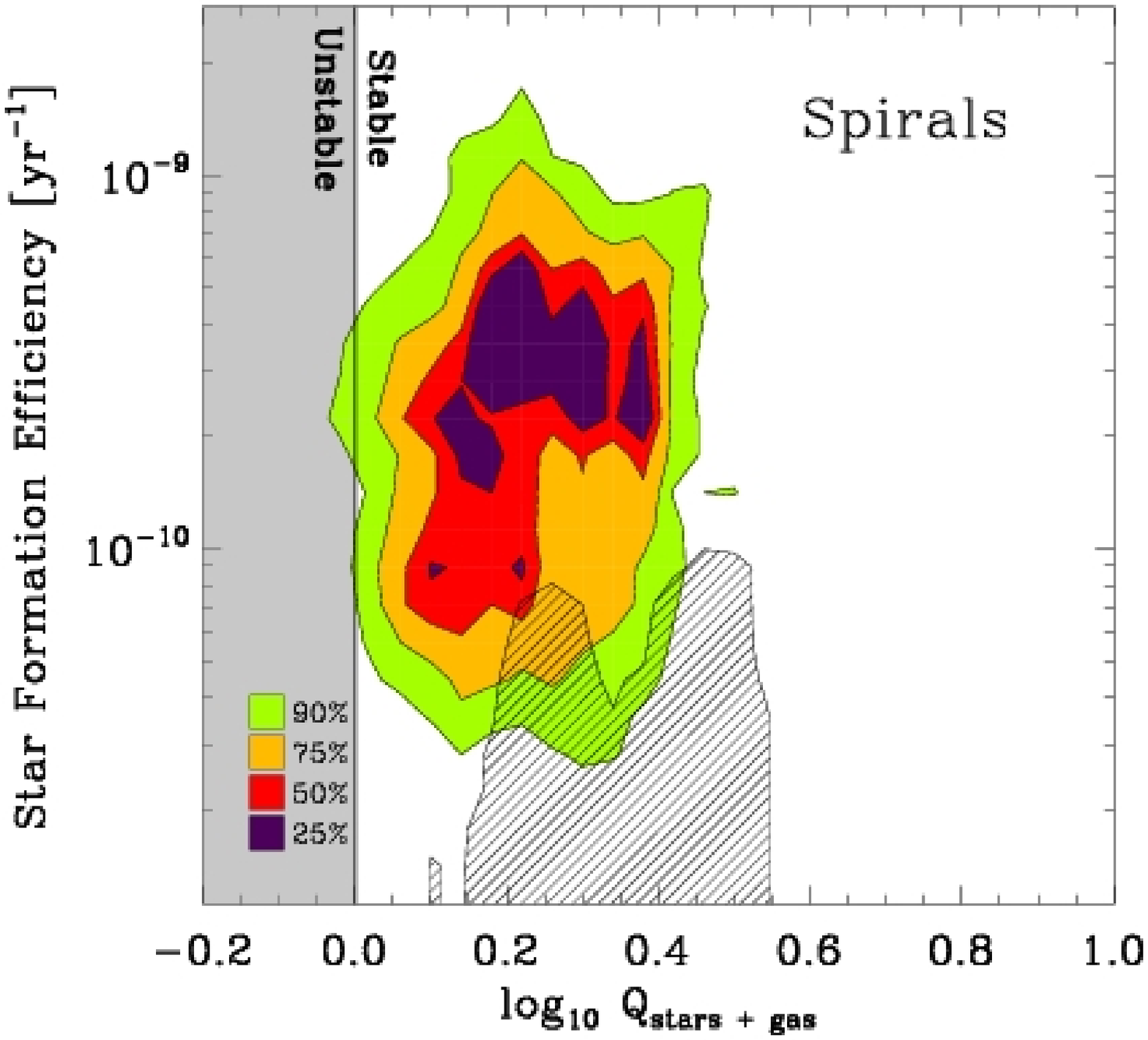}
    \plottwo{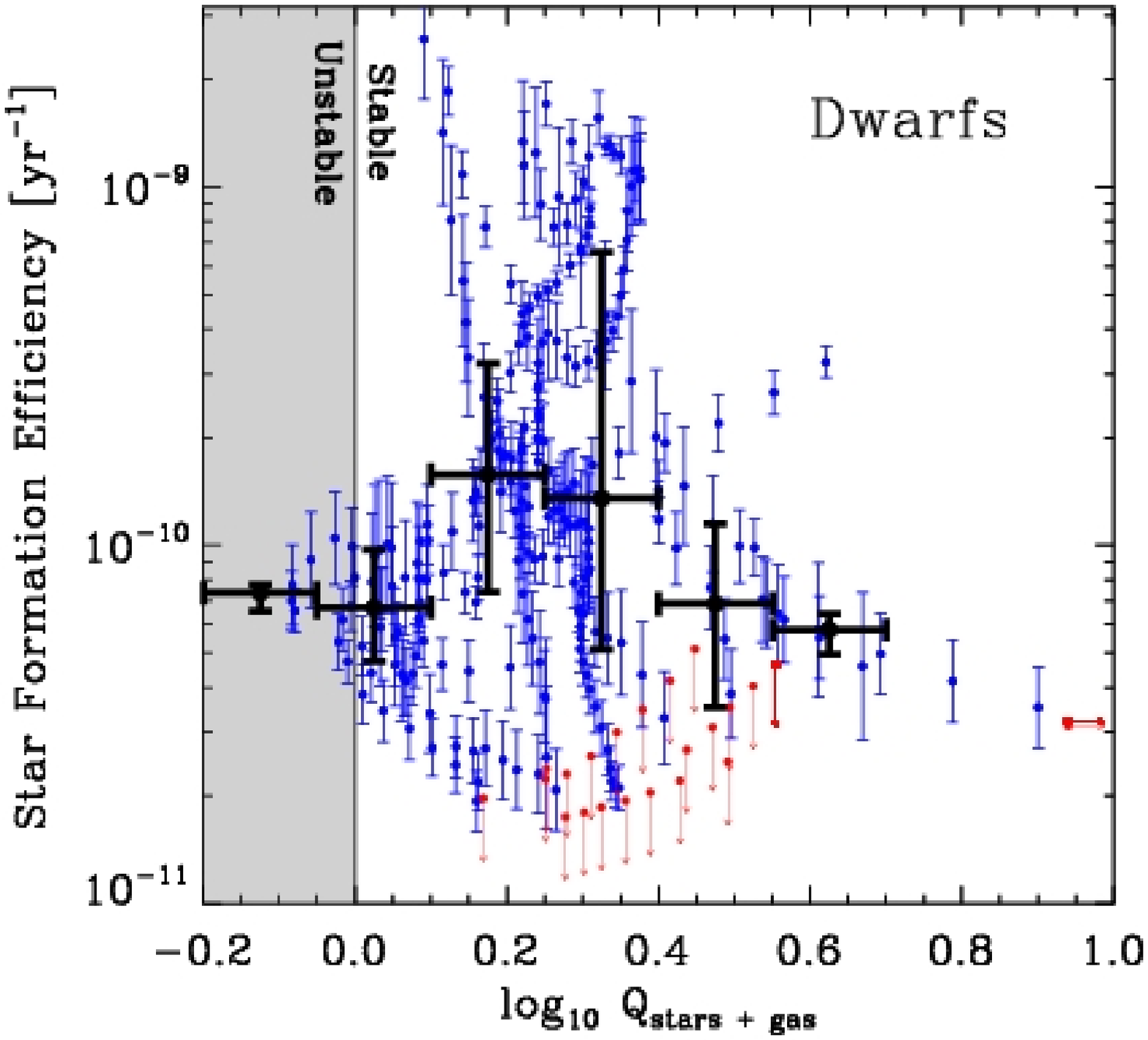}{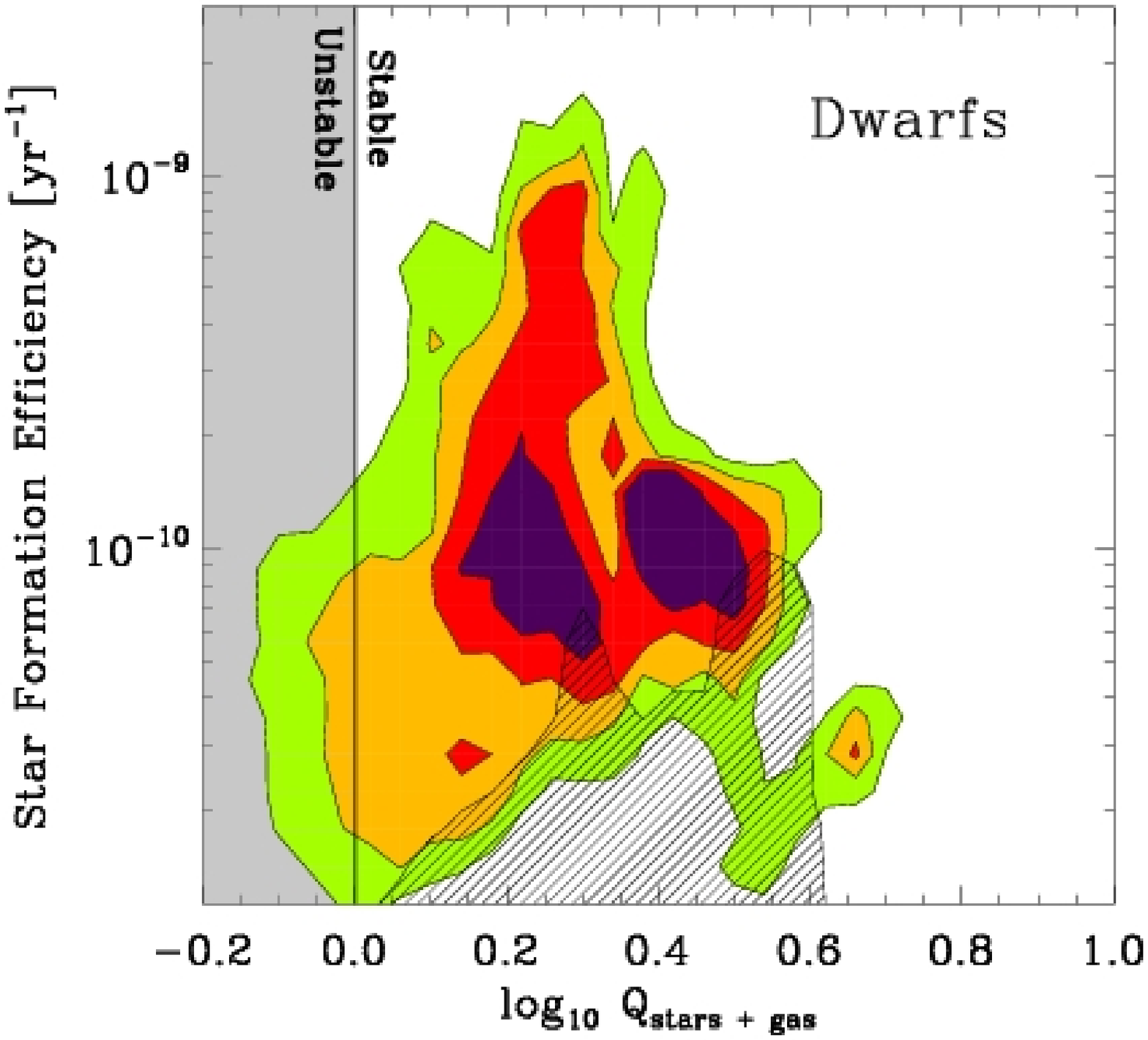}
\end{center}
\caption{\label{SFEVSQSTARGAS} SFE as a function of $Q_{\rm stars+gas}$
  \citep{RAFIKOV01}, which measures instability in a gas disk in the presence
  of a collisionless stellar disk. Symbols and conventions follow Figure
  \ref{SFEVSRAD}. The gray region indicates where gas is unstable. Compared to
  $Q_{\rm gas}$, including stars renders the disk more nearly unstable and
  yields a much lower range of values.}
\end{figure*}

Figure \ref{SFEVSQGAS} shows SFE as a function of $Q_{\rm gas}$, Toomre's $Q$
parameter for a thin gas disk; the top left panels in Figures
\ref{THRESHVSRADSPIRALS} and \ref{THRESHVSRADDWARFS} show $Q_{\rm gas}$ as a
function of radius.

In each plot, a gray area indicates the theoretical condition for instability.
We see immediately that almost no area in our sample is formally unstable.
Rather, most lines of sight are strikingly stable, $Q_{\rm gas} \sim 4$ is
typical inside $\sim 0.8~r_{25}$ and $Q_{\rm gas} > 10$ is common.

We find no clear evidence for a $Q_{\rm gas}$ threshold (at any value) that can
unambiguously distinguish regions with high SFE from those with low SFE. In
spirals, $Q_{\rm gas} \lesssim 2.5$ appears to be a sufficient, but by no
means necessary condition for high SFE; there are also areas where the ISM is
mostly \htwo , SFE is quite high and $Q_{\rm gas} \gtrsim 10$. In dwarfs
$Q_{\rm gas}$ appears, if anything, anti-correlated with SFE, though this may
partially result from incomplete estimates of $\Sigma_{\rm gas}$.

These conclusions appear to contradict the findings by \citet{KENNICUTT89} and
\citet{MARTIN01}, who found marginally stable gas ($Q_{\rm gas}\sim 1.5$)
across the optical disk with a rise in $Q_{\rm gas}$ corresponding to dropping
SFE at large radii. In fact, after correcting for different assumptions, our
median $Q_{\rm gas}$ matches theirs quite well.  Both \citet{KENNICUTT89} and
\citet{MARTIN01} assumed $\xco =2.8 \times 10^{20}$~\xcounits\ and
$\sigma_{\rm gas} =6$~km~s$^{-1}$, while we take $\xco =2.0 \times
10^{20}$~\xcounits\ and $\sigma_{\rm gas} =11$~km~s$^{-1}$. As a result, we
estimate less \htwo\ and more kinetic support than they do for the same
observations. If we match their assumptions, our median $Q_{\rm gas}$ in
spirals and the outer parts of dwarfs agrees quite well with their threshold
value, though we find the central regions of dwarfs systematically above this
value \citep[as did][]{HUNTER98}. We show this in Figures
\ref{THRESHVSRADSPIRALS}, \ref{THRESHVSRADDWARFS}, and \ref{SFEVSQGAS} by
plotting the \citeauthor{MARTIN01} threshold converted to our assumptions
($Q_{\rm gas} \sim 3.9$) as a a dashed line.

The main observational difference between our result and \citet{MARTIN01} is
that $Q_{\rm gas}$ shows much more scatter in our analysis. As a result, a
systematic transition from low to high $Q_{\rm gas}$ near the edge of the
optical disk is not a universal feature of our data, though a subset of spiral
galaxies do show increasing $Q_{\rm gas}$ at large radii (Figure
\ref{THRESHVSRADSPIRALS}).

This discrepancy in $Q_{\rm gas}$ derived from similar data highlights the
importance of assumptions. The largest effect comes from $\sigma_{\rm gas}$,
which we measure to be $\approx 11$~km~s$^{-1}$ and roughly constant in \hi
--dominated outer disks (Appendix \ref{KINAPP}). We assume $\sigma_{\rm gas}$
to be constant everywhere, an assumption that may break down on small scales
and in the molecular ISM. In this case we expect $\sigma_{\rm gas}$ to be
locally lower than the average value, lowering $Q_{\rm gas}$ and making gas
less stable.  Black dots in the upper right panel of Figure \ref{SFEVSQGAS}
show the effect of changing $\sigma_{\rm gas}$ from 11~km~s$^{-1}$ (our value)
to 6~km~s$^{-1}$ (the \citeauthor{MARTIN01} value) and then to 3~km~s$^{-1}$,
the value expected and observed for a cold \hi\ component
\citep[e.g.][]{YOUNG03, SCHAYE04, DEBLOK06}. If most gas is cold then $Q_{\rm
  gas}$ may easily be $\lesssim 1$ for this component (if only a small
fraction of gas is cold, the situation is less clear).

\subsubsection{Gravitational Instability Including Stars}
\label{QSTARSECT}

Stars dominate the baryon mass budget over most of the areas we study and
stellar gravity may be expected to affect the stability of the gas disk. In \S
\ref{SECT_BACKGROUND} we described a straightforward extension of $Q_{\rm
  gas}$ to the case of a disk containing gas and stars \citep{RAFIKOV01}. In
Figure \ref{SFEVSQSTARGAS} we plot SFE as a function of this parameter,
$Q_{\rm stars+gas}$, which we plot as a function of radius in the top right
panels of Figures \ref{THRESHVSRADSPIRALS} and \ref{THRESHVSRADDWARFS}.

The gray region indicates where gas is unstable to axisymmetric collapse.
Including stars does not render large areas of our sample unstable, but it
does imply that most regions are only marginally stable, $Q_{\rm stars+gas}
\sim 1.6$. This in turn suggests that it is not so daunting to induce collapse
as one would infer from only $Q_{\rm gas}$.

In addition to lower values, $Q_{\rm stars+gas}$ exhibits a much narrower
range of values than $Q_{\rm gas}$, mostly areas in both spiral and dwarf
galaxies show $Q_{\rm stars+gas} = 1.3$ -- $2.5$. This may offer support to
the idea of self--regulated star formation, but it also means that $Q_{\rm
  stars+gas}$ offers little leverage to predict the SFE. High SFE, mostly
molecular regions show the same $Q_{\rm stars+gas}$ as low SFE regions from
outer disks (indeed, the highest values we observe come from the central parts
of spiral galaxies).

As with $Q_{\rm gas}$, our assumptions have a large impact on $Q_{\rm
  stars+gas}$.  In addition to $\sigma_{\rm gas}$ and \xco (which affect the
calculation via $Q_{\rm gas}$), the stellar velocity dispersion, $\sigma_*$,
and mass--to--light ratio, \ML , strongly affect our stability estimate. We
assume that $\sigma_* \propto \Sigma_*^{0.5}$ in order to yield a constant
stellar scale height. If we instead fixed $\sigma_*$, we would derive $Q_{\rm
  stars+gas}$ increasing steadily with radius. Radial variations in $\ML$ may
create a similar effect.

\citet{BOISSIER03} find similar results to our own when they incorporate stars
in their stability analysis; they adopt a lower $\sigma_{\rm gas}$ than we do,
but also lower \xco\ and the effects roughly offset.  \citet{YANG07} recently
derived $Q_{\rm stars+gas}$ across the LMC and found widespread instability
that corresponded well with the distribution of star formation. If we match
their adopted $\sigma_{\rm gas}$ (5~km~s$^{-1}$) and assumptions regarding
$\sigma_*$ (constant at 15~km~s$^{-1}$) we also find widespread instability
throughout our dwarf subsample, $Q_{\rm stars+gas}$ decreasing with radius; we
find a similar result for spirals if we fix a typical outer--disk $\sigma_*$.

Our approach is motivated by observations of disk galaxies (see Appendix
\ref{KINAPP}), but direct observations of $\sigma_*$ at large radii are still
sorely needed.

\subsubsection{Shear Threshold}
\label{SHEAR_SECT}

\begin{figure*}
  \begin{center}
    \plottwo{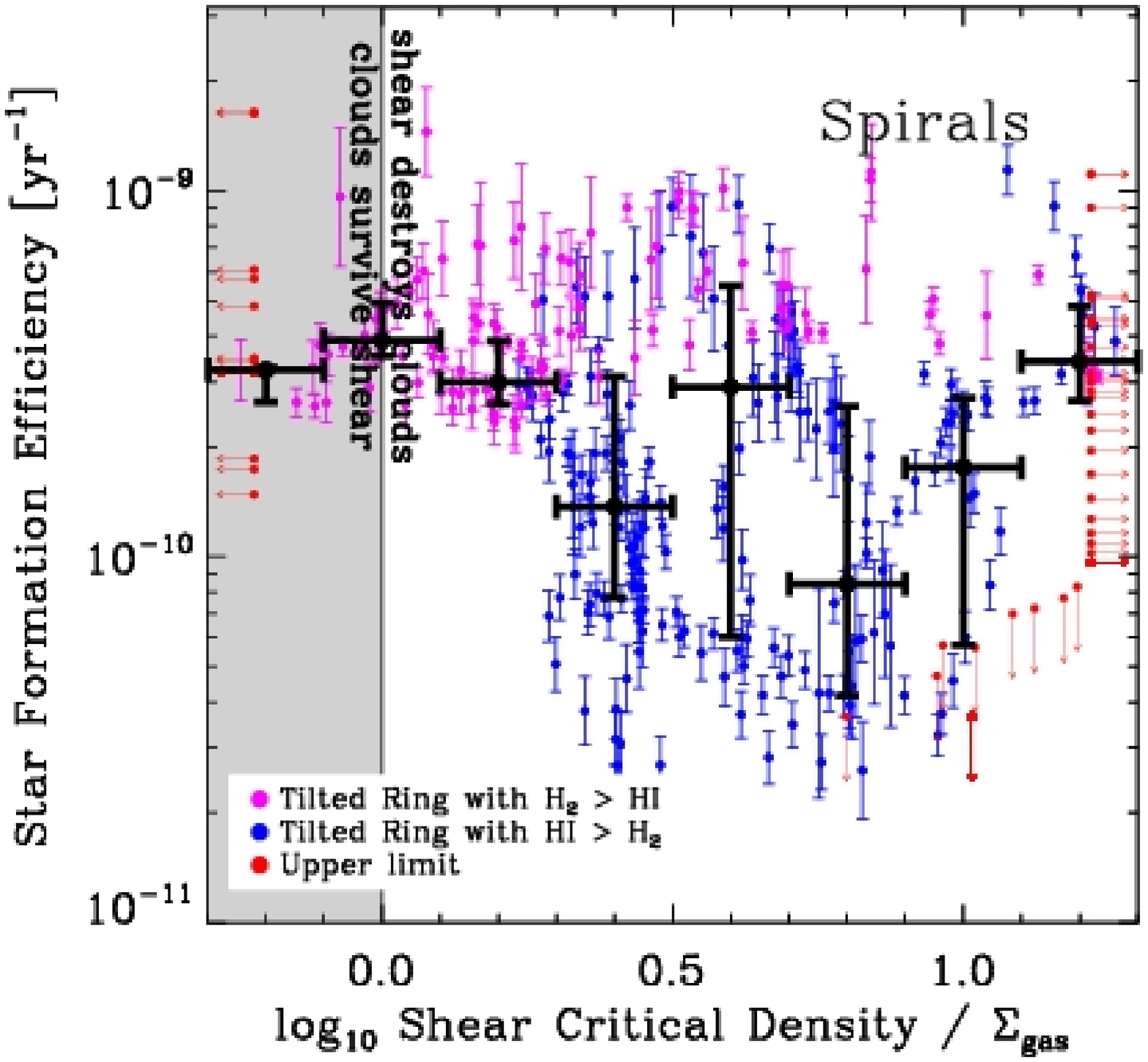}{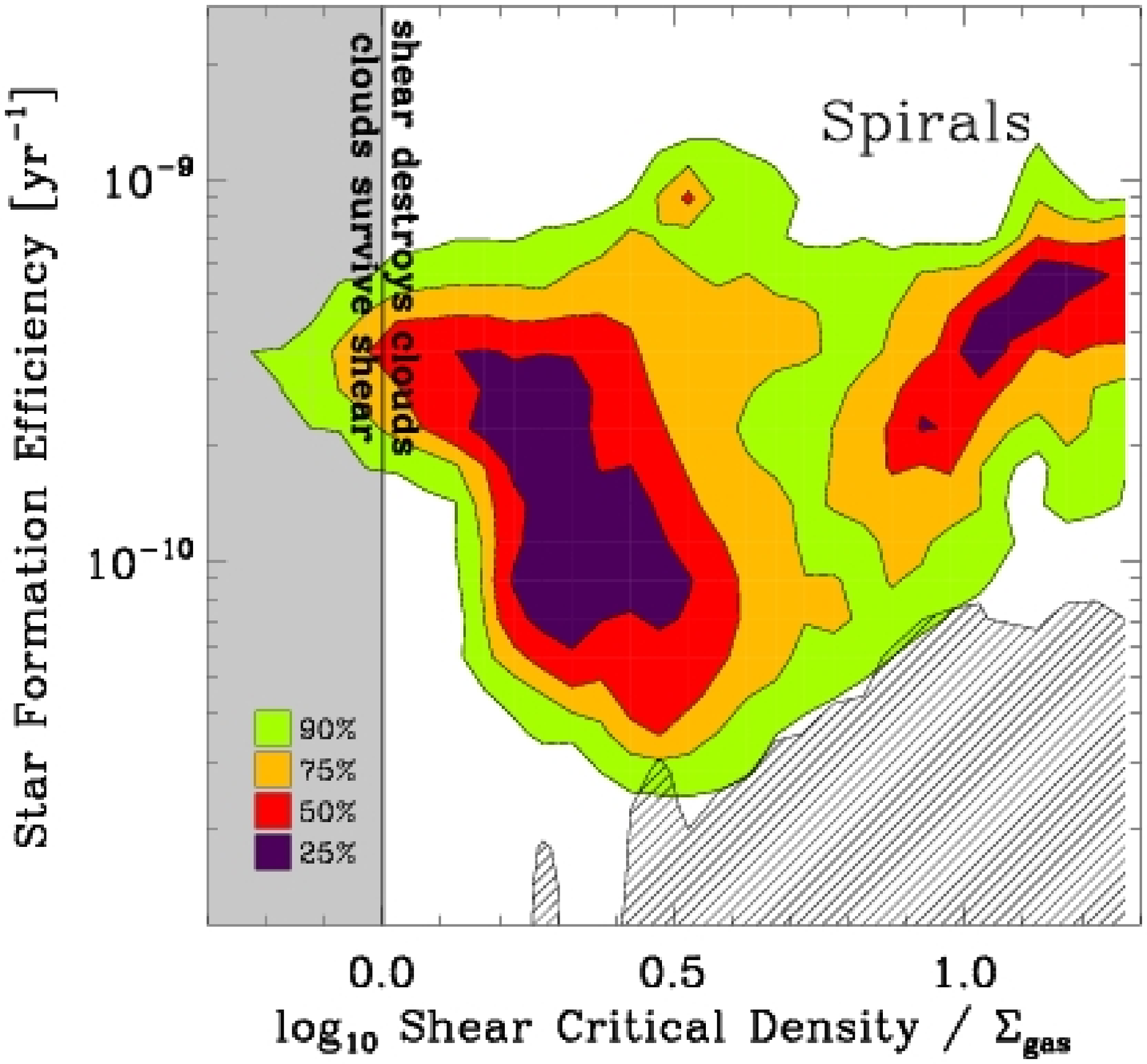}
    \plottwo{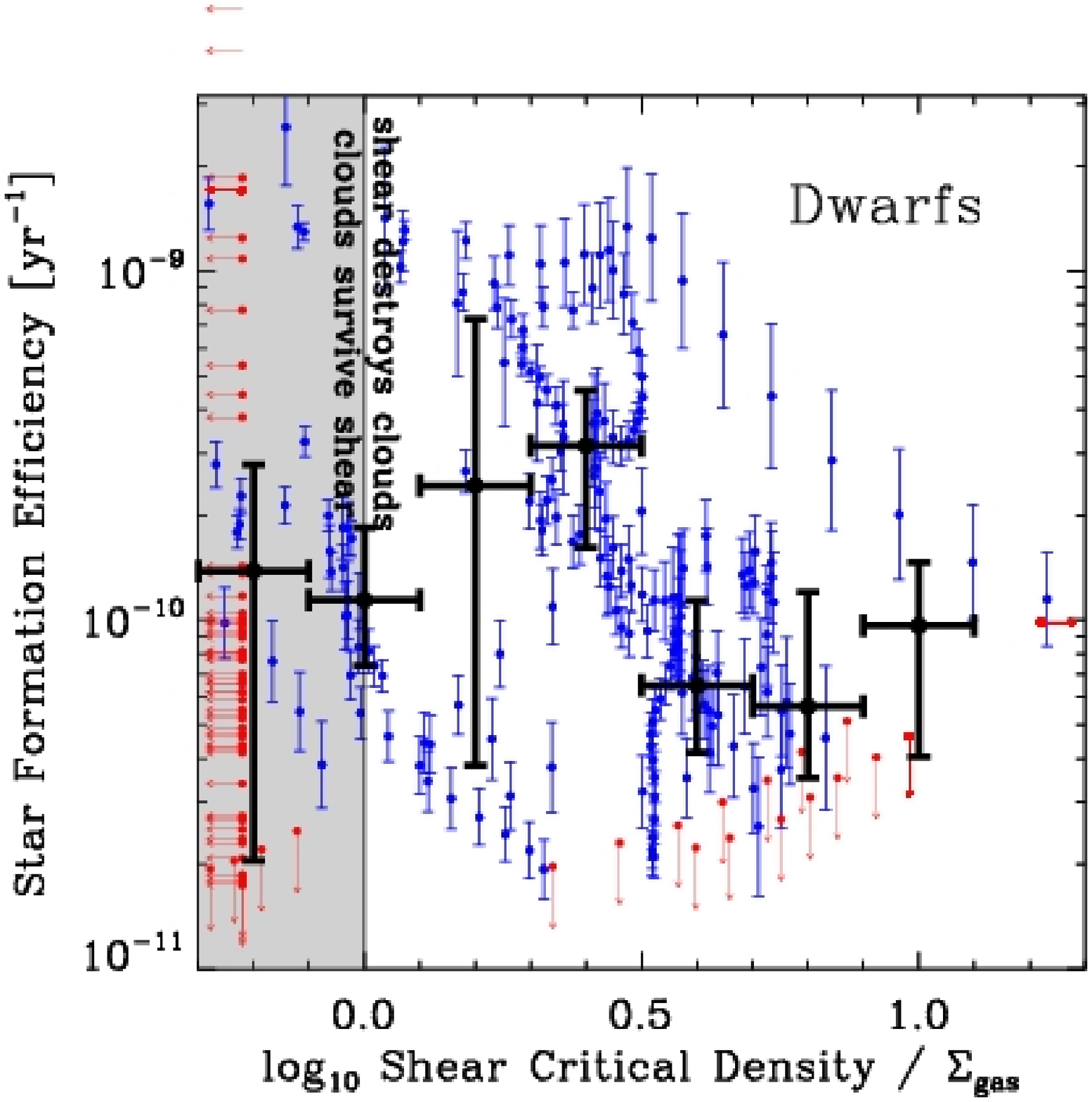}{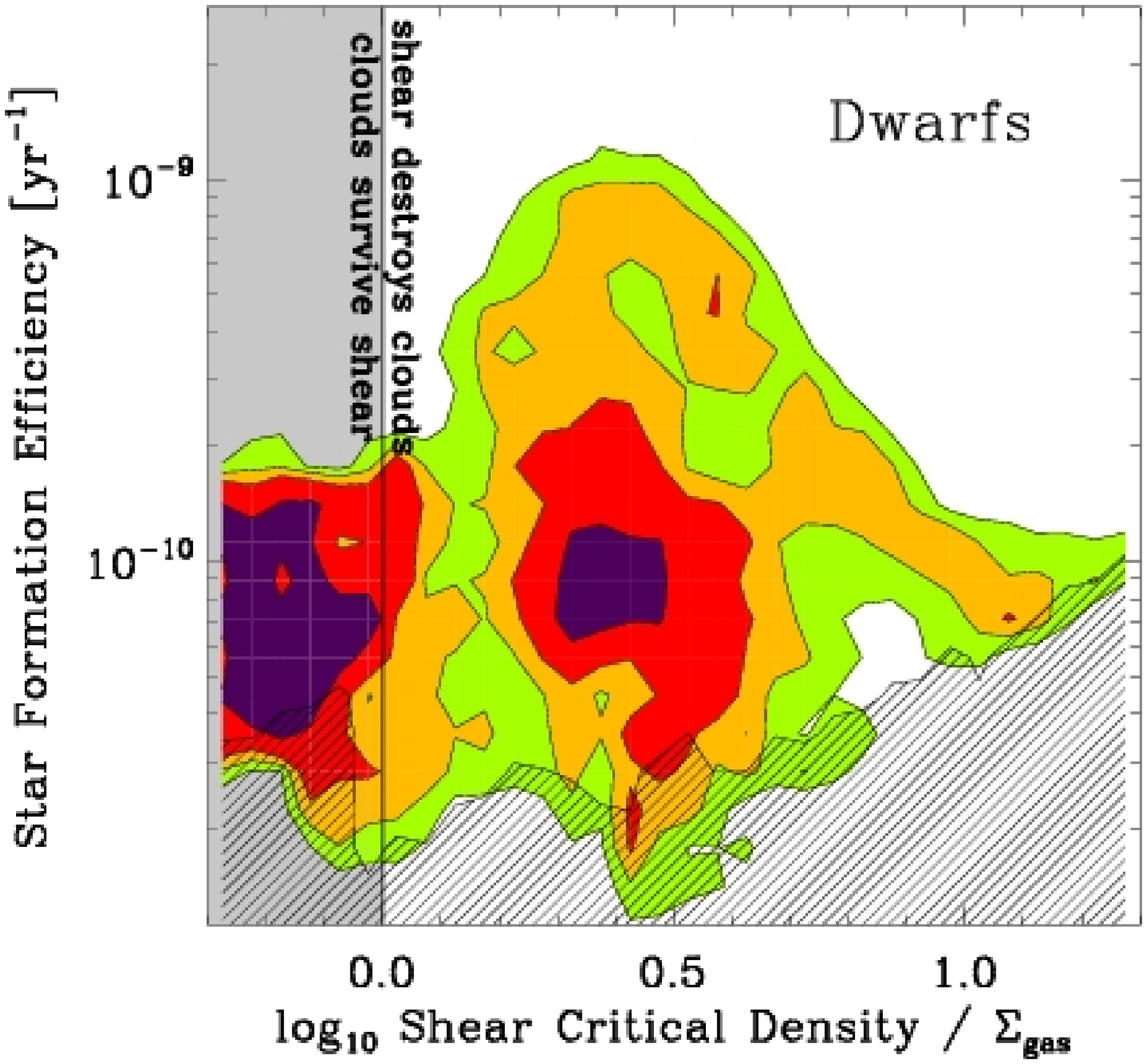}
\end{center}
\caption{\label{SFEVSSHEAR} SFE as a function of $\Sigma_{\rm
    crit,A}/\Sigma_{\rm gas}$, the threshold for cloud growth in the presence
  of shear \citep{HUNTER98} for spiral (top row) and dwarf (bottom row)
  galaxies.  Conventions and symbols follow Figure \ref{SFEVSRAD}. The gray
  area shows where clouds should be able to survive distribution by shear.}
\end{figure*}

If clouds form efficiently, e.g., through the aid of magnetic fields to
dissipate angular momentum, then \citet{HUNTER98} suggest that the time
available for a perturbation to grow in the presence of destructive shear may
limit where star formation is widespread. \citet{KIM01} describe a similar
scenario where magneto-Jeans instabilities can grow in regions with weak shear
or strong magnetic fields. In the bottom left panels of Figures
\ref{THRESHVSRADSPIRALS} and \ref{THRESHVSRADDWARFS}, we plot this shear
threshold as a function of radius and in Figure \ref{SFEVSSHEAR} we compare it
to the SFE.

The gray region shows the condition for instabilities to grow into GMCs,
$\Sigma_{\rm crit,A} / \Sigma_{\rm gas} < 1$. This matches the condition
$Q_{\rm gas} < 1$ where $\beta = 0$, e.g., in outer disks of spirals. In the
inner parts of spirals and in dwarf galaxies, however, $\Sigma_{\rm crit,A} /
\Sigma_{\rm gas}$ is lower than $Q_{\rm gas}$, i.e., the conditions for star
formation are more nearly supercritical (because shear is low in these
regions).  These areas harbor \htwo\ or widespread star formation, so
supercritical values are expected.

This trend of more supercritical data at lower radii agrees with the steady
increase of SFE with decreasing radius that we saw in Figure \ref{SFEVSRAD}.
However, the scatter in $\Sigma_{\rm crit,A} / \Sigma_{\rm gas} < 1$ is as
large as that in $Q_{\rm gas}$ (as one would expect from their forms, see
Table \ref{IDEATAB}).  As a result, a direct plot of SFE against $\Sigma_{\rm
  crit,A} / \Sigma_{\rm gas}$ does not yield clear threshold behavior or a
strong correlation between $\Sigma_{\rm crit,A} / \Sigma_{\rm gas}$ and SFE.
The strongest conclusion we can draw is that the inner parts of both spiral
and dwarf galaxies are marginally stable for the shear threshold (an
improvement over $Q_{\rm gas}$ and $Q_{\rm stars+gas}$ in these regions).

\subsubsection{Cold Phase Formation}
\label{SCHAYE_SECT}

\begin{figure*}
  \begin{center}
  \plottwo{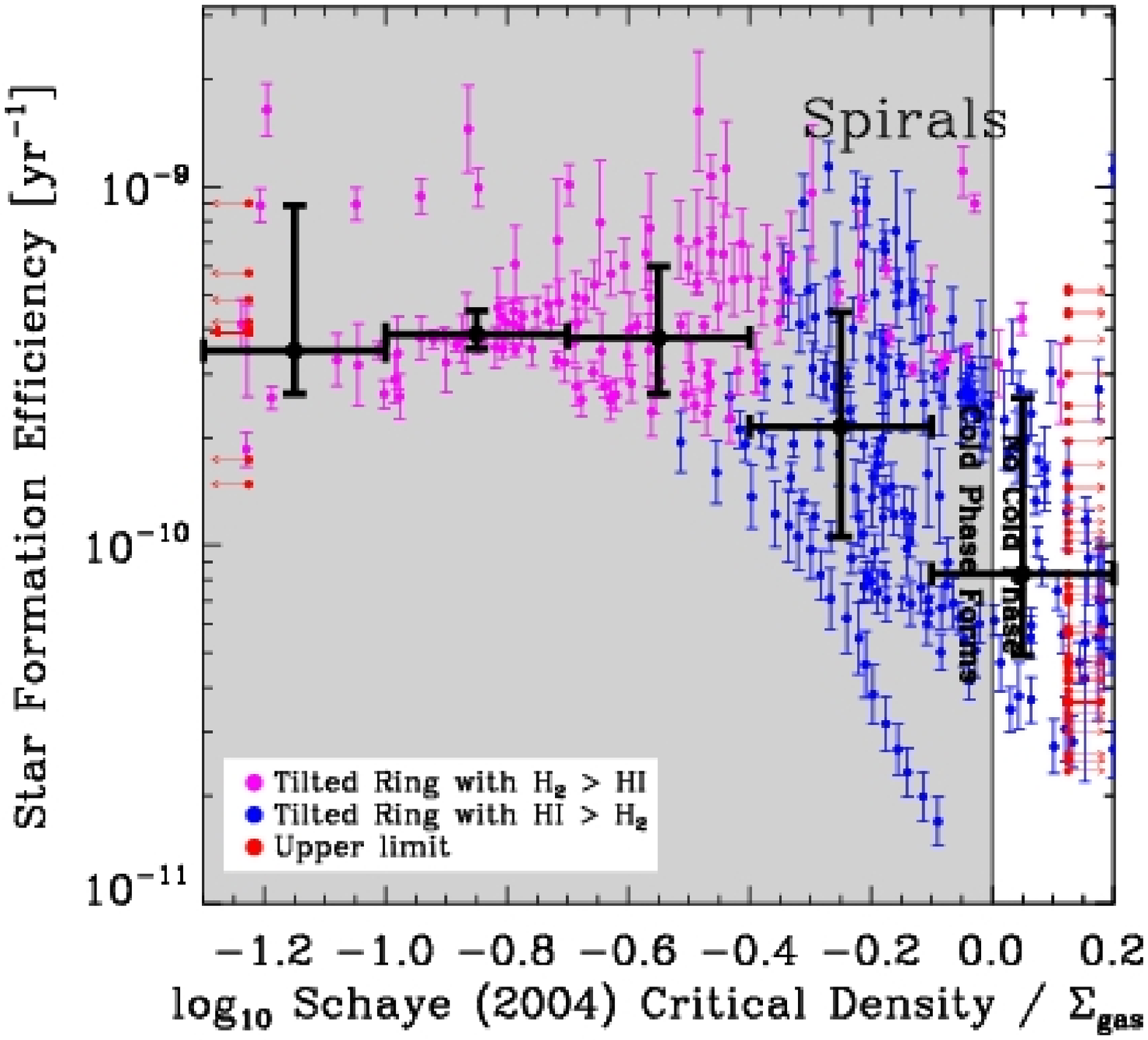}{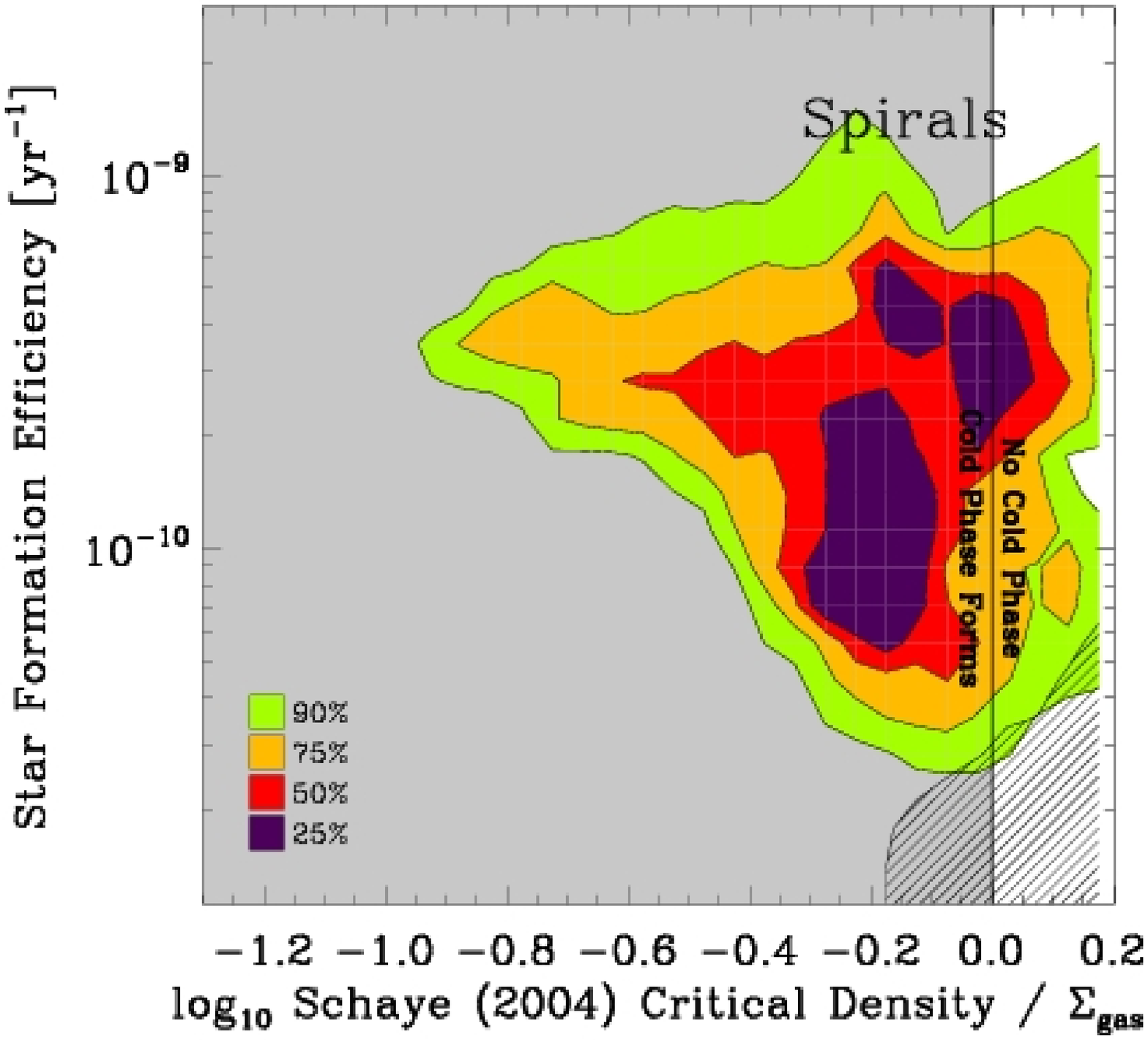}
  \plottwo{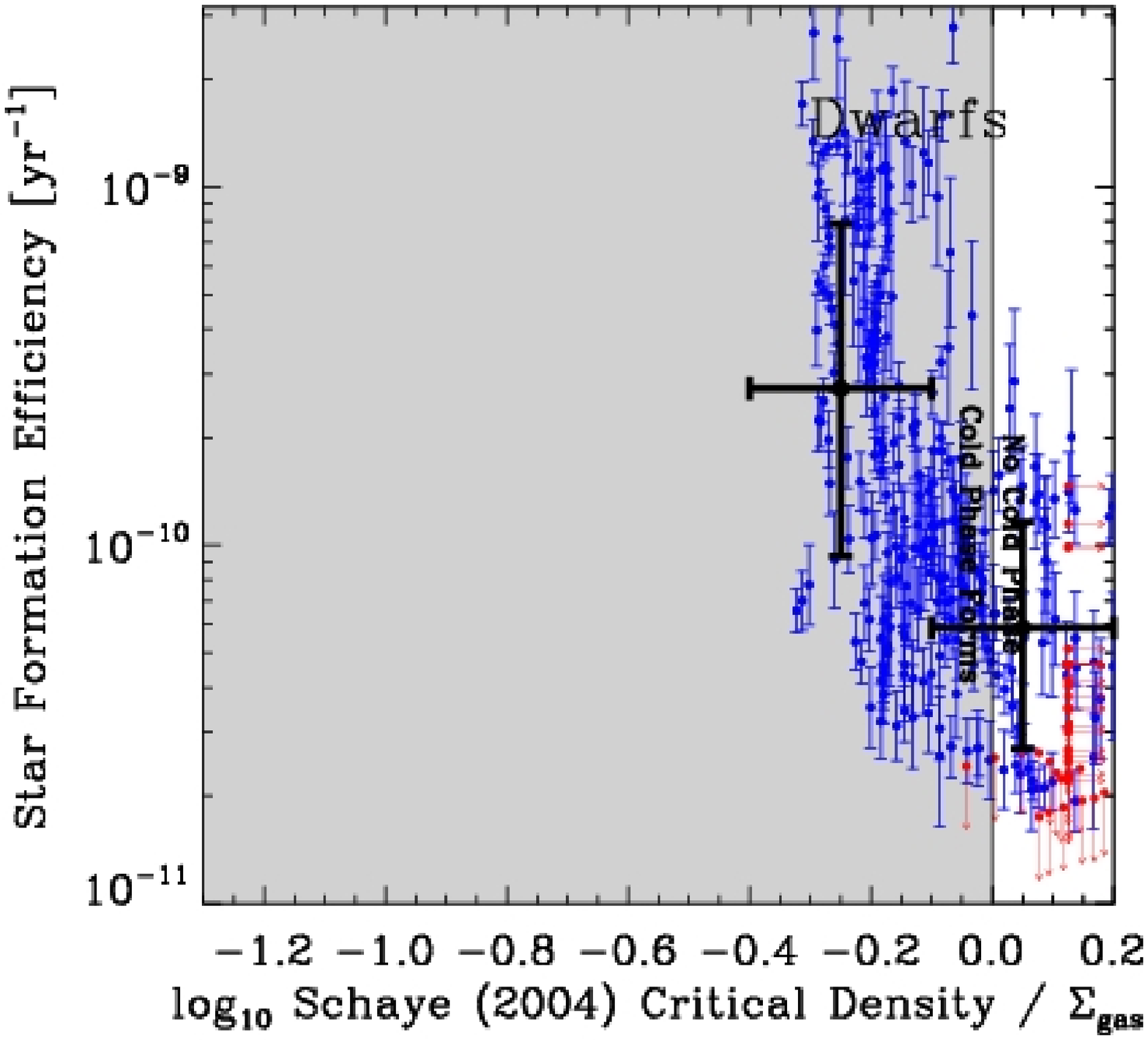}{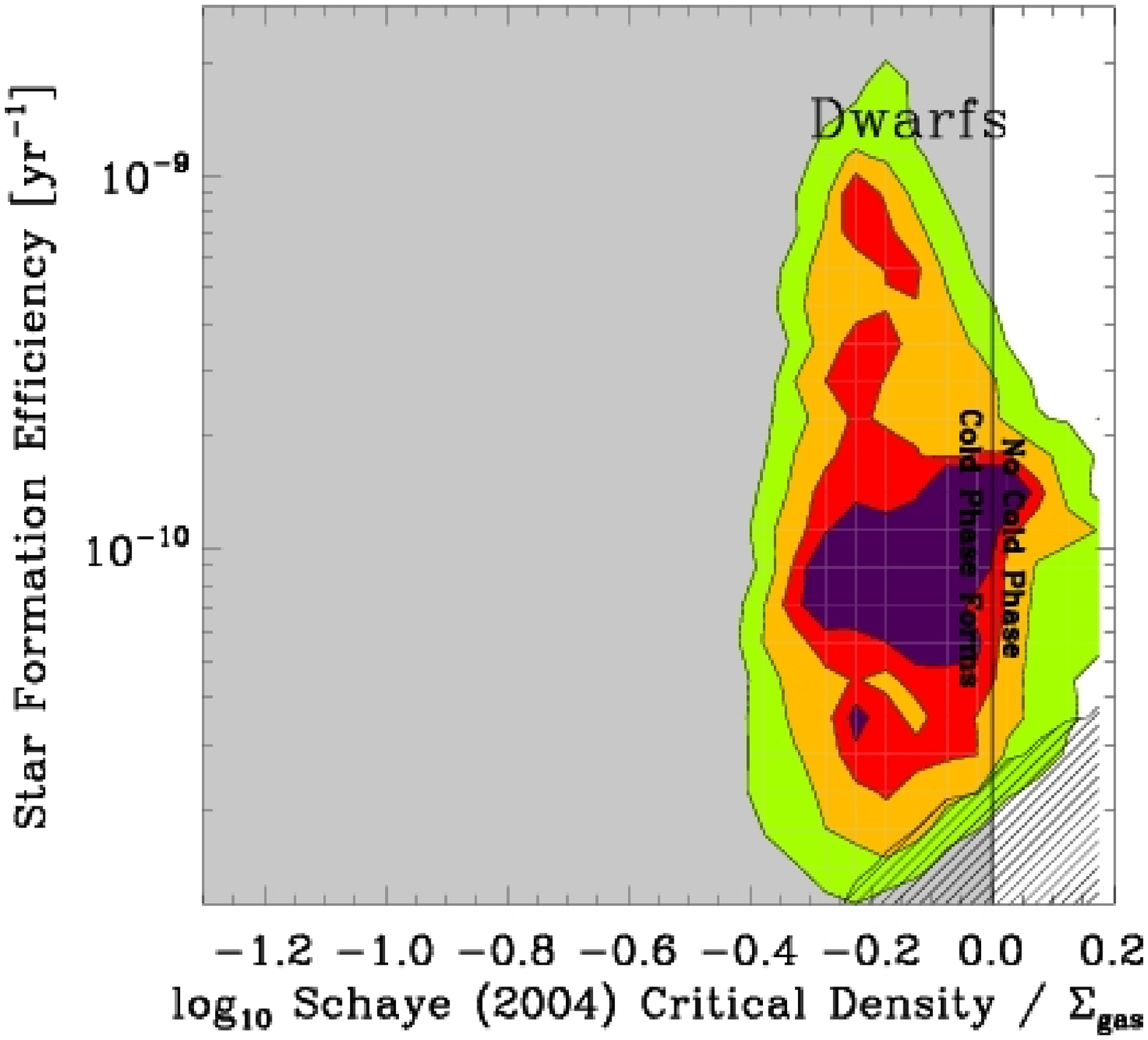}
\end{center}
\caption{\label{SFEVSSCHAYE} SFE as a function of $\Sigma_{\rm
    S04}/\Sigma_{\rm gas}$, the threshold for the formation of a cold phase
  \citet{SCHAYE04}, for spiral (top row) and dwarf galaxies (bottom row).
  Conventions and symbols follow Figure \ref{SFEVSRAD}.  The gray area
  indicates where \citet{SCHAYE04} estimates that a cold phase can form. Most
  areas where we observe star formation meet this criterion and the areas that
  do not tend to have low SFE.}
\end{figure*}

Even where the ISM is stable against gravitational collapse on large scales,
star formation may still proceed if a cold (narrow-line width) phase can form
locally and thus induce gravitational instability in a fraction of the gas
(recall the effect of lower $\sigma_{\rm gas}$ in Figure \ref{SFEVSQGAS}).
\citet{SCHAYE04} argued that this is the usual path to star formation in the
outer parts of galaxies and modeled the critical gas surface density for such
a phase to form, $\Sigma_{\rm S04}$. The bottom right panels in Figures
\ref{THRESHVSRADSPIRALS} and \ref{THRESHVSRADDWARFS} show $\Sigma_{\rm
  S04}/\Sigma_{\rm gas}$ as a function of radius and Figure \ref{SFEVSSCHAYE}
shows the SFE as a function of this ratio. The gray area in both figures shows
where a cold phase can form.

We calculate $\Sigma_{\rm S04}$ from Equation \ref{SCHAYEEQ}, which depends on
$I / [10^6$~cm$^{-2}$~s$^{-1}]$, the flux of ionizing photons. In outer disks,
we assume $I = 10^6$~cm$^{-2}$~s$^{-1}$, \citeauthor{SCHAYE04}'s fiducial
value, and in inner disks we take $I \propto \Sigma_{\rm SFR}$,

\begin{equation}
  I \approx 10^6~{\rm cm^{-2}~s^{-1}}~\left(\frac{\Sigma_{\rm SFR}}{5 \times 10^4~{\rm M}_{\odot}~{\rm
        yr}^{-1}~{\rm kpc}^{-2}}\right)~.
\end{equation}

\noindent The normalization is the average $\Sigma_{\rm SFR}$ between
$0.8$--$1.0~r_{25}$ in our spiral subsample.

Equation \ref{SCHAYEEQ} also accounts for variations about
\citeauthor{SCHAYE04}'s fiducial metallicity $Z=0.1~Z_\odot$, typical for the
outer disk of a spiral. We lack estimates of $Z$ and so neglect this term but
note the sense of the uncertainty. Inner galaxy disks will tend to have higher
metallicities, which will lower $\Sigma_{\rm S04}$.  We already find
$\Sigma_{\rm gas} > \Sigma_{\rm S04}$ over most inner disks; therefore missing
$Z$ seems unlikely to seriously bias our results.

Figures \ref{THRESHVSRADSPIRALS}, \ref{THRESHVSRADDWARFS}, and
\ref{SFEVSSCHAYE} show that we expect a cold phase over most of the disk in
both spiral and dwarf galaxies. In our spiral subsample, most data inside
$r_{\rm gal} \sim 0.9~r_{25}$ meet this criterion. Because most subcritical
data come from large radii, we also find that most lines of sight with
$\Sigma_{\rm gas} < \Sigma_{\rm S04}$ exhibit low SFEs or upper limits.

Because most data are supercritical, the \cite{SCHAYE04} threshold is of
limited utility for predicting the SFE {\em within} a galaxy disk.
\citet{SCHAYE04} does not predict the ratio of \htwo --to--\hi\ where cold gas
forms; he is primarily concerned with the edges of galaxies. Figures
\ref{THRESHVSRADSPIRALS} and \ref{SFEVSSCHAYE} broadly confirm that his
proposed threshold matches both the edge of the optical disk and the typical
threshold found by \citet{MARTIN01}.

This relevance of this comparison to the SFE within the optical disk is that
based on the \citet{SCHAYE04} model, we expect a widespread narrow--line phase
throughout most of our galaxies \citep[][obtain a similar result for the Milky
Way]{WOLFIRE03}. This suggests that cold phase formation followed by collapse
may be a common path to star formation and offers a way to form stars in our
otherwise stable disks.

\section{Discussion}
\label{SECT_DISCUSSION}

In \S \ref{SECT_SFEOBS}, \ref{SECT_SFELAW} and \ref{SECT_SFETHRESH} we
examined the SFE as a function of basic physical parameters, laws, and
thresholds. Here we collect these results into general conclusions
regarding the SFE in galaxies and identify key elements of a
successful theory of star formation in galaxies.

\subsection{Fixed SFE of \htwo}
\label{FIXEDHTWO_SECT}

\begin{figure*}
  \begin{center}
  \plotone{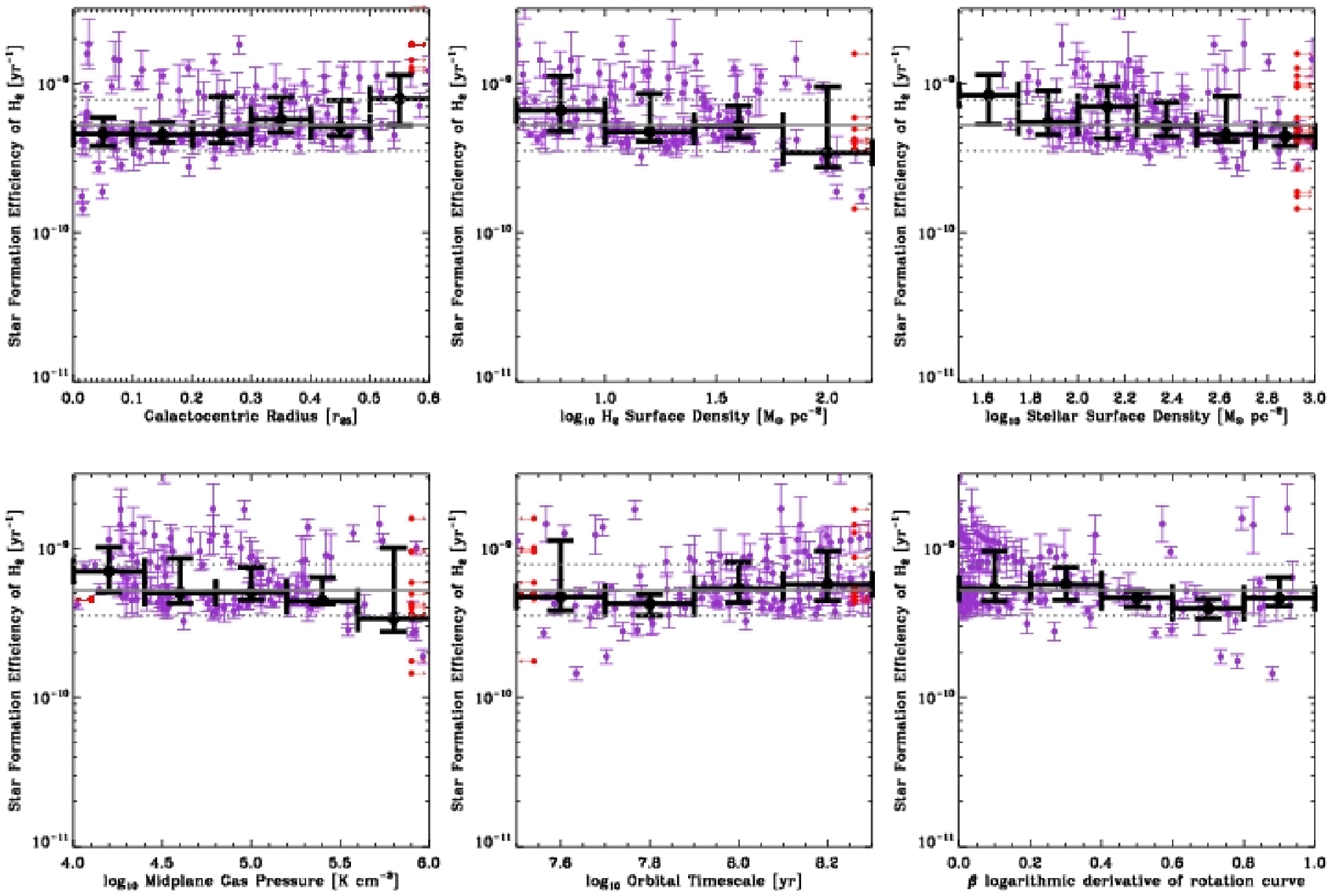}
\end{center}
\caption{\label{SFEH2FIG} SFE~(\htwo ) in individual tilted rings from spiral
  galaxies as a function of: ({\em top left}) galactocentric radius, ({\em top
    middle}) \htwo\ surface density, ({\em top right}) stellar surface
  density, ({\em bottom left}) midplane pressure, ({\em bottom middle})
  orbital timescale, ({\em bottom right}) logarithmic derivative of the
  rotation curve. Gray lines show the median $\log_{10} {\rm SFE~(\htwo )} =
  -9.28 \pm 0.17$ for our data.}
\end{figure*}

Using a data set that overlaps the one presented here, \citet{BIGIEL08} found
a linear relationship between $\Sigma_{\rm SFR}$ and $\Sigma_{\rm H2}$. Here
we extend that finding: {\em Where the ISM is mostly \htwo\ in spiral
  galaxies, the SFE does not vary strongly with any of the quantities that we
  consider}, including radius, $\Sigma_{\rm gas}$, $\Sigma_{\rm *}$, $P_{\rm
  h}$, $\tau_{\rm orb}$, and $\beta$. We plot SFE~(\htwo ) as a function of
each of these quantities in Figure \ref{SFEH2FIG}. The median value for tilted
rings from our spiral subsample is $\log_{10} {\rm SFE~(\htwo )} = -9.28 \pm
0.17$, i.e.,

\begin{equation} 
  {\rm SFE}~(\htwo )~= 5.25 \pm 2.5 \times 10^{-10}~{\rm
    yr}^{-1}~.
\end{equation}

Constant SFE~(\htwo ) might be expected if 1) conditions within a GMC, rather
than the larger scale properties of the ISM, drive star formation
\citep[e.g.,][]{KRUMHOLZ05} and 2) GMC properties are relatively universal
rather than, e.g., a sensitive function of formation mechanism or environment.
This appears to be the case in the inner Milky Way (excluding the Galactic
center) and in M31 and M33, where GMC properties are largely a function of
cloud mass alone
\citep[][]{SOLOMON87,ROSOLOWSKY03,ROSOLOWSKY07,BLITZ07,BOLATTO08}.  The
constancy of SFE~(\htwo ) hints that a similar case holds in our spiral
subsample.

Figure \ref{SFEVSPRESS} illustrates why (relatively) universal GMC properties
may be plausible in our sample. From Equation \ref{PRESSEQ}, the internal
pressure of a starless GMC with $\Sigma_{\rm gas} \approx
170$~M$_{\odot}$~pc$^{-2}$ \citep[][]{SOLOMON87} is $P_{\rm h}/k_{\rm B} \sim
10^6~{\rm K~cm}^{-3}$. This is the highest value we plot in Figures
\ref{SFEVSPRESS} and \ref{SFEH2FIG} and only a small fraction of our data have
higher $P_{\rm h}$ so that even where the ISM is mostly \htwo , $P_{\rm h}$ is
usually well below the typical internal pressure of a GMC. Thus GMCs are not
necessarily pressure-confined, which allows the possibility of bound, isolated
GMCs out of pressure equilibrium with the rest of the ISM.  In this case, the
environmental factors that we consider may never be communicated to GMCs
(though some mechanism may still be needed to damp out any imprint left by
environment during GMC formation).

The range of $P_{\rm h}$ in our sample also underscores that one should not
expect a constant SFE~(\htwo ) to extend to starburst conditions, where
$P_{\rm h}$ and $\Sigma_{\rm gas}$ on kiloparsec scales exceed those found for
individual Galactic GMCs and SFE~(\htwo ) {\em is} observed to vary strongly
with local conditions \citep[e.g.,][]{KENNICUTT98A,RIECHERS07}.

Another important caveat is that the distribution of GMC masses is observed to
vary with environment \citep{ROSOLOWSKY05}, possibly as a result of varying
formation mechanisms. This suggests that either the SFE of a GMC is only a
weak function of its mass (and thus other properties) or that real variations
in SFE~(\htwo ) may exist in dwarf galaxies and the outskirts of spirals.

\subsection{Conditions at the \hi -to--\htwo\ Transition in Spirals}
\label{TRANS_SECT}

In spiral galaxies, the transition between an \hi --dominated ISM and a
mostly--\htwo\ ISM occurs at a characteristic value for most quantities. This
can be seen from Figures \ref{SFEVSRAD}, \ref{SFEVSSTARS}, \ref{SFEVSGAS},
\ref{SFEVSPRESS}, and \ref{SFEVSTORB}, in which \hi -dominated regions (blue
points) typically occupy one region and \htwo -dominated regions (magenta
points) occupy another.

Table \ref{TRANSITIONTAB} gives our estimates of properties where $\Sigma_{\rm
  HI} \approx \Sigma_{\rm H2}$ in spiral galaxies. For each galaxy, we measure
the median of the property in question over all pixels where $\Sigma_{\rm H2}
= 0.8$ -- $1.2~\Sigma_{\rm HI}$. Table \ref{TRANSITIONTAB} lists the median
transition value in our spiral subsample, along with the ($1\sigma$) scatter
and log scatter among galaxies. These values appear as dotted vertical lines
in Figures \ref{SFEVSRAD}, \ref{SFEVSGAS}, \ref{SFEVSSTARS}, \ref{SFEVSPRESS},
and \ref{SFEVSTORB}. Note that methodology --- the choice to use pixels or
rings, to interpolate, use the mean or median, etc. --- affects the values in
Table \ref{TRANSITIONTAB} by $\sim 20\%$.

From Table \ref{TRANSITIONTAB}, we find that physical conditions at the \hi
--to--\htwo\ transition are fairly similar to those found in the solar
neighborhood.  The orbital time is $\approx 1.8 \times 10^8$~years and the
free--fall time in the gas disk is $\approx 4.2 \times 10^{7}$~years. The
midplane gas pressure is $P_{\rm h} / k_{\rm B} \approx 2.3 \times
10^4$~cm$^{-3}$~K, corresponding to a particle density $n \sim 1$~cm$^{-3}$.
The baryon mass budget in the disk is dominated by stars, $\Sigma_{*} \approx
81$~M$_{\odot}$~pc$^{-2}$ while $\Sigma_{\rm gas} \approx
14$~M$_\odot$~pc$^{-2}$. Accordingly, the gas is stable against large scale
gravitational collapse on its own ($Q_{\rm gas} \approx 3.8$), but in the
presence of stars is only marginally stable $Q_{\rm stars+gas} \sim 1.6$.

Approximately $1\%$ of gas is converted to stars per free fall time at the
transition, in agreement with expectations by \citet[][]{KRUMHOLZ05}. About
$6\%$ of gas is converted to stars per $\tau_{\rm orb}$. This agrees well with
the disk--averaged value of $\sim 7\%$ derived by \citet{KENNICUTT98A}
(adapted to our IMF and CO-to-\htwo\ conversion factor) and with the range of
efficiencies found by \citet{WONG02}.

\begin{deluxetable}{l c c c}
  \tabletypesize{\small} \tablewidth{0pt} \tablecolumns{4}
  \tablecaption{\label{TRANSITIONTAB} Conditions at the \hi -to-\htwo\
    Transition} \tablehead{\colhead{Quantity} & \colhead{Median} &
    \colhead{Scatter} & \colhead{Scatter} \\
    & Value\tablenotemark{a} & & in log$_{10}$ } \startdata
  $r_{\rm gal}$ [$r_{25}$] & $0.43$ & $0.18$ & $0.17$ \\
  $\Sigma_{\rm *}$ [M$_{\odot}$~pc$^{-2}$] & $81$ & $25$ & $0.15$ \\
  $\Sigma_{\rm gas}$ [M$_{\odot}$~pc$^{-2}$]  & $14$ & $6$ & $0.18$ \\
  $P_{h}/k_{\rm B}$ [cm$^{-3}$~K] & $2.3 \times 10^4$ & $1.5 \times 10^4$ &
  $0.26$ \\
  $\tau_{\rm ff}$ [yr] & $4.2 \times 10^7$ & $1.2 \times 10^7$ & $0.14$ \\
  $\tau_{\rm orb}$ [yr] & $1.8 \times 10^8$ & $0.4 \times 10^8$ & $0.09$ \\
  $Q_{\rm gas}$ & $3.8$ & $2.6$ & $0.31$ \\
  $Q_{\rm stars+gas}$ & $1.6$ & $0.4$ & $0.09$ \\
  \enddata
  \tablenotetext{a}{Median value in the spiral subsample.}
\end{deluxetable}

\subsection{H$_2$ in Dwarf Galaxies}
\label{SECT_MISSINGH2}

Because of uncertainties in \xco , we do not directly estimate the amount of
\htwo\ in dwarf galaxies . However, indirect evidence suggests that a
significant part of the ISM is \htwo\ in the central parts of these galaxies.
Specifically, we observe very high SFE in the centers of dwarf galaxies ---
higher than SFE~(\htwo ) in spirals --- often under conditions associated with
an \htwo -dominated ISM in spirals (\S \ref{TRANS_SECT}).  It would be
surprising if the SFE of \hi\ in dwarfs indeed exceeds SFE~(\htwo ) in
spirals. We argue that an unaccounted--for reservoir of \htwo\ is a more
likely explanation.

The SFE~(\htwo ) that we observe in spiral galaxies offers an approximate way
to estimate how much \htwo\ may be present. If we assume that SFE~(\htwo ) is
the same in dwarf and spiral galaxies then we can calculate $\Sigma_{\rm H2}$
from the observed $\Sigma_{\rm SFR}$ via

\begin{equation}
  \label{H2FROMSFR}
  \Sigma_{\rm H2} \approx \frac{10^{-6}~\Sigma_{\rm SFR}}{5.25 \times 10^{-10}~{\rm yr}^{-1}}~.
\end{equation}

This treatment suggests that in our typical dwarf galaxies, most of the ISM is
\htwo\ within $\sim 0.25~r_{25}$. This may be seen directly from Equation
\ref{DWARFSFEFIT2}, which translates our fit of SFE to radius to a relation
between $R_{\rm mol}$ and radius assuming Equation \ref{H2FROMSFR}. From
Equation \ref{DWARFSFEFIT2}, $\Sigma_{\rm H2} / \Sigma_{\rm HI}$ in dwarf
galaxies is $1$--$2$ inside $\sim 0.25~r_{25}$, rising as high as $\sim 3$ at
$r_{\rm gal} = 0$.

\subsection{Environment-Dependent GMC/\htwo\ Formation}
\label{GMCFORM_SECT}

\begin{figure*}
  \begin{center}
    \plottwo{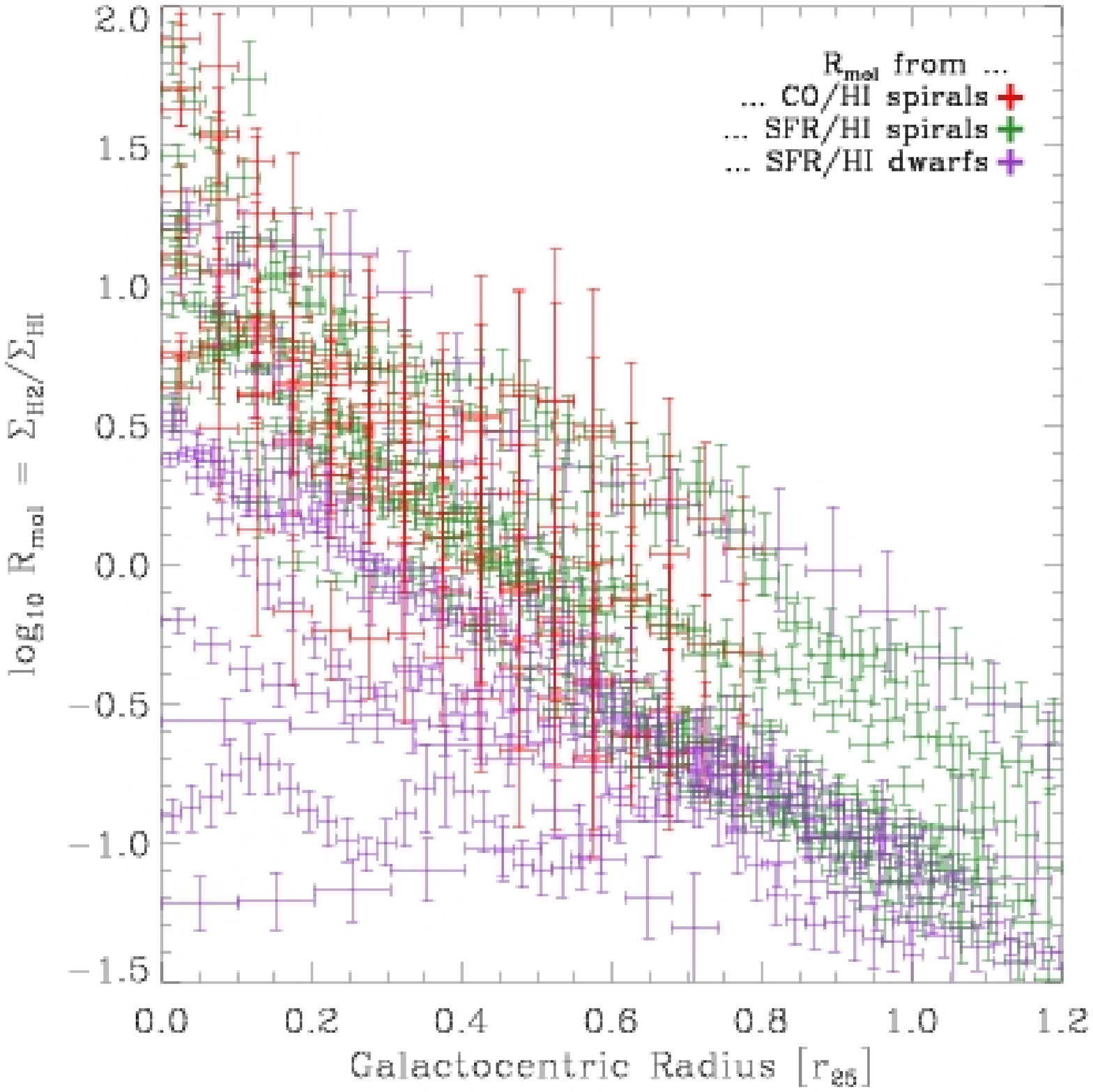}{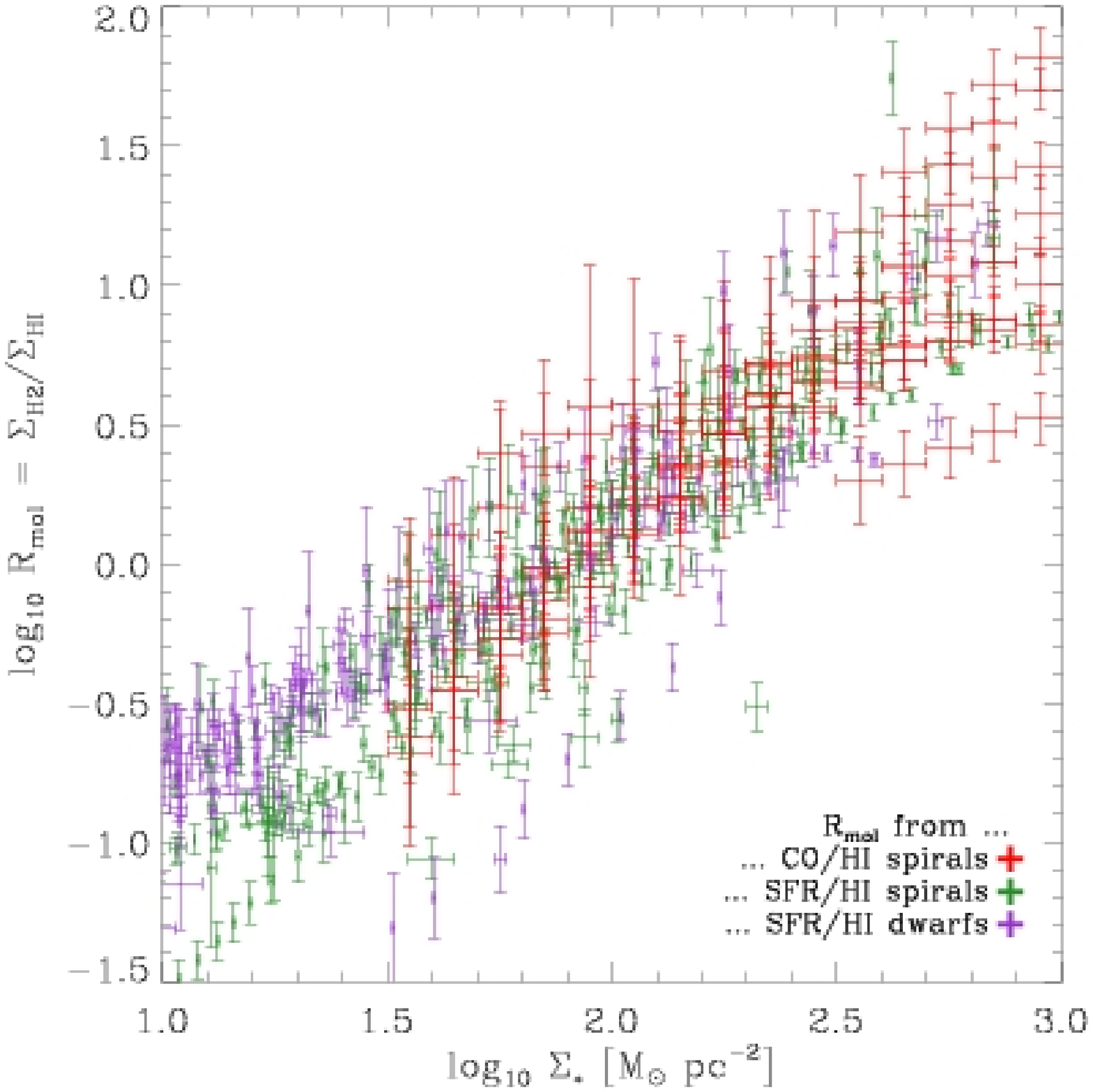}
    \plottwo{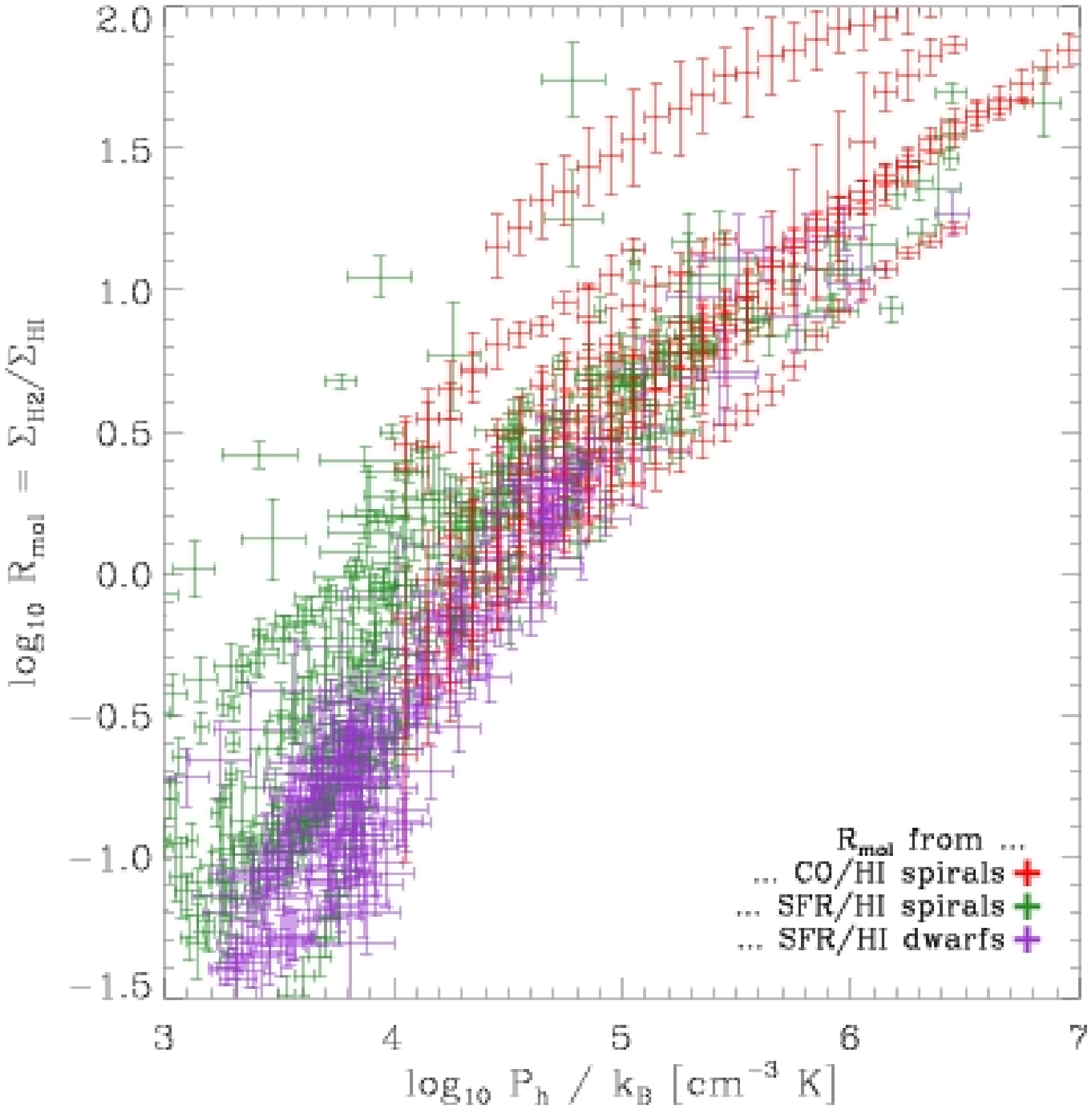}{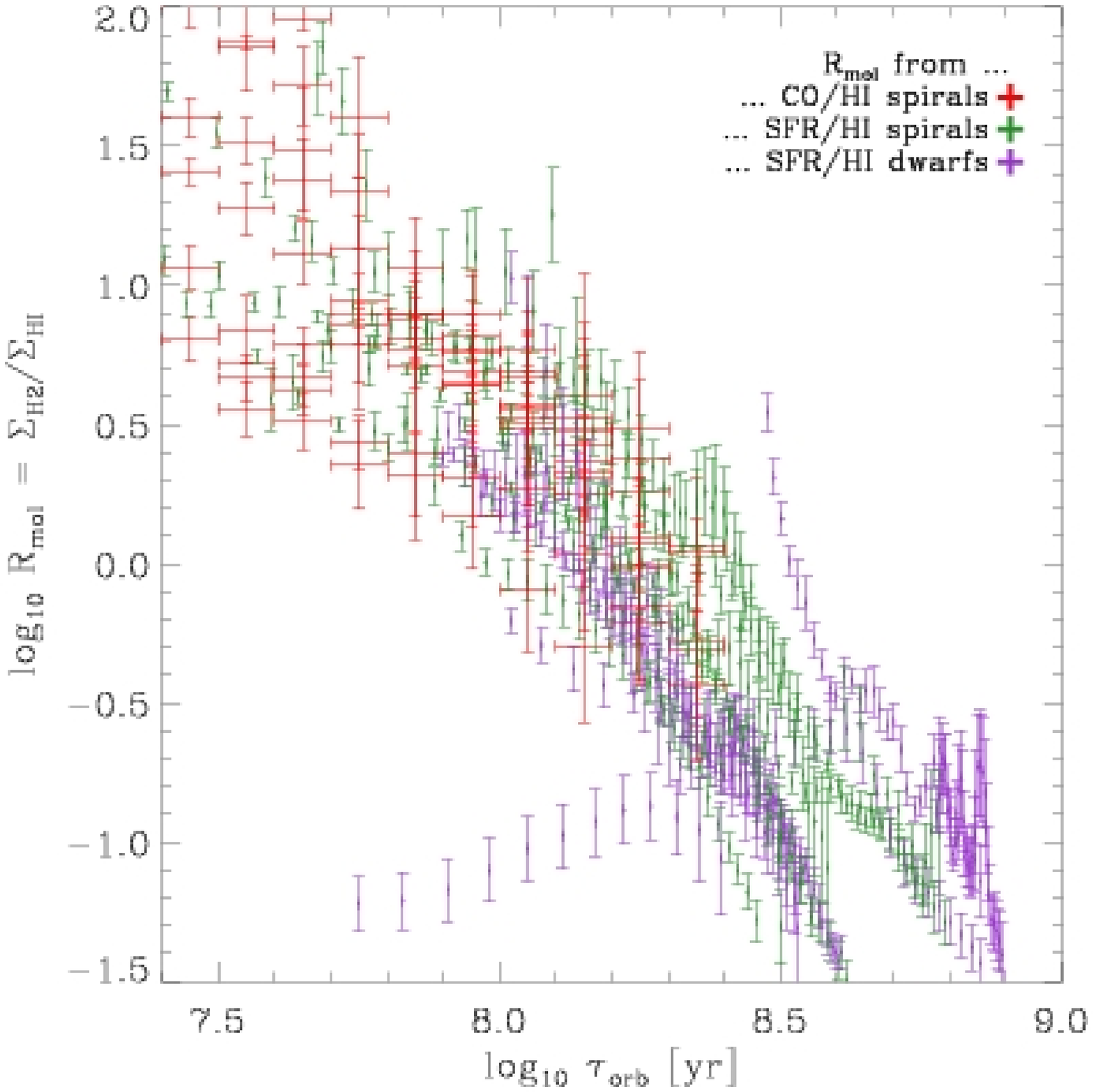}
    \end{center}
    \caption{\label{RMOLVSXXX} The \htwo --to--\hi\ ratio, $R_{\rm mol} =
      \Sigma_{\rm H2} / \Sigma_{\rm HI}$, as a function of ({\em top left})
      radius, ({\em top right}) $\Sigma_*$, ({\em bottom left}) $P_{\rm h}$
      ($\propto \tau_{\rm ff}^{-2}$), and ({\em bottom right}) $\tau_{\rm
        orb}$. Red points are pixel--by--pixel measurements of $R_{\rm mol}$
      in spirals, binned by the quantity on the $x$-axis.  Green and purple
      points show tilted rings in spiral and dwarf galaxies with $R_{\rm mol}$
      inferred from $\Sigma_{\rm SFR}$ and $\Sigma_{\rm HI}$ assuming a fixed
      SFE~(\htwo ). We show the same data, binned, in Figure \ref{RMOLVSFIT}.}
\end{figure*}

\begin{figure*}
  \begin{center}
    \plottwo{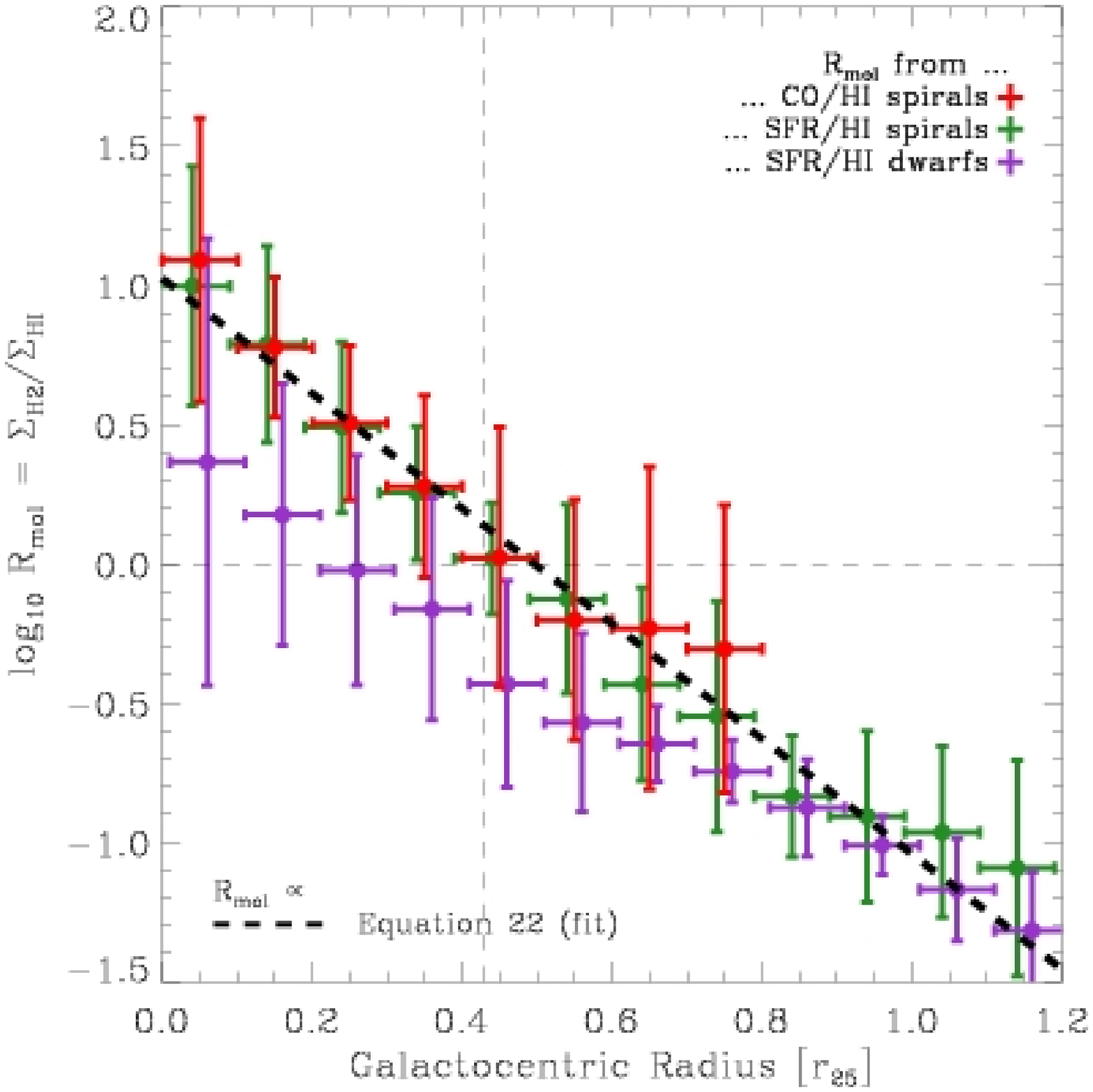}{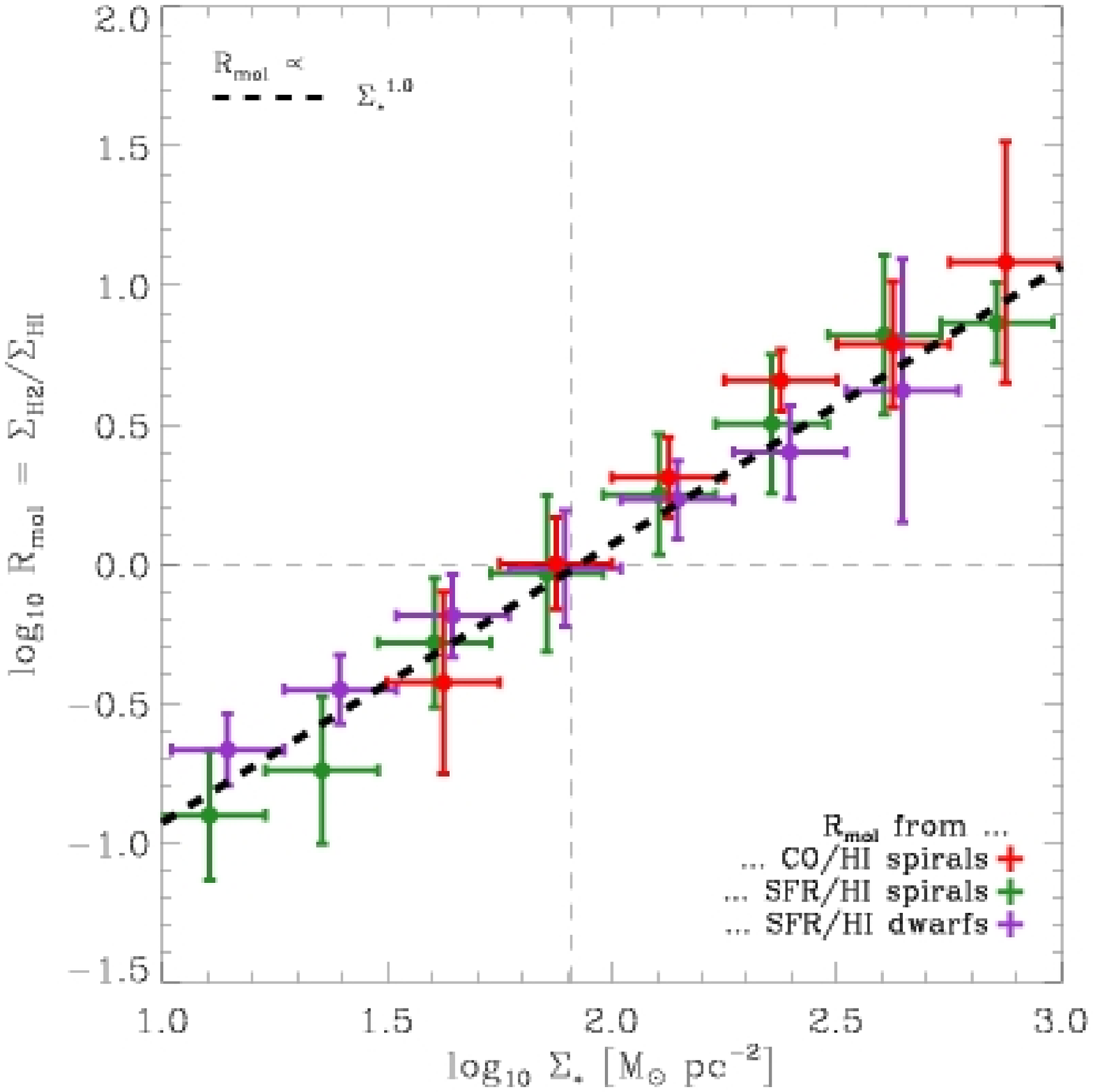}
    \plottwo{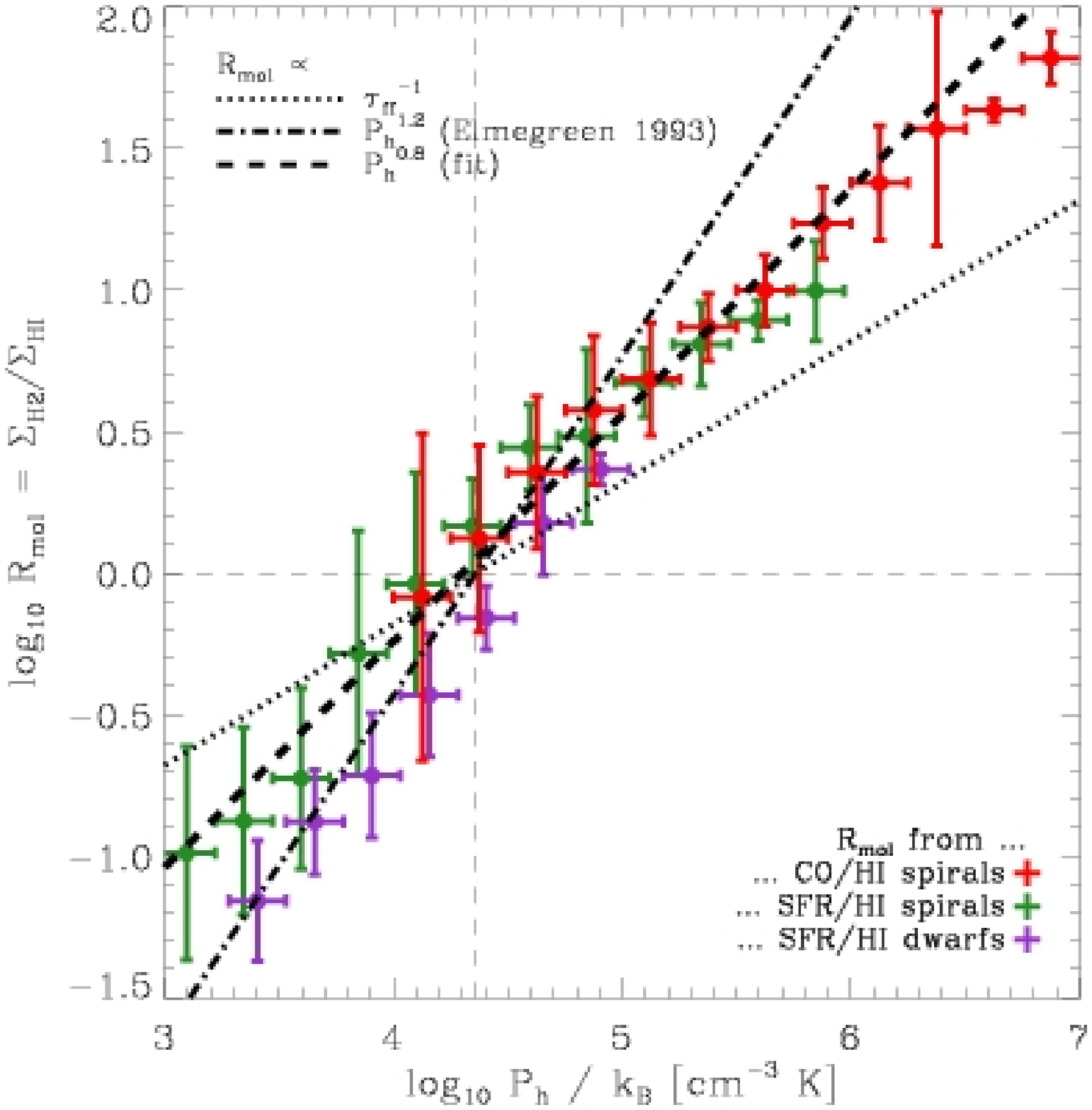}{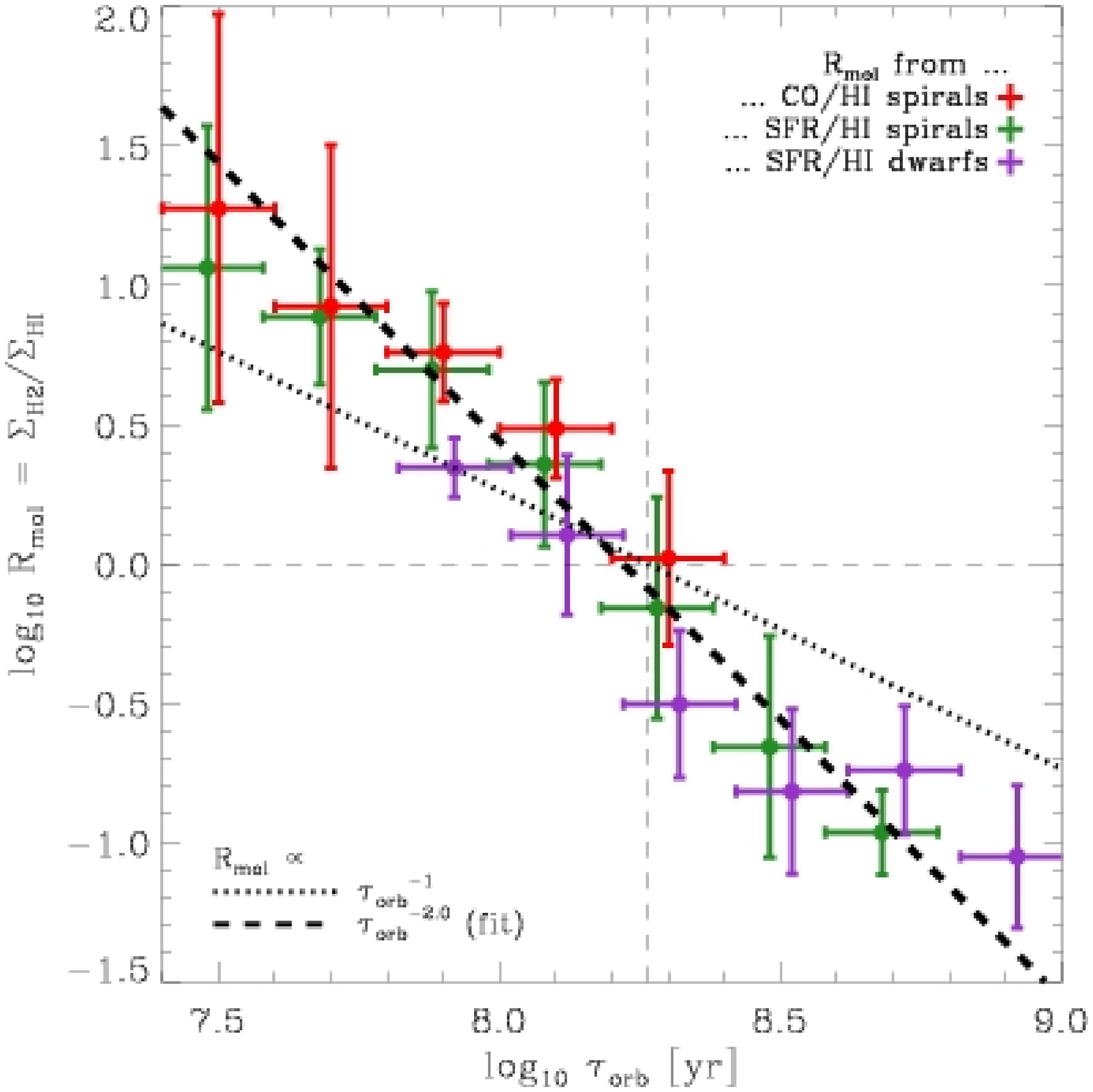}
    \end{center}
    \caption{\label{RMOLVSFIT} The data from Figure \ref{RMOLVSXXX}, binned by
      the quantity on the $x$--axis into three trends: $R_{\rm mol}$ measured
      pixel--by--pixel in spirals (red) and inferred from $\Sigma_{\rm SFR}$
      and $\Sigma_{\rm HI}$ in (green) spiral and (purple) dwarf galaxies.
      Thin dashed lines show $R_{\rm mol} = 1$ (horizontal) and our estimate
      of each quantity at the \hi --to--\htwo\ transition (vertical). Dotted
      lines show $R_{\rm mol} \propto \tau_{\rm ff}^{-1}$ (bottom left) and
      $R_{\rm mol} \propto \tau_{\rm orb}^{-1}$ (bottom right). Thick dashed
      lines show fits of $R_{\rm mol}$ to each quantity.}
\end{figure*}

Where the ISM is \hi --dominated --- in dwarf galaxies and outside the \hi
-to-\htwo\ transition in spirals --- the SFE declines steadily with increasing
radius. In this regime, the SFE is covariant with a number of environmental
factors, including $\Sigma_*$, pressure, density, free fall time, and orbital
timescale. This observation, together with those in \S \ref{FIXEDHTWO_SECT}
and \ref{TRANS_SECT}, implies that while star formation {\em within} GMCs is
largely decoupled from environment, {\em the formation of H$_2$ / GMCs from
  \hi\ depends sensitively on local conditions}.

In this case, we can break the SFE into two parts: star formation
within GMCs and GMC formation, so that

\begin{equation}
  \label{SFESPLITEQ}
  {\rm SFE} = {\rm SFE}\left( {\rm H}_2
  \right)\frac{\Sigma_{\rm H2}}{\Sigma_{\rm gas}} = {\rm SFE}\left( {\rm
      H}_2 \right)\frac{R_{\rm mol}}{R_{\rm mol}+1}~,
\end{equation}

\noindent i.e., the SFE is a product of a constant SFE (\htwo ) and $R_{\rm
  mol} = \Sigma_{\rm H2}/\Sigma_{\rm HI}$, which is a function of local
conditions.

We show this directly in Figure \ref{RMOLVSXXX} and plot the same data,
binned, in Figure \ref{RMOLVSFIT}. We plot $R_{\rm mol} = \Sigma_{\rm
  H2}/\Sigma_{\rm HI}$ on the $y$-axis as a function of radius, $\Sigma_*$,
$P_{\rm h}$ ($\propto \tau_{\rm ff}^{-2}$), and $\tau_{\rm orb}$. Red points
show direct measurements of $R_{\rm mol}$ from CO and \hi\ assembled following
the methodology used by \citet{BLITZ06} to compute $R_{\rm mol}$ as a function
of $P_{\rm h}$:

\begin{enumerate}
\item For each galaxy, we examine scatter plots to estimate a value of $P_{\rm
    h}$ above which our pixel-by-pixel measurements of $\Sigma_{\rm H2}$ are
  approximately complete.
\item Where $P_{\rm h}$ is above this limit, we measure $R_{\rm mol}$ for each
  pixel.
\item We sort pixels into bins based on $P_{\rm h}$ and calculate the average
  and scatter in $\log_{10} R_{\rm mol}$ for the pixels in each bin.
\end{enumerate}

\noindent A red point in Figure \ref{RMOLVSXXX} corresponds to one $P_{\rm h}$
bin in one spiral galaxy; the $x$-- and $y$--error bars indicate the width of
the bin and the scatter in $R_{\rm mol}$ within the bin. We carry out
analogous procedures to compute $R_{\rm mol}$ as a function of $r_{\rm gal}$,
$\Sigma_*$, and $\tau_{\rm orb}$.
 
Because of the limited sensitivity of the CO data, these direct measurements
of $R_{\rm mol}$ seldom probe far below $R_{\rm mol} = 1$ and do not extend to
dwarfs. Therefore we also use $\Sigma_{\rm SFR}$ and $\Sigma_{\rm HI}$ to
estimate $R_{\rm mol}$ by assuming a fixed SFE (\htwo ). For each tilted
ring in both subsamples, we convert $\Sigma_{\rm SFR}$ into $\Sigma_{\rm H2}$
using Equation \ref{H2FROMSFR}. We divide this by the observed $\Sigma_{\rm
  HI}$ to estimate $R_{\rm mol}$ for that ring. We plot the results as green
points for spirals and purple points for dwarf galaxies\footnote{Because
  $P_{\rm h}$ depends on $\Sigma_{\rm gas}$, we make a first--order correction
  to $P_{\rm h}$ in dwarf galaxies based on the estimated $R_{\rm mol}$.}.
This approach --- essentially plotting SFE~(\hi ) in units of $R_{\rm mol}$
--- allows us to estimate $R_{\rm mol}$ far below the sensitivity of our CO
maps.  While this extrapolation of SFE~(\htwo ) may be aggressive, the
quantity $\Sigma_{\rm SFR} / \Sigma_{\rm HI}$ must be closely related to the
ability of \hi\ to assemble into star--forming clouds.

Figure \ref{RMOLVSFIT} shows the data in Figure \ref{RMOLVSXXX} binned by the
quantity on the $x$-axis. Thin dashed lines horizontal show $R_{\rm mol} = 1$,
i.e., $\Sigma_{\rm HI} = \Sigma_{\rm H2}$, and the value of the property on
the $x$-axis that we estimate at the \hi --to--\htwo\ transition (\S
\ref{TRANS_SECT} and Table \ref{TRANSITIONTAB}). Dashed and dotted lines show
fits and expectations that we discuss later in this section.

In spirals, the agreement between direct measurements of $R_{\rm mol}$ and
estimates based on $\Sigma_{\rm SFR}$ and $\Sigma_{\rm HI}$ is quite good.
There is also general agreement between spirals and dwarf galaxies: the two
subsamples sweep out similar, though slightly offset, trends in all four
panels. The magnitude of the offsets between dwarf and spiral galaxies that we
see in Figure \ref{RMOLVSFIT}, typically $0.2$--$0.3$~dex, offers indirect
evidence that differences between the subsamples --- metallicity, radiation fields,
spiral structure (\S \ref{SAMPLE_SECT}) --- affect cloud formation or SFE
(\htwo ) at the factor of $\sim 2$--$3$ level.

Figures \ref{RMOLVSXXX} and \ref{RMOLVSFIT} show explictly what we have
already seen indirectly throughout \S \ref{SECT_RESULTS}. $R_{\rm mol}$ is a
continuous function of environment spanning from the \htwo --dominated
($R_{\rm mol} \sim 10$) to \hi --dominated ($R_{\rm mol} \sim 0.1$) ISM, from
inner to outer galaxy disks, and over a wide range of ISM pressures. This
qualitatively confirms and extends similar findings by \citet{WONG02} and
\citet{BLITZ06}, which were mostly confined to the inner, molecule--dominated
parts of spirals.

\subsubsection{Cloud Formation Timescales}
\label{SECT_TIMES}

In \S \ref{SECT_BACKGROUND}, we discuss two basic ways that $R_{\rm mol}$
might be set by environment. First, the timescale to form GMCs may depend on
local conditions. If \hi\ and \htwo\ are in approximate equilibrium, with the
entire neutral ISM actively cycling between these two phases, then

\begin{equation}
\label{RMOLEQ}
R_{\rm mol} = \frac{\Sigma_{\rm H2}}{\Sigma_{\rm HI}} \approx
\frac{{\rm GMC~lifetime}}{\tau \left( {\rm \hi}\rightarrow{\rm
      \htwo}\right)}~.
\end{equation}

\noindent For constant GMC lifetimes --- perhaps a reasonable extension of
fixed SFE~(\htwo ) --- $R_{\rm mol}$ is set by $\tau \left( {\rm
    \hi}\rightarrow{\rm \htwo}\right)$. In \S \ref{DENSITY_SECT} and \S
\ref{KIN_SECT} we saw that SFE anti-correlates with $\tau_{\rm ff}$ and
$\tau_{\rm orb}$ where $\Sigma_{\rm HI} > \Sigma_{\rm H2}$. If GMCs form over
these timescales then $R_{\rm mol} \propto \tau_{\rm ff}^{-1}$ or $R_{\rm mol}
\propto \tau_{\rm orb}^{-1}$.

However, we found that the SFE decreased more steeply than one
would expect if these timescales alone dictated $R_{\rm mol}$, so that
increasing timescale for GMC formation cannot explain all of the decline in
$R_{\rm mol}$. Figures \ref{RMOLVSXXX} and \ref{RMOLVSFIT} show this directly:
dotted lines in the bottom two panels illustrate $R_{\rm mol} \propto
\tau_{\rm ff}^{-1}$ and $R_{\rm mol} \propto \tau_{\rm orb}^{-1}$. In both
cases the prediction is notably shallower than the data in both the \htwo\ and
\hi --dominated regimes.

\subsubsection{Disk Stability Thresholds}
\label{SECT_THRESHDISC}

Of course, the entire ISM may not participate in cloud formation.  Star
formation thresholds are often invoked to explain the decrease in SFE between
inner and outer galaxy disks. The amount of stable, warm \hi\ may depend on
environment, with a variable fraction of the disk actively cycling between
\hi\ and GMCs. This suggests a straightforward extension of Equation
\ref{RMOLEQ},

\begin{equation}
\label{RMOLPLUSTHRESH}
R_{\rm mol} = \frac{\Sigma_{\rm H2}}{\Sigma_{\rm HI}} \approx
\frac{{\rm GMC~lifetime}}{\tau \left( {\rm \hi}\rightarrow{\rm
      \htwo}\right)} \times f_{\rm GMC~forming}~,
\end{equation}

\noindent which again balances GMC formation and destruction but now includes
the factor $f_{\rm GMC~forming}$ to represent the fact that only a fraction of
the \hi\ is actively cycling between the molecular and atomic ISM.

We considered three thresholds in which large--scale instabilities dictate
$f_{\rm GMC~forming}$ --- $Q_{\rm gas}$, $Q_{\rm stars+gas}$, and shear. One
would naively expect these thresholds to correspond to $f_{\rm GMC~forming}
\sim 1$ for supercritical gas and $f_{\rm GMC~forming} \ll 1$ for subcritical
gas, yielding a step function in SFE or $R_{\rm mol}$. However, we do not
observe such relationships between thresholds and SFE (\S
\ref{SECT_SFETHRESH}), which agrees with \citet{BOISSIER07} who also based
their SFR profiles on extinction-corrected FUV maps and found no evidence for
sharp star formation cutoffs.

If these instabilities regulate star formation but operate below our
resolution, we still expect a correspondence between SFE and the average
threshold value, which should indicate what fraction of the ISM is unstable.
Despite this expectation, $Q_{\rm gas}$ shows little correspondence to the SFE
and almost all of our sample is stable against axisymmetric collapse.
\citet{KIM01} and \citet{KIM07} discuss $Q_{\rm gas}$ thresholds for the
growth of non-axisymmetric instabilities, but these are in the range $Q_{\rm
  gas} \sim 1$--$2$, still lower than the typical values that we observe (\S
\ref{QSECT}). Even independent of the normalization, $Q_{\rm gas}$ shows
little relation to the SFE, particularly in dwarf galaxies \citep[see
also][]{HUNTER98,WONG02,BOISSIER03}.

Including the effects of stellar gravity reduces stability. Over most of our
sample, $Q_{\rm stars+gas} \lesssim 2$ with a much narrower range than $Q_{\rm
  gas}$ \citep[similar improvements were seen by][]{BOISSIER03,YANG07}.  These
values are roughly consistent with the conditions for cloud formation found
from simulations. \citet{LI05} find gas collapses where $Q_{\rm stars+gas}
\lesssim 1.6$ and \citet{KIM01,KIM07} find runaway instabilities where $Q_{\rm
  gas} \lesssim 1.4$ \citep[though this is $Q_{\rm gas}$ and not $Q_{\rm
  stars+gas}$; for a region like the solar neighborhood,][argue that disk
thickness, which tends to increase stability, approximately offsets the effect
of stars on $Q$]{KIM07}.

$Q_{\rm stars+gas}$ increases towards the central parts of spirals, so that
although the ISM in these regions is usually dominated by \htwo , they appear
more stable than gas near the \hi --to--\htwo\ transition. \citet{HUNTER98}
and \citet{KIM01} suggest that because of low shear, instabilities aided by
magnetic fields may grow in these regions despite supercritical $Q$.
Comparing to the shear threshold proposed by \citet{HUNTER98}, we find some
support for this idea: at $\lesssim 0.2~r_{25}$ many dwarf and spiral galaxies
appear unstable or marginally stable. As with $Q_{\rm gas}$, however,
$\Sigma_{\rm crit,A}/\Sigma_{\rm gas}$ shows large scatter and no clear
ability to predict the SFE.

Thus, we find no clear evidence that disk stability at large scales drives the
observed variations in SFE and $R_{\rm mol}$. Improved handling of
second--order effects (disk thickness, $\sigma_{\rm gas}$, \xco, $\sigma_*$,
and $\ML$) may change this picture, but comparing our first--order analysis to
expectations and simulations, disks appear marginally stable more or less
throughout with little correlation between proposed thresholds and SFE.

\subsubsection{Cold Phase Formation}

Timescales and thresholds computed at $400$ (dwarfs) and $800$~pc (spirals)
scales do not offer a simple way to predict $R_{\rm mol}$. An alternative view
is that physics on smaller scales regulates cloud formation. Comparison with
models by \citet{SCHAYE04} suggests that a cold phase can form across the
entire disk of most of our sample, which agrees with results from
\citet{WOLFIRE03} modeling our own Galaxy. High density, narrow--linewidth
clouds may easily be unstable or be rendered so by the passage of spiral arms
or supernova shocks, even where the ISM as a whole is subcritical. Both
\citet{SCHAYE04} and \citet{DEBLOK06} have emphasized the effect of lower
$\sigma_{\rm gas}$ on instability and we have seen that a shift from the
observed $\sigma_{\rm gas} = 11$~km~s$^{-1}$ to $\sigma_{\rm gas} =
3$~km~s$^{-1}$ would render most gas disks in our sample unstable or
marginally stable (of course a proper calculation requires estimating the
density and fraction of the mass in this phase as well).

A narrow--line component is observed from high--velocity resolution \hi\
observations of nearby irregular galaxies \citep{YOUNG03,DEBLOK06}, but an
important caveat is the lack of direct evidence for such a component in
THINGS. With $\sim 2.5$ or $5$~km~s$^{-1}$ velocity resolution, one cannot
distinguish a narrow component directly. Therefore, \citet{USERO08} followed
up on work by \citet{BRAUN97}, who used the peak intensity along each line of
sight to estimate the maximum contribution from a cold phase and found
pervasive networks of high brightness filaments. \citet{USERO08} find no clear
evidence for a cold phase traced by networks of high brightness filaments,
suggesting that a cold phase, if present, is mixed with the warm phase at the
THINGS resolution of several times $\sim 100$~pc.

\subsubsection{$R_{\rm mol}$ and Pressure}
\label{SECT_PRESS}

 \begin{deluxetable}{l c c}
   \tabletypesize{\small} \tablewidth{0pt} \tablecolumns{4}
   \tablecaption{\label{PRESSFITTAB} Fits\tablenotemark{a} of $R_{\rm mol} =
     \Sigma_{\rm H2}/\Sigma_{\rm HI}$ to $\left( P_{\rm h} /
       P_0\right)^{\alpha}$} \tablehead{\colhead{Source} & \colhead{$\log_{10}
       P_0 / k_{\rm B}$} & \colhead{$\alpha$} \\ & (cm$^{-3}$~K) & }
   \startdata
   spiral subsample (CO/\hi ) & 4.19 & 0.73 \\
   spiral subsample (SFR/\hi )\tablenotemark{b} & $4.30$ & $0.79$ \\
   spiral subsample (combined) & $4.23$ & $0.80$ \\
   dwarf subsample (SFE/\hi )\tablenotemark{b} & $4.51$ & $1.05$ \\
   \citet{WONG02} & \nodata & $0.8$ \\
   \citet{BLITZ06} & $4.54 \pm 0.07$ & $0.92 \pm 0.07$ 
   \enddata
   \tablenotetext{a}{Over the range $R_{\rm mol} = 0.1$ -- $10$.}
   \tablenotetext{b}{Estimating $\Sigma_{\rm H2}$ from $\Sigma_{\rm SFR}$.}
 \end{deluxetable}

 Despite this caveat, our results offer significant circumstantial support
 that ISM physics below our resolution dictates $R_{\rm mol}$: the lack of
 obvious threshold behavior, marginal stability of our disks, the ability of a
 cold phase to form, and the continuous variations in SFE and $R_{\rm mol}$ as
 a function of radius, $\Sigma_*$, and $P_{\rm h}$.

 In particular, the relationship between $R_{\rm mol}$ and $P_{\rm h}$ has
 been studied before. Following theoretical work by \citet{ELMEGREEN93} and
 \citet{ELMEGREEN94}, \citet{WONG02} and \citet{BLITZ06} showed that $R_{\rm
   mol}$ and $P_{\rm h}$ correlate in nearby spiral galaxies (mostly at
 $R_{\rm mol} > 1$) and \citet{ROBERTSON08} recently produced a similar
 relationship from simulations that include cool gas and photodissociation of
 \htwo ; they emphasize the importance of the latter to reproduce the observed
 scaling.

 The dash-dotted line in the bottom left panel of Figure \ref{RMOLVSFIT} shows
 $R_{\rm mol} \propto P_{\rm h}^{1.2}$, predicted by \citet{ELMEGREEN93} from
 balancing \htwo\ formation and destruction in a model ISM. This is a
 reasonable description of dwarf galaxies, where we derive a best--fit power
 law with index $\approx 1.05$. Spirals show a slightly shallower relation
 between $R_{\rm mol}$ and $P_{\rm h}$ with best--fit power law index $\approx
 0.80$. The thick dashed line in the bottom left panel of Figure
 \ref{RMOLVSFIT} shows our best fit to the spiral subsample (both CO/\hi\ and
 SFR/\hi ) over the range $R_{\rm mol} = 0.1$ -- $10$. Table \ref{PRESSFITTAB}
 lists this fit along with fits to dwarf galaxies and the results of
 \citet{WONG02} and \citet{BLITZ06}.

 The entry ``spiral subsample (combined)'' in Table \ref{PRESSFITTAB} lists
 the best fit power law like Equation \ref{RMOLPRESS} for our spiral
 subsample.  This fit has an index $\alpha = 0.80$ and normalization
 $\log_{10} P_0 / k_{\rm B} = 4.23$ (this is an OLS bisector fit over the
 range $0.1 < R_{\rm mol} < 10$ giving equal weight to each of the red and
 green points in Figure \ref{RMOLVSXXX}).  Formally, the uncertainty in the
 fit is small because it includes a large number of data points. However, both
 $\log_{10} P_0 / k_{\rm B}$ and $\alpha$ scatter by several tenths when fit
 to individual galaxies.  This agrees well with $\alpha=0.8$ derived by
 \citet{WONG02} and with $\alpha = 0.92 \pm 0.07$ obtained by \citet{BLITZ06}
 given the uncertainties. Fitting the dwarf subsample in the same manner
 yields $\log_{10} P_0 / k_{\rm B} = 4.51$, the pressure at the \hi
 --to--\htwo\ transition. This is $0.2$--$0.3$~dex higher than $\log_{10} P_0
 / k_{\rm B} = 4.23$ in spirals, suggesting that at the same pressure
 (density) GMC/\htwo\ formation in our dwarf subsample is a factor of $\sim 2$
 less efficient than in spirals.

\subsubsection{$R_{\rm mol}$ and Environment}
\label{SECT_FITS}

The fits between $R_{\rm mol}$ and $P_{\rm h}$ in Table \ref{PRESSFITTAB} are
reasonable descriptions of the data, but do not represent a ``smoking gun''
regarding the underlying physics; radius, $\Sigma_*$, $P_{\rm h}$, and
$\tau_{\rm orb}$ are all covariant and each could be used to predict $R_{\rm
  mol}$ with reasonable accuracy in spirals. Therefore we close our discussion
by noting a set of four scaling relations between $R_{\rm mol}$ and
environment that describe our spiral subsample

\begin{eqnarray}
\label{FITS}
  R_{\rm mol} &=& 10.6~\exp \left(-r_{\rm gal} /
    0.21~r_{25}\right)~ \\
  R_{\rm mol} &=& \Sigma_{*} / 81~M_\odot~{\rm pc}^{-2} \\
  R_{\rm mol} &=& \left( P_{\rm h} / 1.7 \times 10^4~{\rm cm}^{-3}~{\rm
      K}~k_{\rm B} \right)^{0.8} \\
  R_{\rm mol} &=& \left( \tau_{\rm orb} / 1.8 \times 10^8~{\rm yr} \right)^{-2.0}
\end{eqnarray}

\noindent these appear as thick dashed lines in Figure \ref{RMOLVSFIT}. 

In particular, we stress the relationship between $R_{\rm mol}$ and $\Sigma_*$
(see also Figure \ref{SFEVSSTARS}). This has several possible interpretations,
the most simple of which is that stars form where they have formed in the
past. There are physical reasons to think relationship may be causal, however.
Considering a similar finding in dwarf irregular galaxies, \citet{HUNTER98}
suggested that stellar feedback may play a critical role in triggering cloud
formation. Recently the importance of the stellar potential well has been
highlighted, either to triggering large-scale instabilities
\citep{LI05,LI06,YANG07} or in bringing gas to high densities in order for
small-scale physics to operate more effectively
\citep{ELMEGREEN93,ELMEGREEN94,WONG02,BLITZ04,BLITZ06}.

\subsection{A Note on Systematics: $\xco$, $\sigma_{\rm gas}$,
  $\sigma_*$, $\ML$}

In this paper, we work ``to first order,'' using the simplest
well-motivated assumptions to convert observations to physical
quantities. These assumptions are described in \S \ref{SECT_DATA} and
Appendices \ref{GASAPP} -- \ref{SFRAPP}. These are not always unique
and here we note differences with the literature and the effect that
they may have on our analysis.

{\em \xco :} In spirals, we adopt a fixed $\xco = 2 \times 10^{20}$~\xcounits
. \citet{WONG02} and \citet{BLITZ06} adopt the same value.
\citet{KENNICUTT89}, \citet{KENNICUTT98A}, \citet{MARTIN01}, and
\citet{KENNICUTT07} also use a fixed value, but a slightly higher one, $\xco =
2.8 \times 10^{20}$~\xcounits . \citet{BOISSIER03} test the effects of a
metallicity--dependent \xco\ that tends to yield lower $\Sigma_{\rm H2}$ than
our values in the inner parts of spirals, but higher in the outer parts.

Variations in the normalization of \xco\ will affect the location of the \hi
--to--\htwo\ transition and the value of SFE~(\htwo ), but not the
observations of fixed SFE~(\htwo ) or steadily varying $R_{\rm mol}$. A strong
dependence of \xco\ on environment in spirals would affect many of our
results, but leave the basic observation of environment--dependent SFE~(\hi )
intact. Variations in the CO $J=2\rightarrow1$/$1\rightarrow0$ line ratio
(Appendix \ref{GASAPP}) will manifest as changes in \xco .

{\em $\sigma_{\rm gas}$:} We adopt $\sigma_{\rm gas} = 11$~km~s$^{-1}$ based
on the THINGS second moment maps (Appendix \ref{KINAPP}). This is almost twice
the commonly used $\sigma_{\rm gas} = 6$~km~s$^{-1}$
\citep{KENNICUTT89,KENNICUTT98A,MARTIN01,BOISSIER03} and also higher than
$\sigma_{\rm gas} = 8$~km~s$^{-1}$, adopted by \citet{BLITZ06}. We emphasize
the importance of $\sigma_{\rm gas}$ to the stability analysis in \S
\ref{QSECT}; observations with velocity and spatial resolution capable of
distenangling different \hi\ components and a multi--phase analysis are needed
to move forward on this topic.

{\em $\sigma_{\rm *}$:} We assume an isothermal stellar disk with a fixed
scale height, as do \citet{BOISSIER03} and \citet{BLITZ06}. \citet{WONG02} and
\citet{YANG07} assume a fixed stellar velocity dispersion. This has a moderate
effect on $P_{\rm h}$ and a strong effect on $Q_{\rm stars+gas}$ (\S
\ref{QSTARSECT}).

{\em \ML:} We adopt $\ML = 0.5 \mlunit$ (Appendix \ref{STARAPP}), consistent
with our adopted IMF and \citet{BELL03A}, the same value used by
\citet{BLITZ04,BLITZ06}. $\ML$ directly affects $\Sigma_*$, $P_{\rm h}$,
$\tau_{\rm ff}$, and $Q_{\rm stars+gas}$. It may vary by $\sim 30\%$ both
within and among galaxies \citep{BELL01}, with larger variations in the bluest
galaxies or from changes to the assumed IMF \citep{BELL03A}.

{\em Star formation rate tracer:} We use FUV+24$\mu$m to estimate recent
$\Sigma_{\rm SFR}$, discussed in detail in Appendix \ref{SFRAPP}. This is
similar to \citet{BOISSIER07} but in contrast to \citet{KENNICUTT89},
\citet{KENNICUTT98A}, \citet{MARTIN01}, \citet{WONG02}, and
\citet{BOISSIER03}, who each used H$\alpha$ emission with various extinction
corrections. \citet{BOISSIER07} considered differences between between
H$\alpha$ and FUV profiles in detail, suggesting that stochasticity leads
H$\alpha$ to show signs of knees and turnoffs while FUV remains smooth. We
work pixel--by--pixel and in radial profile \citep[similar
to][]{MARTIN01,WONG02,BOISSIER03,BLITZ06} rather than attempting to isolate
individual star forming regions \citep[e.g.,][]{KENNICUTT07}. Both differences
mean that we measure ``recent'' rather than ``present'' $\Sigma_{\rm SFR}$,
which may account for some of the smoothness in the trends seen in \S
\ref{SECT_RESULTS}.

\section{Star Formation Recipes}
\label{SECT_RECIPE}

\begin{figure*}
  \begin{center}
  \plotone{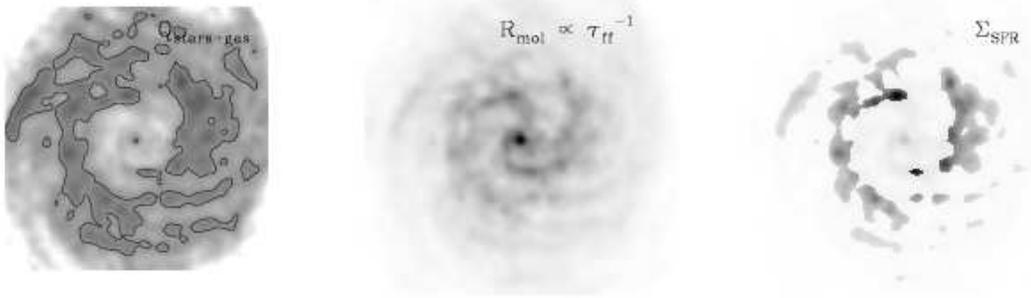}
  \end{center}
  \caption{\label{EGPREDIM} How we predict $\Sigma_{\rm SFR}$, illustrated for
    NGC~3184. We calculate the threshold value ({\em left}, here $Q_{\rm
      stars+gas}$), identify supercritical areas (solid contour). In parallel,
    we estimate $R_{\rm mol} = \Sigma_{\rm H2} / \Sigma_{\rm HI}$ ({\em
      middle}), here from $R_{\rm mol} \propto \tau_{\rm ff}^{-1}$. We combine
    these with the assumption of a fixed SFE (\htwo ) and a (fixed) low SFE in
    subcritical areas to predict $\Sigma_{\rm SFR}$ ({\em right}).}
\end{figure*}

As a final exercise, we compare our galaxies to simple star formation recipes
based on the laws and thresholds discussed in $\S 2$ and normalized to the \hi
--to--\htwo\ transition in spirals (\S \ref{TRANS_SECT}). We predict the SFE
in this way:

\begin{enumerate}
\item We assume SFE (\htwo )$= 5.25 \times 10^{-10}~{\rm yr}^{-1}$
\item We calculate $R_{\rm mol} = \Sigma_{\rm H2} / \Sigma_{\rm HI}$ either:
\begin{enumerate} 
\item By setting $R_{\rm mol} = \tau_{\rm ff,0} / \tau_{\rm ff}$ or $\tau_{\rm
    orb,0} / \tau_{\rm orb}$.
\item From our fits of $R_{\rm mol}$ to radius, $\Sigma_*$, $P_{\rm h}$, and
  $\tau_{\rm orb}$ in spirals galaxies (Equation \ref{FITS}).
\end{enumerate}

\item We derive $\Sigma_{\rm SFR}$ from SFE~(\htwo ), $R_{\rm mol}$,
  and $\Sigma_{\rm HI}$.

\item We calculate the predicted SFE, dividing $\Sigma_{\rm SFR}$ by
  $\Sigma_{\rm gas}$ in spirals and $\Sigma_{\rm HI}$ in dwarf
  galaxies.

\item We combine $R_{\rm mol} \propto \tau_{\rm ff}^{-1}$ or $\tau_{\rm
    orb}^{-1}$ with thresholds. In subcritical areas, we set SFE~$=5 \times
  10^{-11}$~yr$^{-1}$, roughly the observed value at $r_{25}$ in both
  subsamples.
\end{enumerate}

\noindent Figure \ref{EGPREDIM} illustrates the procedure for $R_{\rm mol}
\propto \tau_{\rm ff}^{-1}$ and the $Q_{\rm stars+gas}$ threshold in the
spiral galaxy NGC~3184.

We set $\tau_{\rm ff,0}$ and $\tau_{\rm orb,0}$ equal to the timescale at the
\hi --to--\htwo\ transition in spirals (\S \ref{TRANS_SECT}, Table
\ref{TRANSITIONTAB}), i.e., we predict $R_{\rm mol}$ using the dotted lines in
Figure \ref{RMOLVSXXX}. The predictions will therefore intersect our data
where $R_{\rm mol} = \Sigma_{\rm H2}/\Sigma_{\rm HI} \approx 1$ in spirals.

We adopt the same approach to normalize thresholds. For shear and $Q_{\rm
  stars+gas}$, we define the boundary between supercritical and subcritical
data as $2.3$ and $1.6$, respectively, approximately the values at the \hi
--to--\htwo\ transition in spirals. For the \citet{SCHAYE04} cold phase
threshold we use a critical value of $1$.

We implement thresholds pixel--by--pixel and present our results in radial
average.  Within a tilted ring, some lines of sight can be supercritical and
some can be subcritical, allowing the threshold to damp the average SFE in a
ring without setting it to the minimum value.

The choice to normalize the recipes for both dwarf and spiral galaxies using
values measured for spirals is meant to highlight differences between the
subsamples.

\subsection{Results}
\label{RECIPE_RESULTS}

\begin{figure*}
  \begin{center}
    \plottwo{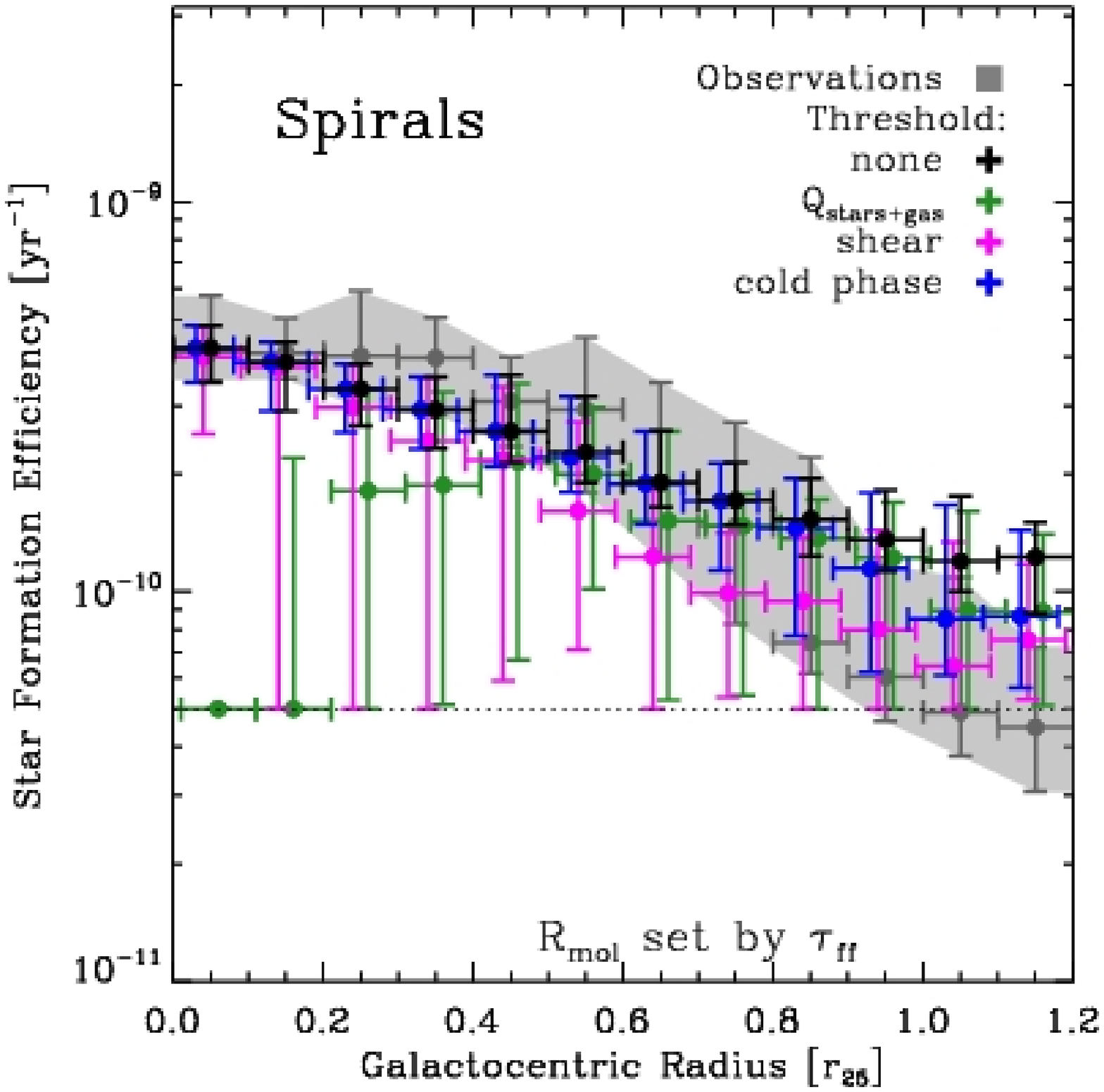}{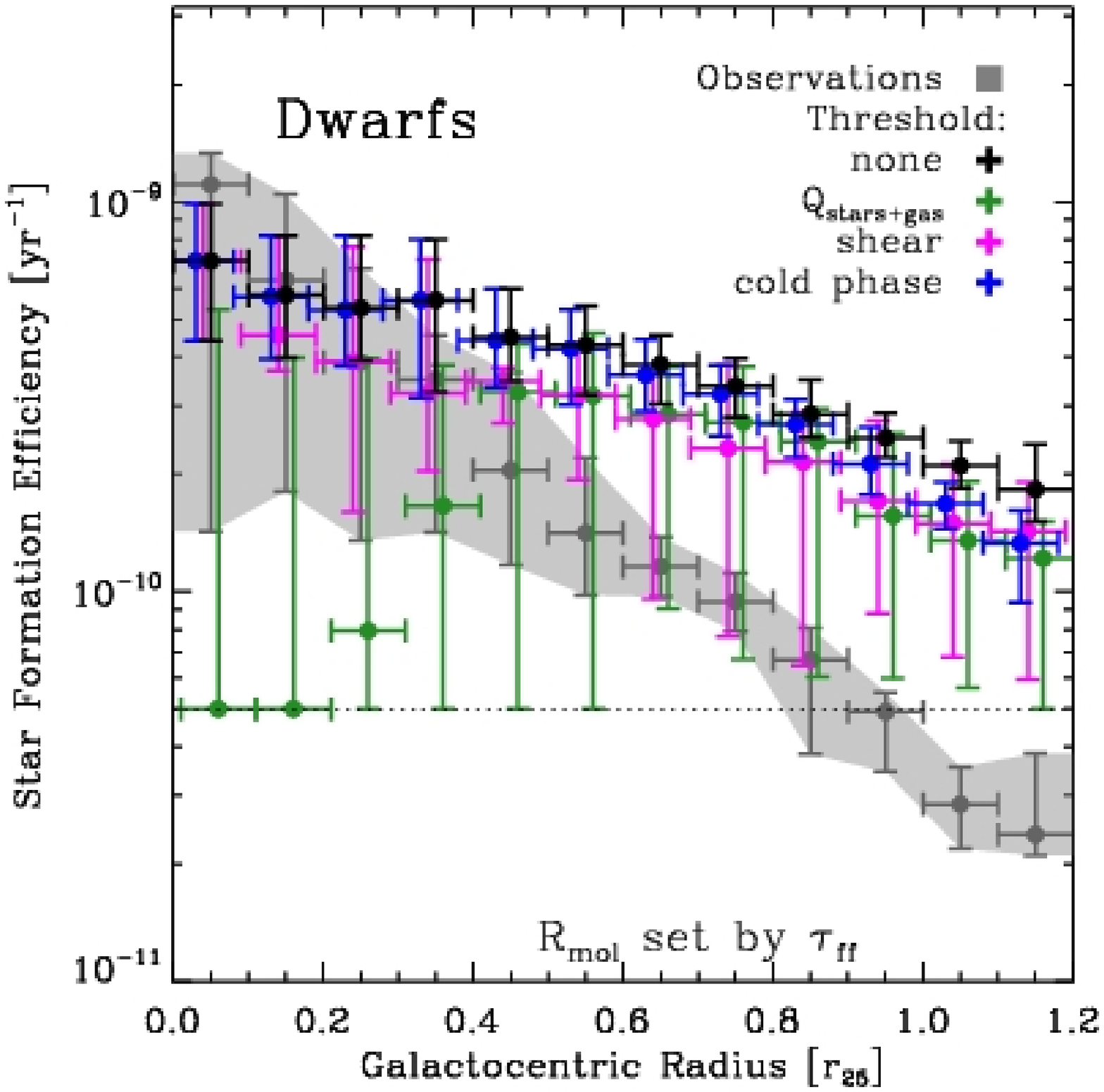}
    \plottwo{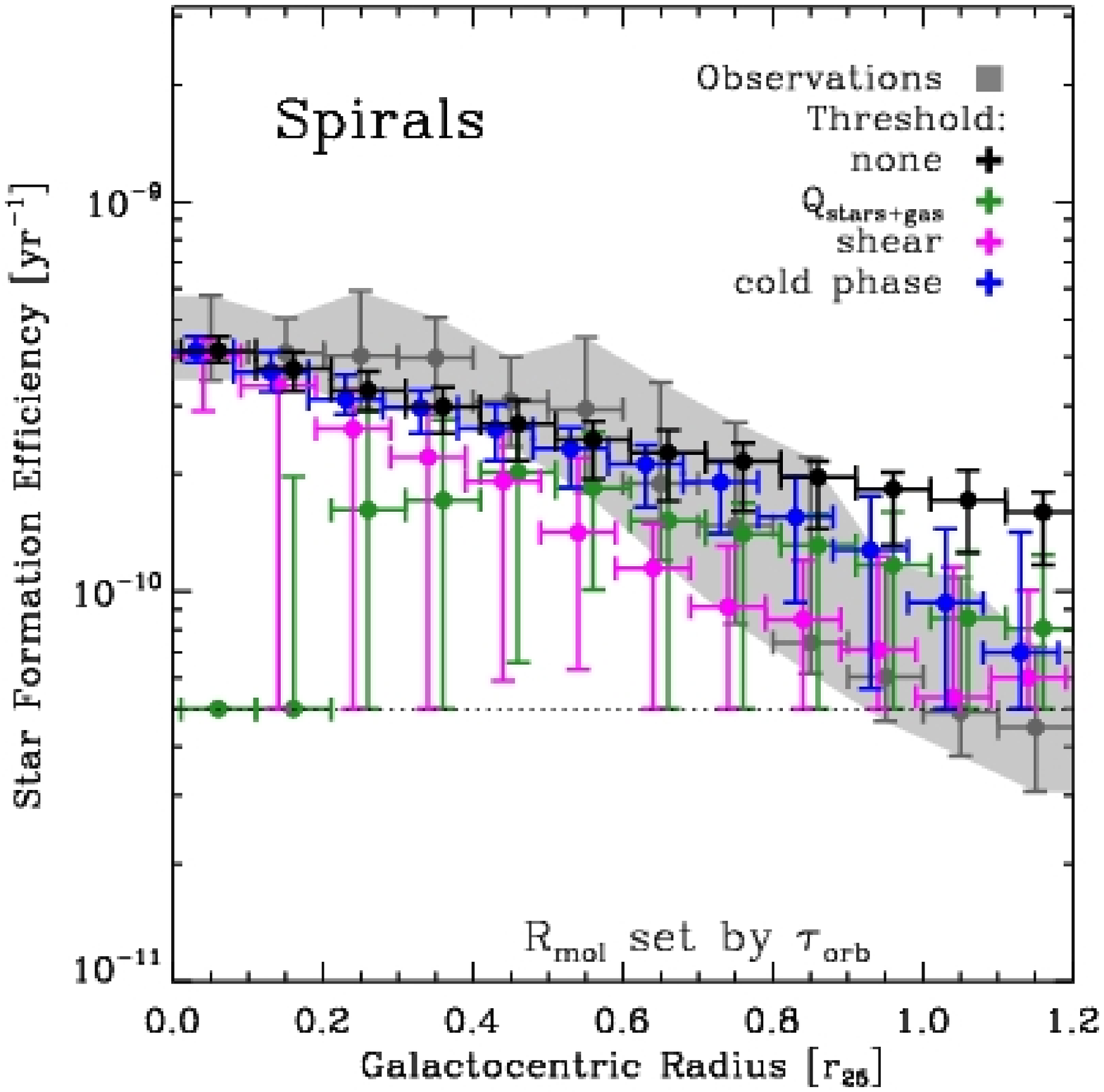}{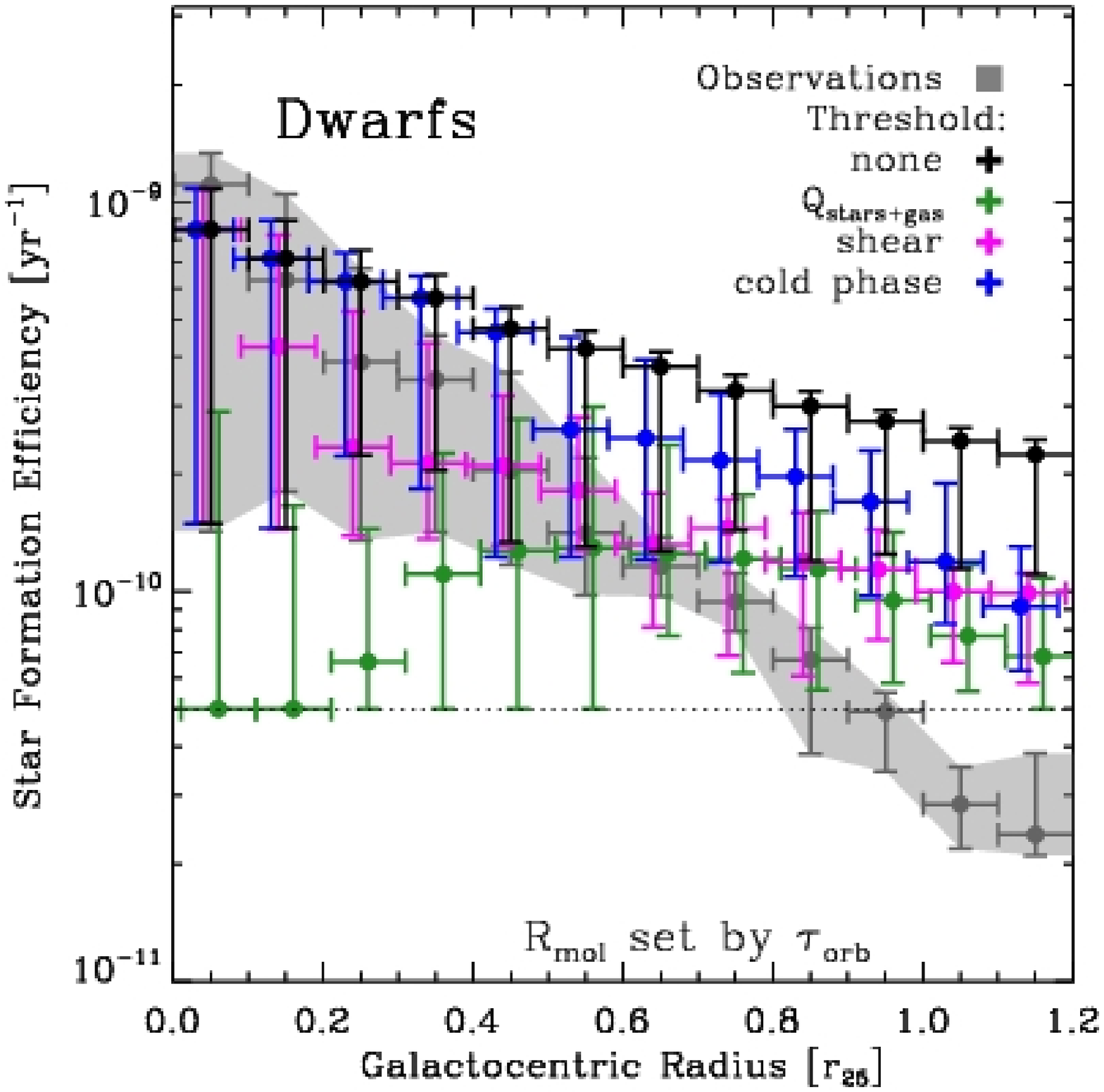}
    \plottwo{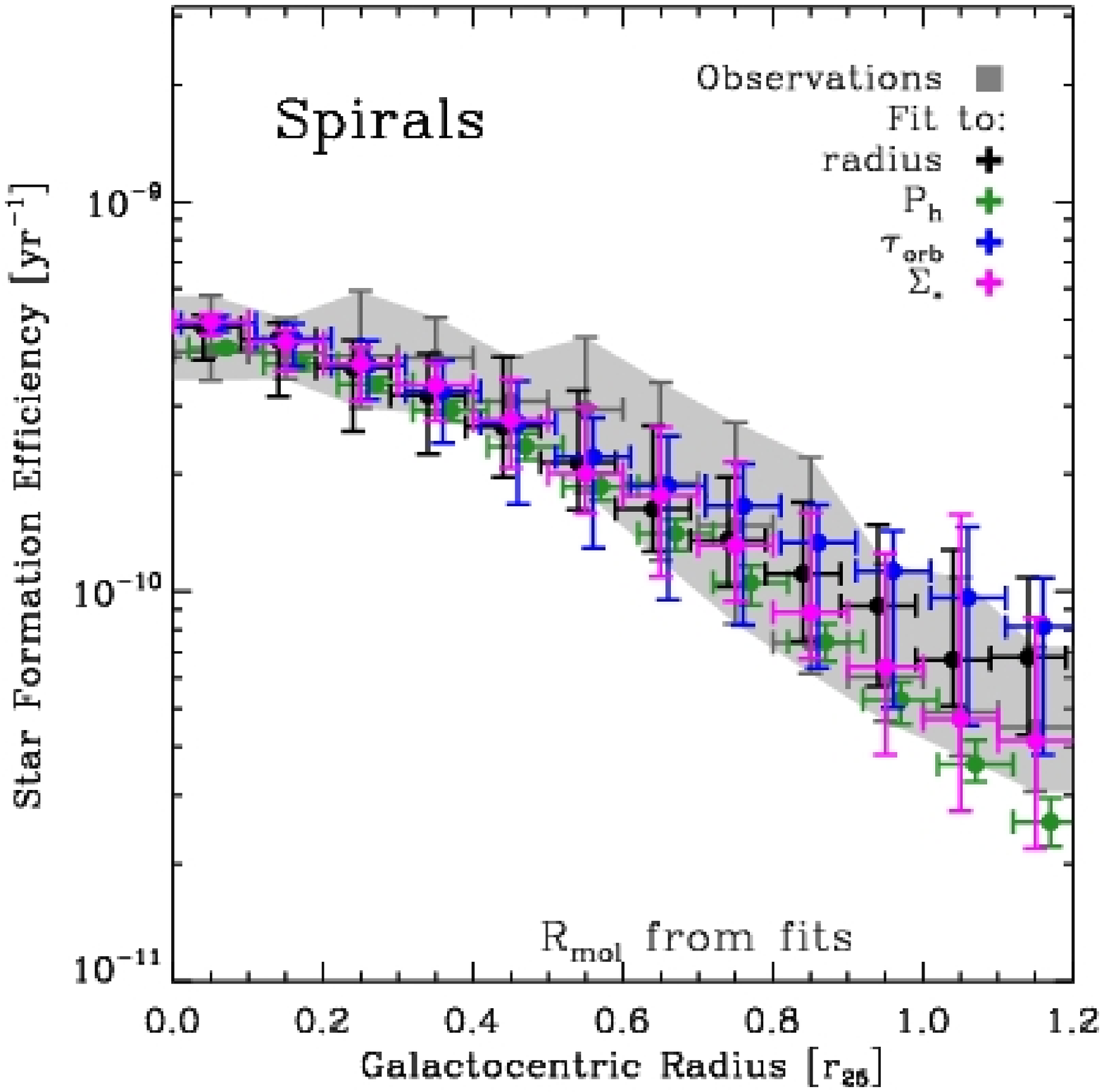}{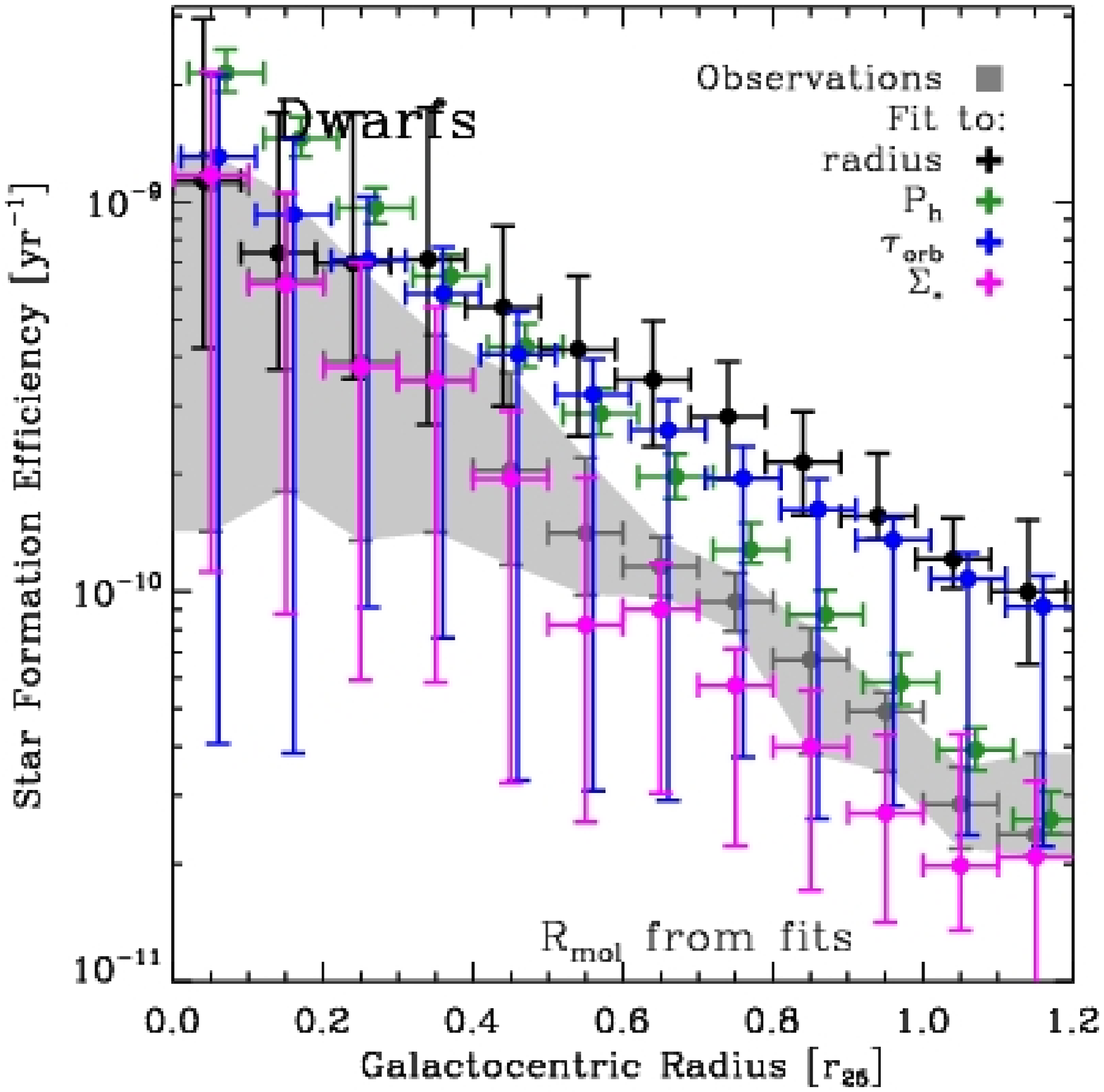}
  \end{center}
\caption{\label{RECIPEFIG} Comparison of predicted (color bins) to observed
  (gray region) SFE in spiral ({\em left}) and dwarf ({\em right}) galaxies.
  We adopt fixed SFE~(\htwo ) and predict $R_{\rm mol}$ from $\tau_{\rm
    ff}^{-1}$ ({\em top row}) and $\tau_{\rm orb}^{-1}$ ({\em middle panel})
  combined with thresholds. We also show four fits of $R_{\rm mol}$ to other
  quantities in spirals ({\em bottom row}). The dotted horizontal line in the
  top two rows shows the SFE that we adopt for subcritical data.}
\end{figure*}

Figure \ref{RECIPEFIG} shows the results of these calculations. The observed
SFE as a function of radius appears as a shaded gray region (based on Figure
\ref{SFEVSRAD}). Radial profiles of SFE compiled from predictions appear in
color (these follow the same methodology used to make the bins in Figure
\ref{SFEVSRAD}). The top row shows results for spiral ({\em left}) and dwarf
({\em right}) galaxies setting $R_{\rm mol} \propto \tau_{\rm ff}^{-1}$
combined with several thresholds, the middle row shows results for $R_{\rm
  mol} \propto \tau_{\rm orb}^{-1}$ combined with the same stable of
thresholds, and the bottom row shows $R_{\rm mol}$ set by fits to radius,
$\Sigma_*$, $P_{\rm h}$, and $\tau_{\rm orb}$.

Figure \ref{RECIPEFIG} illustrates much of what we saw in \S
\ref{SECT_RESULTS} and \ref{SECT_DISCUSSION}. First, adopting fixed SFE~(\htwo
) ensures that we match the observed SFE with reasonable accuracy in the inner
parts of spirals regardless of how we predict $R_{\rm mol}$. Using fits to
predict $R_{\rm mol}$ (bottom row) offers a small refinement over the
timescales in this regime, but as long as $R_{\rm mol} \gtrsim 1$ then SFE
$\sim$ SFE~(\htwo ). As a result, the available gas reservoir sets the SFR
in this regime.

Setting $R_{\rm mol} \propto \tau_{\rm ff}^{-1}$ or $\tau_{\rm orb}^{-1}$
smoothly damps the SFE with increasing radius, but not by enough to match
observations. Without a threshold, $\tau_{\rm ff}$ and $\tau_{\rm orb}$
overpredict the SFE at large radii in spiral galaxies and at $r \gtrsim
0.4~r_{25}$ in dwarf galaxies (black bins, top two rows).

Thresholds damp the SFE with mixed success (green, magenta, and blue bins in
the top two panels). Each somewhat lowers the SFE in the outer parts of
galaxies. In the process, however, both $Q_{\rm stars+gas}$ and shear predict
suppressed star formation at low or intermediate radii in both dwarf and
spiral galaxies, areas where we observe ongoing star formation (the vertical
error bars show that the 50\% range includes completely subcritical galaxies
in both cases).  We saw in \S \ref{SECT_SFETHRESH} that the radial variation
in these thresholds is often less than the scatter among galaxies at a given
radius and that the step function behavior that we implement here is not clear
in our data.

The \citet{SCHAYE04} threshold predicts that a cold phase can form almost
everywhere in our sample and so only comes into play in the outer parts of
spirals and in dwarf galaxies, where it damps the predicted SFE, but not by
enough to match observations.

The bottom left panel shows that the fits mostly do a good job of reproducing
the SFE in spirals, which is expected because they are fits to these data.

The same fits (to spirals) yield mixed results when applied to dwarf galaxies.
The fits to $\Sigma_*$ and $\tau_{\rm orb}$ show very large scatter and fits
to radius, $P_{\rm h}$, and $\tau_{\rm orb}$ all overpredict SFE by varying
amounts (similar discrepancies are evident comparing spirals and dwarfs in the
top two panels). The scaling relations relating $R_{\rm mol}$ to environment
in spirals apparently do not apply perfectly to dwarfs. Likely drivers for the
discrepancy are the lower abundance of metals and dust and more intense
radiation fields, which affect phase balance in the ISM and the rate of \htwo\
formation and destruction. Focusing on the pressure fit (green), we can phrase
the observation this way: for the same pressure (density), cloud formation in
our dwarf subsample is suppressed relative to that in spirals by a factor of
$\sim 2$.

\section{Summary}
\label{CONCLUSIONS}

We combine THINGS, SINGS, the GALEX NGS, HERACLES, and BIMA SONG to study what
sets the star formation efficiency in 12 nearby spirals and 11 nearby dwarf
galaxies.

We use these data to estimate the star formation rate surface density, gas
kinematics, and the mass surface densities of \hi, \htwo, and stars
(Appendices \ref{GASAPP} and \ref{STARAPP}). To trace recent star formation,
we use a linear combination of GALEX FUV and {\em Spitzer} 24$\mu$m (Appendix
\ref{SFRAPP}).  We suggest that this combination represent a useful tool given
the outstanding legacy data sets now available from these two observatories
(e.g., SINGS and the GALEX NGS).

We focus on the star formation efficiency (SFE), $\Sigma_{\rm SFR} /
\Sigma_{\rm gas}$, and the \htwo -to-\hi\ ratio, $R_{\rm mol}$. These
quantities remove the basic scaling between gas and SFR, allowing us to focus
on where gas forms stars quickly/efficiently (SFE) and the phase of the
neutral ISM ($R_{\rm mol}$).  We measure the SFE out to $\sim 1.2~r_{25}$,
compare it to a series of variables posited to influence star formation, and
test the ability of several predictions to reproduce the observed SFE.

\subsection{Structure of Our Typical Spiral and Dwarf Galaxy}

We deliberately avoid discussing individual galaxies in the main text (these
data appear in Appendices \ref{RPROFAPP} and \ref{ATLASAPP}). Instead, we
study ``stacked'' versions of a spiral and dwarf galaxy. We sketch their basic
structure here.

The spiral galaxy has a roughly constant distribution of \hi , $\Sigma_{\rm
  HI} \sim 6$~M$_{\odot}$~pc$^{-2}$ out to $\sim r_{25}$. \hi\ surface
densities seldom exceed $\Sigma_{\rm HI} \sim 10$~M$_{\odot}$~pc$^{-2}$; gas
in excess of this surface density tends to be molecular.  We observe no
analogous saturation in $\Sigma_{\rm H2}$, finding $\Sigma_{\rm H2} \gtrsim
100$~M$_{\odot}$~pc$^{-2}$ in the very central parts of many galaxies.

Molecular gas, star formation, and stellar surface density all decline with
nearly equal exponential scale lengths, $\sim 0.2~r_{25}$, giving the
appearance of a long--lived star--forming disk embedded in a sea of \hi. The
ISM is mostly \htwo\ within $\sim 0.5~r_{25}$ and where $\Sigma_* \gtrsim
80$~M$_{\odot}$.

Over a wide range of conditions the SFR per unit H$_2$, SFE (\htwo ), $=5.25
\pm 2.5 \times 10^{-10}$~yr$^{-1}$ at scales of 800~pc. This is a ``limiting
efficiency'' in the sense that we do not observe the average SFE in spirals to
climb above this value.  Where the ISM is mostly \hi , the SFE is lower than
this limiting value and declines radially with an exponential scale length
$\sim 0.2$--$0.25~r_{25}$. In this regime, the star formation rate per unit
stellar mass remains nearly fixed at a value about twice the cosmologically
average rate (i.e., the stellar assembly time is $\sim$ twice the Hubble
time).

Dwarf galaxies also exhibit flat \hi\ distributions, declining SFE with
increasing radius, and a nearly constant stellar assembly time. Normalized to
$r_{25}$, the scale length of the decline in the SFE is identical to that
observed in spirals within the uncertainties. The stellar assembly time is
half that found in spirals, corresponding to roughly a Hubble time.  Dwarfs
exhibit only the crudest relationship between $\Sigma_{\rm SFR}$ and
$\Sigma_{\rm HI}$ and, as a result, $\Sigma_*$ is a much better predictor of
the SFR than $\Sigma_{\rm gas}$ \citep[in good agreement with][]{HUNTER98}.
The lack of a clear relationship between $\Sigma_{\rm SFR}$ and $\Sigma_{\rm
  gas}$ is at least partially due to an incomplete census of the ISM:
conditions in the central parts of dwarf galaxies often match those where we
find \htwo\ in spirals and in these same regions the SFE is (unexpectedly)
higher than we observe anywhere in spirals (where \htwo\ is included).

\subsection{Conclusions for Specific Laws and Thresholds}

We compare the observed SFE to proposed star formation laws and thresholds
described in \S \ref{SECT_BACKGROUND}. For star formation laws we find:

\begin{itemize} 

\item The SFE varies dramatically over a small range of $\Sigma_{\rm HI}$ and
  very little with changing $\Sigma_{\rm H2}$. Therefore, the {\em disk
    free-fall time for a fixed scale height disk} or any other weak dependence
  of SFE on $\Sigma_{\rm gas}$ is of little use to predict the SFE (\S
  \ref{FIXEDSH_SECT}).

\item {\em The disk free-fall time accounting for a changing scale height},
  $\tau_{\rm ff}$, correlates with both SFE and $R_{\rm mol}$ (\S
  \ref{DENSITY_SECT}). Setting SFE proportional to $\tau_{\rm ff}$ broadly
  captures the drop in SFE in spirals, but predicts variations in SFE~(\htwo )
  that we do not observe and is a poor match to dwarf galaxies. Taking
  $\tau_{\rm ff}$ to be the relevant timescale for \hi\ to form GMCs (i.e.,
  $R_{\rm mol} \propto \tau_{\rm ff}^{-1}$), fails to capture the full drop in
  the SFE in either subsample.

\item The {\em orbital timescale}, $\tau_{\rm orb}$, also correlates with both
  SFE and $R_{\rm mol}$, but in outer spirals and dwarf galaxies both SFE and
  $R_{\rm mol}$ drop faster than $\tau_{\rm orb}$ increases (\S
  \ref{KIN_SECT}).  As with $\tau_{\rm ff}$, $\tau_{\rm orb}$ alone cannot
  describe cloud or star formation in our sample.

\item In spirals, we observe no clear relationship between SFE~(\htwo ) and
  the logarithmic derivative of the rotation curve, $\beta$ (\S
  \ref{KIN_SECT}). In dwarf galaxies, SFE correlates with $\beta$.  Both
  observations are contrary to the anti-correlation between SFE and $\beta$
  expected if {\em cloud--cloud collisions} set the SFE \citep{TAN00}.

\item {\em Fixed GMC efficiency} appears to be a good description of our
  spiral subsample \citep[\S \ref{SECT_SFEOBS} and][]{BIGIEL08}. SFE (\htwo)
  is constant as a function of a range of environmental parameters. This
  observation applies only to the disks of spiral galaxies, not starbursts or
  low metallicity dwarf galaxies.

\item We observe a correspondence between hydrostatic {\em pressure and ISM
    phase} (\S \ref{DENSITY_SECT} and \S \ref{SECT_PRESS}). In spirals our
  results are consistent with previous work \citep{WONG02,BLITZ06}. In dwarf
  galaxies and the outer parts of spirals, inferring $R_{\rm mol}$ from
  SFE~(\hi ) yields results roughly consistent with predictions by
  \citet{ELMEGREEN93}.
\end{itemize}

\noindent For thresholds we find:

\begin{itemize}
\item Despite a suggestion of increased stability at large radii in spirals,
  there is no clear relation between $Q_{\rm gas}$ --- which measures
  stability against axisymmetric collapse due to {\em self-gravity in the gas
    disk alone} --- and SFE.  Most regions are quite stable and $Q_{\rm gas}$
  has large scatter, even appearing weakly {\em anti}--correlated with the SFE
  in dwarfs (\S \ref{QSECT}).

\item When the effects of stars are included, most disks are only marginally
  stable: $Q_{\rm stars+gas}$ \citep{RAFIKOV01}, which measures {\em
    gravitational instability in a disk of gas and stars}, lies mostly in the
  narrow range $1.3$--$2.5$, increasing slightly towards the centers and edges
  of galaxies. We emphasize that adopted parameters --- \xco , $\sigma_{\rm
    gas}$, $\ML$, and $\sigma_*$ --- strongly affect both $Q_{\rm gas}$ and
  $Q_{\rm stars+gas}$ (\S \ref{QSTARSECT}).

\item The ability of instabilities to survive {\em competition with shear}
  \citep{HUNTER98} shows the same large scatter and high stability as $Q_{\rm
    gas}$ in the outer disks of spirals, but identifies most areas in dwarf
  galaxies and inner spirals as only marginally stable, an improvement over
  $Q_{\rm gas}$ (\S \ref{SHEAR_SECT}).

\item Most areas in both dwarf and spiral galaxies meets the condition needed
  {\em for a cold phase to form} (\S \ref{SCHAYE_SECT}) \citep{SCHAYE04}.
  Regions that do not meet this criterion tend to come from outer disks and
  have low SFE. Because this criterion is met over such a large area, it is of
  little use on its own to predict variation in the SFE within galaxy disks.
\end{itemize}

Finally, we distinguish three different {\em critical surface densities}.
First, in spirals $\Sigma_{\rm gas} \sim 14$~M$_{\odot}$~pc$^{-2}$ at the \hi
-to-\htwo\ transition. We find no evidence that this is a real threshold for
cloud formation: $R_{\rm mol} = \Sigma_{\rm H2}/\Sigma_{\rm HI}$ varies
continuously across $R_{\rm mol} = 1$ as a function of other quantities.
However, it is useful to predict the SFE, which will be nearly constant above
this $\Sigma_{\rm gas}$. A related (but not identical) value, $\Sigma_{\rm HI}
\sim 10$~M$_{\odot}$~pc$^{-2}$, is the surface density at which \hi\
``saturates.'' Gas in excess of this surface density is in the molecular phase
\citep[][]{MARTIN01,WONG02,BIGIEL08}. This presumably drives the observation
that most vigorous star formation takes place where $\Sigma_{\rm HI} \gtrsim
10$~M$_{\odot}$~pc$^{-2}$ \citep[e.g.,][]{SKILLMAN87}. Last, lower values,
$\Sigma_{\rm gas} \sim 3$--$4$~M$_{\odot}$~pc$^{-2}$
\citep[e.g.][]{KENNICUTT89,SCHAYE04}, may correspond to the edge of the
star--forming disk. At our resolution such values are relatively rare inside
$1.2~r_{25}$ and we draw no conclusion regarding whether this ``outer disk
threshold'' corresponds to a real shift in the mode of star formation.

\subsection{General Conclusions}

Our general conclusions are:

\begin{enumerate}

\item In the disks of spiral galaxies, the SFE of \htwo\ is roughly constant
  as a function of: galactocentric radius, $\Sigma_*$, $\Sigma_{\rm gas}$,
  $P_{\rm h}$, $\tau_{\rm orb}$, $Q_{\rm gas}$, and $\beta$ (\S
  \ref{FIXEDHTWO_SECT}). This fixed SFE~(\htwo )$=5.25 \pm 2.5 \times
  10^{-10}$~yr$^{-1}$ ($\tau_{\rm Dep} ({\rm H2}) = 1.9 \times 10^9$~yr) sets
  the SFE of total gas across the \htwo --dominated inner parts ($r_{\rm gal}
  \lesssim 0.5~r_{25}$) of spiral galaxies.

\item In spiral galaxies, the transition between a mostly--\hi\ and a
  mostly--\htwo\ ISM is a well--defined function of local conditions (\S
  \ref{TRANS_SECT}). It occurs at a characteristic radius ($0.43 \pm
  0.18~r_{25}$), $\Sigma_*$ ($81 \pm 25$~M$_{\odot}$~pc$^{-2}$), $\Sigma_{\rm
    gas}$ ($14 \pm 6$~M$_{\odot}$~pc$^{-2}$), $P_{\rm h}$ ($2.3 \pm 1.5 \times
  10^4~k_{\rm B}$~cm$^{-3}~K$), and $\tau_{\rm orb}$ ($1.8 \pm 0.4 \times
  10^8$~yr).

\item We find indirect evidence for abundant \htwo\ in the central parts of
  many dwarf galaxies, where SFE~(\hi ) exceeds SFE (\htwo ) found in spirals.
  The simplest explanation is that \htwo\ accounts for a significant fraction
  of the ISM along these lines of sight (\S \ref{SECT_SFERAD} and \S
  \ref{SECT_MISSINGH2}). The implied central $\Sigma_{\rm H2}/\Sigma_{\rm HI}$
  is $\sim 2.5$ with $\Sigma_{\rm H2} = \Sigma_{\rm HI}$ at $\sim
  0.25~r_{25}$.

\item Where $\Sigma_{\rm HI} > \Sigma_{\rm H2}$ --- in the outer parts of
  spirals and throughout dwarf galaxies (by assumption) --- we observe the SFE
  to decline steadily with increasing radius, with scale length $\sim
  0.2$--$0.25~r_{25}$ in both subsamples (\S \ref{SECT_SFEOBS}). We also
  observe a decline in SFE with decreasing $\Sigma_*$, decreasing $P_{\rm h}$,
  and increasing $\tau_{\rm orb}$, which are all covariant with radius.

\item Where $\Sigma_{\rm HI} > \Sigma_{\rm H2}$, we find little relation
  between SFE and $\Sigma_{\rm gas}$ (\S \ref{GAS_SECT}) but a strong
  relationship between SFE and $\Sigma_*$ (\S \ref{SECT_SFESTARS}). The
  simplest explanation is that present day star formation roughly follows past
  star formation. A more aggressive interpretation is that the stellar
  potential well or feedback are critical to bring gas to high densities.

\item The \htwo --to--\hi\ ratio, $R_{\rm mol}= \Sigma_{\rm H2}/\Sigma_{\rm
    HI}$, and by extension cloud formation, depends strongly on environment.
  $R_{\rm mol}$ correlates with radius, $P_{\rm h}$, $\tau_{\rm ff}$,
  $\tau_{\rm orb}$, and $\Sigma_*$ in spirals.  We find corresponding
  correlations between these quantities and $\Sigma_{\rm SFR} / \Sigma_{\rm
    HI}$, a proxy for the efficiency of cloud formation in dwarfs and the
  outer parts of spirals. At our resolution, $R_{\rm mol}$ appears to be a
  continuous function of environment from the \hi --dominated ($R_{\rm mol}
  \sim 0.1$) to \htwo --dominated ($R_{\rm mol} \sim 10$) regime (\S
  \ref{GMCFORM_SECT}).

\item The variation in $R_{\rm mol}$ is too strong to be reproduced only by
  varying $\tau_{\rm orb}$ or $\tau_{\rm ff}$ (\S \ref{SECT_SFELAW} and
  \ref{SECT_TIMES}). Physics other than these timescales must also play an
  important role in cloud formation (points 8 -- 11).

\item Thresholds for large scale stability do not offer an obvious way to
  predict $R_{\rm mol}$. We find no clear relationship (continuous or
  step--function) between SFE and $Q_{\rm gas}$, $Q_{\rm stars+gas}$, or the
  shear threshold. The threshold values we find suggest disks that are stable
  or marginally stable throughout once the effects of stars are included (\S
  \ref{SECT_SFETHRESH} and \ref{SECT_THRESHDISC}).

\item We derive a power law relationship between $R_{\rm mol}$ and hydrostatic
  pressure \citep{ELMEGREEN89} that is roughly consistent with expectations by
  \citet[][]{ELMEGREEN93}, observations by \citet{WONG02} and \citet{BLITZ06},
  and simulations by \citet{ROBERTSON08}. In its simplest form, this is a
  variation on the classical Schmidt law, i.e., $R_{\rm mol}$ set by gas
  volume density (\S \ref{SECT_SFELAW} and \ref{SECT_PRESS}).

\item Power law fits of $R_{\rm mol}$ to $P_{\rm h}$ ($\tau_{\rm ff}$),
  radius, $\tau_{\rm orb}$, and $\Sigma_{*}$ reproduce observed SFE reasonably
  in spiral galaxies but yield large scatter or higher--than--expected SFE in
  the outer parts of dwarf galaxies, offering indirect evidence that the
  differences between our two subsamples --- metallicity (dust), radiation
  field, and strong spiral shocks --- play a role in setting these relations
  (\S \ref{SECT_FITS} and \ref{SECT_RECIPE}).

\item Our data do not identify a unique driver for the SFE, but suggest that
  ISM physics below our resolution --- balance between warm and cold \hi\
  phases, \htwo\ formation, and perhaps shocks and turbulent fluctuations
  driven by stellar feedback --- govern the ability of the ISM to form GMCs
  out of marginally stable galaxy disks (\S \ref{SECT_FITS}).

\end{enumerate}

\acknowledgements We thank the anonymous referee for helpful suggestions that
improved the paper. We thank Daniela Calzetti, Robert Kennicutt, Erik
Rosolowsky, Tony Wong, Leo Blitz, and Helene Roussel for suggestions and
discussion during this project. We gratefully acknowledge the hard work of the
SINGS, GALEX NGS, and BIMA SONG teams and thank them for making their data
publicly available.  EB gratefully acknowledges financial support through an
EU Marie Curie International Reintegration Grant (Contract No.
MIRG-CT-6-2005-013556). FB acknowledges support from the Deutsche
Forschungsgemeinschaft (DFG) Priority Program 1177. The work of WJGdB is based
upon research supported by the South African Research Chairs Initiative of the
Department of Science and Technology and National Research Foundation. We have
made use of: the NASA/IPAC Extragalactic Database (NED) which is operated by
the Jet Propulsion Laboratory, California Institute of Technology, under
contract with the National Aeronautics and Space Administration; the HyperLeda
catalog, located on the World Wide Web at
http://www-obs.univ-lyon1.fr/hypercat/intro.html; NASA's Astrophysics Data
System (ADS); and data products from the Two Micron All Sky Survey, which is a
joint project of the University of Massachusetts and the Infrared Processing
and Analysis Center/California Institute of Technology, funded by the National
Aeronautics and Space Administration and the National Science Foundation.

\begin{appendix}

  These appendices describe how we assemble the database of radial profiles
  and maps that are used in the main text.  We discuss the data and methods
  that we use to derive gas surface densities (Appendix \ref{GASAPP}),
  kinematics (Appendix \ref{KINAPP}), stellar surface densities (Appendix
  \ref{STARAPP}), and star formation rate surface densities (Appendix
  \ref{SFRAPP}). Finally, we present a table containing radial profiles of key
  quantities (Appendix \ref{RPROFAPP}) and an atlas showing maps,
  profiles, and basic results for each galaxy  (Appendix \ref{ATLASAPP}) . \\

\section{Maps of \hi\ and \htwo\ Surface Density}
\label{GASAPP}

\subsection{$\Sigma_{\rm HI}$ from THINGS 21cm Maps}

The \hi\ Nearby Galaxy Survey \citep[THINGS,][]{WALTER08} mapped 21-cm line
emission from all of our sample galaxies using the Very Large Array. We
calculate atomic gas mass surface density, $\Sigma_{\rm{HI}}$, from
natural-weighted data that have mean angular resolution $11\arcsec$ and mean
velocity resolution $5$~\kmpers . THINGS includes data from the most compact
VLA configuration and therefore comfortably recovers extended structure (up to
$15\arcmin$) in our sources. At $30\arcsec$ resolution, THINGS maps are
sensitive to $\Sigma_{\rm HI}$ as low as $\sim 0.5$~M$_{\odot}$~pc$^{-2}$;
here we adopt a working sensitivity of $\Sigma_{\rm HI} =
1$~M$_{\odot}$~pc$^{-2}$. In practice the sensitivity and field of view of the
THINGS maps are sufficient to measure $\Sigma_{\rm HI}$ to $\gtrsim r_{25}$ in
almost every galaxy. For detailed description and presentation of THINGS, we
refer the reader to \citet{WALTER08}.

To convert from integrated intensity to $\Sigma_{\rm HI}$ we use

\begin{equation}
  \Sigma_{\rm HI}~\left[{\rm M_\odot~pc}^{-2}\right] = 0.020~\cos i~I_{\rm 21cm}~\left[{\rm K~km~s}^{-1}\right]~.
\end{equation}

\noindent which accounts for inclination and includes a factor of $1.36$ to
reflect the presence of helium.

\subsection{$\Sigma_{\rm H2}$ from HERACLES (IRAM 30-m) and BIMA SONG CO Maps}

We estimate the surface density of molecular hydrogen, $\Sigma_{\rm H2}$, from
CO emission, the most commonly used tracer of \htwo. Along with
\citet{BIGIEL08}, this study presents the first scientific results from
HERACLES, a large project that used the HERA focal plane array
\citep{SCHUSTER04} on the IRAM~30--m telescope to map CO $J=2\rightarrow1$
emission from the full optical disk in $18$ THINGS galaxies \citep{LEROY08}.
These data have an angular resolution of $11\arcsec$ and a velocity resolution
of $2.6$~km~s$^{-1}$. The typical noise in an individual channel map is
40--80~mK, yielding (masked) integrated intensity maps that are sensitive to
$\Sigma_{\rm H2} \gtrsim 4$~M$_{\odot}$~pc$^{-2}$ at our working resolution
and adopted conversion factor.

HERA maps are not available for NGC~3627 and NGC~5194. In these galaxies, we
use CO~$J=1\rightarrow0$ maps from the BIMA Survey of Nearby Galaxies
\citep[BIMA SONG,][]{HELFER03} to estimate $\Sigma_{\rm H2}$. These data have
angular resolution $\sim 7\arcsec$ and include zero-spacing data from the Kitt
Peak 12m, ensuring sensitivity to extended structure.

We derive $\Sigma_{\rm H2}$ from integrated CO intensity, $I_{\rm CO}$ by
adopting a constant CO-to-H$_2$ conversion factor, $\xco = 2 \times
10^{20}$~\xcounits . Based on comparison to $\gamma$-ray and FIR observations,
this value is appropriate in the Solar Neighborhood
\citep[][]{STRONG96,DAME01}. For CO $J=1\rightarrow0$ emission, the conversion
to $\Sigma_{\rm H2}$ is

\begin{equation}
  \Sigma_{\rm H2}~\left[{\rm M}_\odot~{\rm pc}^{-2}\right] = 4.4~\cos i~I_{\rm CO}~\left(1\rightarrow0\right)~\left[{\rm K~km~s}^{-1}\right]~.
\end{equation}

To relate CO $J=2\rightarrow1$ to CO $J=1\rightarrow0$ intensity, we further
assume a line ratio of $I_{\rm CO} (2\rightarrow1) = 0.8~I_{\rm CO}
(1\rightarrow0)$. Based on direct comparison of HERACLES and previous surveys,
thisis a typical value in our sample \citep{LEROY08} and is intermediate in
the range ($\sim 0.6 -- 1.0$) observed for the Milky Way and other spiral
galaxies \citep[, e.g.,][]{BRAINE93,SAWADA01,SCHUSTER07}. Thus, for the
HERACLES maps we derive $\Sigma_{\rm H2}$ via

\begin{equation}
\Sigma_{\rm H2}~\left[{\rm M}_\odot~{\rm pc}^{-2}\right] = 5.5~\cos i~I_{\rm CO}~\left(2\rightarrow1\right)~\left[{\rm K~km~s}^{-1}\right]~.
\end{equation}

\subsection{The CO--to--\htwo\ Conversion Factor}
\label{XCOAPP}

The CO--to--\htwo\ conversion factor is presumably a source of significant
systematic uncertainty in $\Sigma_{\rm H2}$. $\xco$ almost certainly varies:
it is likely to be lower than Galactic (yielding lower $\Sigma_{\rm H2}$) in
overwhelmingly molecular, heavily excited regions; it is likely to be higher
(yielding higher $\Sigma_{\rm H2}$) in regions with low dust content and
intense radiation fields, such as dwarf irregular galaxies. There is
compelling evidence for both senses of variation, but it is our assessment
that no reliable calibration of $\xco$ as a function of metallicity, radiation
field, and $\Sigma_{\rm H2}$ yet exists.  A useful calibration must reflect
all of these quantities, which all affect $\xco$ and are not universally
covariant.

In light of this uncertainty, our approach is: 1) to treat $\xco$ as unknown
in low mass, low--metallicity galaxies, where different approaches to measure
$\Sigma_{\rm H2}$ yield results that differ by an order of magnitude or more
and 2) to assume that variations in $\xco$ within spiral galaxies are
relatively small. The second point might be expected based on theoretical
modeling of GMCs \citep{WOLFIRE93} and the observed uniformity of GMC
properties across a wide range of environments \citep{BOLATTO08}. We emphasize
that even if present, the most extreme variations are likely to contribute
primarily to the central resolution element, which is not the focus of this
study, and the far outer disk, where $\Sigma_{\rm H2}$ is not the dominant
mass component.

Variations aside, estimates of ``typical'' values of $\xco$ in the Milky Way
and other spiral galaxies span the range $\sim 1.5$--$4 \times
10^{20}$~\xcounits\ ~\citep[e.g.][in addition to the references already
given]{BLITZ07,DRAINE07}.  The choice of mean \xco\ within this range can have
a large impact on, e.g., assessing gravitational stability or conditions at
the \htwo -to-\hi\ transition.  We refer the reader to \citet{BOISSIER03} for
a quantitative exploration of how different assumptions regarding \xco\ affect
a stability analysis.

\subsection{Masking the \hi\ and CO Data Cubes}

The \hi\ and CO data cubes have large bandwidth, only a small part of
which contains signal of the spectral line. In order to produce
integrated intensity maps with good signal-to-noise ratio, we blank
signal-free regions of the \hi\ and CO cubes.  \citet{WALTER08}
describe this process for THINGS. We apply an analogous procedure to
the HERACLES and BIMA SONG data. We convolve the cubes to $30\arcsec$
resolution, identify regions with significant emission, and then blank
the original data cubes outside these regions. We integrate these
blanked cubes to create intensity maps. For HERACLES, we require
$I_{\rm CO} > 2\sigma_{\rm RMS}$ in 3 consecutive (2.6~km~s$^{-1}$)
channels at $30\arcsec$ resolution. Note that our use of masking
drives the small ($\sim 10\%$) numerical differences with the HERACLES
survey paper (which uses a different approach to create integrated
intensity maps). For BIMA SONG, we require either $I_{\rm CO} >
3\sigma_{\rm RMS}$ in a single (10~km~s$^{-1}$) channel at $30\arcsec$
resolution or $I_{\rm CO} > 2\sigma_{\rm RMS}$ in consecutive velocity
channels, similar to the original masking by \citet{HELFER03}. In both
cases we consider only CO emission within $\sim 100$~km~s$^{-1}$ of
the mean \hi\ velocity.

\section{Kinematics}
\label{KINAPP}

\subsection{Rotation Curves from THINGS}

\begin{figure*}
  \plotone{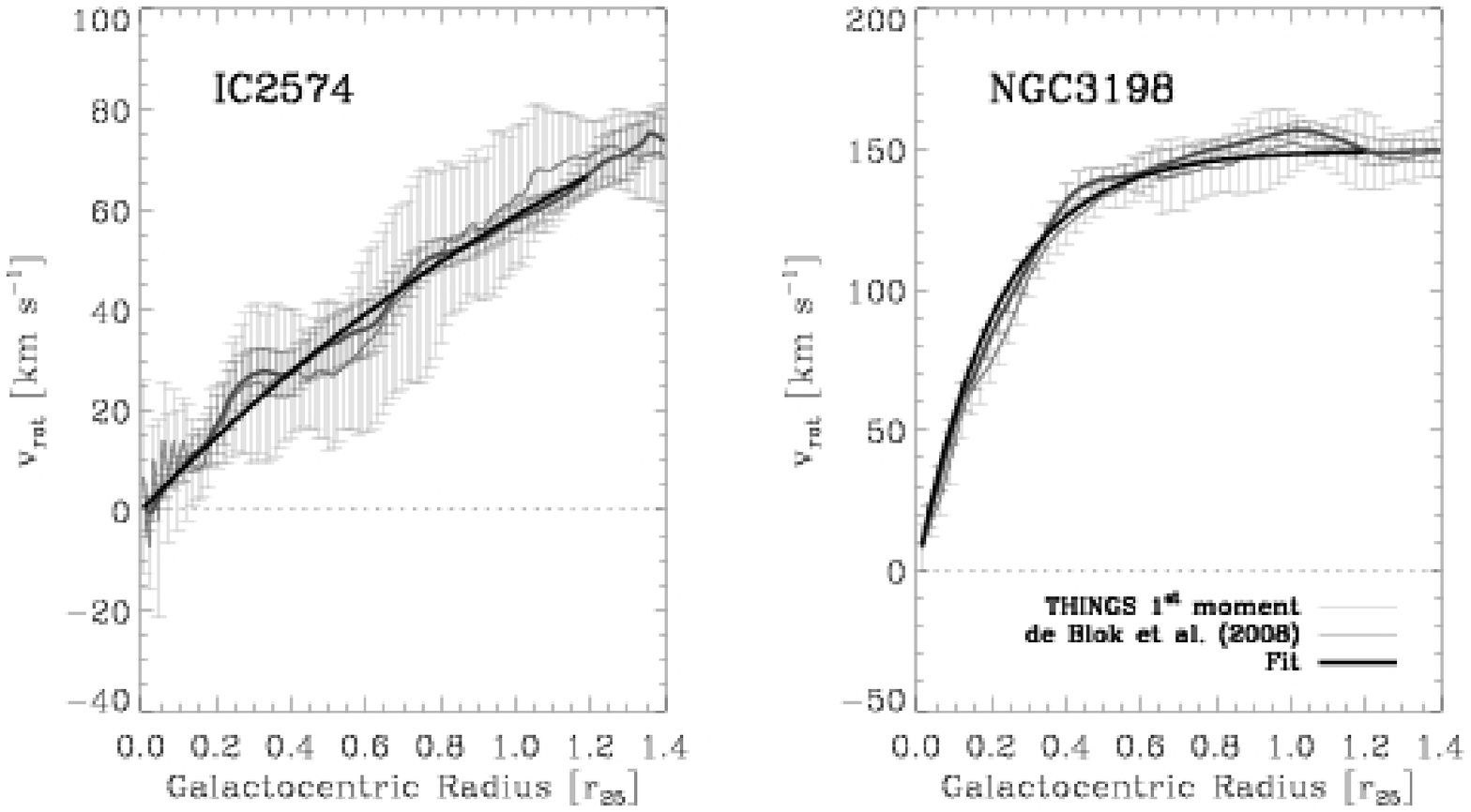} \figcaption{\label{RCEXAMPLE} Illustration of our
    rotation curve treatment. Light gray profiles show median rotation
    velocity and scatter measured directly from the THINGS first moment maps;
    dark gray profiles and scatter show the higher quality \citep{DEBLOK08}
    rotation curves. The thick black lines show the fit that we use to
    approximate the rotation curve. This simple function (Equation
    \ref{ROTCUREQ}) does a good job of capturing both the steadily rising
    rotation curves typical of dwarf galaxies (e.g., IC~2574, left panel) and
    the rapidly rising then flat curves seen in more massive spiral galaxies
    (e.g., NGC~3198, right panel).}
\end{figure*}

We approximate all galaxies to have rotation curves with the following
functional form \citep{BOISSIER03}:

\begin{equation}
\label{ROTCUREQ}
v_{\rm rot} \left(r \right) = v_{\rm flat}~\left[ 1 - \exp \left( \frac{-
      r}{l_{\rm flat}} \right) \right]
\end{equation}

\noindent where $v_{\rm rot}$ is the circular rotation speed of the galaxy at
a radius $r$ and $v_{\rm flat}$ and $l_{\rm flat}$ are free parameters that
represent the velocity at which the rotation curve is flat and the length
scale over which it approaches this velocity. For a continuously rising
rotation curve, common for low--mass galaxies, we expect large $l_{\rm flat}$,
while the almost flat rotation curves of massive spiral galaxies will have
small $l_{\rm flat}$ and then remain nearly constant at $v_{\rm flat}$.

In most cases, Equation \ref{ROTCUREQ} captures the basic behavior of the
rotation curve well. Small scale variations are lost, but these may be due to
streaming motions near spiral arms or warps in the gas disk as easily as real
variations in the circular velocity. On the other hand, Equation
\ref{ROTCUREQ} offers the distinct advantage of having a smooth, analytic
derivative.  Our analysis uses the rotation curve to estimate the orbital
timescale, shear, and coriolis force (see \S \ref{SECT_BACKGROUND}). The
former is quite reasonably captured by Equation \ref{ROTCUREQ} and the latter
two depend critically on the derivative of the rotation curve $\beta = d
\log~v(r_{\rm gal}) / d \log~r_{\rm gal}$.

For each galaxy, we derive $v_{\rm flat}$ and $l_{\rm flat}$ from a non-linear
least squares fit using Equation \ref{ROTCUREQ} and profiles of $v_{\rm rot}$
measured from the THINGS data cubes. We calculate $v_{\rm rot}$ from the
intensity--weighted first moment, $v_{\rm r}$, via

\begin{equation}
\label{VROTEQ}
v_{\rm rot} = \frac{v_{\rm r} - v_{\rm sys}}{\sin i~\cos \theta}~.
\end{equation}

\noindent Here $v_{\rm sys}$ is the systemic velocity, $i$ is the inclination,
and $\theta$ is the azimuthal angle relative to the receding major axis
measured in the plane of the galaxy. We calculate maps of $v_{\rm rot}$ and
then convert these into profiles of the median and $1\sigma$ scatter in
$v_{\rm rot}$ within $60\arcdeg$ of the major axis in a series
$5\arcsec$--wide tilted rings. We fit Equation \ref{ROTCUREQ} to the profile
of median $v_{\rm rot}$ weighted by the scatter in that ring.

For many of our galaxies, high quality rotation curves are available from the
analysis of \citet[][see Table \ref{SAMPLETAB}]{DEBLOK08}. Wherever possible,
we include these in our fit with very high weight, so that they drive the
best--fit $v_{\rm flat}$ and $l_{\rm flat}$ for these galaxies. For the $7$
low--inclination galaxies in our sample that are not part of the study by
\citet{DEBLOK08} (see Column 4 of Table \ref{SAMPLETAB}) we only fit the first
moment data.

Figure \ref{RCEXAMPLE} shows examples of this procedure for two galaxies: the
dwarf irregular IC~2574, which has a steadily rising rotation curve, and the
spiral galaxy NGC~3198, which has a quickly rising rotation curve that remains
flat over most of the disk. We plot $v_{\rm rot}$ and associated scatter, the
\citet{DEBLOK08} rotation curve, and the best--fit version of Equation
\ref{ROTCUREQ}. The best--fit values of $v_{\rm flat}$ and $l_{\rm flat}$ for
all galaxies are given in Table \ref{STRUCTURETAB}; for the three galaxies
that overlap the sample of \citet{BOISSIER03}, we match their fitted
parameters well.

The dynamics of the irregular galaxies NGC~3077 and NGC~4449 are not
well--described by Equation \ref{ROTCUREQ}; the former is disturbed by an
ongoing interaction with M81 and the latter has a counter--rotating core,
perhaps due to a recent interaction \citep{HUNTER98B}. We neglect both
galaxies in the kinematic analyses.

\subsection{Gas Velocity Dispersion}

Throughout this paper, we adopt a single gas velocity dispersion, $\sigma_{\rm
  gas} = 11$~km~s$^{-1}$. This is typical of the outer (\hi -dominated) parts
of THINGS galaxies and agrees well with values derived by Tamburro et al. (in
prep.), who are conducting a thorough study of $\sigma_{\rm gas}$ in THINGS.
The left panel in Figure \ref{VDISPANDSTARS} motivates this choice. We plot
the median and $1\sigma$ range of $\sigma_{\rm gas}$ over the range
$0.5~r_{25}$--$1.0~r_{25}$ for each galaxy in THINGS as a function of the
inclination of the galaxy. We restrict ourselves to the outer disk because
over this regime \hi\ usually dominates the ISM. This figure shows that a
fixed $\sigma_{\rm gas} = 11 \pm 3$~km~s$^{-1}$ is a good description of the
outer disk for galaxies with $i < 60\arcdeg$; variations both within and among
galaxies are comparatively small, typically $25\%$. On the other hand, highly
inclined galaxies show large scatter and systematically high velocity
dispersions, likely because the velocity dispersion is significantly affected
by projection effects.

Variations in the gas velocity dispersion inside $0.5~r_{25}$ could be
expected to take two forms: $\sigma_{\rm gas}$ in the warm neutral medium may
increase in regions of active star formation due to stellar feedback
\citep[e.g.][]{DIB06} and the fraction of gas in a narrow--line width (cold)
\hi\ phase may increase towards the centers of galaxies
\citep[e.g.][]{SCHAYE04}. The first effect may be observed in THINGS: the
second moment maps show a gradual increase in $\sigma_{\rm gas}$ from the
outskirts to the centers of galaxies. The second effect can, in principle, be
observed using 21--cm line observations \citep[][]{DEBLOK06}, but doing so is
very challenging, requiring better spatial and velocity resolution and a
higher signal--to--noise ratio than is achieved in most THINGS targets.
Further, we know that a large fraction of the ISM is \htwo\ in the central
parts of our spiral galaxies, making it even more complicated to interpret
measurements based only on \hi . Because measuring the detailed behavior of
$\sigma_{\rm gas}$ inside $\sim 0.5~r_{25}$ is beyond the limit of our current
data, and because $\sigma_{\rm gas}$ varies only gradually in the outer parts
of galaxies, we adopt a fixed $\sigma_{\rm gas}$ \citep[an almost universal
approach in this field,
following][]{KENNICUTT89,HUNTER98,MARTIN01,WONG02,BOISSIER03}.

\subsection{Stellar Velocity Dispersion}

Direct measurements of the stellar velocity dispersion, $\sigma_*$, across the
disks of nearby galaxies are extremely scarce. In lieu of such observations
for our sample, we make four assumptions to estimate $\sigma_*$. First, we
assume that the exponential stellar scale height, $h_*$, of a galaxy does not
vary with radius. This is generally observed for edge--on disk galaxies
\citep{VANDERKRUIT81,DEGRIJS97,KREGEL02}. Second, we assume that $h_*$ is
related to the stellar scale length, $l_*$, by $l_* / h_* = 7.3 \pm 2.2$, the
average flattening ratio measured by \citet{KREGEL02}. Because we measure
$l_*$, this yields an estimate of $h_*$. Third, we assume that our disks are
isothermal in the $z$-direction, so that hydrostatic equilibrium yields $h_* =
1/2~\left( \sigma_{*,z}^2 / 2 \pi G \rho_*\right)^{0.5}$
\citep[][]{VANDERKRUIT81}, where $\rho_*$ is the midplane stellar volume
density and $\Sigma_* = 4 \rho_* h_*$ \citep{VANDERKRUIT88}.  Eliminating
$\rho_*$, then in terms of measured quantities, $\sigma_{*,z} =
\sqrt{2~\pi~G~\Sigma_*~h_*}$ \citep{VANDERKRUIT88} and

\begin{equation}
  \sigma_{*,z} = \sqrt{\frac{2~\pi~G~l_*}{7.3}}~\Sigma_*^{0.5}~.
\end{equation}

\noindent Finally, we assume a fixed ratio $\sigma_{*,z} = 0.6~\sigma_{*,r}$
to relate the radial and vertical velocity dispersions, which is reasonable
for most late-type galaxies based on the limited available evidence
\citep[e.g.,][]{SHAPIRO03}.

These assumptions yield disk--averaged $Q_{\rm stars}$ (Equation
\ref{QSTAREQ}) mostly in the range $\sim 2$--$4$, in reasonable agreement with
estimates in the Milky Way \citep[][]{JOG84,RAFIKOV01} and the expectation
that stellar disks are marginally stable against collapse, $Q_{\rm stars} \sim
2$ \citep[][and references therein]{KREGEL05}.  Our fixed flattening ratio
yields nearly identical results to the fit used by \citet{BLITZ06} to derive
$h_*$ from $l_*$. The scaling between $\sigma_*$ and maximum rotation velocity
observed by \citet{BOTTEMA93} and \citet{KREGEL05} yields roughly similar
scale heights but is more sensitive to adopted structural parameters (a
problem for several face--on galaxies). The scatter among the various methods
to estimate $\sigma_*$ or $h_*$ and observations remains $\sim 50\%$ and this
is clearly an area where more observations are needed \citep[particularly
measuring $\sigma_*$ as a function of radius, though
see][]{CIARDULLO04,MERRETT06}.

\section{Stellar Surface Densities from the IRAC 3.6 $\mu{\rm m}$ Band}
\label{STARAPP}

\begin{figure}
  \plottwo{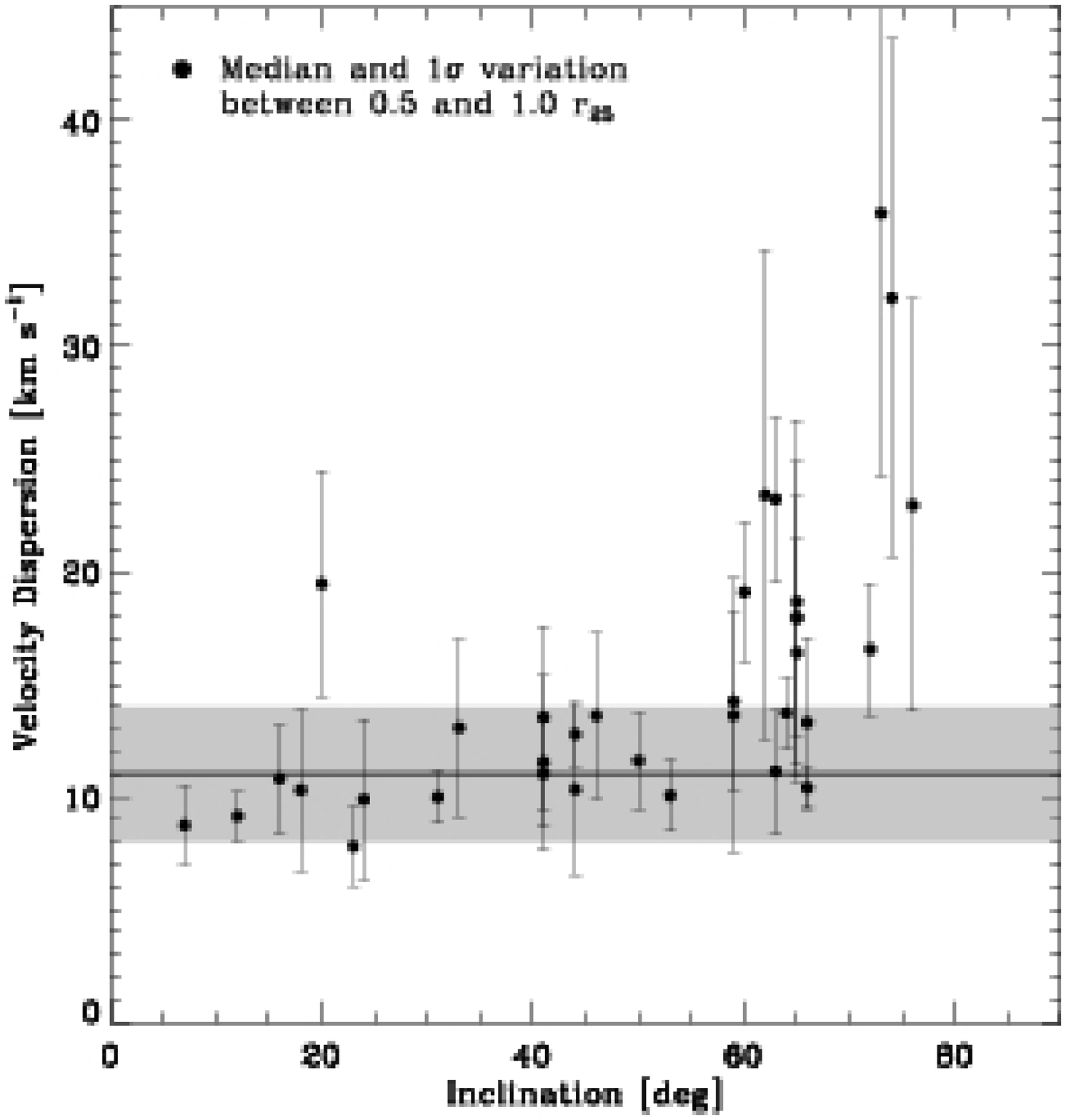}{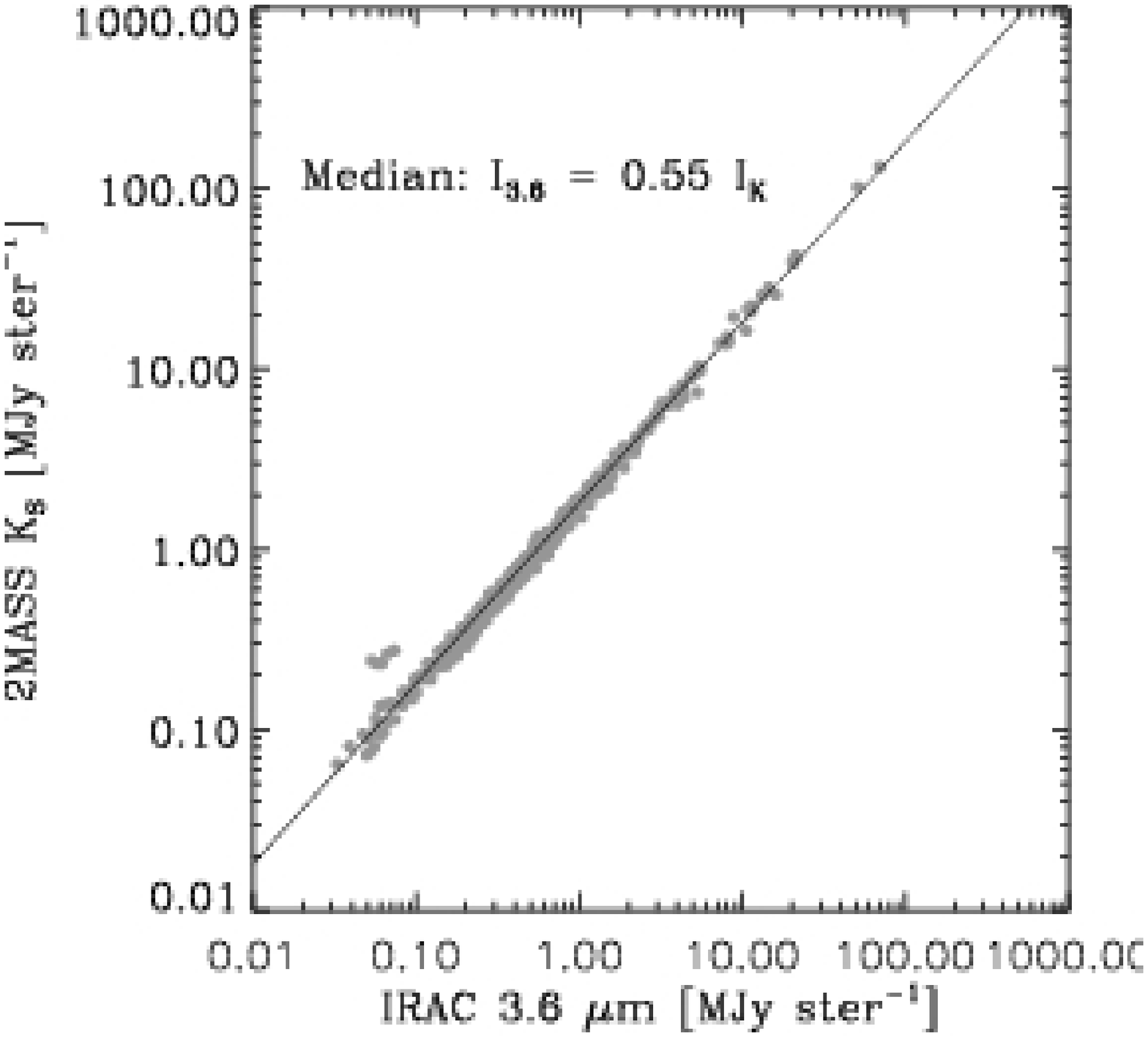} \figcaption{\label{VDISPANDSTARS}
    (left) Motivation for our adopted gas velocity dispersion, $\sigma_{\rm
      gas}$. We plot the median and $1\sigma$ range of $\sigma_{\rm gas}$ over
    the outer disk ($0.5$--$1.0~r_{25}$) of each THINGS galaxy (including
    galaxies that are not part of this study) as a function of inclination.
    We see that variations within a galaxy are relatively small and that
    $\sigma_{\rm gas} = 11 \pm 3$~km~s$^{-1}$ is a good description of
    galaxies with inclination $i \lesssim 60\arcdeg$. Above this inclination,
    $\sigma_{\rm gas}$ is systematically higher with more scatter as a likely
    consequence of projection effects in these systems.  We adopt a fixed
    $\sigma_{\rm gas}$ (gray line) throughout this work.  (right) Median
    $K$-band intensity vs.  median $3.6\mu$m intensity in $10\arcsec$-wide
    tilted rings. A single scaling, $I_{3.6} = 0.55~I_{K}$ (shown by the solid
    line), relates the two well.}
\end{figure}

SINGS \citep{KENNICUTT03} imaged most of our sample using the IRAC instrument
on {\em Spitzer} \citep{FAZIO04}. Emission from old stellar photospheres
accounts for most of the emission seen in the IRAC 3.6~$\mu$m band
\citep[e.g.,][]{PAHRE04}, although we note that there may be some contribution
from very hot dust and PAH features. Therefore we use these data to estimate
radial profiles of the stellar surface density, $\Sigma_*$. To convert from
3.6$\mu$m intensity, $I_{3.6}$, to $\Sigma_*$ we apply an empirical conversion
from 3.6$\mu$m to $K$-band intensity and then adopt a standard $K$-band
mass-to-light ratio.

We work with median profiles of $I_{3.6}$, taken over a series of
$10\arcsec$-wide tilted rings using the structural parameters in Table
\ref{STRUCTURETAB}. Real azimuthal variations, e.g., due to bars or spiral
arms, are lost. This is balanced by three major advantages from the median: 1)
we avoid contamination by hot dust or PAH emission near star forming regions,
a potential issue with the 3.6$\mu$m band; 2) we filter out foreground stars;
and 3) we increase our sensitivity by averaging over the ring. The first
advantage avoids a serious possible bias due to confusing $\Sigma_*$ and
$\Sigma_{\rm SFR}$. The latter two allow us to measure $\Sigma_*$ out to large
radii.

To calibrate the ratio of $I_{3.6}$ to $K$-band intensity, $I_{K}$, we compare
$I_{3.6}$ profiles to $I_{K}$ profiles from the 2MASS Large Galaxy Atlas
\citep[LGA][]{JARRETT03}. The profiles from the LGA are not sensitive enough
to reach $\gtrsim r_{25}$ in most cases, but they yield sufficient data to
measure a typical $I_{K}$-to-$I_{3.6}$ ratio. The right panel in Figure
\ref{VDISPANDSTARS} shows this measurement. We plot $I_{K}$ as a function of
$I_{3.6}$; each point gives median intensities in one 10$\arcsec$-wide tilted
ring in one galaxy. The solid line shows a fixed ratio $I_{3.6} = 0.55~I_K$
(both in MJy ster$^{-1}$), which matches the data very well. This agrees with
results from \citet{OH08}, who investigated the $K$-to-$3.6\mu$m ratio using
stellar population modeling and found only very weak variations.

To convert $I_{K}$ to $\Sigma_*$ we apply a fixed $K$-band mass-to-light
ratio, $\ML = 0.5$~M$_{\odot} / $L$_{\odot,K}$. This is near the mean expected
for our sample: applying the \citet{BELL03A} relation between $B-V$ color and
mean $\ML$, we find $\ML = 0.48$--$0.60$~M$_{\odot} / $L$_{\odot,K}$
\citep[using global $B-V$ colors and assuming a][IMF to match our star
formation rate]{KROUPA01}. This small range in mean $\ML$ motivates our
decision to adopt a constant value.

With our $K$-to-$3.6\mu$m ratio, $\ML = 0.5$~M$_{\odot} / $L$_{\odot,K}$, and
the $K$-band magnitude of the Sun $=3.28$ mag \citep{BINNEY98}, the conversion
from 3.6~$\mu$m intensity to stellar surface density is

\begin{equation}
\label{APPSIGSTAREQ}
\Sigma_* = \ML~\left<\frac{I_{K}}{I_{3.6}}\right>~\cos i~I_{3.6}~= 280~\cos i~I_{3.6},
\end{equation}

\noindent with $\Sigma_*$ in M$_\odot$~pc$^{-2}$ assuming a \citet{KROUPA01}
IMF and $I_{3.6}$ in MJy~ster$^{-1}$.

The major uncertainty in Equation \ref{APPSIGSTAREQ} is the mass--to--light
ratio, which depends on the star formation history, metallicity, and IMF. The
mass--to--light ratio varies less in the NIR than the optical but it does
vary, showing $\sim 0.1$~dex scatter for redder galaxies and $0.2$~dex for
bluer galaxies \citep{BELL01,BELL03A}. Because metallicity and star formation
history exert different influences on galaxy colors and $\ML$, these
variations are not readily inferred from colors \citep[unlike in the optical,
e.g.,][]{BELL03A}.

In their analysis of the THINGS rotation curves, \citet{DEBLOK08} also
derive $\Sigma_*$ from $I_{3.6}$. They use $J-K$ colors from the 2MASS
LGA to estimate variations in $\ML$. Their Figure 21 compares our
integrated masses to those that they derive using color--dependent
$\ML$ for both a \citep{KROUPA01} and ``diet Salpeter''
\citep[see][]{BELL01} IMF. Because they use the \citet{BELL01}
results, which have a fairly strong dependence on NIR color, they find
$\ML$ $\sim 30$--$40\%$ higher than we do in massive (red) spiral
galaxies, even for matched \citep{KROUPA01} IMFs.


\section{Star Formation Rate Surface Density Maps}
\label{SFRAPP}

We combine GALEX FUV and {\em Spitzer} 24$\mu$m maps to estimate the star
formation rate surface density, $\Sigma_{\rm SFR}$, along each line of sight.
FUV maps show mostly photospheric emission from O and B stars and thus trace
unobscured star formation over a timescale of $\tau_{\rm FUV} \sim
10$--$100$~Myr \citep[e.g.,][]{KENNICUTT98B,CALZETTI05,SALIM07}. Emission at
24$\mu$m originates from small dust grains mainly heated by UV photons from
young stars. It has been shown to directly relate to ongoing star formation
over a timescale of $\tau_{24} \sim 10$~Myr
\citep[e.g.,][]{CALZETTI05,PEREZGONZALEZ06,CALZETTI07}. We adopt this tracer
because: 1) the resolution and sensitivity of the GALEX FUV and {\em Spitzer}
24$\mu$m maps are both good (and well--matched), 2) these data are available
for our whole sample, and 3) the combination is directly sensitive to both
exposed and embedded star formation.

In this section, we take a practical approach, calibrating our tracer by
comparing it to other estimates of $\Sigma_{\rm SFR}$. For a more thorough
discussion of the relationship between extinction, UV, and IR emission, we
refer the reader to, e.g.,
\citet[][]{CALZETTI95,BUAT02,BELL03B,CORTESE06,BOISSIER07}. Our tracer builds
mainly on two recent results: 1) for entire galaxies, \citet{SALIM07} showed
that FUV emission can be used to accurately measure star formation rates (with
typical $\tau_{\rm FUV} \sim 20$~Myr) if extinction is properly accounted for
and 2) \citet[][]{CALZETTI07} and \citet{KENNICUTT07} demonstrated that
24$\mu$m data could be used to accurately correct H$\alpha$ for extinction. We
combine these results using a method similar to that of \citet{CALZETTI07}:
via comparisons to other estimates of extinction--corrected $\Sigma_{\rm
  SFR}$, we derive a linear combination of FUV and 24$\mu$m intensity that we
use to estimate $\Sigma_{\rm SFR}$,

\begin{eqnarray}
\label{SFREQ}
\Sigma_{\rm SFR} = \left(~8.1 \times 10^{-2}~I_{\rm FUV} + 3.2_{-0.7}^{+1.2} \times
  10^{-3}~I_{24}~\right)~\cos i~.
\end{eqnarray}

\noindent Here $\Sigma_{\rm SFR}$ has units of ${\rm M}_\odot~{\rm
  kpc}^{-2}~{\rm yr}^{-1}$ and FUV and 24$\mu$m intensity are each in ${\rm
  MJy~ster}^{-1}$. The first term measured unobscured SFR using the FUV-to-SFR
calibration found by \citet{SALIM07}; the second term measures embedded SFR
from 24$\mu$m and is $30\%$ higher than the matching term in the
H$\alpha$+24$\mu$m calibration of \citet[][]{CALZETTI07}.  The additional
weight reflects the fact that FUV is more heavily absorbed than H$\alpha$.
More detailed motivation for Equation \ref{SFREQ} is given in Appendix
\ref{APP_SFRMOTIVATION}.

Following \citet{CALZETTI07}, Equation \ref{SFREQ} assumes the default initial
mass function (IMF) of STARBURST99 \citep{LEITHERER99}, the broken power law
given by \citet{KROUPA01} with a maximum mass of $120$~M$_{\odot}$. This
yields $\Sigma_{\rm SFR}$ a factor of $1.59$ lower than a
$0.1$--$100$~M$_{\odot}$ \citet{SALPETER55} IMF
\citep[e.g.][]{KENNICUTT89,KENNICUTT98A}. Our FUV term is Equation 10 from
\citet{SALIM07} divided by this value ($1.59$); the calibration is the same
found for the \citet{CHABRIER03} IMF over the range 0.1--100~M$_\odot$ (their
Equations 7 and 8).

\subsection{Data}

\subsubsection{GALEX NGS FUV Maps}

We use FUV maps obtained by the GALEX satellite \citep{MARTIN05} as part of
the GALEX Nearby Galaxies Survey \citep[NGS,][]{GILDEPAZ07}.  The GALEX FUV
band covers $\lambda = 1350$--$1750~$\AA\ with a resolution of $5.6\arcsec$
and a $1.25\arcdeg$ diameter field of view.  These maps have excellent
sensitivity and well-behaved backgrounds over a large field of view.  GALEX
simultaneously observes in a near--UV (NUV) band ($\lambda = 1750$--$2750~$\AA
). We use these data to measure UV colors and to identify foreground stars.


We correct the FUV maps for Galactic extinction using the dust map of
\citet{SCHLEGEL98}. We subtract a small background, measured away from the
galaxy. We identify and remove foreground stars using their UV color: any
pixel with a NUV--to--FUV intensity ratio $\gtrsim 15$ (varying $\pm 5$ from
galaxy--to--galaxy) that is also detected in the NUV map with $> 5\sigma$
significance is blanked. In convolution to our working resolution, blank
pixels are replaced with the average of nearby data.  We also blank a few
regions with obvious artifacts. These include bright stars (e.g., in NGC~3198
and NGC~6946) that are usually beyond the optical radius of the galaxy and
M51b, the companion of M51a.

\subsubsection{SINGS 24$\mu$m Maps}

We use maps of 24$\mu$m emission obtained as part of the SINGS Legacy program
\citep{KENNICUTT03} using the MIPS instrument on {\em Spitzer}
\citep{RIEKE04}. \citet{GORDON05} describe the reduction of these scan maps,
which have $6\arcsec$ resolution. The sensitivity and background subtraction
are both very good, and it is typical to find $3\sigma$ emission at $\sim
r_{25}$ in a spiral. The MIPS PSF at 24$\mu$m is complex at low levels, but
our working resolution of $\sim20\arcsec$ makes this only a minor concern.

NGC~3077, NGC~4214, and NGC~4449 are not part of SINGS. For these galaxies we
use 24$\mu$m (and IRAC) maps from the {\em Spitzer} archive. We use the post
basic calibrated data produced by the automated {\em Spitzer} pipeline.

As with the FUV maps, we subtract a small background from the 24~$\mu$m maps,
which we measure away from the galaxy. We blank the same set of foreground
stars as in the FUV maps. In convolution to our working resolution, these
pixels are replaced with the average of nearby data. We also blank the edges
of the 24$\mu$m maps perpendicular to the scan, which are noisy (and outside
the optical radius) and the same artifacts blanked in the FUV maps.

\subsubsection{SINGS H$\alpha$}

The SINGS fourth data release includes H$\alpha$ maps, which we use to compare
$\Sigma_{\rm SFR}$ derived from H$\alpha$, FUV, and 24$\mu$m emission in $13$
galaxies. We convert H$\alpha$ to $\Sigma_{\rm SFR}$ following
\citet{CALZETTI07} and the SINGS data release documentation.  We correct for
[NII] contamination following \citet{CALZETTI05} and \citet{LEE06} and correct
for Galactic extinction using the \citet{SCHLEGEL98} dust maps. We check the
flat-fielding by eye and mask regions or fit backgrounds where necessary.

\subsection{Motivation for the FUV+24 to $\Sigma_{\rm SFR}$ Relation}
\label{APP_SFRMOTIVATION}

Because 24$\mu$m lies well short of the FIR peak for a typical galaxy SED
\citep[e.g.,][]{DALE02}, a measurement in this band does not directly trace
the total IR luminosity. Therefore using 24$\mu$m to measure embedded SFR
relies on modeling of the IR SED or empirical calibration against other
estimates of extinction.  \citet{CALZETTI07} compared H$\alpha$ and 24$\mu$m
to Paschen $\alpha$ (Pa$\alpha$) emission, a tracer of ionizing photons
largely unaffected by extinction. They showed that a linear combination of
H$\alpha$ and 24$\mu$m,

\begin{eqnarray}
\label{CALZETTISFR}
\nonumber {\rm SFR}_{\rm Tot} &=& {\rm SFR}_{\rm H\alpha}^{\rm unobscured}
\left( {\rm H}\alpha \right) + {\rm SFR}_{\rm H\alpha}^{\rm embedded} \left(
  24\mu{\rm m}\right) \\
{\rm SFR}_{\rm Tot} &=& 5.3 \times 10^{-42}
\left( L \left( H\alpha \right) + 0.031 ~ L \left(  24\mu{\rm m} \right) \right)~,
\end{eqnarray}

\noindent matches ${\rm SFR}_{\rm Tot}$ inferred from Pa$\alpha$ for 220
individual star forming regions in 33 nearby galaxies and that the same
calibration also works well when integrated over a large fraction of a galaxy
disk. Here $L\left( H\alpha \right)$ is the luminosity of H$\alpha$ emission
from the region in erg~s$^{-1}$ and $L \left( 24\mu{\rm m} \right) = \nu_{\rm
  24 \mu m}~L_{\nu} \left( 24\mu{\rm m}\right)$, also in erg~s$^{-1}$, is the
specific luminosity of the region times the frequency at 24$\mu$m. SFR is the
star formation rate in that region in ${\rm M}_\odot~{\rm yr}^{-1}$.

We require an analogous formula to combine FUV and 24$\mu$m data,

\begin{equation}
  \label{OURGOAL}
  {\rm SFR}_{\rm Tot} = {\rm SFR}^{\rm unobscured}_{\rm FUV}\left({\rm FUV}\right) +
  {\rm SFR}^{\rm embedded}_{\rm FUV}\left({\rm FUV}, 24\mu{\rm m}\right)~.
\end{equation}

\noindent The first term is the SFR implied by a particular FUV luminosity
taking no account of internal extinction. The second term is the SFR that can
be attributed to FUV light that does not reach us --- i.e., the extinction
correction for the first term ---, which we infer from the 24$\mu$m luminosity
and may also depend on the ratio of FUV to 24$\mu$m intensity.

We adopt the first term in Equation \ref{OURGOAL} from \citet{SALIM07}, who
studied the relationship between FUV emission and SFR in $\sim 50,000$
galaxies, combining multi-band photometry with population synthesis modeling
and comparing to H$\alpha$ emission. They found

\begin{equation}
  \label{SALIMSFR}
  {\rm SFR}^{\rm unobscured}_{\rm FUV} = 0.68 \times 10^{-28}~L_{\nu}\left({\rm  FUV}\right)~,
\end{equation}

\noindent with SFR in ${\rm M}_{\odot}~{\rm yr}^{-1}$ and $L_{\nu}\left({\rm
    FUV}\right)$ in ${\rm erg~s}^{-1}{\rm~Hz}^{-1}$. This yields SFRs $\sim
30\%$ lower than the relation given by \citet{KENNICUTT98B} because of
metallicity, model, and star formation history differences between their
sample and \citeauthor{KENNICUTT98B}'s model.

We calibrate the second term in Equation \ref{OURGOAL} in two ways: 1) we use
simple assumptions to extrapolate SFR$_{\rm FUV}^{\rm embedded} \left({\rm
    FUV}, 24\mu{\rm m}\right)$ from ${\rm SFR}_{\rm H\alpha}^{\rm embedded}
\left( 24\mu{\rm m}\right)$, which was measured by \citet{CALZETTI07}; and 2)
we make several independent estimates of SFR$_{\rm Tot}$ --- by comparing
SFR$_{\rm Tot}$ with FUV emission, we directly measure the second term in
Equation \ref{OURGOAL}. We phrase both analyses in terms of the factor $W_{\rm
  FUV}$, defined as

 \begin{equation}
\label{DEFINEWEIGHT}
W_{\rm FUV} \left({\rm FUV}, 24\mu{\rm m}\right) = \frac{{\rm SFR}^{\rm embedded}_{\rm FUV}}{{\rm SFR}^{\rm
    embedded}_{\rm H\alpha} \left( 24\mu{\rm m} \right)}~.
\end{equation}

\noindent The numerator is the second term in Equation \ref{OURGOAL} and the
denominator is the relation between embedded H$\alpha$ and 24$\mu$m emission
measured by \citet[][Equation \ref{CALZETTISFR}]{CALZETTI07}. To measure
$W_{\rm FUV}$, we combine Equations \ref{OURGOAL} and \ref{DEFINEWEIGHT} to
obtain

\begin{equation} {\rm SFR}_{\rm Tot} = {\rm SFR}_{\rm FUV}^{\rm
    unobscured}\left({\rm FUV} \right) + W_{\rm FUV} \left({\rm FUV},
    24\mu{\rm m}\right){\rm SFR}^{\rm embedded}_{\rm H\alpha}\left( 24\mu{\rm
      m} \right)
\end{equation}

\noindent and solve for $W_{\rm FUV}$ in terms of measurable quantities

\begin{equation} 
  \label{WFUVEQ}
  W_{\rm FUV} = \frac{{\rm SFR}_{\rm Tot} - {\rm SFR}_{\rm FUV}^{\rm
      unobscured} \left({\rm FUV}\right)}{
    {\rm SFR}^{\rm embedded}_{\rm H\alpha}\left( 24\mu{\rm
        m} \right)}~.
\end{equation}

\noindent So that to estimate $W_{\rm FUV}$ over a line of sight we require
FUV and 24$\mu$m intensities and an estimate of ${\rm SFR}_{\rm Tot}$.

\subsubsection{Simple Extrapolation}

In conjunction with a direct measurement, it is helpful to have a basic
expectation for $W_{\rm FUV}$. We calculate this by combining a Galactic
extinction law and a typical nebular-to-stellar extinction ratio. In terms of
H$\alpha$ extinction, $A_{\rm H\alpha}$, and FUV extinction, $A_{\rm FUV}$,
Equations \ref{CALZETTISFR} and \ref{OURGOAL} are

\begin{eqnarray}
  \label{HAEXTEQN} {\rm SFR}_{\rm Tot} &=& {\rm SFR}^{\rm unobscured}_{\rm
    H\alpha}~10^{A_{\rm H\alpha} / 2.5} \\
  \nonumber {\rm SFR}_{\rm Tot} &=& {\rm SFR}_{\rm FUV}^{\rm unobscured}~10^{A_{\rm FUV} / 2.5}~.
 \end{eqnarray}

 \noindent For a Galactic extinction law $A_{\rm FUV} / A_R \approx 8.24 /
 2.33$ \citep{CARDELLI89,WYDER07}. We may also expect that FUV originates from
 a slightly older and more dispersed population than H$\alpha$. If we assume a
 typical nebular-to-stellar extinction ratio of $A_{\rm H\alpha} / A_R \approx
 2$ \citep{CALZETTI94,ROUSSEL05}, then we expect $A_{\rm FUV} / A_{\rm
   H\alpha} \approx 1.8$ (if FUV comes mostly from a very young population
 coincident with H$\alpha$, we instead expect $A_{\rm FUV} / A_{\rm H\alpha}
 \sim 3.6$). Combined with Equations \ref{OURGOAL} -- \ref{HAEXTEQN} these
 assumptions yield

 \begin{eqnarray}
 \label{SIMPLECASE}
 \frac{1}{f + W_{\rm FUV} - 1} + 1 = \left(\frac{W_{\rm FUV}}{f} +
   1\right)^{1/1.8}~ {\rm where}~f = \frac{{\rm SFR}_{\rm
     FUV}^{\rm unobscured}\left({\rm FUV}\right)}{{\rm SFR}_{\rm H\alpha}^{\rm embedded} \left(24\mu{\rm m}\right)}~,
\end{eqnarray}

\noindent which we may solve for $W_{\rm FUV}$ given $f$, the ratio of
observed FUV-to-24$\mu$m intensities (in SFR units). 

For $A_{\rm H\alpha} = 1.1$~mag, a typical value in disk galaxies
\citep{KENNICUTT98B}, $f \approx 0.26$ and Equation \ref{SIMPLECASE} suggests
$W_{\rm FUV} \approx 1.3$. For higher $A_{\rm H\alpha}$, expected for the
inner parts of spiral galaxies or arms, $f$ will be lower and we expect lower
values of $W_{\rm FUV}$, approaching $W_{\rm FUV} = 1$ where both FUV and
H$\alpha$ are almost totally absorbed (and SFR$_{\rm Tot}$ is determined
totally from 24$\mu$m emission). For lower $A_{\rm H\alpha}$, e.g., expected
in dwarf galaxies or the outer parts of spirals, we expect $W_{\rm FUV}$ to
approach the ratio of extinctions, $1.8$, in the optically thin case.

\subsubsection{Measuring $W_{\rm FUV}$}

We measure $W_{\rm FUV}$ directly from observations by comparing FUV and
24$\mu$m emission to various estimates of SFR$_{\rm Tot}$.  We perform these
tests in the $13$ galaxies with SINGS H$\alpha$ data. Over a common set of
lines of sight where we estimate H$\alpha$, FUV, and 24$\mu$m to all be
complete, we estimate $\Sigma_{\rm SFR}$ and $W_{\rm FUV}$ (from Equation
\ref{WFUVEQ}) in 5 ways:

\begin{enumerate}
\item Combining H$\alpha$+24$\mu$m using Equation \ref{CALZETTISFR}
  \citep[][]{CALZETTI07}.
\item From $24\mu$m emission alone, using the (nonlinear) relation found by
  \citet[][their Equation 8]{CALZETTI07}.
\item From H$\alpha$ alone, taking $A_{\rm H\alpha} = 1.1$~mag, a typical
  extinction averaged over disk galaxies, though not necessarily a good
  approximation for each line of sight \citep{KENNICUTT98B}.
\item From H$\alpha$ emission, estimating $A_{\rm H\alpha}$ from $\Sigma_{\rm
    HI}$ and $\Sigma_{\rm H2}$ following \citet{WONG02}. We assume a Galactic
  dust--to--gas ratio and treat dust associated with \hi\ as a foreground
  screen obscuring the H$\alpha$ while treating dust associated with \htwo\ as
  evenly mixed with H$\alpha$ emission.
\item From FUV emission, estimating $A_{\rm FUV}$ for every line of sight
  applying the relationship between FUV--to--NUV color and $A_{\rm FUV}$
  measured for nearby galaxies by \citet[][]{BOISSIER07}.
\end{enumerate}

\noindent In principal, the first method is superior to the others because
\citet{CALZETTI07} directly calibrated it against Pa$\alpha$, and because it
incorporates both H$\alpha$ and IR emission, offering direct tracers of both
ionizing photons and dust--absorbed UV light.  The other four methods offer
checks on SFR$_{\rm Tot}$ that are variously independent of 24$\mu$m, FUV, or
H$\alpha$ emission, allowing us to estimate the plausible range of both
$W_{\rm FUV}$ and the uncertainty in $\Sigma_{\rm SFR}$.

\subsubsection{Derived Relation}

\begin{figure*}
  \plotone{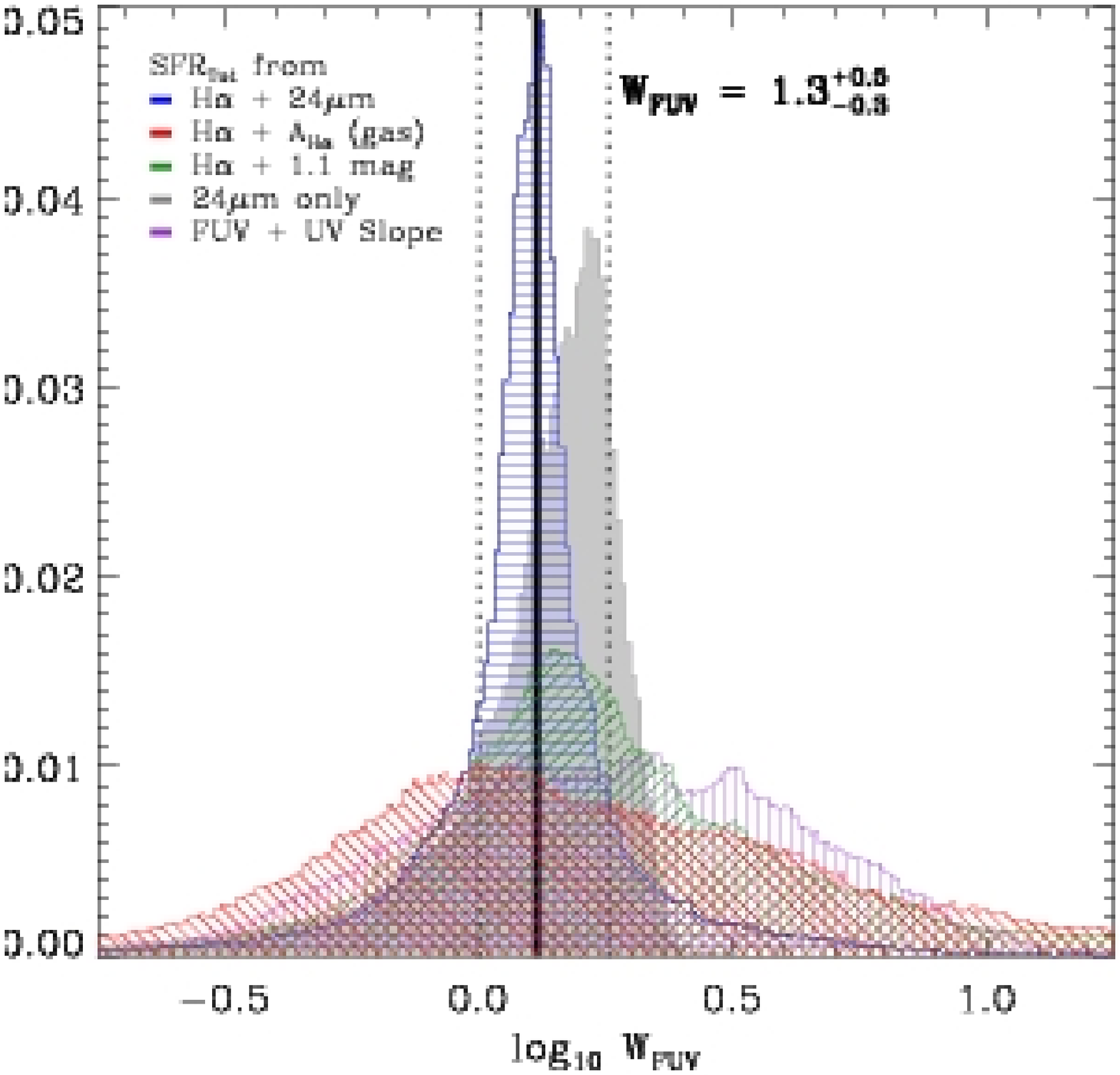}
  \plottwo{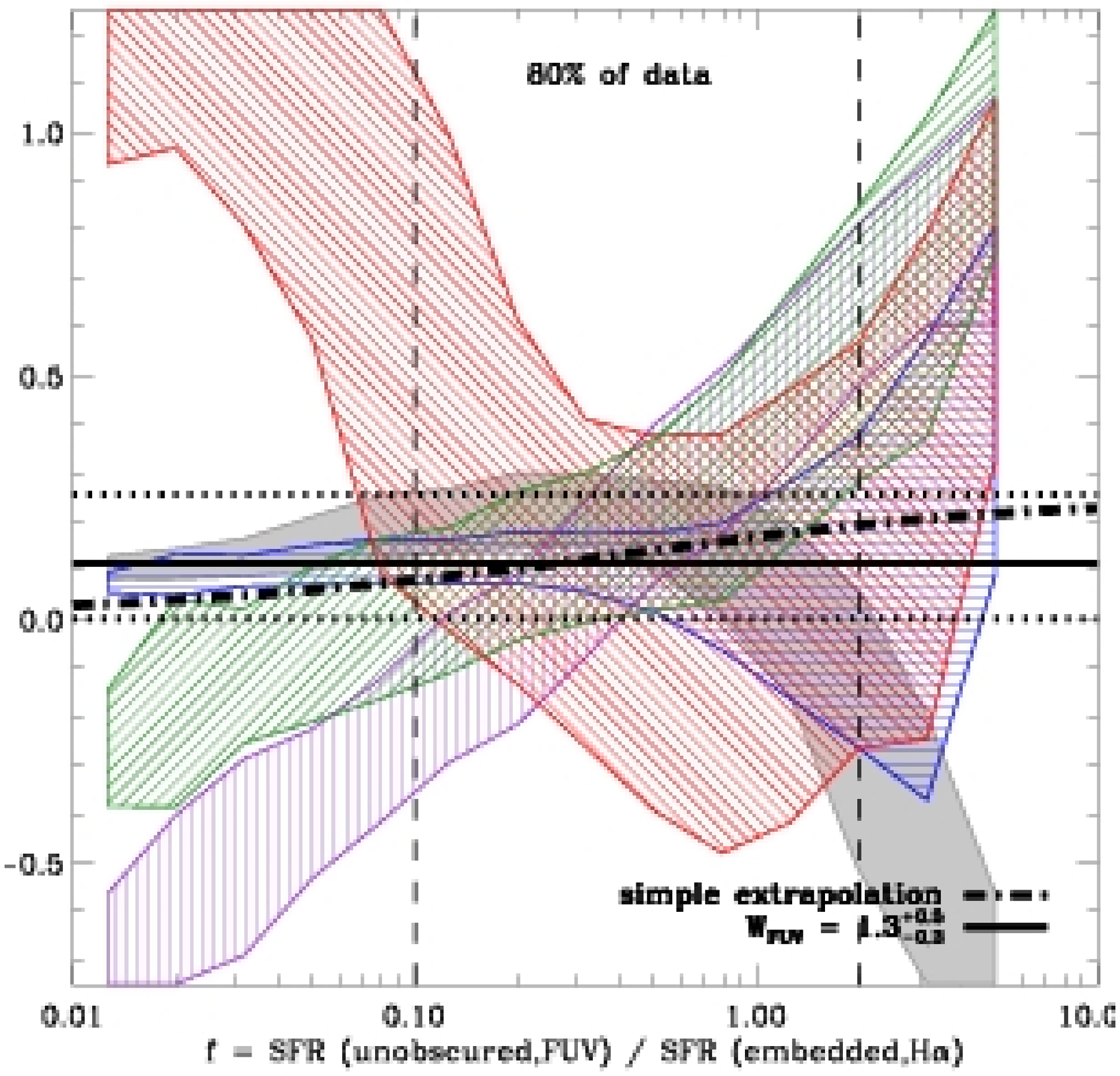}{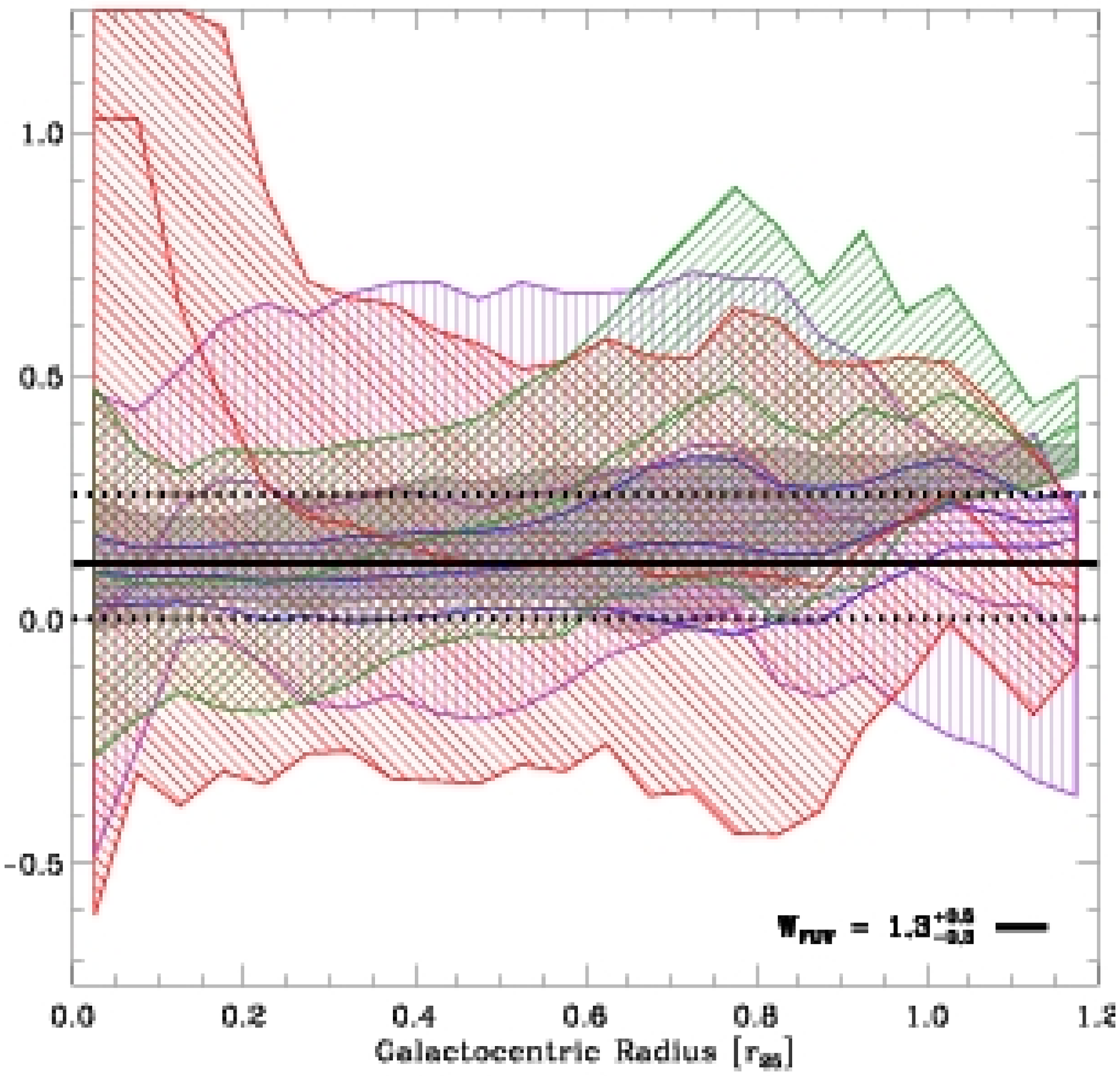}  
  \figcaption{\label{SFMETHODS} $W_{\rm FUV}$, the calibration of the 24$\mu$m
    term to estimate the SFR from a linear combination of FUV and 24$\mu$m
    emission. We measure $W_{\rm FUV}$ pixel--by--pixel by comparing FUV and
    24$\mu$m intensity to $\Sigma_{\rm SFR}$ estimated in five ways: ({\em
      blue}) combining H$\alpha$ and 24$\mu$m; ({\em gray}) using only
    24$\mu$m; ({\em red}) using H$\alpha$, estimating extinction from the gas;
    ({\em green}) using H$\alpha$, assuming a typical extinction; and ({\em
      purple}) using FUV emission, estimating $A_{\rm FUV}$ from the UV color.
    We plot the resulting $W_{\rm FUV}$ in three ways: ({\em top}) as
    normalized histograms; ({\em bottom left}) as a function of $f$, the ratio
    FUV to 24$\mu$m emission along a line of sight (see Equation
    \ref{SIMPLECASE}); and ({\em bottom right}) as a function of
    galactocentric radius normalized by $r_{25}$.  The hatched regions in the
    bottom panels show the median trend $\pm1\sigma$ for each case.  In each
    panel, we indicate our adopted $W_{\rm FUV} = 1.3^{+0.5}_{-0.3}$.  The
    dash--dotted curve in the bottom left panel shows the expectation for a
    typical extinction law and nebular-to-stellar extinction ratio and the
    vertical dashed lines show the range of $f$ that includes $80\%$ of the
    data.}
\end{figure*}

Figure \ref{SFMETHODS} shows the results of these calculations. In the top
panel, we plot the normalized distribution of $W_{\rm FUV}$ for each estimate
of $\Sigma_{\rm SFR}$. The bottom left panel shows how each distribution of
$W_{\rm FUV}$ depends on the FUV--to--24$\mu$m ratio, $f$ (Equation
\ref{SIMPLECASE}). The bottom right panel shows how $W_{\rm FUV}$ varies with
normalized galactocentric radius.

The median $W_{\rm FUV}$ derived in various ways spans a range from $\sim 1.0$
-- $1.8$. The two 24$\mu$m--based methods (blue and gray) both yield $W_{\rm
  FUV} \sim 1.3$ with relatively narrow distributions. Using FUV and
UV--colors yields the highest expected $W_{\rm FUV}$, $\sim 1.8$; estimating
$A_{\rm H\alpha}$ from gas yields the lowest $W_{\rm FUV}$, peaked near $\sim
1.0$, though the distribution is very wide. This range of values agrees
reasonably with our extrapolation (seen as a dash--dotted curve in the top
right panel), which also lead us to expect a typical $W_{\rm FUV}$ of $1.3$
and a reasonable range of $1.0$--$1.8$.

The bottom panels show that while individual methods to estimate $W_{\rm FUV}$
do exhibit significant systematics (particularly at very high and low of $f$),
simply fixing $W_{\rm FUV} = 1.3$ is a reasonable description of most data
(the dashed lines in the center panel bracket $\sim 80\%$ of the measured
$f$).  $W_{\rm FUV}$ does not have to be constant. Indeed, we expect it to
vary with $f$ based on simple assumptions and very basic arguments.  However,
a constant $W_{\rm FUV}$ is consistent with the data and is also the simplest,
most conservative approach. Therefore, this is how we proceed: taking $W_{\rm
  FUV} = 1.3^{+0.5}_{-0.3}$, Equation \ref{OURGOAL} becomes

\begin{eqnarray}
  \label{FINALSFR}
  {\rm SFR}_{\rm Tot} &=& 0.68 \times 10^{-28}~L_{\nu}\left( {\rm FUV}
  \right) + 2.14^{+0.82}_{-0.49} \times 10^{-42}~L \left( {\rm 24}\mu{\rm m}\right)~,
\end{eqnarray}

\noindent We convert Equation \ref{FINALSFR} from luminosity to intensity
units using $\nu_{24\mu{\rm m}} = 1.25 \times 10^{13}$~Hz, $1$~MJy$ =
10^{-17}$~erg~s$^{-1}$~Hz$^{-1}$~cm$^{-2}$, and $L_{\nu} = 4\pi~A~I_{\nu}$,
where $A$ is the physical area subtended by the patch of sky being considered.
This yields Equation \ref{SFREQ},

 \begin{eqnarray}
\label{FINALSD}
 \Sigma_{\rm SFR} &=& 8.1 \times 10^{-2} I_{\rm FUV}+3.2^{+1.2}_{-0.7}\times 10^{-3} I_{24}~,
 \end{eqnarray}

 \noindent with $I_{\rm FUV}$ and $I_{24}$ in units of MJy~ster$^{-1}$ and
 $\Sigma_{\rm SFR}$ in units of ${\rm M}_\odot~{\rm kpc}^{-2}~{\rm yr}^{-1}$.

 \subsubsection{Uncertainty in $\Sigma_{\rm SFR}$}

\begin{figure*}
  \begin{center}
  \leavevmode \epsfxsize=.31\columnwidth \hbox{\epsfbox{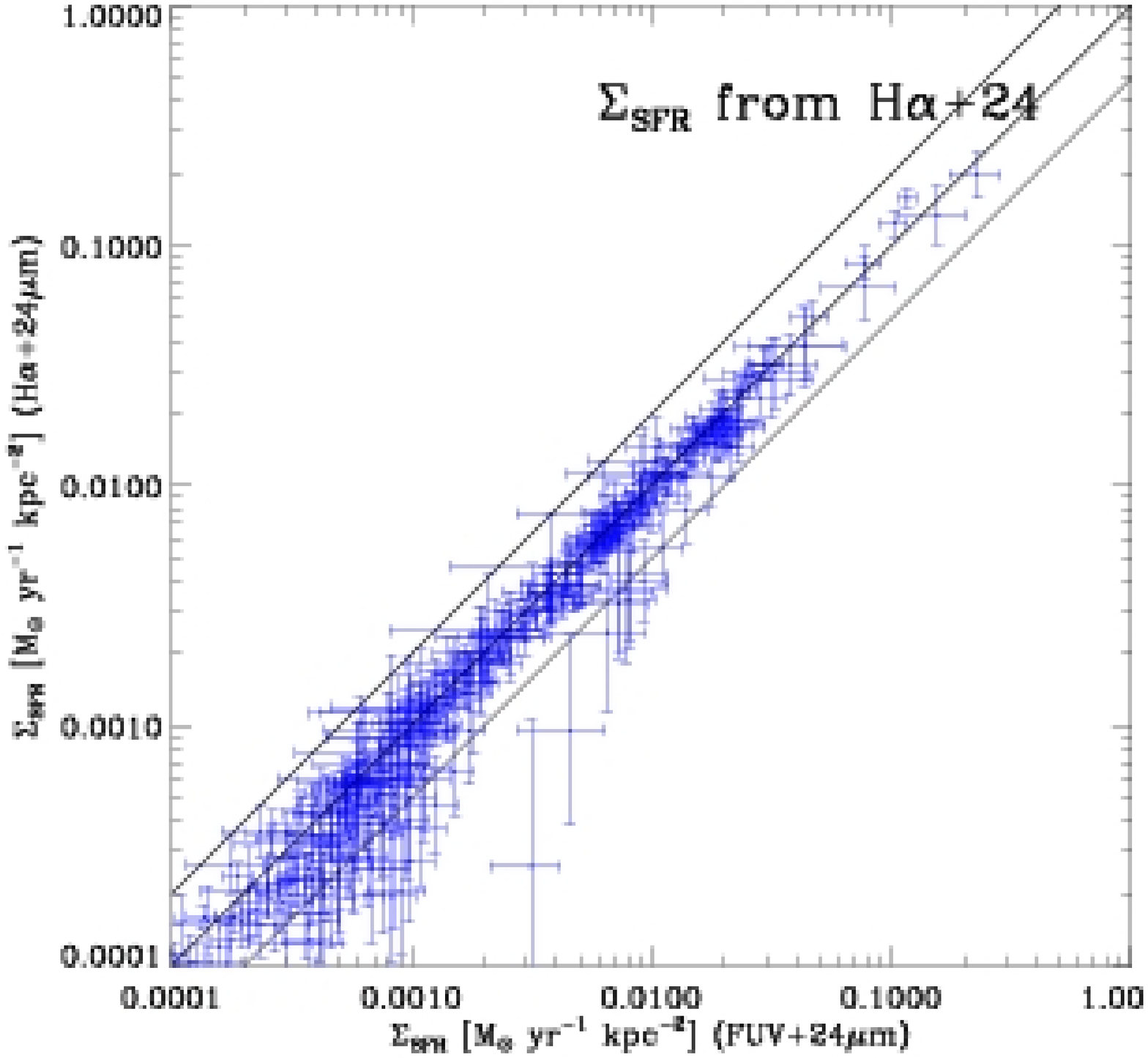}}\hfil
  \epsfxsize=.31\columnwidth \hbox{\epsfbox{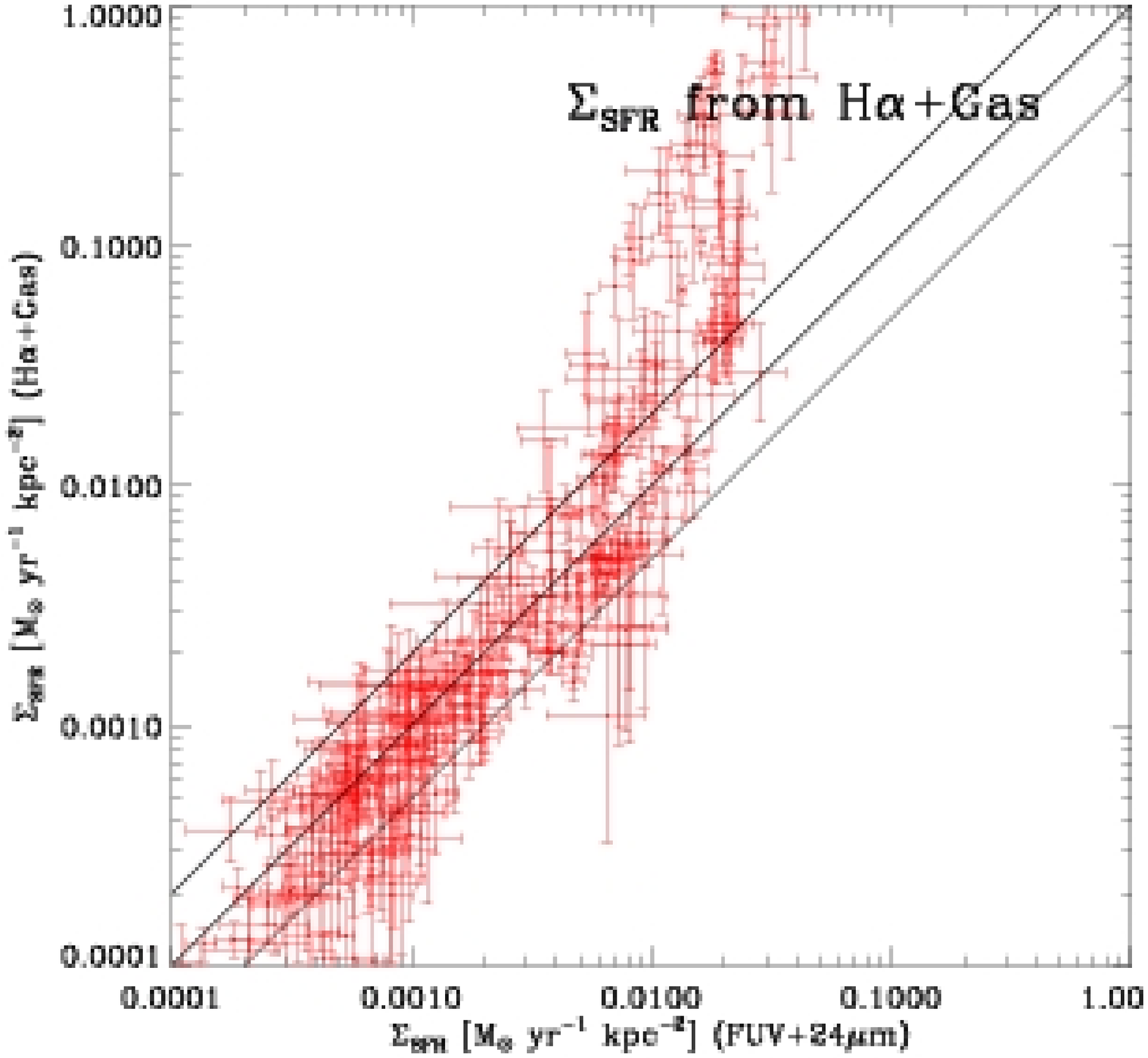}}\hfil
  \epsfxsize=.31\columnwidth \hbox{\epsfbox{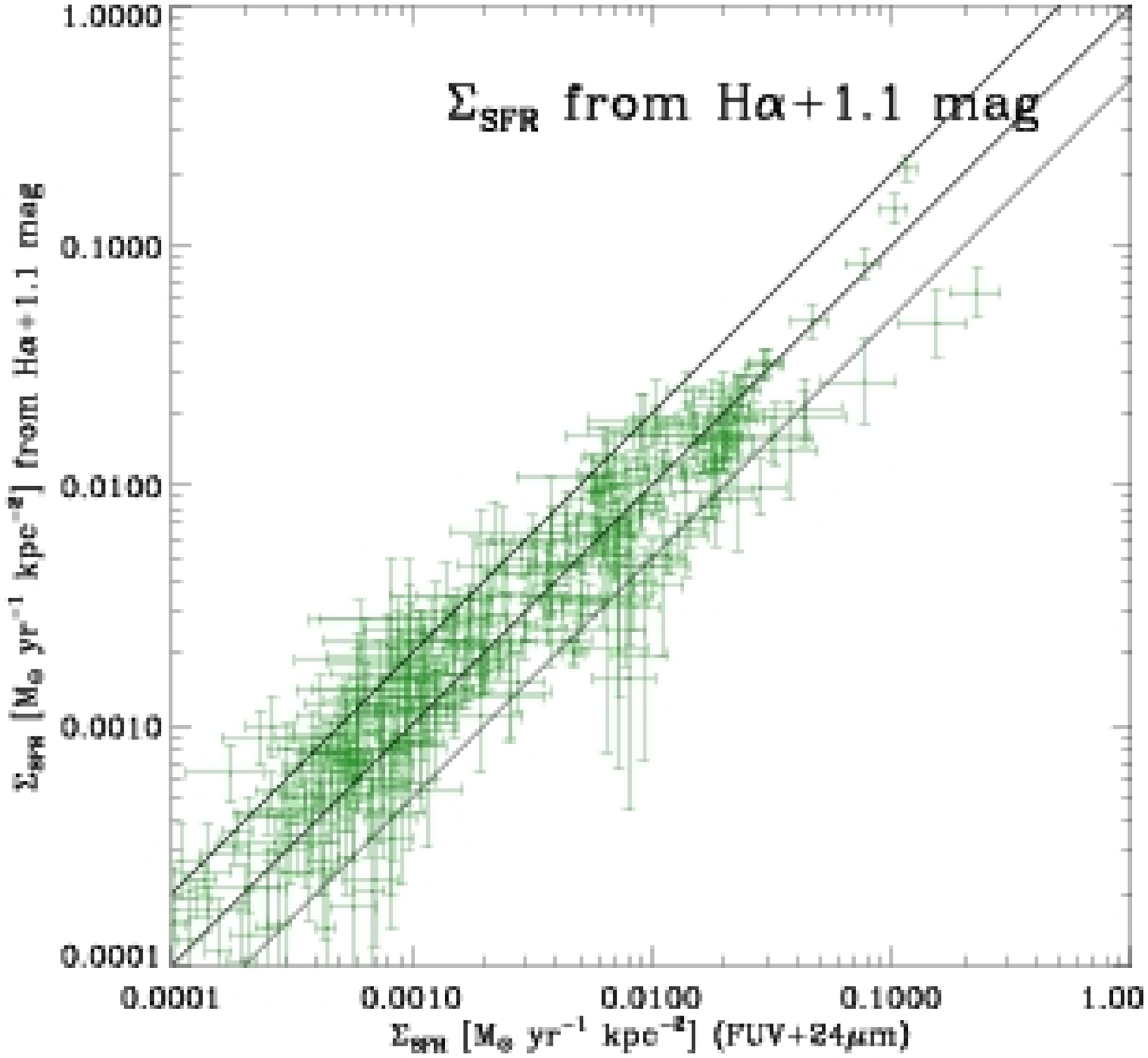}}\hfil
  \leavevmode \epsfxsize=.31\columnwidth \hbox{\epsfbox{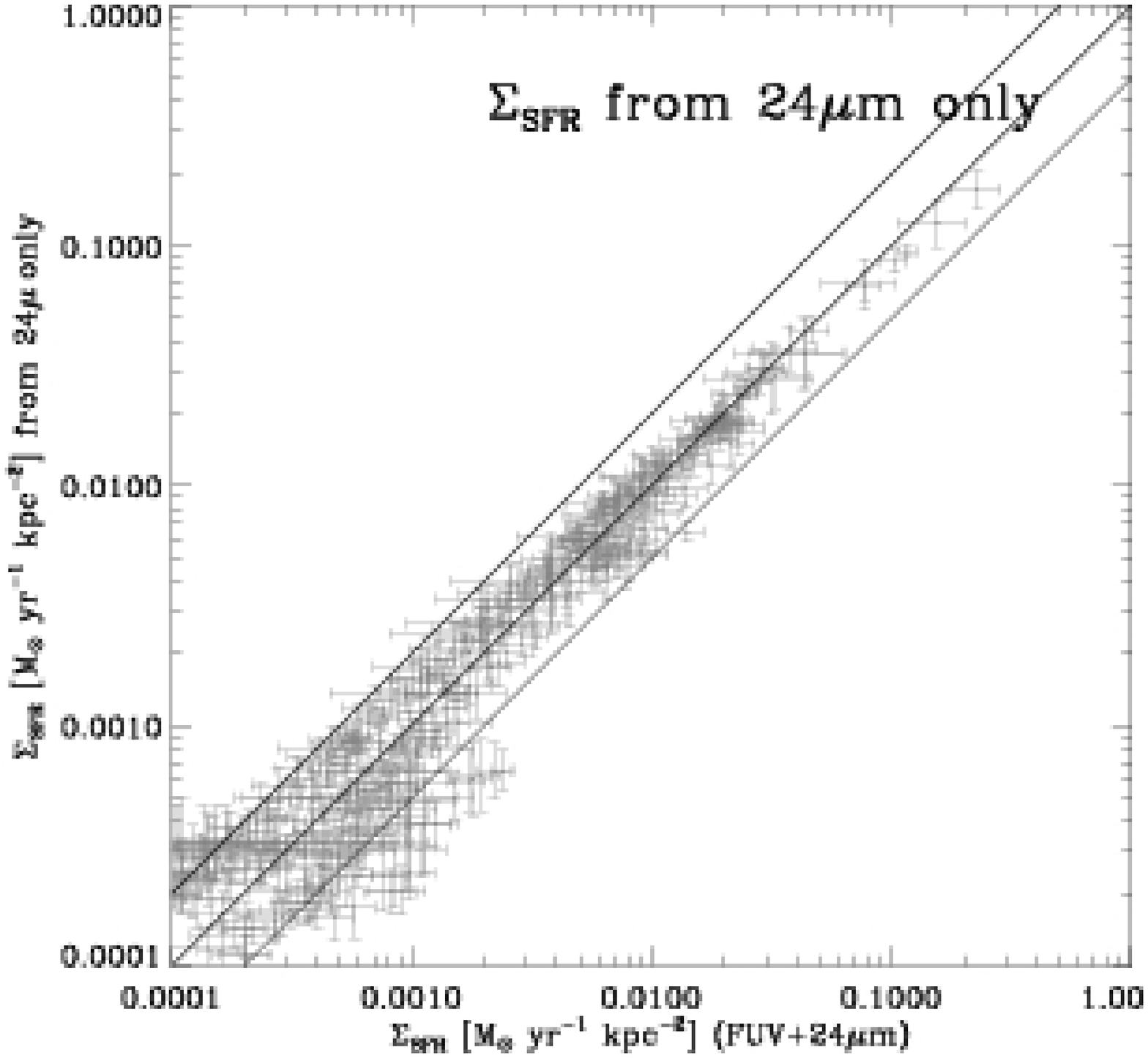}}\hfil
  \epsfxsize=.31\columnwidth \hbox{\epsfbox{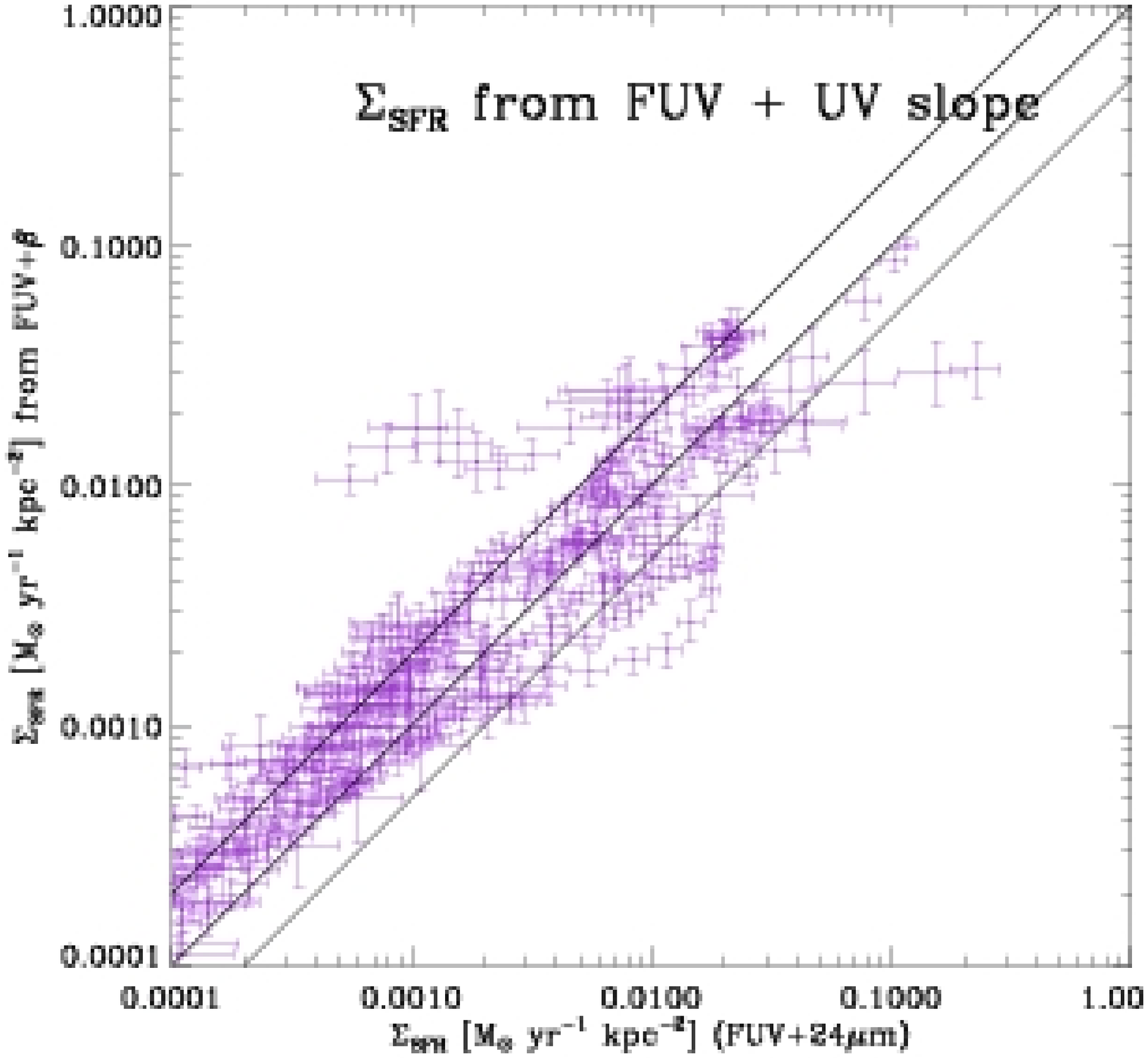}}\hfil
  \epsfxsize=.31\columnwidth \hbox{\epsfbox{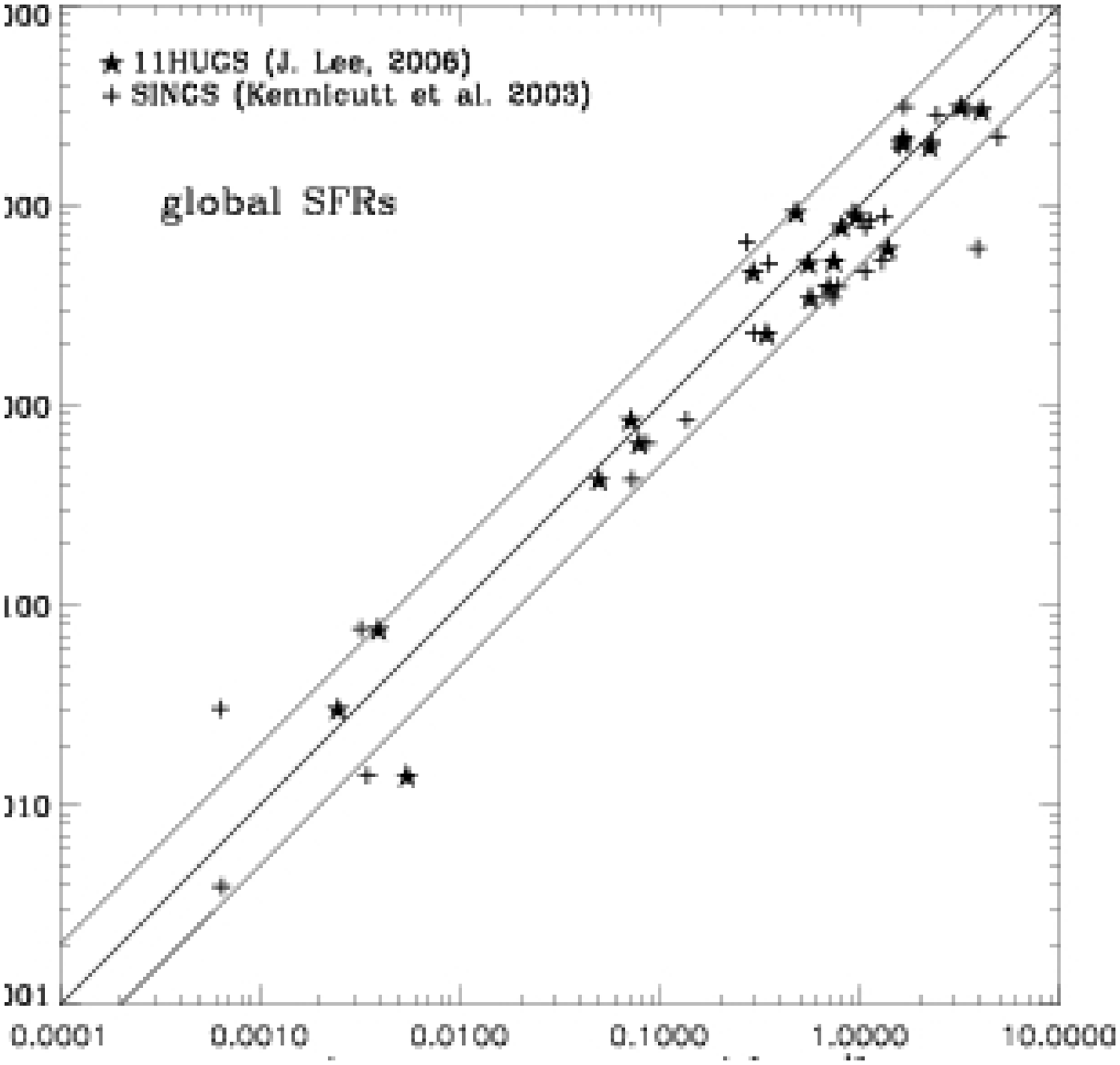}}\hfil
  \end{center}
  \figcaption{\label{SFCHECKS} Five estimates of $\Sigma_{\rm SFR}$ ($y$-axis)
    as a function of $\Sigma_{\rm SFR}$ predicted by our combination of FUV
    and 24$\mu$m. The color scheme is the same as Figure \ref{SFMETHODS} and
    the methodology used to derive $\Sigma_{\rm SFR}$ for comparison is
    labeled in each plot. In the bottom right panel, each point shows the
    integrated SFR for a galaxy derived from H$\alpha$ as a function of the
    SFR derived from FUV+24$\mu$m emission. In all plots, solid lines show
    slopes of $0.5$, $1$, and $2$.}
\end{figure*}

In Figure \ref{SFCHECKS}, we plot our 5 alternate estimates of $\Sigma_{\rm
  SFR}$ ($y$--axis) as a function of $\Sigma_{\rm SFR}$ derived from Equation
\ref{FINALSD} ($x$--axis). Each point corresponds to a $10\arcsec$-wide tilted
ring. In the bottom right panel, we plot the SFR integrated over the disk
(over $r_{\rm gal} < r_{25}$) as a function of SFR estimated from nebular line
emission by \citet{KENNICUTT03} and H$\alpha$ by \citet{LEE06}. Solid lines in
all six panels show the line of equality plus or minus a factor of 2.

If we adopt the naive tack of treating all approaches as equal, the aggregate
data in Figure \ref{SFCHECKS} yield a median ratio $\Sigma_{\rm SFR}~({\rm
  other}) / \Sigma_{\rm SFR} ({\rm FUV+24}) \approx 1.05$ with $\approx
0.22$~dex (i.e., $\sim 65\%$) $1\sigma$ scatter. The dominant sources of this
scatter are the choice of ``other'' $\Sigma_{\rm SFR}$ and galaxy--to--galaxy
variations. Once a galaxy and methodology are chosen, the data tend to follow
a fairly well--defined and often nearly linear relation.  For comparison, the
24$\mu$m part of the H$\alpha$+24$\mu$m calibration has $\approx 20$--$30\%$
uncertainty \citep[][]{CALZETTI07,KENNICUTT07} only considering star forming
peaks. In light of the wider range of star formation histories and geometries
encountered working pixel--by--pixel or averaging over whole rings, the
estimate of $\sim 65\%$ seems quite reasonable.  Comparing our {\em
  integrated} SFRs (Figure \ref{SFCHECKS}, bottom right) to those estimated by
11HUGS \citet{LEE06} and SINGS \citet{KENNICUTT03} bears out this estimate; we
match these estimates with a similar scatter. Another view of this comparison
may be seen in Appendix \ref{ATLASAPP}, where we present radial profiles of
$\Sigma_{\rm SFR}$ based on H$\alpha$ on the same plots as our FUV+24$\mu$m
profiles.

Despite the overall good agreement between our $\Sigma_{\rm SFR}$ and other
estimates, Figures \ref{SFMETHODS} and \ref{SFCHECKS} do show systematic
differences among tracers. We note several of these before moving on:

 \begin{enumerate}
 \item Using only 24$\mu$m emission (gray) yields a low estimate of
   $\Sigma_{\rm SFR}$ for the two low metallicity galaxies in our comparison
   sample: Holmberg~II and IC~2574. Dust is known to be deficient in these
   galaxies \citep{WALTER07}, which is likely to lead to a breakdown in the
   fit between 24$\mu$m emission and Pa$\alpha$. This effect, already
   recognized by \citet{CALZETTI07}, highlights the importance of including a
   non--IR component in a SFR tracer.
 \item Estimating $A_{\rm H\alpha}$ from gas \citep[red,][]{WONG02} yields
   very high $\Sigma_{\rm SFR}$ (and high $W_{\rm FUV}$) in the inner parts of
   galaxies.  This underscores the complexity of the geometry and timescale
   effects at play; it is extremely challenging to reverse engineer the true
   luminosity of a heavily obscured source knowing only the amount of nearby
   interstellar matter. These high values are almost certainly overestimates;
   stellar feedback, turbulence, or simply favorable geometry likely always
   allows at least some light from deeply embedded \hii\ regions to escape.
 \item Particularly at low $\Sigma_{\rm SFR}$, inferring $A_{\rm FUV}$ from UV
   colors (purple) yields higher embedded SFR than using 24$\mu$m emission
   (and this method appears to completely fail in NGC~6946, the horizontal row
   of points). A possible explanation is that where $\Sigma_{\rm SFR}$ is
   relatively low, the UV originates from a somewhat older (and thus redder)
   population \citep[e.g.][]{CALZETTI05}; the FUV--UV color relation depends
   on the recent star formation history \citep[e.g., differing between
   starbursts and more quiescent galaxies,][]{BOISSIER07,SALIM07}.

   This discrepancy (and the close association between our SFR tracer and
   stellar mass seen in the main text) argues for a comparison among
   metallicity, stellar populations, and mid--IR emission that is beyond the
   scope of this paper.  We restrict ourselves to a first--order check: we
   compare the ratio of 24$\mu$m--to--$3.6\mu$m and FUV--to--$3.6\mu$m
   emission in our sample to those in elliptical galaxies, which should be
   good indicators of how much an old population contributes to 24$\mu$m or
   FUV emission. Very approximately, in ellipticals $I_{24} / I_{3.6} \sim
   0.1$ \citep[][]{TEMI05,DALE07,JOHNSON07}, with $\sim 0.03$ expected from
   stellar emission alone \citep{HELOU04}, while $I_{\rm FUV} /I_{3.6} \sim
   2$--$4 \times 10^{-3}$ \citep[][taking the oldest bin from the
   latter]{DALE07,JOHNSON07}. We measure $I_{24} / I_{3.6}$ and $I_{\rm FUV} /
   I_{3.6}$ for each ring in our sample galaxies and compare these to the
   elliptical colors. In both cases, only $\sim 5\%$ of individual tilted
   rings have ratios lower than those seen in elliptical galaxies and the mean
   color is $\sim 10$ times that found in elliptical galaxies, though the
   ratio $I_{\rm FUV} /I_{3.6}$ shows large scatter due to the effects of
   extinction. Both the 24$\mu$m and FUV bands do appear to be dominated by a
   young stellar population almost everywhere in our sample.  Discrepancies
   among various tracers thus seem likely to arise from the different
   geometries and age sensitivities of FUV ($\tau \sim 100$~Myr), H$\alpha$
   ($\tau \sim 10$~Myr), and 24$\mu$m (likely intermediate) emission.
\end{enumerate}

Finally, we emphasize that uncertainties inferred via these comparisons mainly
reflect the ability to accurately infer the total UV light or ionizing photon
production from young stars. They do not include uncertainty in the IMF,
ionizing photon production rate (e.g., at low metallicity), or any of the
other factors involved in converting an ionizing photon count or FUV intensity
into a SFR.

\section{Radial Profiles}
\label{RPROFAPP}

\begin{center}
\begin{deluxetable*}{l c c c c c c c c}
  \tabletypesize{\tiny} \tablewidth{0pt} \tablecolumns{9}
  \tablecaption{\label{RPROFTAB} Table of Radial Profiles\tablenotemark{a}}
  \tablehead{\colhead{Galaxy} & \colhead{$r_{\rm gal}$} & \colhead{$r_{\rm
        gal}$} & \colhead{$\Sigma_{\rm HI}$} & \colhead{$\Sigma_{\rm H2}$} &
    \colhead{$\Sigma_*$} & \colhead{FUV+24} & \colhead{FUV part} &
    \colhead{$24\mu$m part} \\
    & $\left({\rm kpc}\right)$ & $\left(r_{25}\right)$ & $\left({\rm
        M_{\odot}~pc}^{-2}\right)$ & $\left({\rm M_{\odot}~pc}^{-2}\right)$ &
    $\left({\rm M_{\odot}~pc}^{-2}\right)$ & \multicolumn{3}{c}{$\left(10^{-4}~{\rm
          M_{\odot}~yr}^{-1}~{\rm kpc}^{-2}\right)$} } \startdata
  NGC0628 &  0.2 & 0.02 & $ 1.6 \pm  0.3$ & $ 22.7 \pm   1.2$ & $ 1209.4 \pm   18.3$ & $ 105.1 \pm   14.0$ & $  19.3$ & $  85.8$ \\
NGC0628 &  0.5 & 0.05 & $ 2.1 \pm  0.3$ & $ 20.2 \pm   1.3$ & $  557.8 \pm    4.8$ & $  92.3 \pm    9.9$ & $  17.1$ & $  75.1$ \\
NGC0628 &  0.9 & 0.08 & $ 2.6 \pm  0.4$ & $ 16.1 \pm   1.2$ & $  313.6 \pm    1.0$ & $  76.7 \pm    5.1$ & $  15.1$ & $  61.6$ \\
NGC0628 &  1.2 & 0.12 & $ 3.1 \pm  0.4$ & $ 12.7 \pm   0.8$ & $  231.9 \pm    0.5$ & $  65.5 \pm    4.2$ & $  14.2$ & $  51.3$ \\
NGC0628 &  1.6 & 0.15 & $ 3.7 \pm  0.3$ & $ 11.4 \pm   1.1$ & $  194.3 \pm    0.5$ & $  62.2 \pm    3.7$ & $  13.8$ & $  48.4$ \\
NGC0628 &  1.9 & 0.19 & $ 4.6 \pm  0.3$ & $ 11.1 \pm   1.2$ & $  163.5 \pm    0.7$ & $  72.4 \pm   12.2$ & $  15.3$ & $  57.1$ \\
NGC0628 &  2.3 & 0.22 & $ 5.3 \pm  0.4$ & $ 11.1 \pm   1.7$ & $  143.9 \pm    0.8$ & $  90.2 \pm   23.5$ & $  18.0$ & $  72.2$ \\
NGC0628 &  2.7 & 0.25 & $ 5.8 \pm  0.5$ & $ 10.6 \pm   1.9$ & $  123.5 \pm    0.5$ & $  90.7 \pm   21.3$ & $  19.1$ & $  71.6$ \\
NGC0628 &  3.0 & 0.29 & $ 6.1 \pm  0.5$ & $  8.9 \pm   1.5$ & $  107.5 \pm    0.4$ & $  71.9 \pm   11.9$ & $  17.7$ & $  54.2$ \\
NGC0628 &  3.4 & 0.32 & $ 6.5 \pm  0.5$ & $  7.2 \pm   1.2$ & $  151.0 \pm   10.5$ & $  57.9 \pm    8.5$ & $  15.7$ & $  42.2$ \\
NGC0628 &  3.7 & 0.36 & $ 7.3 \pm  0.7$ & $  6.2 \pm   1.5$ & $   81.6 \pm    0.4$ & $  55.8 \pm   11.3$ & $  14.3$ & $  41.6$ \\
NGC0628 &  4.1 & 0.39 & $ 7.9 \pm  0.8$ & $  5.9 \pm   1.7$ & $   68.0 \pm    0.4$ & $  59.6 \pm   14.1$ & $  13.5$ & $  46.1$ \\
NGC0628 &  4.4 & 0.42 & $ 8.1 \pm  0.8$ & $  5.4 \pm   1.5$ & $   61.6 \pm    0.4$ & $  59.9 \pm   15.2$ & $  13.9$ & $  46.0$ \\
NGC0628 &  4.8 & 0.46 & $ 7.9 \pm  0.9$ & $  4.3 \pm   1.1$ & $   48.3 \pm    0.2$ & $  48.8 \pm   11.0$ & $  13.6$ & $  35.2$ \\
NGC0628 &  5.1 & 0.49 & $ 8.2 \pm  1.0$ & $  3.1 \pm   0.8$ & $   41.8 \pm    0.2$ & $  37.4 \pm    6.6$ & $  12.7$ & $  24.7$ \\
NGC0628 &  5.5 & 0.53 & $ 8.5 \pm  1.0$ & $  2.1 \pm   0.7$ & $   37.0 \pm    0.2$ & $  33.5 \pm    8.7$ & $  12.3$ & $  21.2$ \\
NGC0628 &  5.8 & 0.56 & $ 8.6 \pm  0.8$ & $  1.2 \pm   0.5$ & $   33.2 \pm    0.4$ & $  30.2 \pm   10.0$ & $  11.9$ & $  18.4$ \\
NGC0628 &  6.2 & 0.59 & $ 8.6 \pm  0.7$ & $ \leq   1.0$ & $   37.0 \pm    2.3$ & $  23.5 \pm    6.5$ & $  10.1$ & $  13.5$ \\
NGC0628 &  6.5 & 0.63 & $ 8.8 \pm  0.6$ & $ \leq   1.0$ & $   52.9 \pm    6.1$ & $  17.4 \pm    3.1$ & $   8.0$ & $   9.4$ \\
NGC0628 &  6.9 & 0.66 & $ 8.8 \pm  0.5$ & $ \leq   1.0$ & $   19.5 \pm    0.1$ & $  13.6 \pm    1.9$ & $   6.6$ & $   7.0$ \\
NGC0628 &  7.3 & 0.69 & $ 8.6 \pm  0.5$ & $ \leq   1.0$ & $   18.9 \pm    0.1$ & $  11.6 \pm    2.3$ & $   5.7$ & $   5.9$ \\
NGC0628 &  7.6 & 0.73 & $ 8.2 \pm  0.6$ & $ \leq   1.0$ & $   18.7 \pm    0.7$ & $   9.8 \pm    2.2$ & $   5.1$ & $   4.8$ \\
NGC0628 &  8.0 & 0.76 & $ 7.6 \pm  0.6$ & $ \leq   1.0$ & $   12.9 \pm    0.1$ & $   7.5 \pm    1.5$ & $   4.1$ & $   3.4$ \\
NGC0628 &  8.3 & 0.80 & $ 7.1 \pm  0.6$ & $ \leq   1.0$ & $   17.6 \pm    1.3$ & $   5.4 \pm    0.9$ & $   3.1$ & $   2.3$ \\
NGC0628 &  8.7 & 0.83 & $ 6.7 \pm  0.5$ & $ \leq   1.0$ & $   17.0 \pm    1.6$ & $   4.1 \pm    0.6$ & $   2.4$ & $   1.7$ \\
NGC0628 &  9.0 & 0.86 & $ 6.5 \pm  0.4$ & $ \leq   1.0$ & $   10.8 \pm    0.4$ & $   3.2 \pm    0.4$ & $   2.0$ & $   1.3$ \\
NGC0628 &  9.4 & 0.90 & $ 6.0 \pm  0.5$ & $ \leq   1.0$ & $    8.0 \pm    0.1$ & $   2.5 \pm    0.3$ & $   1.7$ & $   0.8$ \\
NGC0628 &  9.7 & 0.93 & $ 5.2 \pm  0.4$ & $ \leq   1.0$ & $    7.5 \pm    0.2$ & $   1.8 \pm    0.3$ & $   1.3$ & $   0.5$ \\
NGC0628 & 10.1 & 0.97 & $ 4.5 \pm  0.4$ & $ \leq   1.0$ & $    5.0 \pm    0.1$ & $   1.2 \pm    0.2$ & $   0.9$ & $   0.3$ \\
NGC0628 & 10.4 & 1.00 & $ 4.1 \pm  0.3$ & $ \leq   1.0$ & $    4.1 \pm    0.0$ & $ \leq    1.0$ & \nodata & \nodata \\
NGC0628 & 10.8 & 1.03 & $ 3.9 \pm  0.3$ & $ \leq   1.0$ & $    3.6 \pm    0.0$ & $ \leq    1.0$ & \nodata & \nodata \\
NGC0628 & 11.1 & 1.07 & $ 3.9 \pm  0.4$ & $ \leq   1.0$ & $    3.9 \pm    0.1$ & $   1.0 \pm    0.4$ & $   0.7$ & $   0.3$ \\
NGC0628 & 11.5 & 1.10 & $ 4.0 \pm  0.4$ & $ \leq   1.0$ & $    4.4 \pm    0.2$ & $ \leq    1.0$ & \nodata & \nodata \\
NGC0628 & 11.9 & 1.13 & $ 4.3 \pm  0.5$ & $ \leq   1.0$ & $    9.5 \pm    0.9$ & $ \leq    1.0$ & \nodata & \nodata \\
NGC0628 & 12.2 & 1.17 & $ 4.6 \pm  0.5$ & $ \leq   1.0$ & $    5.8 \pm    0.2$ & $ \leq    1.0$ & \nodata & \nodata \\

  \enddata 
\tablenotetext{a}{The full table of radial profile data is available in
  electronic form.}
\end{deluxetable*}
\end{center}

Table \ref{RPROFTAB} presents radial profiles of $\Sigma_{\rm HI}$,
$\Sigma_{\rm H2}$, $\Sigma_*$, and $\Sigma_{\rm SFR}$. Combined with
kinematics, which may be calculated using Equation \ref{ROTCUREQ} taking
$v_{\rm flat}$ and $l_{\rm flat}$ from Table \ref{STRUCTURETAB}, these
profiles are intended to provide a database that can be used to test theories
of galaxy--wide star formation or to explore the effects of varying our
assumptions. Results for all galaxies are available in an electronic table
online. Table \ref{RPROFTAB} in the print edition shows the results for our
lowest--mass spiral galaxy, NGC~628, as an example.

The individual columns are as follows. {\em Ring identifiers:} (1) galaxy
name; galactocentric radius of ring center (2) in kpc and (3) normalized by
$r_{25}$. {\em Mass surface densities} (in M$_{\odot}$~pc$^{-2}$) along with
associated uncertainty of (4--5) \hi ; (6--7) \htwo ; and (8--9) stars. {\em
  Star formation rate surface density}, $\Sigma_{\rm SFR}$, with associated
uncertainty (10--11) from combining FUV and 24$\mu$m emission in units of
$10^{-4}$~M$_{\odot}$~yr$^{-1}$~kpc$^{-2}$; and the individual contributions
to $\Sigma_{\rm SFR}$ from (12) FUV and (13) 24$\mu$m emission (i.e., the left
and right terms in \ref{SFREQ}) in the same units.

We derive radial profiles from maps using the mean (for $\Sigma_{\rm HI}$,
$\Sigma_{\rm H2}$, $\Sigma_{\rm SFR}$) or median ($\Sigma_*$) value within
$10\arcsec$-wide tilted rings (so that the rings are spaced by half of our
typical working resolution). The rings use the position angle and inclination
in Table \ref{STRUCTURETAB}, adopted from \citet{WALTER08}. We adopt the
THINGS center for each galaxy \citep{WALTER08,TRACHTERNACH08} except for
Holmberg~I, where we use the dynamical center derived by \citet{OTT01} rather
than the photometric center. We consider only data within $60\arcdeg$ of the
major axis, measured in the plane of the galaxy.  This minimizes our
sensitivity to the adopted structural parameters, which most strongly affect
the deprojection along the minor axis. Where there are no data, we take
$\Sigma_{\rm HI} = 0$ and $\Sigma_{\rm H2} = 0$. These are regions that have
been observed but masked out because no signal was identified. We ignore
pixels with no measurement of $\Sigma_{\rm SFR}$, these are simply missing
data.

We take the uncertainty in a quantity averaged over a tilted ring to be

\begin{equation}
\sigma = \frac{\sigma_{\rm RMS}}{\sqrt{N_{\rm pix,ring}/N_{\rm pix,beam}}}
\end{equation}

\noindent where $\sigma_{\rm RMS}$ is the RMS scatter within the tilted ring,
$N_{\rm pix,ring}$ is the number of pixels in the ring, and $N_{\rm pix,beam}$
is the number of pixels per resolution element. This $\sigma$ captures both
random scatter in the data and variations due to azimuthal structure within
the ring. It does not capture systematic uncertainties, e.g., due to choice of
\xco\ or star formation tracer, discussed in these appendices.

\section{Atlas of Maps and Profile Plots}
\label{ATLASAPP}

\begin{figure*}
  \plotone{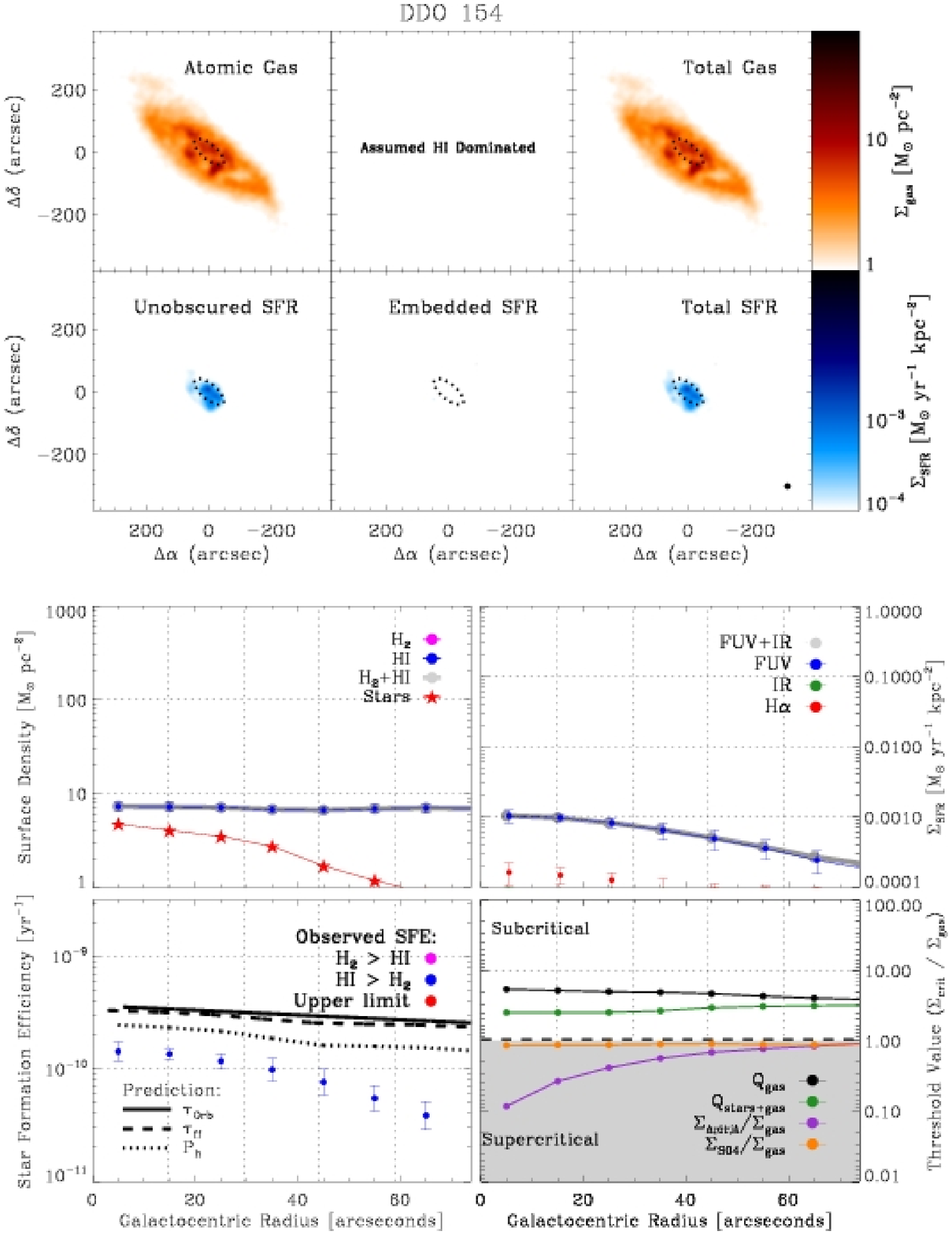}
  \label{ATLAS}
  \caption{Atlas of data and calculations for DDO 154.}
\end{figure*}

In Figure \ref{ATLAS}, we present maps, profiles, and calculations for
individual galaxies. Each page shows results for one galaxy. The top
row shows maps of atomic gas ($\Sigma_{\rm HI}$), molecular gas
($\Sigma_{\rm H2}$), and total gas ($\Sigma_{\rm gas} = \Sigma_{\rm
  HI} + \Sigma_{\rm H2}$). The second row shows unobscured (FUV),
dust-embedded (24$\mu$m), and total star formation surface density
($\Sigma_{\rm SFR}$).  These maps use a color scheme based on the
modified magnitude system described by \citet[][]{LUPTON99}; a bar to
the right of each row of plots illustrates the scheme. The gas maps
and star formation maps for each galaxy use a single color scheme, but
the scheme does vary from galaxy to galaxy, so care should be taken
when comparing different galaxies. Also note that we construct the
table to show empty values below our working sensitivity (i.e., any
data below $\Sigma_{\rm gas} = 1$~M$_{\odot}$~pc$^{-2}$ or
$\Sigma_{\rm SFR} = 10^{-4}$~M$_{\odot}$~yr$^{-1}$~kpc$^{-2}$ appear
as white) but the data (especially THINGS) often show evidence of real
emission below this value. We refer the readers to the original data
papers for more information on each data set.

The dotted circle indicates the optical radius, $r_{25}$, in the plane
of the galaxy for the structural parameters given in Table
\ref{STRUCTURETAB}. A small black circle in the bottom right panel
shows our working resolution.

In the left panel on the third row, we plot mass surface density
profiles. We show \hi\ (blue), \htwo\ (magenta, where available),
stars (red stars), and total gas (thick gray profile). Vertical dotted
lines indicate $0.25$, $0.5$, $0.75$, and $1.0$ times
$r_{25}$. Horizontal dotted lines show fixed mass surface density. The
thick gray vertical line shows where the intensity scale for the
images is set, i.e., $0.1~r_{25}$.

In the right panel on the third row, we plot star formation rate
surface density profiles. We show the total $\Sigma_{\rm SFR}$ (thick
gray profile) and the separate contributions from dust-embedded
(green, 24$\mu$m) and unobscured (blue, FUV) star formation, which add
up to $\Sigma_{\rm SFR}$.  Where they are available, we plot
$\Sigma_{\rm SFR}$ from the SINGS DR4 H$\alpha$ (red) and points
measured from the H$\alpha$ profiles of \citet{MARTIN01} (magenta) and
\citet{WONG02} (purple). All H$\alpha$ profiles assume 1.1~mag of
extinction \citep[a typical average value in disk
galaxies,][]{KENNICUTT98B}. The other markings are as in the left
panel.

In the left panel of the fourth row, we show the observed SFE for the
galaxy. We use the same color scheme as in \S \ref{SECT_RESULTS},
i.e., magenta points indicate rings where $\Sigma_{\rm H2} >
\Sigma_{\rm HI}$, blue points show rings where $\Sigma_{\rm H2} <
\Sigma_{\rm HI}$, and red arrows indicate upper limits. The ensemble
of points in this panel combine to form Figure \ref{SFEVSRAD}. Dashed,
dotted, and dash-dotted lines show the SFE predicted following the
method described in \S \ref{SECT_RECIPE} with no threshold applied
(the thresholds appear in the right panel). The other markings are as
in the panels on the third row.

In the right panel of the fourth row we show azimuthally averaged
values for thresholds described in \S \ref{SECT_THRESHBACK}. We expect
widespread star formation (conditions are ``supercritical'') where the
value of a profile is below $1$ (the shaded area) and isolated or
nonexistent star formation (conditions are ``subcritical'') above
$1$. We plot the Toomre $Q$ parameter for a gas disk, $Q_{\rm gas}$
(black), and for a gas disk in the presence of stars, $Q_{\rm
  stars+gas}$ (green). We show the shear criterion described by
\citet{HUNTER98}, $\Sigma_{\rm crit,A} / \Sigma_{\rm gas}$ in purple
and the condition for the formation of a cold phase given by
\citet{SCHAYE04}, $\Sigma_{\rm S04} / \Sigma_{\rm gas}$ in orange.
The other markings are as in the panels on the third row.

\clearpage

\clearpage

\begin{figure*}
  \plotone{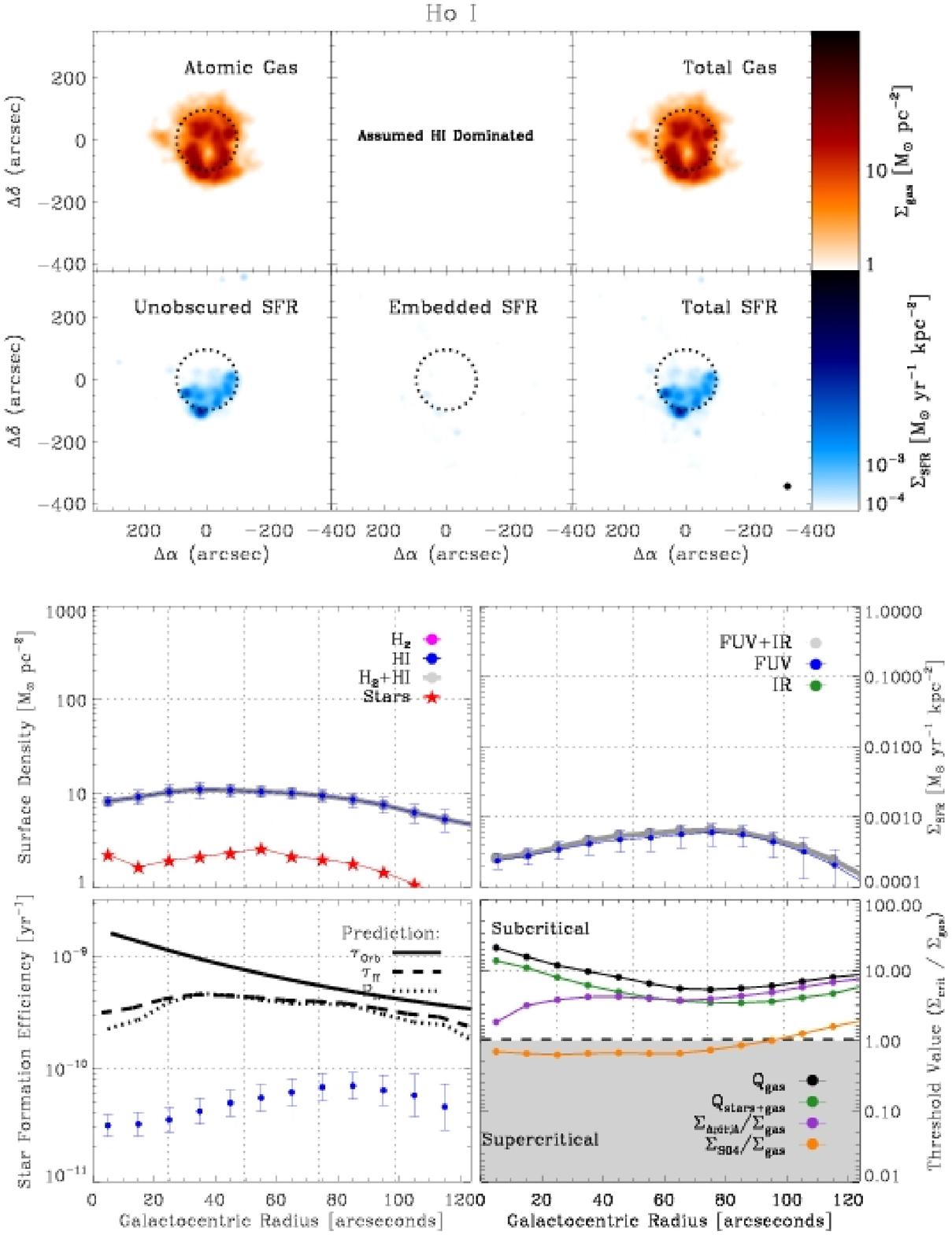}
  \figurenum{\ref{ATLAS}}
  \caption{Atlas of data and calculations for Holmberg I.}
\end{figure*}

\clearpage

\begin{figure*}
  \plotone{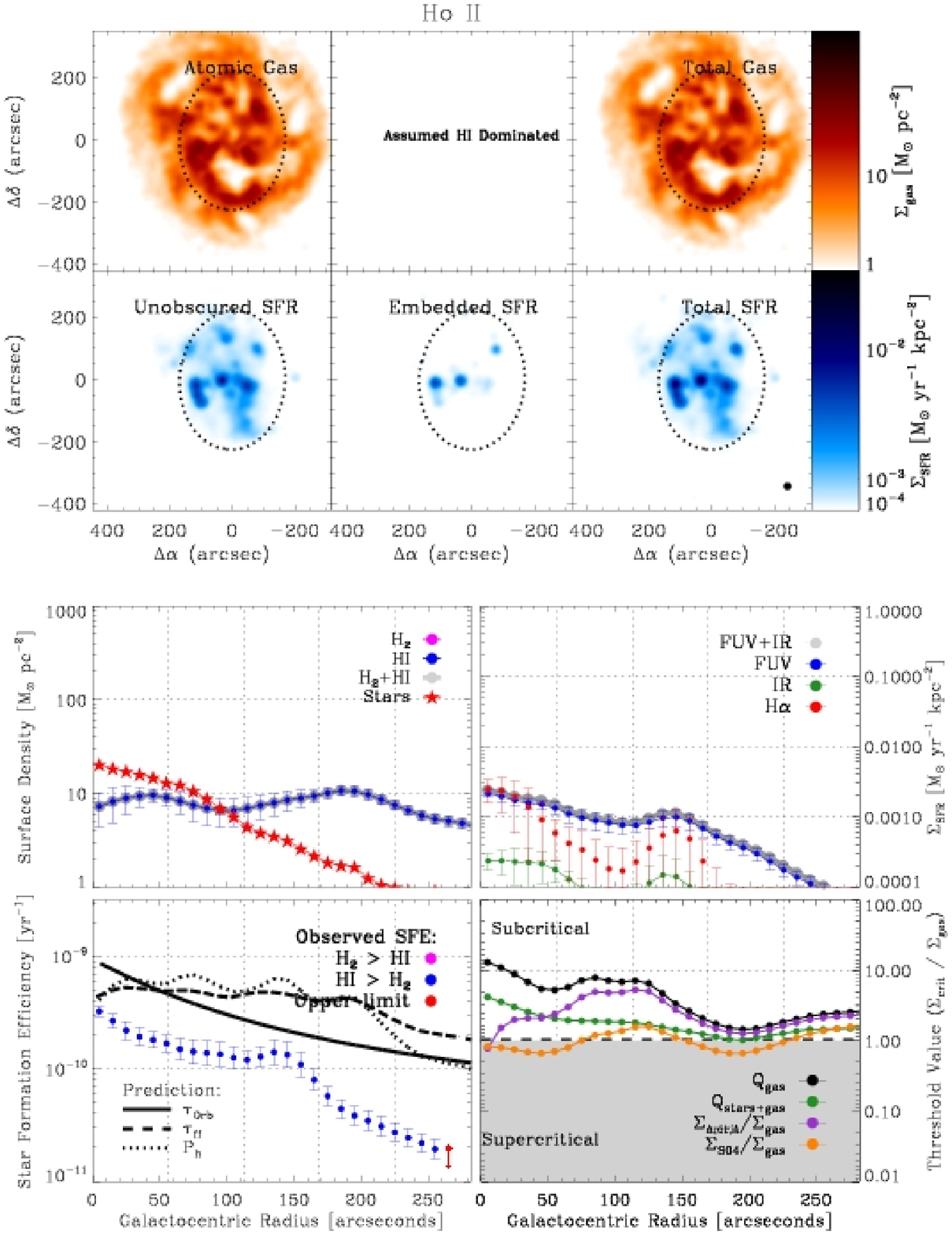}
  \figurenum{\ref{ATLAS}}
  \caption{Atlas of data and calculations for Holmberg II.}
\end{figure*}

\clearpage

\begin{figure*}
  \plotone{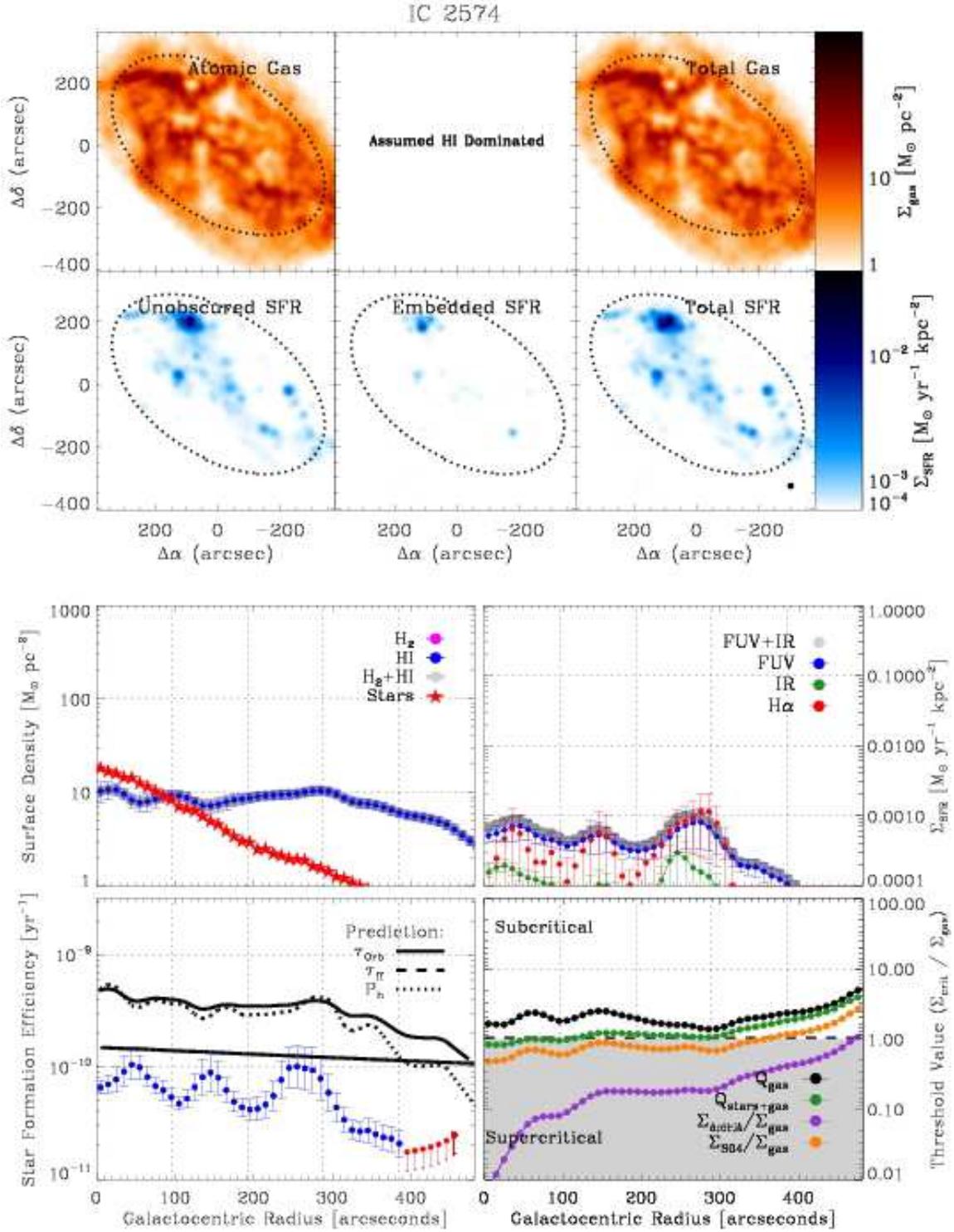}
  \figurenum{\ref{ATLAS}}
  \caption{Atlas of data and calculations for IC 2574.}
\end{figure*}

\clearpage

\begin{figure*}
  \plotone{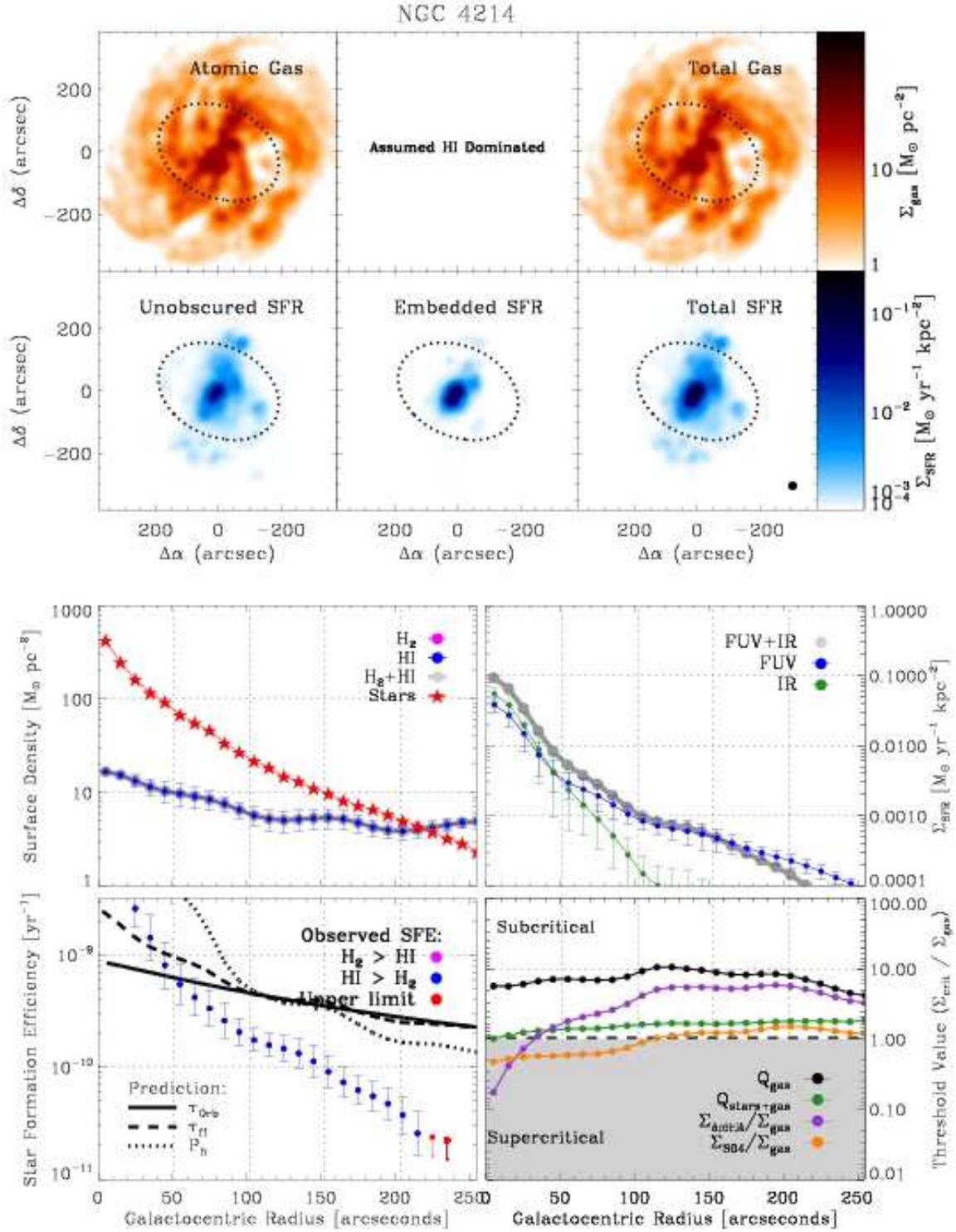}
  \figurenum{\ref{ATLAS}}
  \caption{Atlas of data and calculations for NGC 4214.}
\end{figure*}

\clearpage

\begin{figure*}
  \plotone{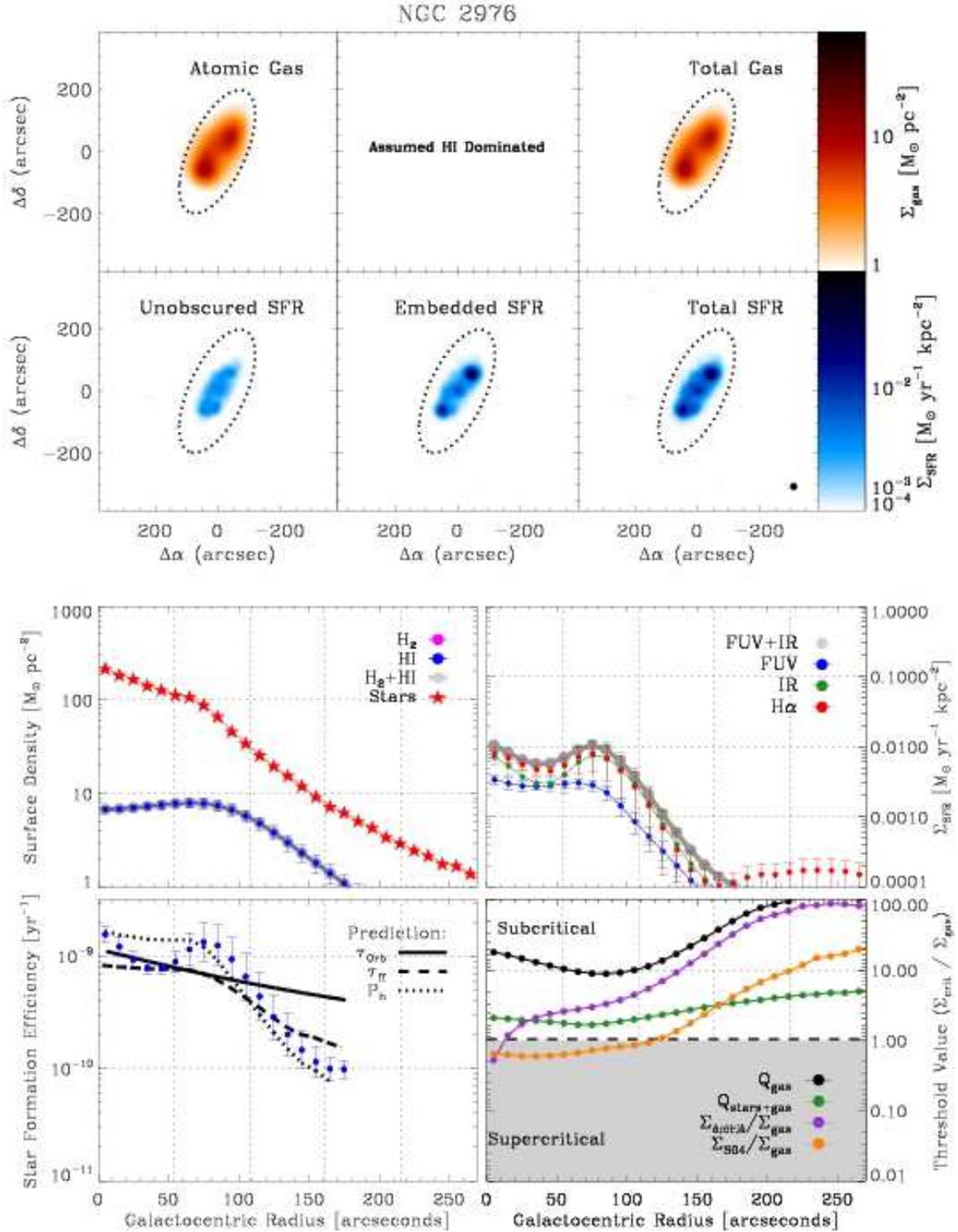}
  \figurenum{\ref{ATLAS}}
  \caption{Atlas of data and calculations for NGC 2976.}
\end{figure*}

\clearpage

\begin{figure*}
  \plotone{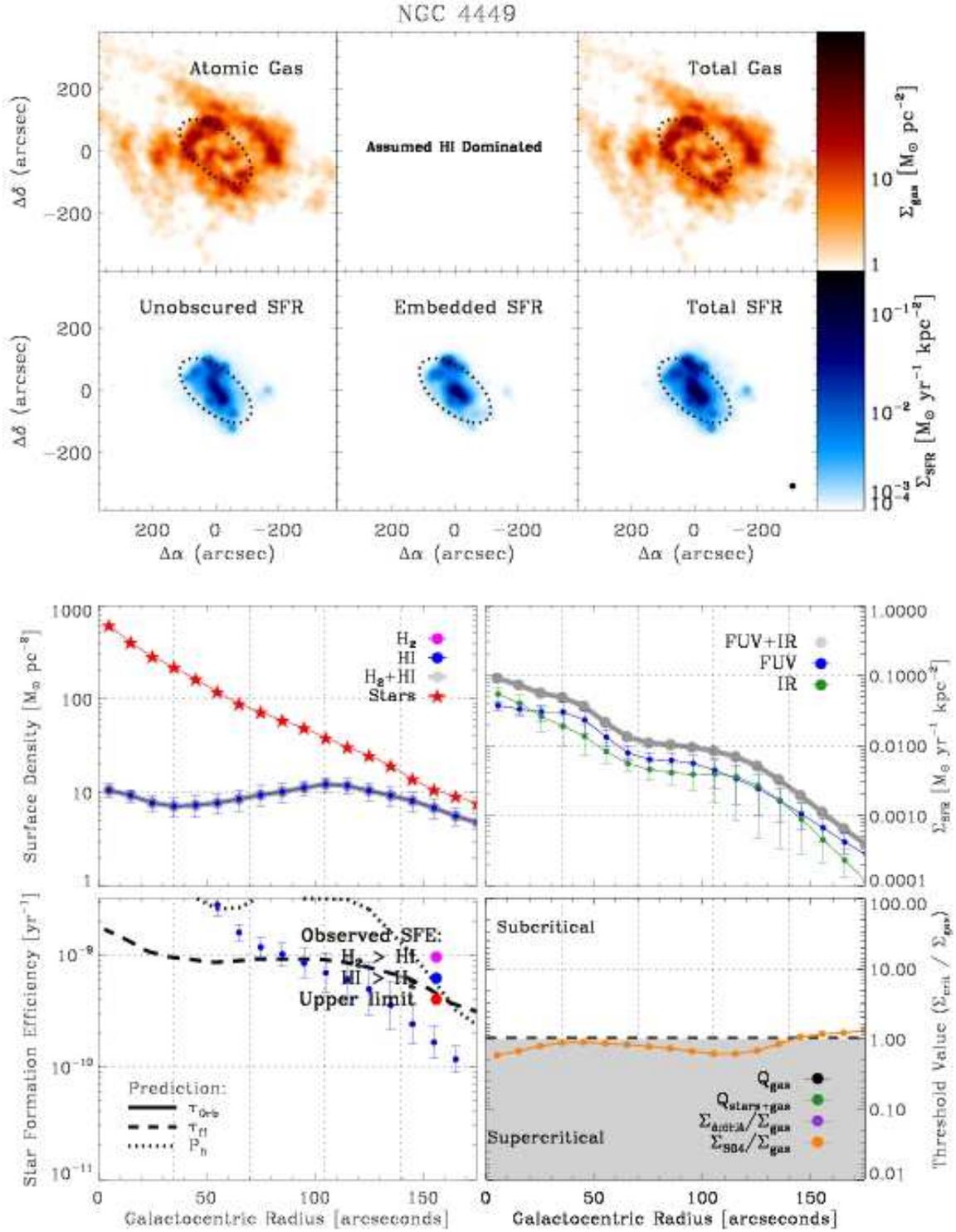}
  \figurenum{\ref{ATLAS}}
  \caption{Atlas of data and calculations for NGC 4449.}
\end{figure*}

\clearpage

\begin{figure*}
  \plotone{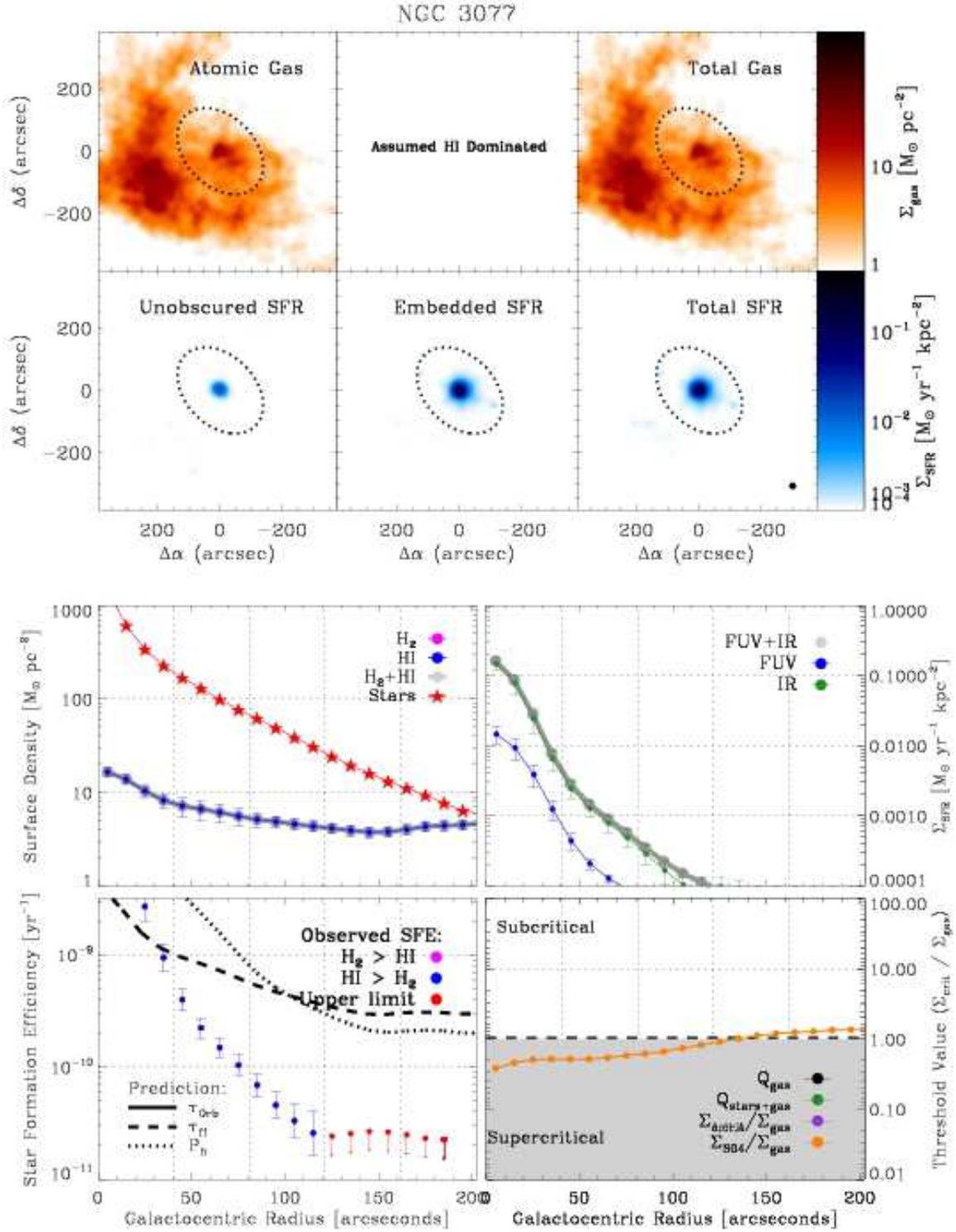}
  \figurenum{\ref{ATLAS}}
  \caption{Atlas of data and calculations for NGC 3077.}
\end{figure*}

\clearpage

\begin{figure*}
  \plotone{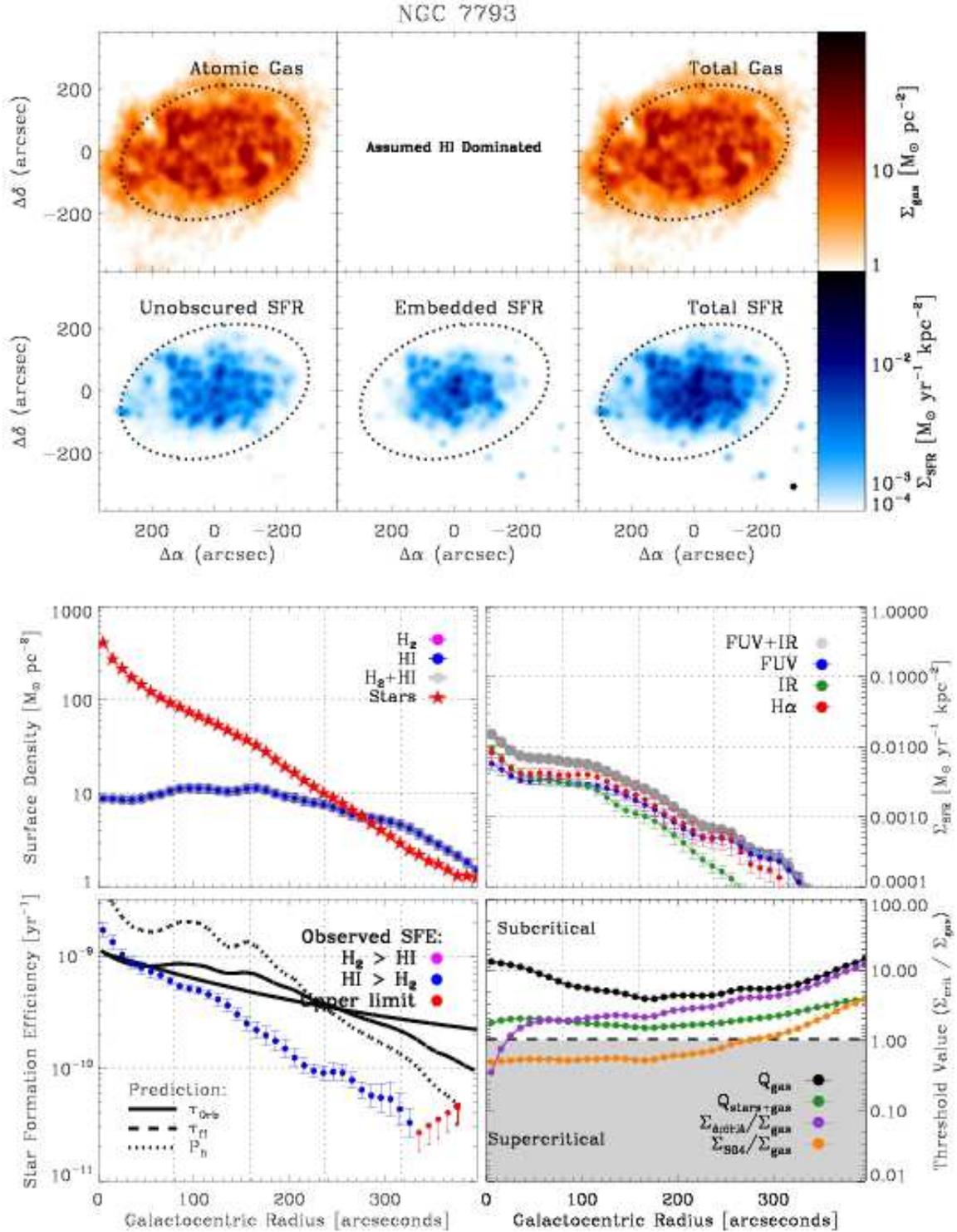}
  \figurenum{\ref{ATLAS}}
  \caption{Atlas of data and calculations for NGC 7793.}
\end{figure*}

\clearpage

\begin{figure*}
  \plotone{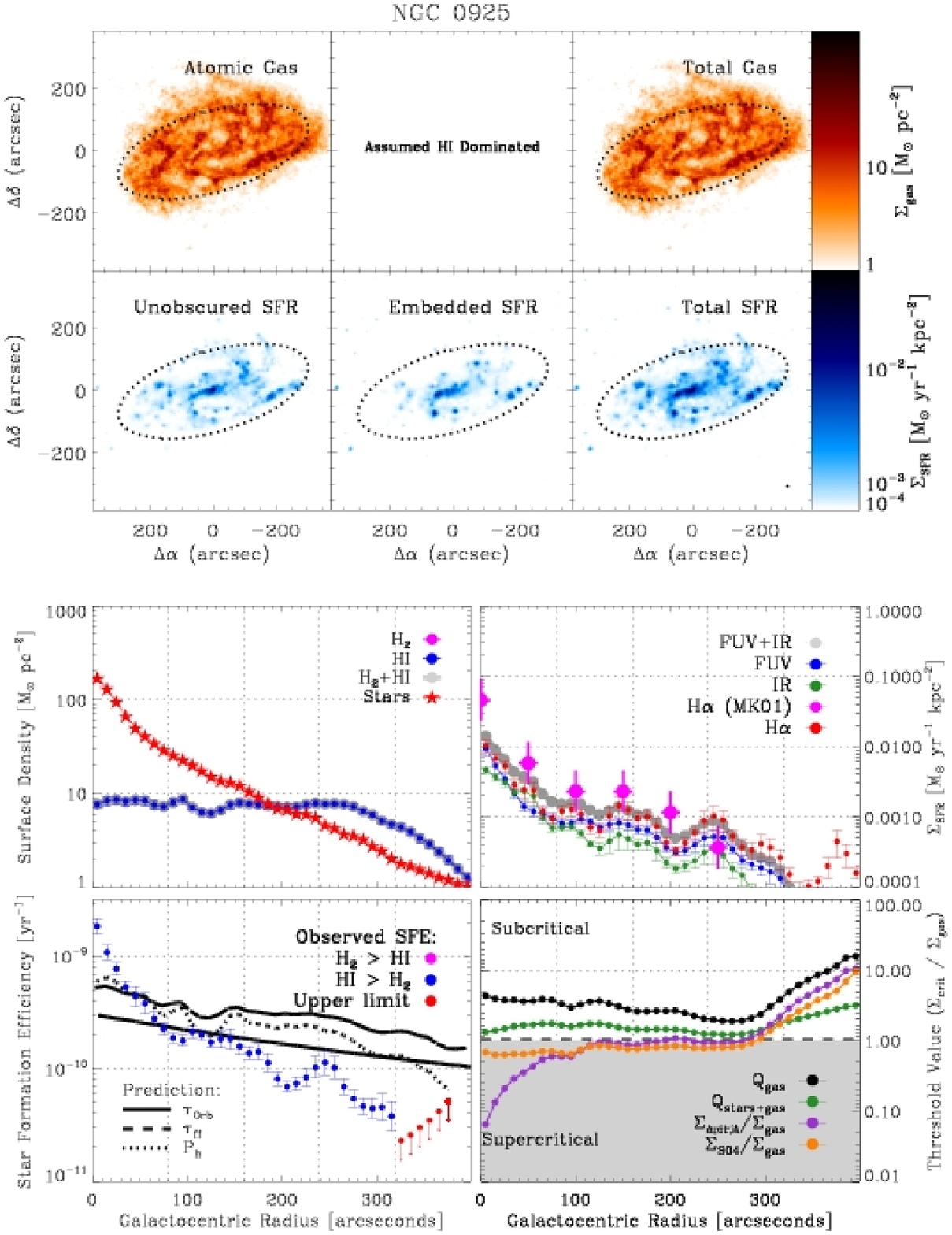}
  \figurenum{\ref{ATLAS}}
  \caption{Atlas of data and calculations for NGC 925.}
\end{figure*}

\clearpage

\begin{figure*}
  \plotone{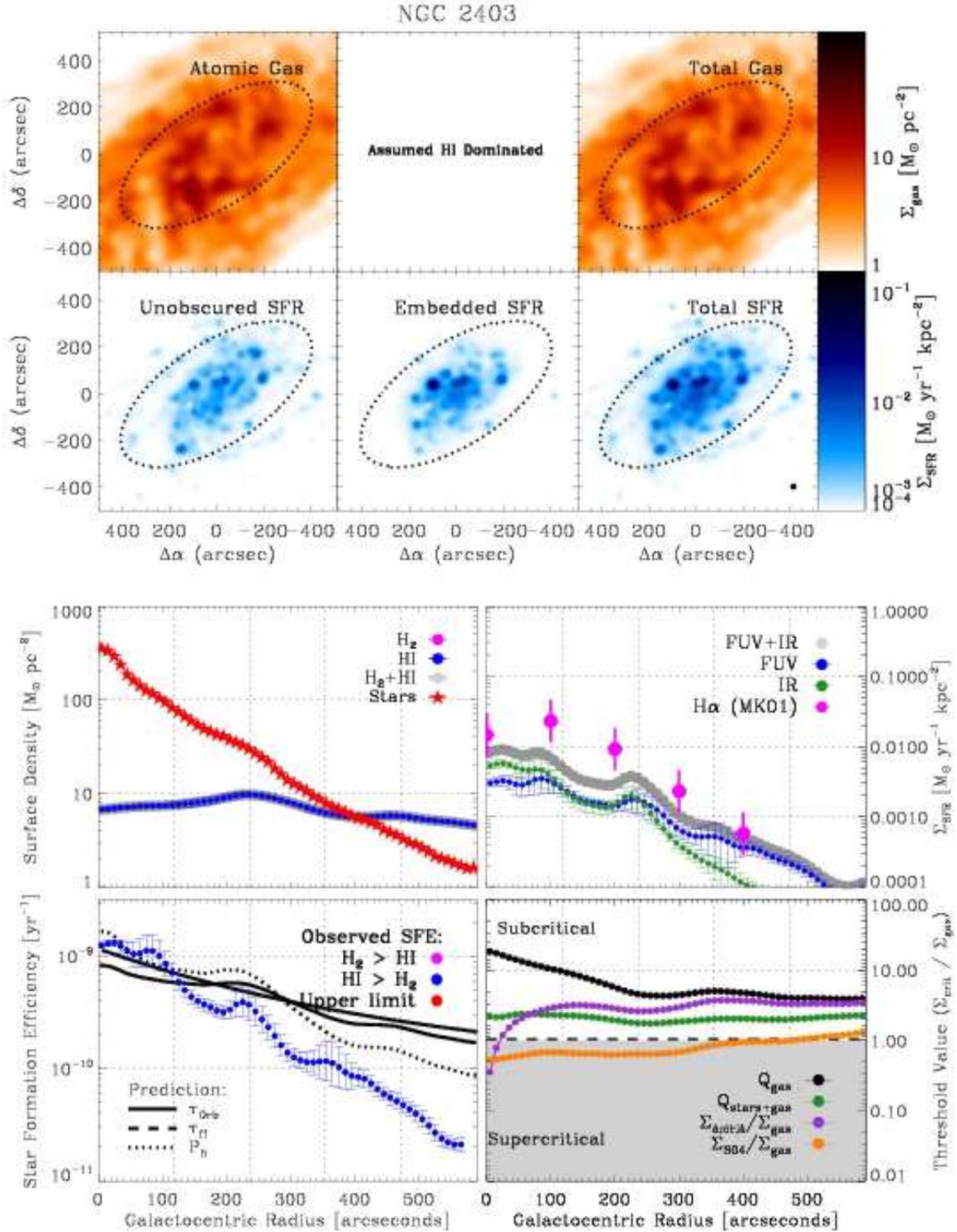}
  \figurenum{\ref{ATLAS}}
  \caption{Atlas of data and calculations for NGC 2403.}
\end{figure*}

\clearpage

\begin{figure*}
  \plotone{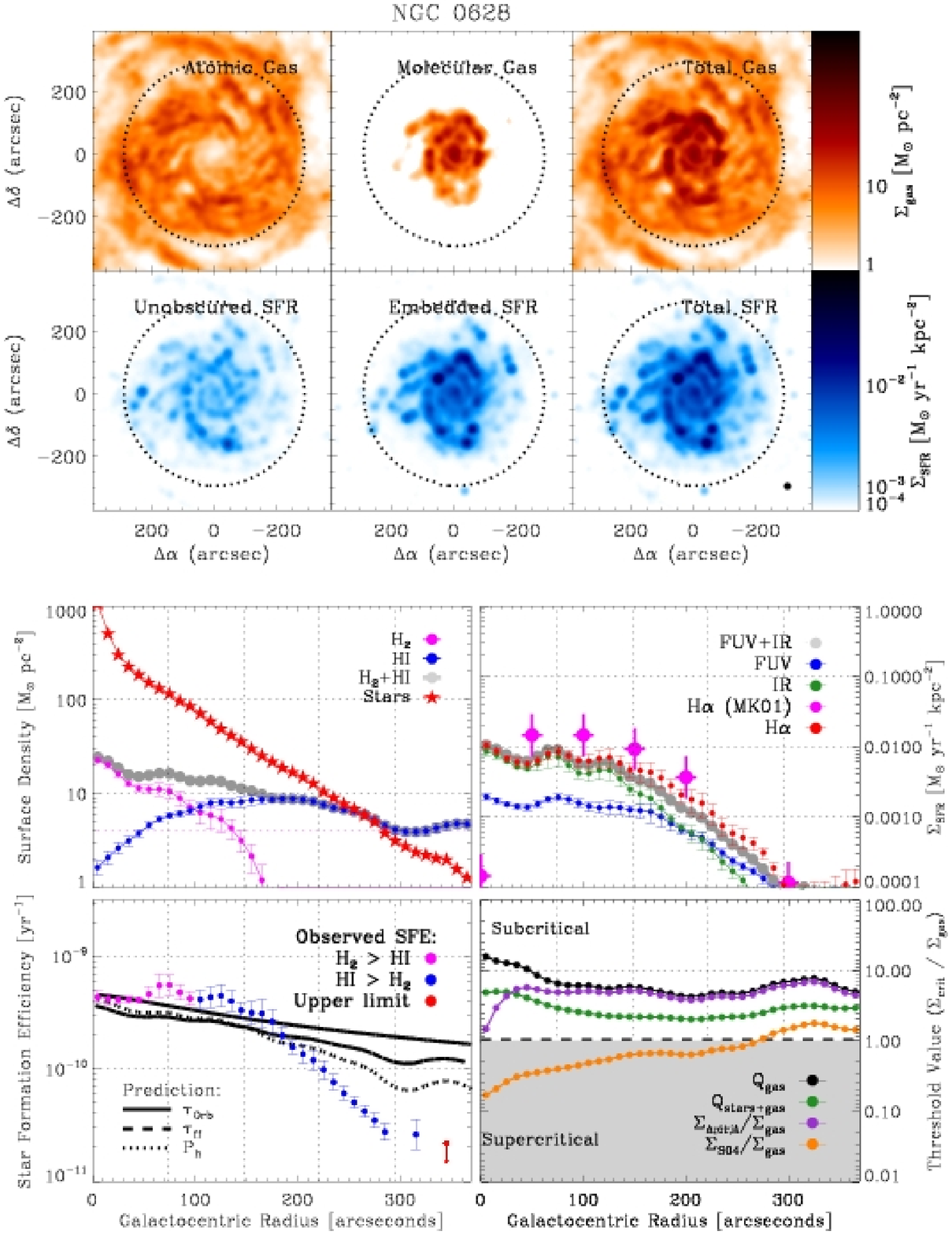}
  \figurenum{\ref{ATLAS}}
  \caption{Atlas of data and calculations for NGC 628.}
\end{figure*}

\clearpage

\begin{figure*}
  \plotone{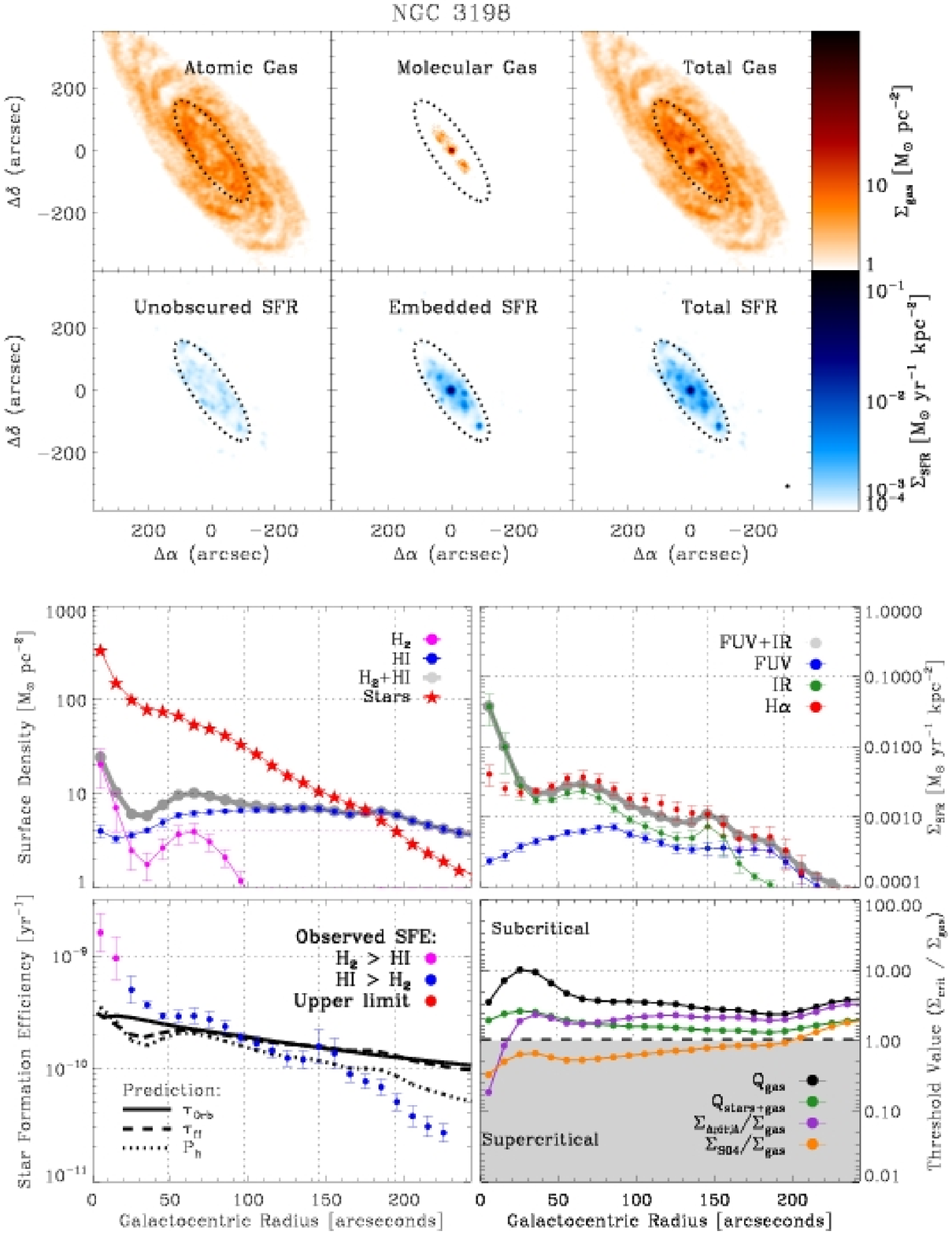}
  \figurenum{\ref{ATLAS}}
  \caption{Atlas of data and calculations for NGC 3198.}
\end{figure*}

\clearpage

\begin{figure*}
  \plotone{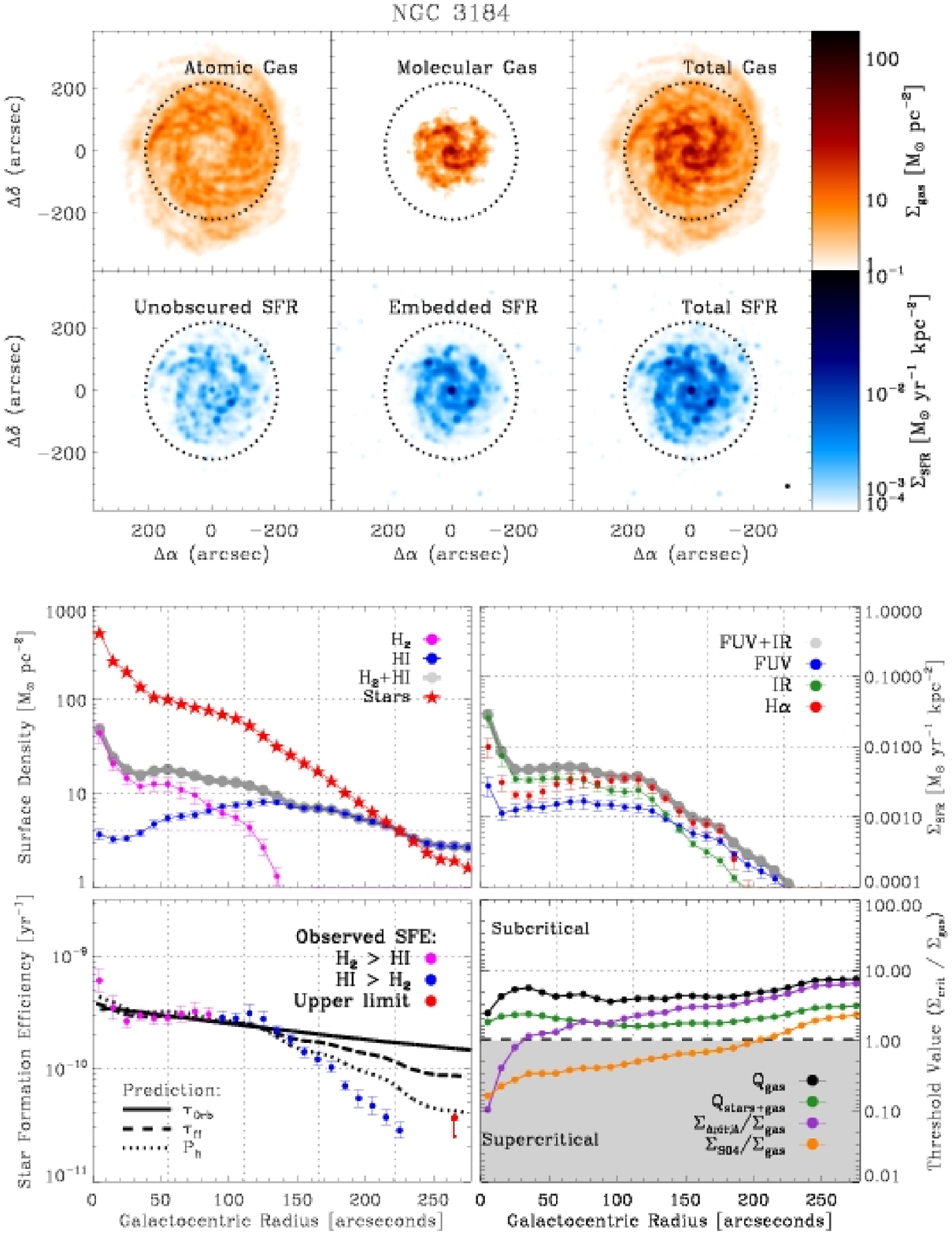}
  \figurenum{\ref{ATLAS}}
  \caption{Atlas of data and calculations for NGC 3184.}
\end{figure*}

\clearpage

\begin{figure*}
  \plotone{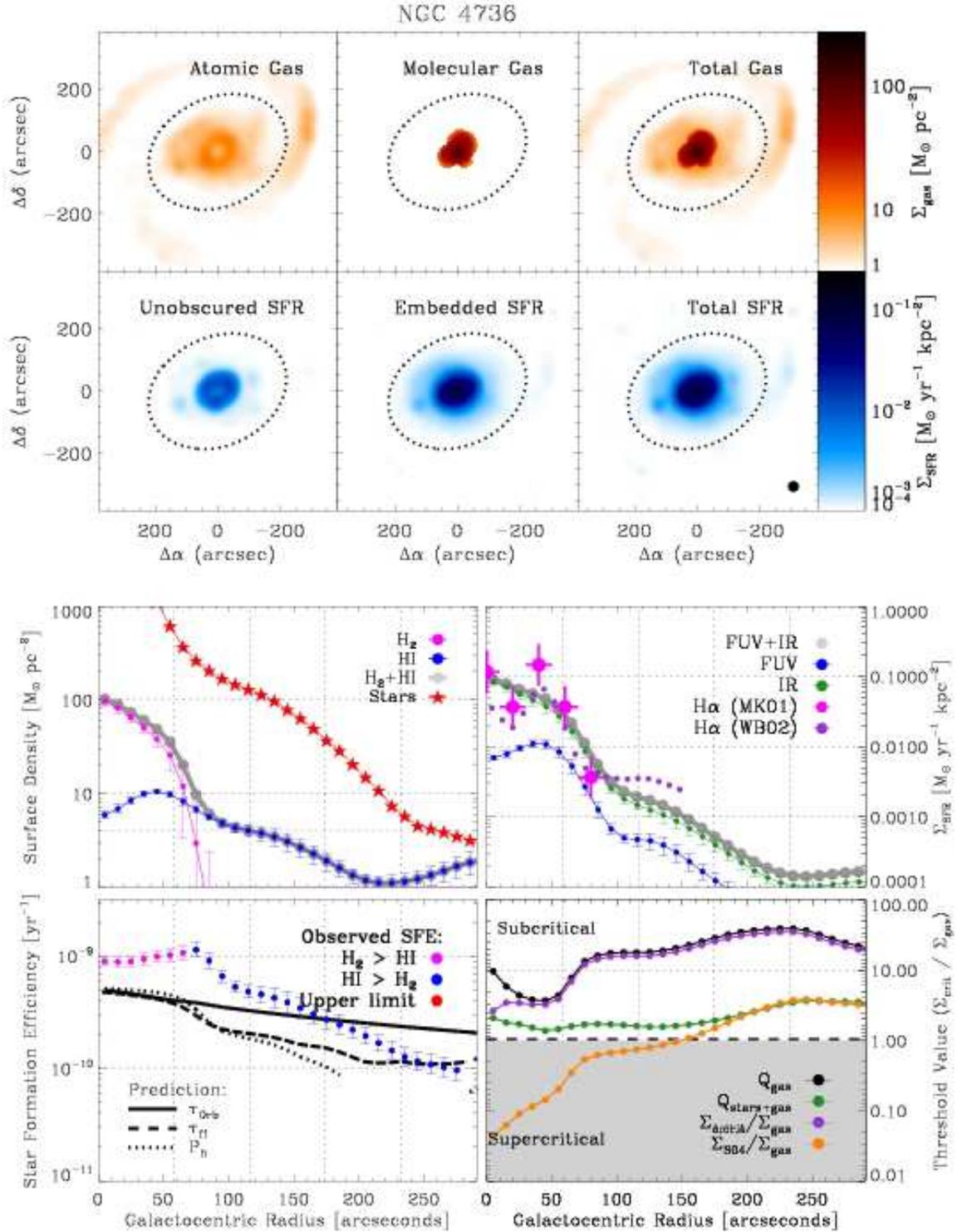}
  \figurenum{\ref{ATLAS}}
  \caption{Atlas of data and calculations for NGC 4736.}
\end{figure*}

\clearpage

\begin{figure*}
  \plotone{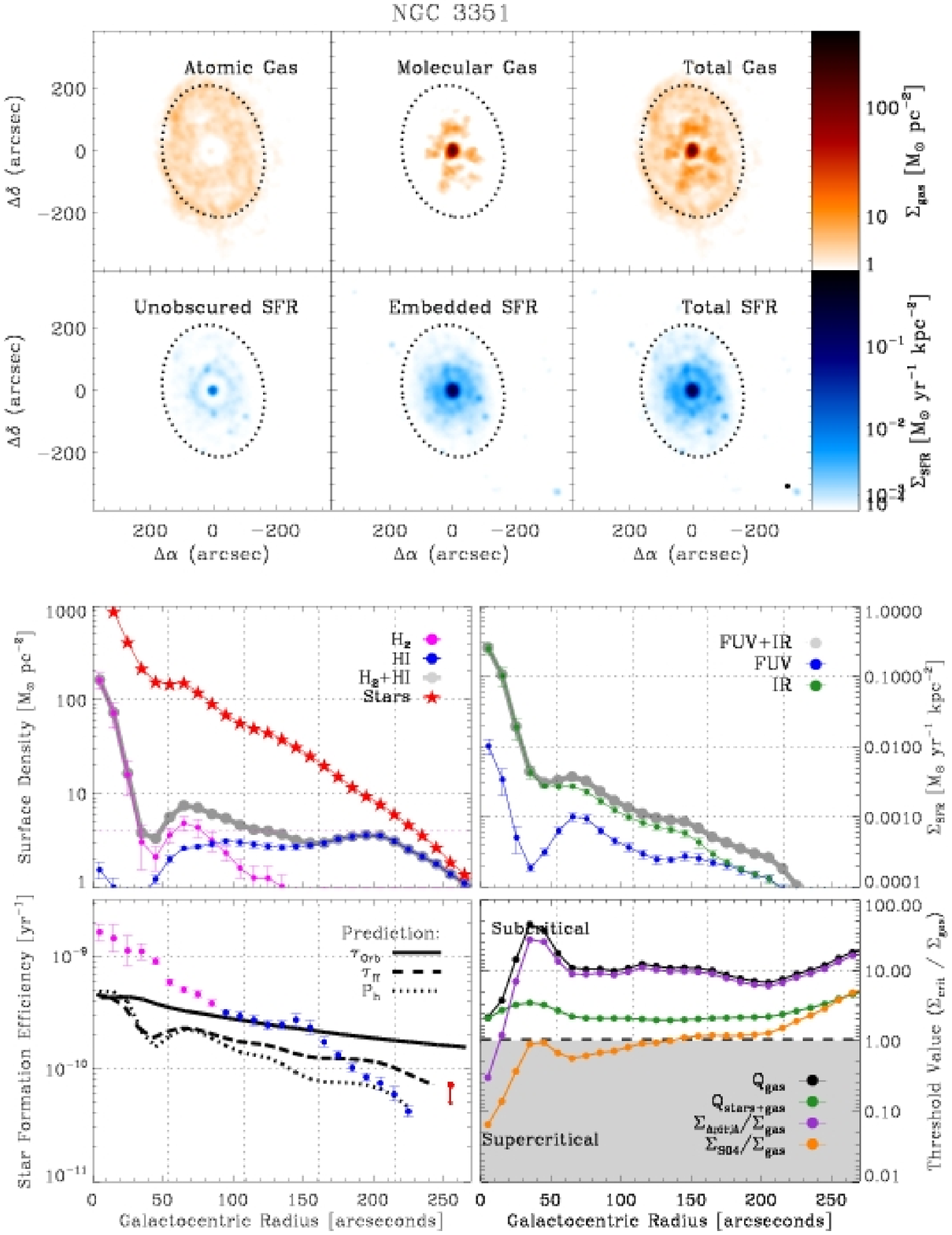}
  \figurenum{\ref{ATLAS}}
  \caption{Atlas of data and calculations for NGC 3351.}
\end{figure*}

\clearpage

\begin{figure*}
  \plotone{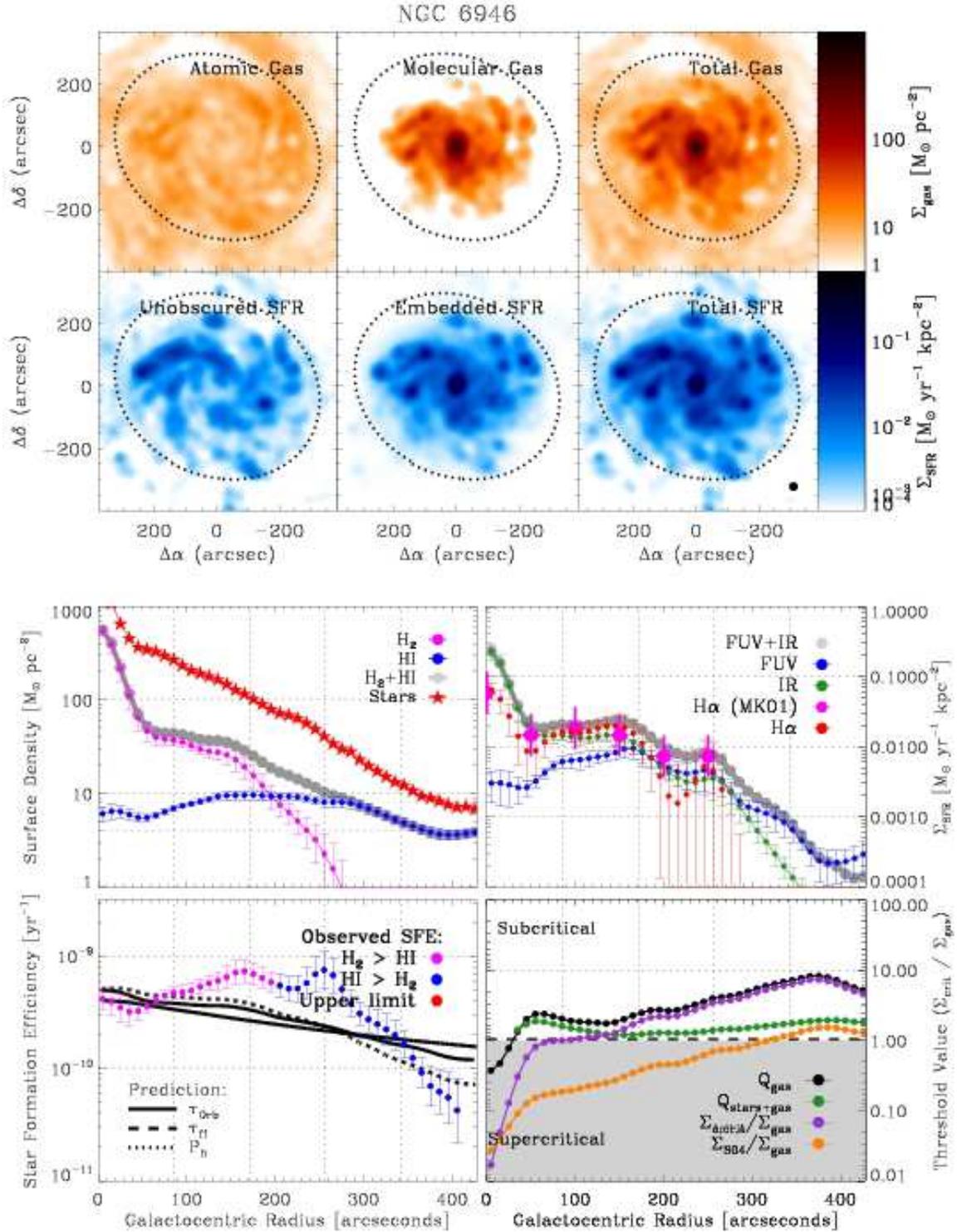}
  \figurenum{\ref{ATLAS}}
  \caption{Atlas of data and calculations for NGC 6946.}
\end{figure*}

\clearpage

\begin{figure*}
  \plotone{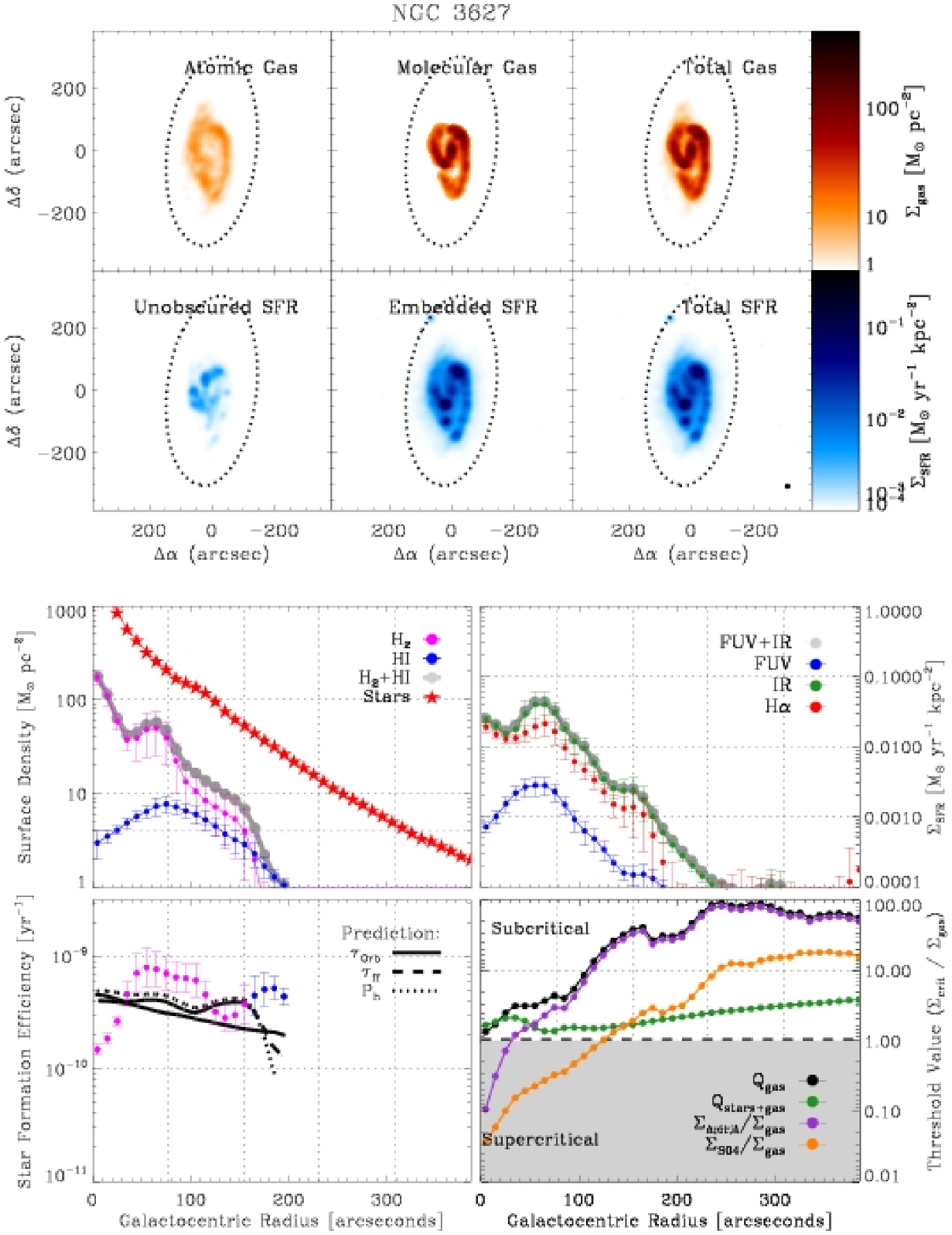}
  \figurenum{\ref{ATLAS}}
  \caption{Atlas of data and calculations for NGC 3627.}
\end{figure*}

\clearpage

\begin{figure*}
  \plotone{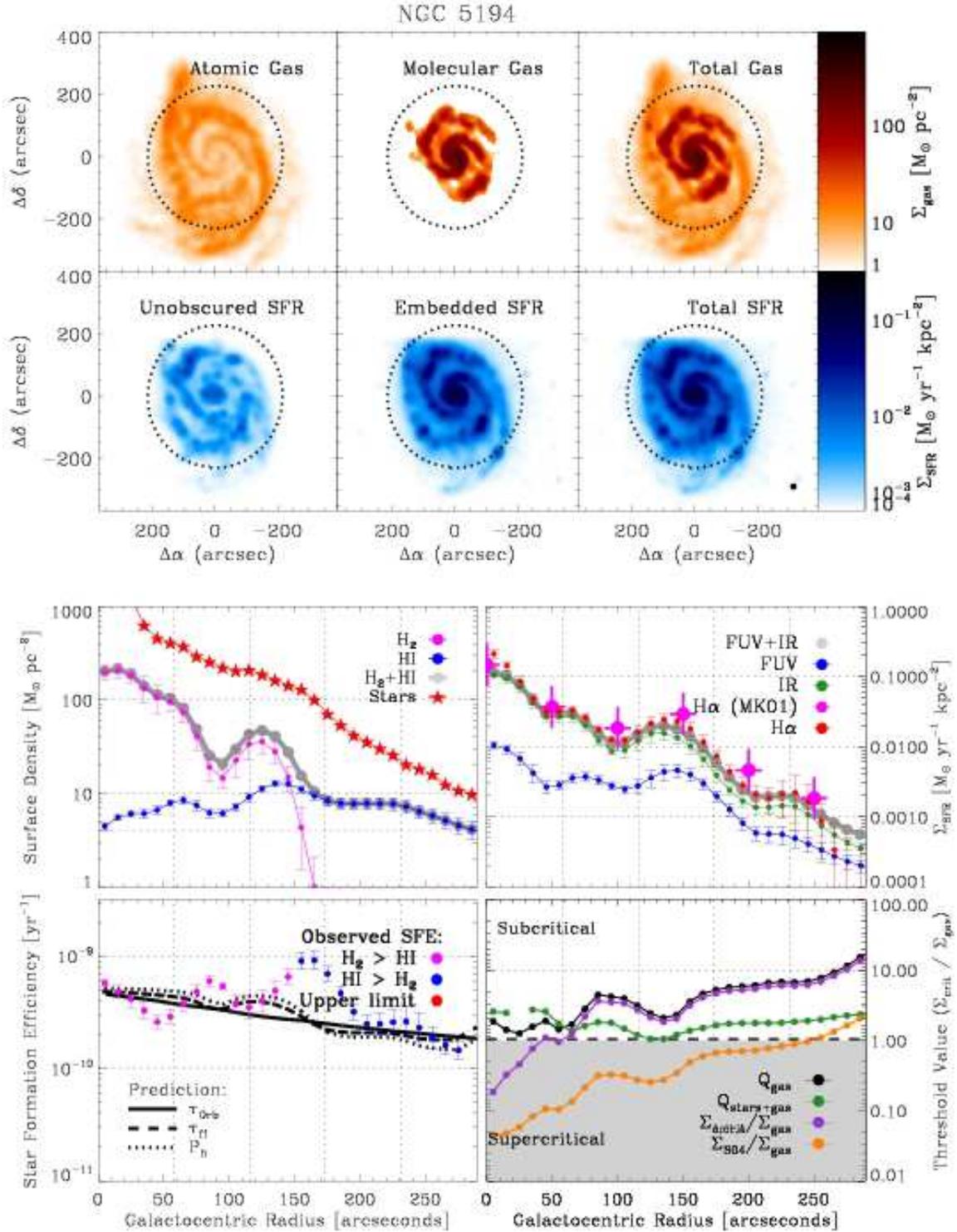}
  \figurenum{\ref{ATLAS}}
  \caption{Atlas of data and calculations for NGC 5194.}
\end{figure*}

\clearpage

\begin{figure*}
  \plotone{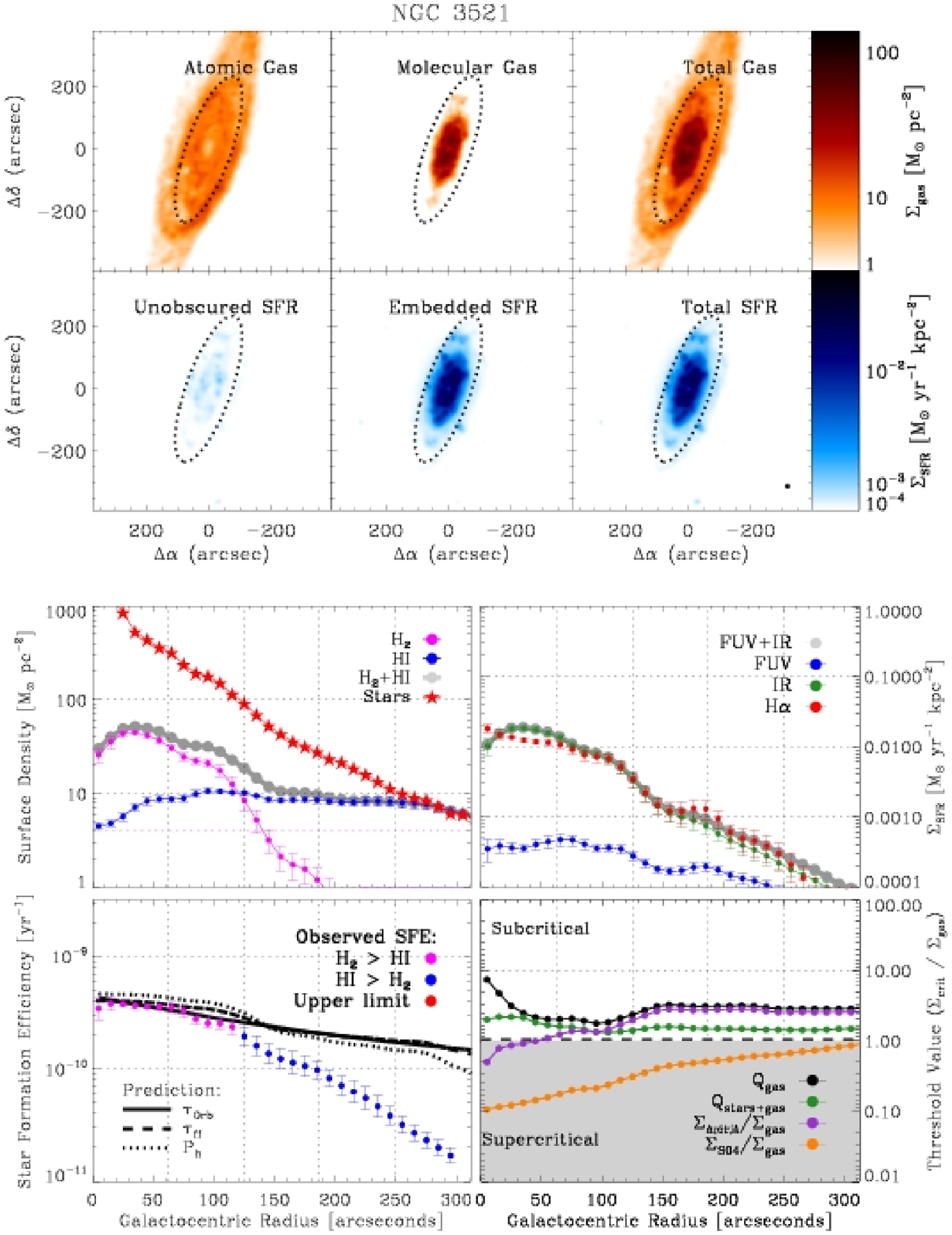}
  \figurenum{\ref{ATLAS}}
  \caption{Atlas of data and calculations for NGC 3521.}
\end{figure*}

\clearpage

\begin{figure*}
  \plotone{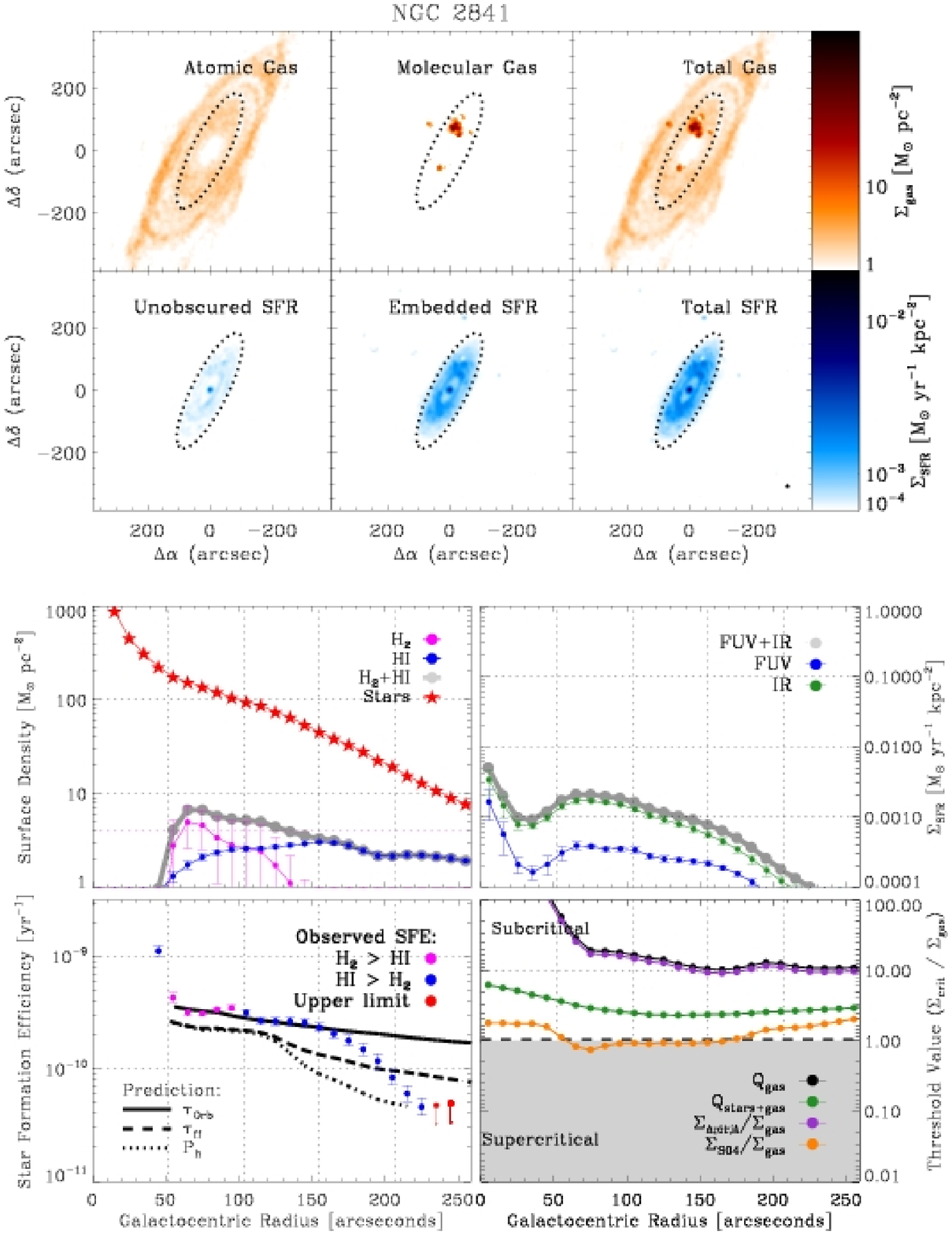}
  \figurenum{\ref{ATLAS}}
  \caption{Atlas of data and calculations for NGC 2841.}
\end{figure*}

\clearpage

\begin{figure*}
  \plotone{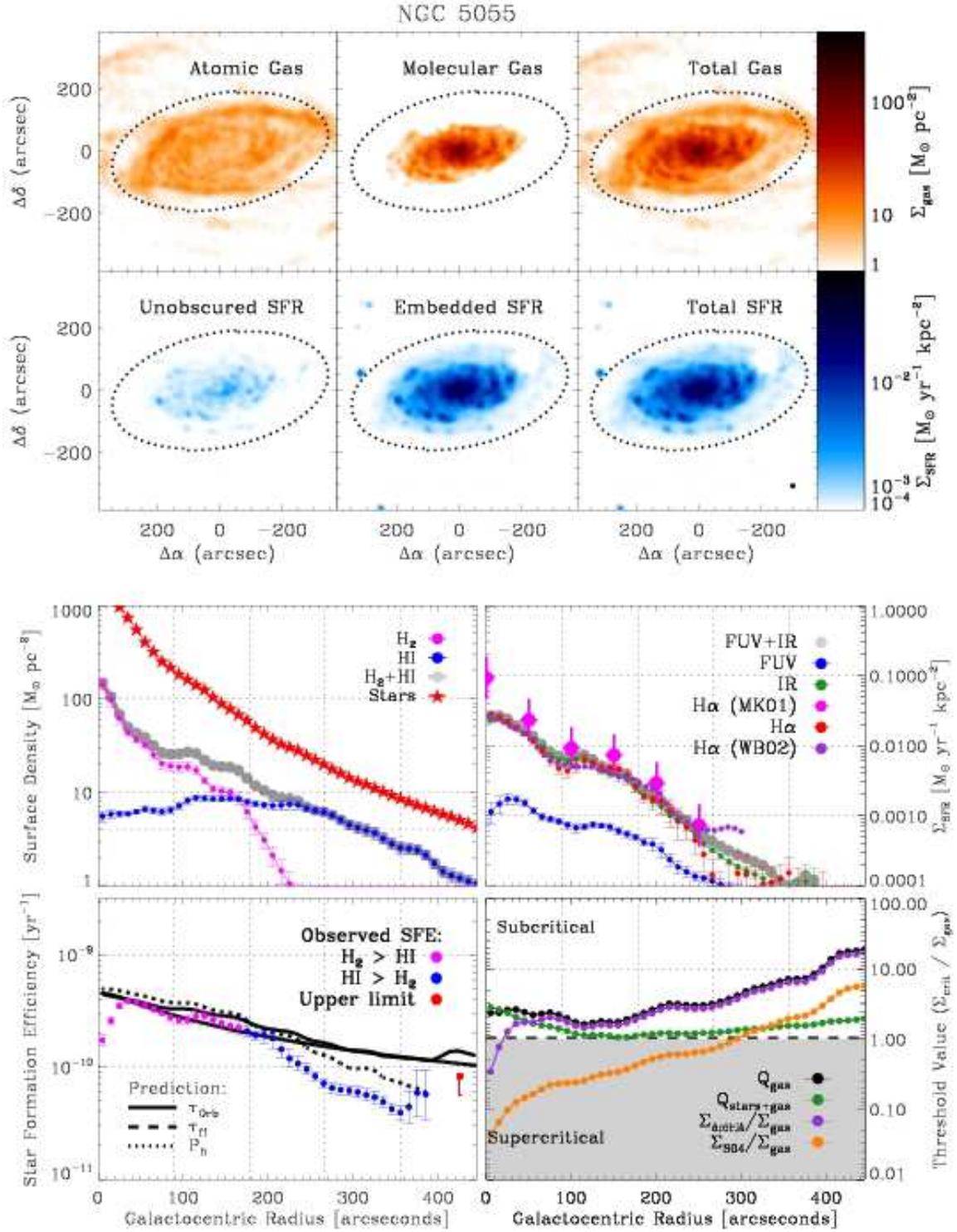}
  \figurenum{\ref{ATLAS}}
  \caption{Atlas of data and calculations for NGC 5055.}
\end{figure*}

\clearpage

\begin{figure*}
  \plotone{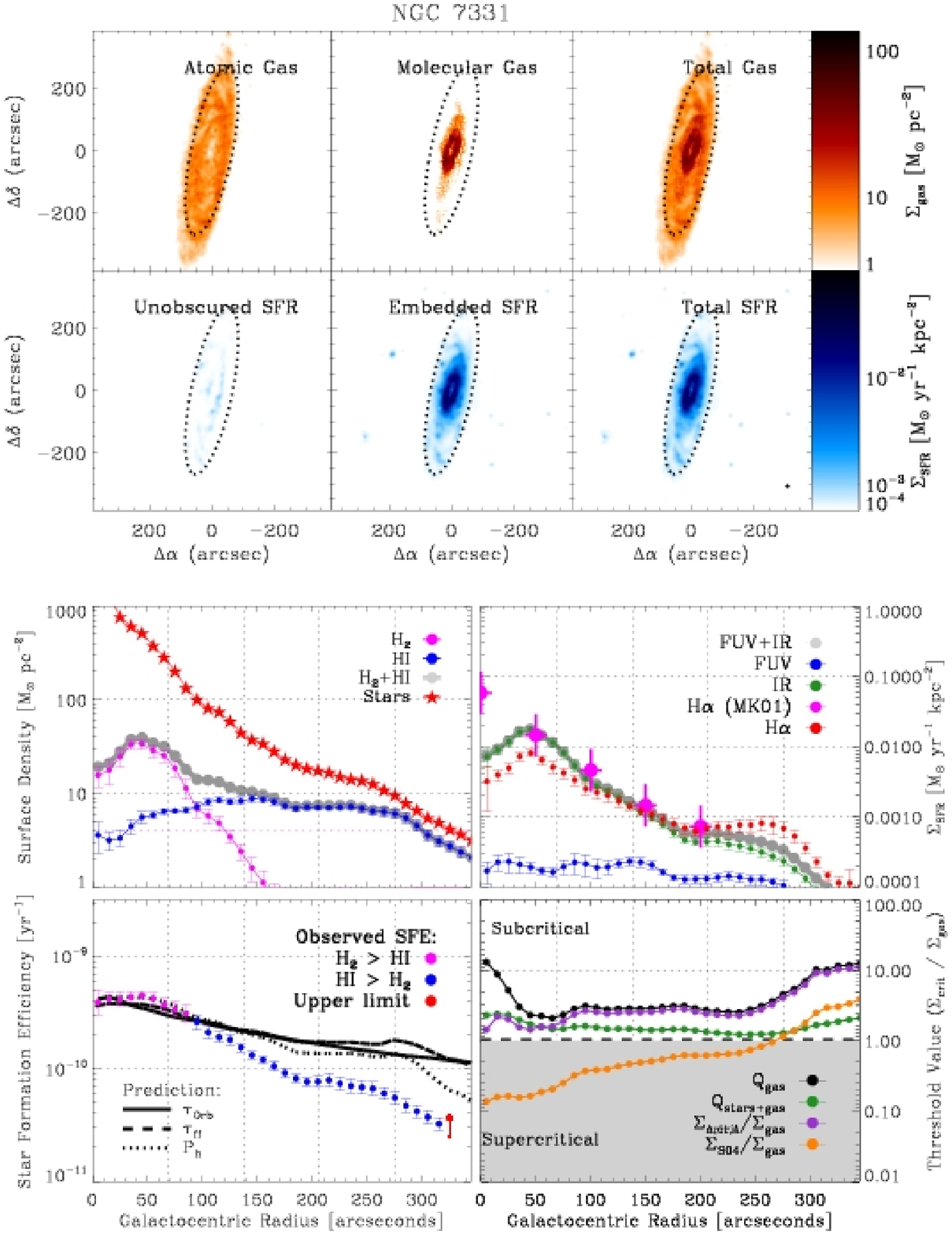}
  \figurenum{\ref{ATLAS}}
  \caption{Atlas of data and calculations for NGC 7331.}
\end{figure*}

\clearpage

\end{appendix}

\end{document}